%% file: revised.tex
\documentclass[draft]{agujournal2019}
\usepackage{url} %this package should fix any errors with URLs in refs.
\usepackage{lineno}
\usepackage[inline]{trackchanges} %for better track changes. finalnew option will compile document with changes incorporated.
\usepackage{soul}
\usepackage{amssymb}
\usepackage{comment}
\usepackage{natbib}
\setcitestyle{citesep={,}}
\usepackage{totcount}

\newtotcounter{citnum} %From the package documentation
\def\oldbibitem{} \let\oldbibitem=\bibitem
\def\bibitem{\stepcounter{citnum}\oldbibitem}

\input{bibsymlist.tex}

\draftfalse

\journalname{JGR: Planets}

\begin{document}

%% ------------------------------------------------------------------------ %%
%  Title
%
% (A title should be specific, informative, and brief. Use
% abbreviations only if they are defined in the abstract. Titles that
% start with general keywords then specific terms are optimized in
% searches)
%
%% ------------------------------------------------------------------------ %%

\title{Hot Jupiters: Origins, Structure, Atmospheres}

%% ------------------------------------------------------------------------ %%
%
%  AUTHORS AND AFFILIATIONS
%
%% ------------------------------------------------------------------------ %%

% Authors are individuals who have significantly contributed to the
% research and preparation of the article. Group authors are allowed, if
% each author in the group is separately identified in an appendix.)

% List authors by first name or initial followed by last name and
% separated by commas. Use \affil{} to number affiliations, and
% \thanks{} for author notes.
% Additional author notes should be indicated with \thanks{} (for
% example, for current addresses).

% Example: \authors{A. B. Author\affil{1}\thanks{Current address, Antartica}, B. C. Author\affil{2,3}, and D. E.
% Author\affil{3,4}\thanks{Also funded by Monsanto.}}

\authors{Jonathan J. Fortney$^{1}$, Rebekah I. Dawson$^{2}$, and Thaddeus D. Komacek$^{3}$}

% \affiliation{1}{First Affiliation}
% \affiliation{2}{Second Affiliation}
% \affiliation{3}{Third Affiliation}
% \affiliation{4}{Fourth Affiliation}

\affiliation{1}{Department of Astronomy and Astrophysics, University of California, Santa Cruz, CA, USA}
\affiliation{2}{Department of Astronomy \& Astrophysics, Center for Exoplanets and Habitable Worlds, The Pennsylvania State University, University Park, PA, 16802, USA}
\affiliation{3}{Department of the Geophysical Sciences, The University of Chicago, Chicago, IL, 60637, USA}
%(repeat as many times as is necessary)

%% Corresponding Author:
% Corresponding author mailing address and e-mail address:

% (include name and email addresses of the corresponding author.  More
% than one corresponding author is allowed in this LaTeX file and for
% publication; but only one corresponding author is allowed in our
% editorial system.)

% Example: \correspondingauthor{First and Last Name}{email@address.edu}

\correspondingauthor{Jonathan J. Fortney}{jfortney@ucsc.edu}

%% Keypoints, final entry on title page.

%  List up to three key points (at least one is required)
%  Key Points summarize the main points and conclusions of the article
%  Each must be 100 characters or less with no special characters or punctuation and must be complete sentences

% Example:
% \begin{keypoints}
% \item	List up to three key points (at least one is required)
% \item	Key Points summarize the main points and conclusions of the article
% \item	Each must be 100 characters or less with no special characters or punctuation and must be complete sentences
% \end{keypoints}

\begin{keypoints}
\item The origins of hot Jupiter exoplanets likely involve more than one formation pathway.
\item Explanations for the anomalously large radii of hot Jupiters need a connection to atmospheric temperature.
\item Hot Jupiters have complex atmospheres featuring ions, atoms, molecules, and condensates, where radiation and advection both play significant roles in controlling the 3D temperature structure.
\end{keypoints}

%% ------------------------------------------------------------------------ %%
%
%  ABSTRACT and PLAIN LANGUAGE SUMMARY
%
% A good Abstract will begin with a short description of the problem
% being addressed, briefly describe the new data or analyses, then
% briefly states the main conclusion(s) and how they are supported and
% uncertainties.

% The Plain Language Summary should be written for a broad audience,
% including journalists and the science-interested public, that will not have 
% a background in your field.
%
% A Plain Language Summary is required in GRL, JGR: Planets, JGR: Biogeosciences,
% JGR: Oceans, G-Cubed, Reviews of Geophysics, and JAMES.
% see http://sharingscience.agu.org/creating-plain-language-summary/)
%
%% ------------------------------------------------------------------------ %%

%% \begin{abstract} starts the second page

\begin{abstract}
We provide a brief review of many aspects of the planetary physics of hot Jupiters.  Our aim is to cover most of the major areas of current study while providing the reader with additional references for more detailed follow-up.   We first discuss giant planet formation and subsequent orbital evolution via disk-driven torques or dynamical interactions.  More than one formation pathway is needed to understand the population.  Next, we examine our current understanding of the evolutionary history and current interior structure of the planets, where we focus on bulk composition as well as viable models to explain the inflated radii of the population.  Finally we discuss aspects of their atmospheres in the context of observations and 1D and 3D models, including atmospheric structure and escape, spectroscopic signatures, and complex atmospheric circulation.  The major opacity sources in these atmospheres, including alkali metals, water vapor, and others, are discussed.  We discuss physics that control the 3D atmospheric circulation and day-to-night temperature structures.  We conclude by suggesting important future work for still-open questions. 
\end{abstract}

\section*{Plain Language Summary}
``Hot Jupiters'' are gas giant planets, thought to be akin to Jupiter and Saturn, that orbit their parent stars with typical orbital periods of only a few days.  These perplexing planets under strong stellar irradiation, found around 1\% of Sun-like stars, have been extensively studied.  Here we review many aspects of the physics of hot Jupiters.  First, we discuss the leading scenarios for the formation and orbital evolution of the planets, including the dominant ideas that these planets originally form much further from their parent stars.  Next, we describe models to assess their interior structure and thermal evolution and how strong stellar irradiation leads to radii that are significantly larger than that of Jupiter itself.  Finally, we discuss many aspects of their atmospheres, including the opacity sources that control the temperature structure, the mass-loss processes that drive a planetary wind, and the dynamical processes that control atmospheric circulation and day-to-night temperature contrasts.

%% ------------------------------------------------------------------------ %%
%
%  TEXT
%
%% ------------------------------------------------------------------------ %%

%%% Suggested section heads:
% \section{Introduction}
%
% The main text should start with an introduction. Except for short
% manuscripts (such as comments and replies), the text should be divided
% into sections, each with its own heading.

% Headings should be sentence fragments and do not begin with a
% lowercase letter or number. Examples of good headings are:

% \section{Materials and Methods}
% Here is text on Materials and Methods.
%
% \subsection{A descriptive heading about methods}
% More about Methods.
%
% \section{Data} (Or section title might be a descriptive heading about data)
%
% \section{Results} (Or section title might be a descriptive heading about the
% results)
%
% \section{Conclusions}

\section{Discovery of Hot Jupiters}
In 1995 \citeauthor{Mayor95} shocked the scientific world with the discovery of a Jupiter-mass planet, 51 Pegasi b, in a 4-day orbit around a Sunlike star, found via high-precision stellar radial velocity monitoring.  Soon after, \citet{Butler97} found additional close-in orbiting giant planets, and the astrophysical study of ``hot Jupiter" exoplanets was born.  While it had been suggested on theoretical grounds that planetary orbital migration could occur due to interaction with the planet-forming disk \citep{gold80, Lin86}, there was no particular prediction that gas giant planets could be found on such extreme orbits.  There were immediate questions regarding all aspects of these planets, including their past and ongoing tidal evolution \citep{Guillot96,tril98}, pathways through which the planets could settle in so close to their stars \citep{Lin96,rasi96,weid96}, atmospheric composition \citep{SS98}, and how strong stellar forcing could alter the structure of the planets \citep{Guillot96} and lead to mass loss \citep{Burrows95}.

% Paragraph below is cut-able, since most of this shows up later.

%Five years later, the detection of the transits of planet HD 209458b \citep{Charb00,Henry00}, brought forth a new era in the physics of hot Jupiters, since planetary radii and densities could now be measured.  This led to new models for the structural evolution of hot Jupiters \citep{Burrows00} and the suggestion that HD 209458b was larger in radius than could be accommodated by these models \citep{Bodenheimer01,Guillot02}.  At the same time, it was realized that stellar light passing through the planet's thin outer atmosphere could imprint planetary atmosphere absorption features onto this stellar spectrum \citep{Seager2000,Brown01,Hubbard01}.  Concurrent with this theoretical work, \citet{Charb02} detected sodium atoms in the planet's ``transmission spectrum," opening the study of exoplanetary atmospheres. 

Five years later, the detection of the transits of planet HD 209458b \citep{Charb00,Henry00}, brought forth a new era in the physics of hot Jupiters, since planetary radii and densities could now be measured.  The detection of transits motivated researchers to set up wide-field ground-based surveys \citep{Udalski02,Bakos:2005aa,Mandushev05}, to complement radial velocity detections and increase hot Jupiter detections.  Those transiting planets found around relatively bright nearby stars have proven to be best for detailed atmospheric followup.  
%RID: suggested new sentence:
Occurrence rate studies from ground-based radial velocity and space-based transit surveys have revealed that of order 1/100 Sun-like stars host hot a Jupiter (e.g., \citealt{Howard12,wrig12}).
%RID: Original sentence:
%The launch of NASA's \emph{Kepler} Mission \citep{Borucki10} has enabled more precise constraints on the inherent frequency of hot Jupiters, which is currently estimated to be $0.5 \%$ \citep{Howard12} for Sunlike (F,G,K-type) parent stars. 
In this contribution we will first explore the origins of these planets, before examining planetary structure and evolution, and then moving on to atmospheres, along the way pointing towards the many open questions still in need of exploration.

\section{Origins and orbital evolution}

Over the past 25 years astronomers have made significant progress in understanding how hot Jupiters came to reside so close to their host stars. Here we summarize hypotheses for their origins (Section \ref{subsec:hyp}), constraints from orbital properties (Section \ref{subsec:orb}) and stellar and companion (Section \ref{subsec:com}) properties, and take aways (Section \ref{subsec:sum}).

\subsection{Origins hypotheses}
\label{subsec:hyp}
\begin{figure}[h!]
    \centering
    \includegraphics[width=.9\textwidth]{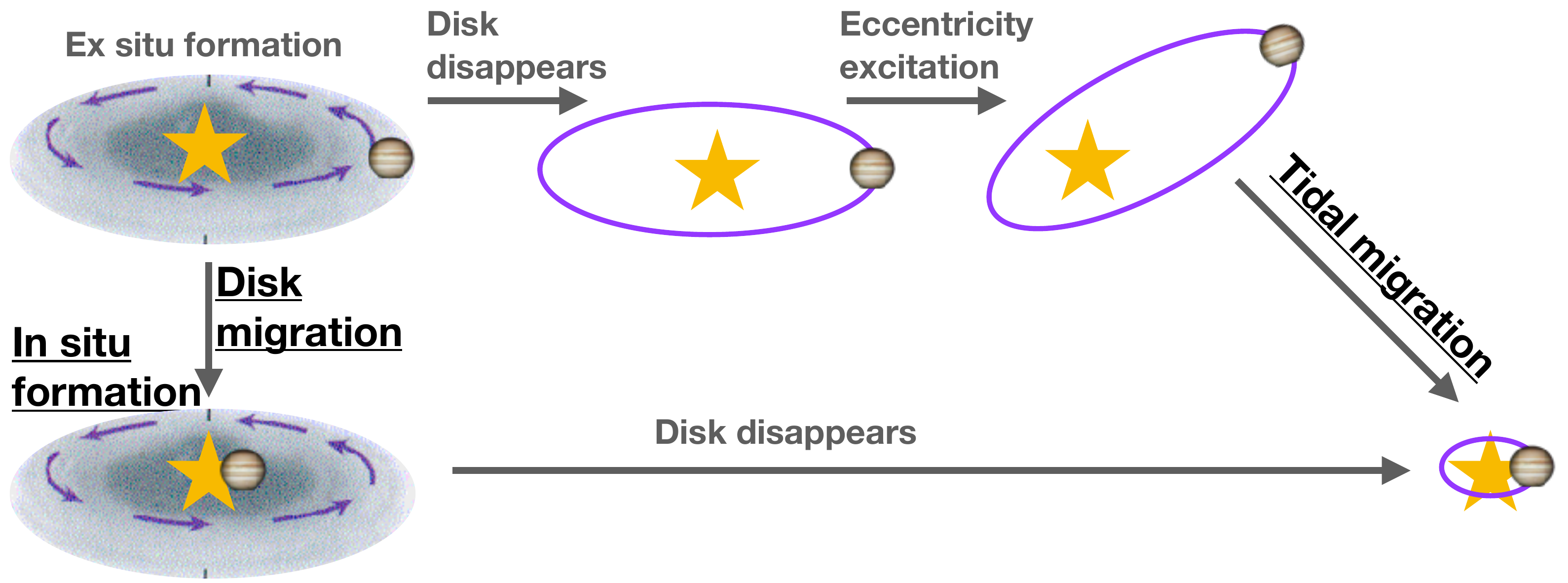}
    \caption{Origins hypotheses for hot Jupiters: in situ formation, disk migration, and high eccentricity tidal migration.}
    \label{fig:cartoon}
\end{figure}

There are three main hypotheses for the origins of hot Jupiters: in situ formation, disk migration, and high eccentricity migration (Fig. \ref{fig:cartoon}). In situ formation posits that hot Jupiters grew or assembled their cores and accreted gaseous envelopes at their present day locations. Alternatively, hot Jupiters may have formed much further out and migrated through the gaseous disk that surrounds a young star. After the gas disk dissipates, a hot Jupiter far from its star could be disturbed onto a highly elliptical orbit and migrate through tidal dissipation. Investigations of hot Jupiters' origins have run parallel to those of giant planets in the Solar System, for which we also debate the roles of planetesimal (e.g., \citealt{malh93}) and gas disk migration (e.g., \citealt{Walsh11}) and scattering to elliptical orbits (e.g., \citealt{thom99}). From a theory standpoint, all three mechanisms are viable but lead to different expectations for hot Jupiter properties.

Giant planets are thought to form either by core accretion, in which a rocky proto-planet core accretes many times its mass in gas from the proto-planetary disk (e.g., \citealt{peri74,poll96}; see \citealt{chab14} for a review), or gravitational instability, in which part of the proto-planetary disk fragments into bound clumps (e.g., \citealt{boss97}; see \citealt{duri07} for a review). Close to the star, gas conditions prevent formation by gravitational instability \citep{rafi05}. Core accretion can operate close to the star (e.g., \citealt{lee14,baty15}), but building a sufficiently large core ($\sim 10 M_\oplus$) is challenging. Feeding zones are small so the local available solids are insufficient (e.g., \citealt{schl14}), mergers of multiple smaller cores are prevented by the disk (e.g., \citealt{Lee16}), and accretion of radially transported pebbles stalls at a much lower mass (see \citealt{joha17} for a review). 

Hot Jupiters may form further out -- where conditions for core accretion and/or gravitational instability are more achievable -- and migrate in through torques from the gaseous disk (e.g., \citealt{gold80,Lin86}; \citealt{baru14} review). The migration rate and direction are sensitive to disk conditions (e.g., \citealt{paar06,duff14}). If the planet reaches its short orbital period before the disk dissipates, it could be tidally disrupted or engulfed by the star. However, tidal interactions with the star (e.g., \citealt{tril98,vals15}) or stalling by a magnetocavity in the innermost disk  (e.g., \citealt{rice08,chan10}) may preserve the hot Jupiter.

Tidal migration begins when a Jupiter is perturbed onto a highly elliptical orbit. Mechanisms include  planet-planet scattering (e.g., \citealt{rasi96,weid96}), cyclic secular interactions (e.g., Kozai-Lidov cycles; \citealt{koza62,lido62,wu03,fabr07} see \citealt{naoz16} for a review), or chaotic secular interactions (e.g., \citealt{wu11}). Tides raised by the star on the Jupiter shrink and circularize its orbit (e.g., \citealt{eggl98}).

\subsection{Constraints on origins hypotheses from orbital properties}
\label{subsec:orb}

Observed orbital properties of hot Jupiters can test hypotheses for their origins. They are found at host star separations where formation and migration can effectively deposit them, and their eccentricities and orbital alignment are relics of their orbital evolution. However, tidal effects complicate our interpretation of these properties. 

Most hot Jupiters have typical orbital periods of $\sim$ 3 days. For in situ formation, we expect hot Jupiters to be located at or beyond the disk edge, corresponding to $\sim$ 10 days (e.g., \citealt{lee17}); therefore observed hot Jupiters are a factor of several closer to their stars than expected. Hot Jupiters may arrive through disk migration to half the corotation period, more consistent with hot Jupiters' observed orbital periods. High eccentricity tidal migration should deliver surviving planets at or beyond twice $a_{\rm Roche}$, the tidal disruption limit. Although many hot Jupiters are beyond $2 a_{\rm Roche}$, high eccentricity migration alone cannot easily account for those between 1--2 $a_{\rm Roche}$. However, in all origins scenarios, hot Jupiters can later raise tides on their stars, further shrinking their semi-major axes (e.g., \citealt{Jackson08a,vals14}). 

Hot Jupiter eccentricities provide evidence that some arrived through the high eccentricity tidal migration channel. Out to $\sim$3 day orbital periods, most hot Jupiters have circular orbits; if these closest-in hot Jupiters ever had elliptical orbits, they would likely quickly tidally circularize because of the strong dependence of tidal dissipation on host star separation (e.g., \citealt{hut81}). In the 3--10 day orbital period range, we observe a mixture of moderately elliptical  ($0.2 < e < 0.6$) and circular orbits for hot Jupiters. If hot Jupiters form in situ or undergo disk migration, they would acquire these elliptical orbits at their present short orbital periods. However, proposed mechanisms for exciting eccentricities close to the star cannot account for many of the observed eccentricities (e.g., gas-disk interactions, \citealt{duff15}; planet-planet scattering, \citealt{petr14}; secular forcing from an outer planet). Under the high eccentricity tidal migration hypothesis, moderately elliptical hot Jupiters are expected: they are in the process of tidal circularization. We may observe a mixture of circular and eccentric hot Jupiters at the same host star separations because of different timescales for the initial eccentricity excitation, diverse tidal dissipation properties, or multiple formation channels (i.e., a mix of high eccentricity tidal migration with disk migration and/or in situ formation).

Many hot Jupiters have orbits aligned to their host stars' spin, but others are severely misaligned (e.g., \citealt{albr12} and references therein). Planets form in circumstellar disks, so we might naively expect hot Jupiters that originate in situ or via disk migration to keep their orbits aligned. In contrast, the gravitational interactions that trigger high eccentricity tidal migration can also misalign planets' orbits (e.g., \citealt{fabr07,chat08}). However, tidal interactions may realign the hot Jupiter \citep{winn10,albr12} -- erasing evidence of high eccentricity tidal migration -- and other physical processes can misalign the stellar spin from the circumstellar disk (e.g., \citealt{baty12,roge12,stor14a}), so that Jupiters form misaligned. Therefore spin-orbit alignments are not necessarily indicative of hot Jupiters' origins channel.

\subsection{Constraints from stellar and companion properties}
\label{subsec:com}

Characteristics of hot Jupiters' host stars and the presence and properties of planetary or stellar companions in the system can test theories for hot Jupiters' origins. The general giant planet occurrence rate increases with host star metallicity \citep{gonz97,sant01,fisc05}, which is interpreted as solid-rich disks facilitating core accretion. There is some evidence that the correlation for hot Jupiters is even stronger \citep{jenk16}, which most directly supports in situ formation (mitigating the challenge of forming a massive core close to the star) or high eccentricity migration triggered by a planetary companion (which may be more abundant in disks that nurture core accretion). Hosts of moderately eccentric hot Jupiters -- which in Section \ref{subsec:orb} we attributed to high eccentricity migration -- tend to have higher metallicities \citep{daws13}.

Hot Jupiters have an occurrence rate of about 1\% (e.g., \citealt{wrig12}), which is about 10\% of the overall giant planet occurrence rate (e.g., \citealt{cumm08}). The occurrence rate of hot Jupiters in the \emph{Kepler} transit survey is about a factor of three lower than radial velocity surveys \citep{howa12,wrig12}. The discrepancy cannot be fully explained by differences in host star metallicity \citep{guo17} but may be the result of an overall suppression of planet formation by close binaries in the \emph{Kepler} sample \citep{moe19}. High eccentricity tidal migration scenarios have not been demonstrated to be efficient enough to produce all hot Jupiters (e.g., \citealt{muno16}).  The efficiencies of in situ formation and disk migration could be high enough but are dependent on uncertain disk conditions (e.g., \citealt{cole16,Lee16}).

Host star ages have the potential to distinguish the contributions of different origins channels: hot Jupiters that form in situ or migrate through the gas disk should already be present during the gas disk stage. In contrast, high eccentricity tidal migration typically operates after the gas disk stage, spawning hot Jupiters throughout a star's lifetime; however, timescales for eccentricity excitation and circularization enable most to arrive early. Therefore we expect hot Jupiters produced by high eccentricity migration to be largely absent during the gas disk stage and somewhat less common around younger stars than older stars. Measuring the contribution of high eccentricity tidal migration to the hot Jupiter population requires a sample of very young stars or a large sample of main sequence stars with precise ages. Recently two hot Jupiters were discovered around T Tauri stars, which are still in the gas disk stage \citep{dona16,yu17}. Ongoing T Tauri occurrence surveys will help evaluate what fraction of hot Jupiters must arrive during the gas disk stage.

In situ formation often leads to hot Jupiters with other planets nearby (e.g., \citealt{bole16}). Disk migration can deliver hot Jupiters in resonance with nearby planets \citep{malh93,lee02} and/or with small planets that form nearby after the hot Jupiter's migration. In contrast, tidal migration destroys small planets within the giant planet's initial orbit (e.g., \citealt{must15}). Hot Jupiters generally have an absence of nearby planets \citep{lath11,stef12}, supporting the tidal migration hypothesis, but a notable exception is WASP-47b \citep{beck15}.

While high eccentricity migration initiated by planet-planet scattering could eject the other planet (e.g., \citealt{rasi96}), secular eccentricity excitation retains the companion, which must have the right properties to have excited the eccentricity. Most hot Jupiters have planetary companions consistent with driven high eccentricity migration \citep{knut14,brya16}, but do not have a capable stellar companion \citep{ngo16}. These survey results point to fellow planets as the triggers of high eccentricity migration.

\subsection{Summary}
\label{subsec:sum} 

Multiple origins channels are needed to explain the observed range of orbital and companion properties for hot Jupiter. We refer readers to \citet{daws18}'s review for a more detailed discussion of the evidence described here and constraints from radius inflation, atmospheric properties (e.g., \citealt{Oberg11,Madhu14}), occurrence rates, and comparisons to warm Jupiters and smaller planets. High eccentricity tidal migration triggered by planet-planet secular interactions is likely  one of the main origins channels, accounting for moderately eccentric hot Jupiters (and their increased host star metallicities), their general lack of small nearby planets, and the presence of distant planet companions capable of triggering high eccentricity migration. A second channel -- disk migration or in situ formation -- can supplement the inefficiency of producing hot Jupiters through tidal migration and account for hot Jupiters with rare small nearby planets, orbiting T Tauri stars and lower metallicity stars, and within $1$--$2 a_{\rm Roche}$ (without requiring subsequent tidal evolution). 

\section{Internal structure}
Under intense stellar forcing, at a level of incident flux 10,000 times higher than Jupiter receives from the Sun, one may expect that the evolution of a hot Jupiter's interior structure may differ significantly compared to Jupiter itself.  Indeed, it was suggested early on that hot Jupiters would not cool off as efficiently, leading to hotter interiors, and larger planetary radii, at old ages \citep{Guillot96}.  While this prediction has proven true, it is the magnitude of these very large radii, or the planetary \emph{radius anomaly} that has proven difficult to understand.  In this section, we introduce constraints on, and models for, the internal structure of hot Jupiters (see \citealp{Baraffe10,fortney_2009}, and \citealp{Baraffe:2014} for detailed reviews in the context of giant planet structure and evolution, outside and inside the solar system). We describe the observed radius distribution of hot Jupiters in \Sec{sec:radiusinflation}, proposed mechanisms to enhance the radii of hot Jupiters in \Sec{sec:mechanisms}, and the bulk composition of hot Jupiters in \Sec{sec:metallicities}.

\subsection{Radius anomaly} %% working title
\begin{figure}
    \centering
    \includegraphics[width=0.9\textwidth]{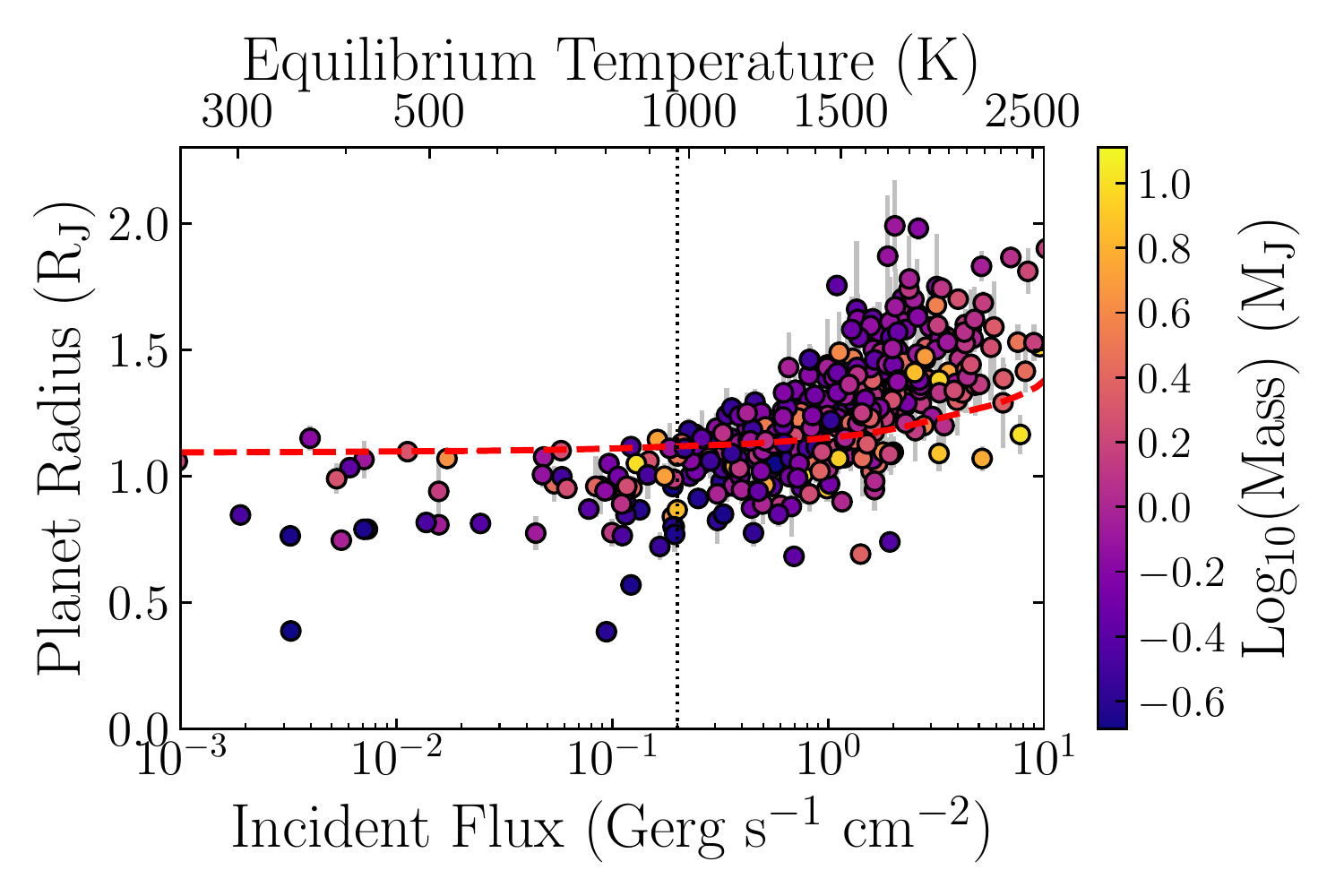}
    \caption{The observed radius of extrasolar gas giants (from 0.1 \mj\ to 13 \mj) plotted as a function of incident stellar flux and colored by planetary mass. The red dashed line shows an evolutionary model for a Jupiter-mass planet without any additional inflation effects at an age of 4.5 Gyr, and the vertical dashed line displays the flux cutoff below which no radius inflation is found.  Figure courtesy of Daniel Thorngren.}
    \label{fig:thorngren2018}
\end{figure}

\label{sec:radiusinflation}
One can assess the observed radius distribution of hot Jupiters for clues regarding planetary structure. Figure \ref{fig:thorngren2018}, updated from \citet{Thorngren:2017}, shows planetary radii vs.~incident stellar flux (bottom x-axis) and the zero Bond albedo equilibrium temperature (top x-axis). There is a clear rise in the distribution of radii, starting at $\sim 2 \times 10^8~\mathrm{Gerg}~\mathrm{s}^{-1}~\mathrm{cm}^{-2}$ in flux, or around 1000 K in equilibrium temperature.  For comparison, planetary evolution model predictions for a 4.5 Gyr, pure H/He, 1 \mj\ planet that undergoes various levels of incident stellar flux are shown in the slowly rising dashed red curve. Many hot Jupiters hotter than the the $\sim 1000~\mathrm{K}$ threshold have radii ``above'' the curve -- this is the radius anomaly. Meanwhile, planets that lie below the red dashed curve likely just indicate planets that are relatively more enriched in heavy elements \citep{Fortney07a,Burrows07}, often termed ``metals'' in an astrophysical context. 

%Thorngren figure here? 
%The sample of transiting hot Jupiters 

\Fig{fig:thorngren2018} shows two important properties of the distribution of hot Jupiter radii that constrain models proposed to explain the radius anomaly (see \citealp{Laughlin:2015} and \citealp{Laughlin:2018aa} for recent reviews). First, Jupiter-mass planets that have equilibrium temperatures cooler than $1000~\mathrm{K}$ (termed ``warm Jupiters'') are not observed to be inflated \citep{Miller:2011,Demory:2011,Laughlin_2011}.
%,Thorngren:2017}.
As a result, the mechanism(s) that cause the radius anomaly are likely not at work on these cooler Jupiters. Second, the radii of hot Jupiters appear to scale with their equilibrium temperature \citep{Laughlin_2011,weis13,Thorngren:2017}. This implies that the mechanism that causes the radius anomaly is linked to the incident flux from the host star. 

%Twenty years after the first radius anomaly was found for HD 209458b \citep{Burrows00, Bodenheimer01}, it is still not known if a specific heating mechanism can explain the entire hot Jupiter radius anomaly, or if a combination of mechanisms or novel explanation is required. However, 
Recent work has found avenues for better constraining the mechanism(s) that enhance radii. \cite{Thorngren:2017} used a statistical analysis of hot Jupiter radii coupled to planetary structure models and found that the required conversion of incident flux to interior heating needed to explain the radius anomaly of the full sample of hot Jupiters peaks at an intermediate value of $T_\mathrm{eq} \sim 1600~\mathrm{K}$. 
%This peak in the conversion of incident flux to heat matches well with the expectations of Ohmic dissipation models, but could also be due to a reduced efficiency of mechanical dissipation due to magnetic drag. 
Additionally, \cite{hart16} found that the radii of hot Jupiters increase with increasing fractional main-sequence age of their host stars (and corresponding higher stellar luminosity as stars brighten on the main sequence), providing further evidence that the mechanism(s) that affect hot Jupiter radii are tied to the incident stellar flux. As predicted by \cite{Lopez:2015}, it has been found that planets that were warm Jupiters when their host stars were on the main sequence ``re-inflate''  as their host stars evolve off of the main sequence and their equilibrium temperatures exceed $1000~\mathrm{K}$ \citep{Grunblatt2016,Grunblatt:2017aa,Grunblatt:2019aa}. Further studies of both main-sequence and post-main-sequence re-inflation will improve constraints on the mechanism(s) that cause the radius anomaly of hot Jupiters \citep{Komacek:2020ab}.

\subsection{Proposed mechanisms to explain the observed radius distribution}
\label{sec:mechanisms}
A wide range of mechanisms have been proposed to explain the hot Jupiter radius anomaly (for previous summaries, see \citealp{weis13} and \citealp{Baraffe:2014}). These mechanisms can be separated into two main categories: keeping the interior of the planet warm by reducing the rate of cooling to space, and utilizing incident stellar radiation or tidal energy to heat the planet from within.
%%not sure if "tidal energy" is the best term here, but it's hard to briefly explain g-mode dissipation

\subsubsection{Reduction of internal cooling}
Gas giant planets are expected to form hot with large initial radii and contract over time as their interiors cool off. 
%The exact initial radius distribution of gas giants is unknown, and depends on details of planet formation, for example the properties of the shock that forms during the core accretion process \citep{Berardo:2017aa}. 
As a result, mechanisms that reduce the cooling rate of the planet can keep the interior hot. Because the interior temperature of a gas giant is directly linked to its radius \citep{Hubbard77,Arras06,Ginzburg:2015}
reducing the interior cooling rate is equivalent to reducing the rate that the planet radius shrinks during its evolution. Gas giants cool off rapidly, and without a reduction in the cooling rate hot Jupiters would not appear inflated after $\sim$ Gyr of evolution \citep{Guillot_2002}. Two mechanisms have been proposed to reduce the internal cooling. However, these mechanisms that rely on the reduction of 
%atmospheric and/or internal 
cooling struggle to explain the increase in radius with increasing incident stellar flux shown in \Fig{fig:thorngren2018}.
%,ginz16}.

%There are two flavors of proposed mechanisms to reduce the interior cooling rate of hot Jupiters (without invoking additional heating, which we discuss in the next section). 
The first mechanism to reduce the rate of cooling is invoking an enhanced atmospheric opacity that leads to longer timescales for radiative heat transport in the stably stratified envelope \citep{Burrows07}.  However, detailed calculations lead to only modest radius increases that cannot explain the entire population \citep{Burrows07}.
The second flavor of proposed mechanism to slow the cooling of hot Jupiters
%interior cooling of hot Jupiters is to directly inhibit cooling of the interior of the planet. One possibility to reduce the ability of the interior of the planet to cool 
is for their interiors to not be fully convective, but experience double-diffusive layered convection \citep{Chabrier:2007,Kurokawa:2015aa}. Double-diffusive convection occurs in the case of a molecular weight gradient that increases with increasing depth and stabilizes the envelope against convection. 
%This results in a form of ``semi-convection'' in which the envelope is comprised of a set of well-mixed layers separated by strong mean molecular weight gradients \citep{Chabrier:2007}. 
%It is thought that Jupiter and Saturn experienced double-diffusive convection due to helium phase separation, core erosion, and/or incomplete mixing of the material delivered from planetesimals \citep{Leconte12,Nettelmann15,Mankovich:2016aa,Vazan:2018aa},
%,Mankovich:2020aa}, 
%which alter cooling, but do not have a major effect on planetary radius evolution. %Furthermore, \cite{Kurokawa:2015aa} showed that only extremely thin (and likely dynamically unstable) layers of width $1-100~\mathrm{m}$ can provide a sufficient reduction in cooling rate to explain the observed radii of hot Jupiters. 
%Perhaps most importantly, mechanisms that rely on the reduction of atmospheric and/or internal cooling struggle to explain the increase in radius with increasing incident stellar flux shown in \Fig{fig:thorngren2018}.
Both mechanisms above cannot explain the observed dependence of radius on incident flux for hot Jupiters.  However semi-convection likely affects the interior evolution of a range of gas giant planets, as both Jupiter and Saturn experienced double-diffusive convection due to helium phase separation or post-formation composition gradients \citep{Leconte12,Nettelmann15,Mankovich:2016aa,Vazan:2018aa}.
\subsubsection{Internal deposition of heat}
%Maybe add Batygin Ohmic dissipation schematic figure, useful for explaining both hydro and MHD dissipation and circulation. %Thorngren inferred heating rate figure and/or RCB depth impact here (maybe ask for data and simplify it somewhat or do some editing in keynote/adobe). Mention re-inflation as a way to test these ideas.
Energy deposited into the interior of the planet can act to replace the cooling of the convective interior, reducing or even reversing the net loss of energy to space. The effect of internal heating on evolution is sensitive to both the depth and amount of energy deposited \citep{Gu:2004aa,Spiegel:2013,Komacek:2017a}, and only heating deep within the convective interior can fully offset interior cooling \citep{Guillot_2002}. There are two classes of deposited heating mechanisms: tidal dissipation and conversion of incident stellar flux to deposited heat. 

Tidal dissipation was the first mechanism proposed to heat the interior of hot Jupiters \citep{Bodenheimer01}. Two types of mechanisms have been proposed to lead to tidal heating of hot Jupiters. The first is eccentricity tides due to the tidal circularization of hot Jupiters that formed through high eccentricity migration \citep{Bodenheimer01,Gu:2003aa,Jackson:681,Ibgui:2009}.
%,Gu2004aa
%,Miller:2009,arra10,Ibgui:2010,Leconte:2010a}.
%due to the $\sim$ Myr  timescales over which the rotation period of hot Jupiters are expected to synchronize with their orbital period \citep{Guillot96,Lubow:1997aa}, 
It is unlikely that eccentricity tides can explain the inflated radii of all hot Jupiters, but they may play a key role for certain systems \citep{Miller:2009,Leconte:2010a}. 

The second type of tidal heating mechanism is thermal tides, which are tides driven by the spatial and temporal inhomogeneity of atmospheric column mass (which is linked to temperature), originally proposed to explain the slow spin rate of Venus \citep{Gold:1969aa}. These atmospheric tides transfer angular momentum and energy from the orbit to the planetary rotation \citep{Thompson:1882aa}, increasing the rotation rate and pushing hot Jupiters away from a synchronous state, leading to gravitational tidal dissipation. Thermal tides have been shown to provide a sufficient rotational perturbation and resulting dissipation to explain the radius anomaly \citep{arra10,socr13},
%,Gu:2019aa,Yu:2020aa}, 
but the depth at which thermal tides deposit heat is uncertain \citep{Gu:2019aa}.

\begin{figure}
    \centering
    \includegraphics[width=0.6\textwidth]{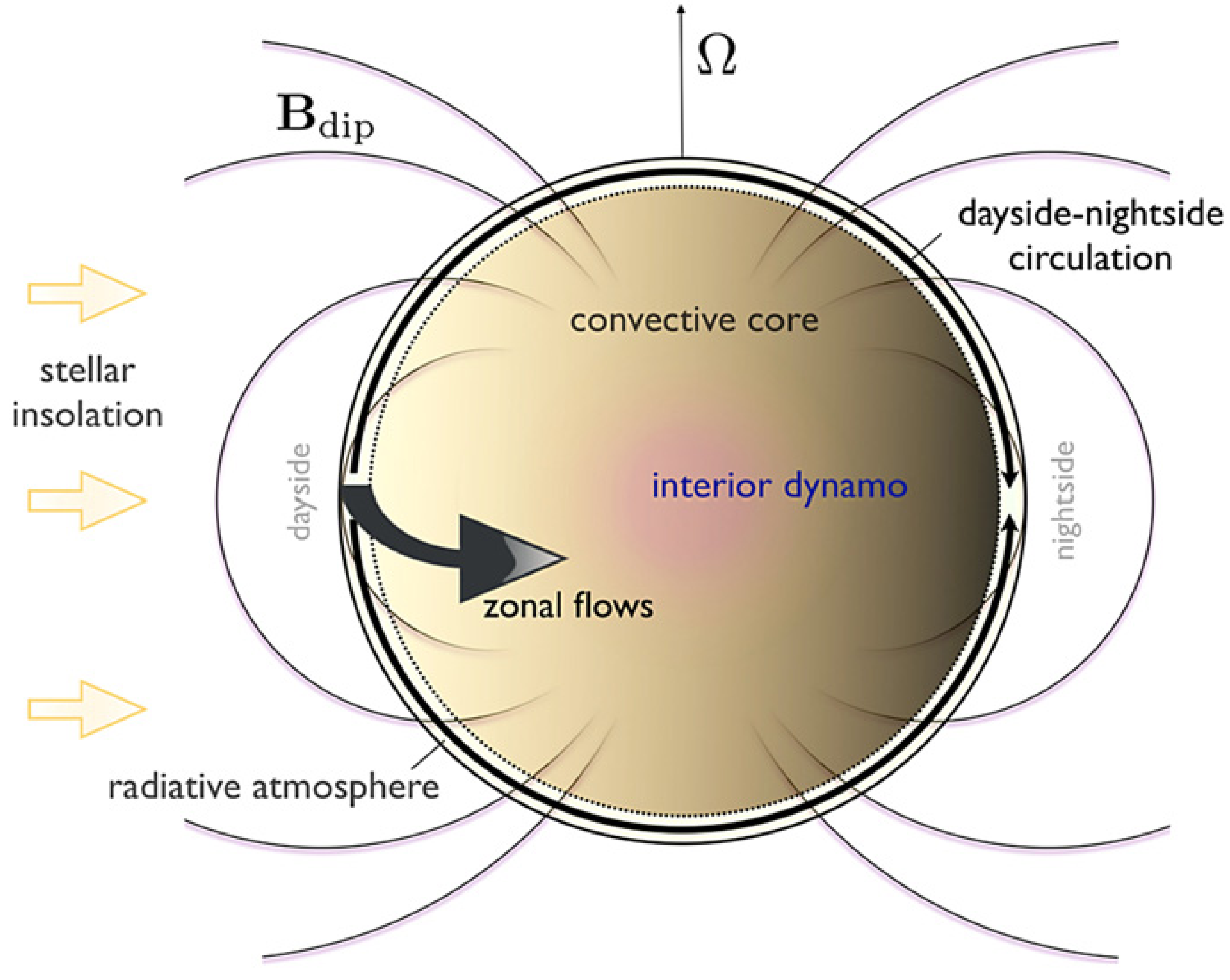}
    \caption{Schematic diagram showing the coupling between the atmospheric circulation and interior magnetic field of hot Jupiters. From \cite{batygin_2013}.}
    \label{fig:batygin2013}
\end{figure}
The second class of mechanism proposed to heat hot Jupiters relies on converting a fraction of the incident stellar flux into deposited heating. Shear instabilities in the deep atmosphere leading to deposited heat were the first of this class to be considered, and rely on dissipation of the $\sim$ km s$^{-1}$ winds in hot Jupiters (see Section \ref{sec:atm}) through fluid dynamic instabilities at interfaces where the winds quickly decay \citep{Guillot_2002,Bodenheimer:2003}.
%showman_2002
Another mechanism relating to the atmospheric circulation is large-scale vertical mixing in the deep atmosphere acting to force a downward heat flux that carries a fraction of incident stellar power to higher pressures, where it can affect the interior evolution \citep{youd10,Tremblin:2017,Sainsbury-Martinez:2019aa}. 

As the atmospheres of hot Jupiters are hot enough that species with low ionization potentials (e.g., Na, K) are partially ionized, \cite{baty10} proposed that the atmospheric current driven by the fast motions of a partially ionized fluid can penetrate into the interior of the planet, leading to ``Ohmic dissipation'' (i.e., electrical resistive heating) at depth. A variety of studies have examined Ohmic dissipation and found varying degrees of success in explaining the radius anomaly (e.g., \citealp{Perna_2010_2,Batygin_2011,Huang_2012,Menou:2012fu,Rauscher_2013,Wu:2013,Rogers:2014,ginz16}). Hot Jupiters are expected to have strong internal magnetic fields \citep{Yadav:2017,Cauley:2019aa}, which couple with the atmospheric currents to induce magnetic forces that act against the circulation (see \Fig{fig:batygin2013}). These magnetic effects naturally reduce the Ohmic dissipation rate for hot planets with equilibrium temperatures $\gtrsim 1500~\mathrm{K}$ \citep{Menou:2012fu,Rogers:2014,ginz16}, in agreement with the findings of \cite{Thorngren:2017}. However, Ohmic dissipation is not thought to provide sufficiently deep heating to lead to re-inflation \citep{Wu:2013,ginz16}. 

While more work is required to fully determine the heating mechanism(s) that cause the radius anomaly of hot Jupiters, \citet{Thorngren:2017} have shown that the additional radius inflation power has a characteristic magnitude as a function of incident flux.  This strongly suggests that mechanisms in the three proceeding paragraphs, that are coupled to this stellar heating, including thermal tides, shear instabilities or vertical mixing, and Ohmic dissipation, are the key contenders.  Indeed, very recently \citet{Sarkis20} have suggested that multiple mechanisms must be at play to explain the population.

\subsection{Bulk planetary metallicities}
\label{sec:metallicities}
While significant theoretical work has gone into examining mechanisms that lead to large radii and low densities for this class of planets, much less work has gone into understanding planets that are not inflated, or are ``over-dense.''  The discovery of HD 149026b \citep{Sato05}, with a very high density compared to other known transiting planets at that time, suggested that the planet must be more metal-rich than Jupiter and Saturn \citep{Sato05,Fortney06,Burrows07}.  The fact that the planet's parent star is quite metal-rich led to the first discussion of a potential star-planet composition connection \citep{Guillot06}.

Only relatively recently has a large enough sample size of transiting giant planets become available that the bulk metallicities of the planets (as diagnosed from a combination of their bulk densities and planet thermal evolution models) can be evaluated.  \citet{Thorngren16}, following up on preliminary work of \citet{Miller11}, find a clear relation between giant planet mass and bulk metallicity. This both confirms a key prediction of core-accretion planet formation \citep{Mordasini14}, in a generic sense,
%RID: Not sure if I would consider this a key prediction
but also gives a well-defined slope and scatter to this relation that future models of planet formation must aim to reproduce.  The more massive giant planets are less metal-rich (reproducing the solar system's trend), but the amount of metals within planets more massive than Jupiter is surprising, suggesting that massive ``super-Jupiters" may often accrete well over 100 \me\ of metals.  A recent new spin on this work is that the metal-richness of the planets does \emph{not} appear to correlate with the metallicity of the parent star \citep{Teske19}.

These kinds of investigations to assess bulk composition are entirely complementary to studies of hot Jupiter atmospheres, where one seeks the abundances of particular atoms and molecules.  Interior structure and evolution modeling has the advantage of having a larger sample size to examine, and it does not require further characterization observations.  However, these atmospheric abundance details are potentially a treasure trove of information, and it is to atmospheres that we next turn. 

\section{Atmospheres}
\label{sec:atm}
The atmospheres of hot Jupiters lie in a unique thermal, chemical, and dynamical regime characterized by strong incident radiation, large horizontal temperature contrasts, species in the ionic, atomic, molecular, and  condensate phases, and fast winds that approach or exceed the speed of sound. Due to their large radii and hot temperatures, hot Jupiters host the only class of exoplanet atmosphere that can currently be observationally characterized in detail with spectroscopic observations over an entire orbital phase.

These observations serve as a testbed for models of hot Jupiter atmospheres, which have largely been successful in explaining the large-scale observed climate of hot Jupiters. Here we describe observational characterization of hot Jupiter atmospheres in \Sec{sec:obs}, summarize our understanding of the atmospheric structure of hot Jupiters in \Sec{sec:tempstructure}, and describe theory for and modeling of the atmospheric circulation of hot Jupiters in \Sec{sec:atmcirc}. 
%^TDK: Figured a brief summary of each section might be useful. 
\subsection{Observational characterization}
\label{sec:obs}
Observations of hot Jupiter atmospheres have been an exciting aspect of exoplanet science over nearly two decades.  Much of this work has utilized the \emph{Hubble} and \emph{Spitzer} Space Telescopes, which were designed and built before the dawn of the exoplanet atmosphere era.  Observers have needed to be extremely creative and resourceful to use these telescopes for high precision photometry and spectroscopy.  See \citet{Winn10book}, \citet{Crossfield15}, \citet{Sing18}, and \citet{Deming19} for a comprehensive introduction and review of these characterization techniques. In the past decade high-resolution atmospheric characterization from the ground has been enabled with spectrographs on the world's largest telescopes, which is covered in more detail in \citet{Birkby18} and \citet{Brogi19}.

\subsubsection{Transmission spectroscopy}
Transmission spectroscopy was the first method used to characterize transiting planet atmospheres, and was first modeled by \citet{SS00}, \citet{Brown01}, and \citet{Hubbard01} and was observed by \citet{Charb02} with \emph{Hubble} for HD 209458b.  With this method, during the transit some flux from the parent star passes through the planet's thin outer atmosphere, imprinting small planetary absorption features on the stellar spectrum, during the transit.  Precisions of $10^{-4}$ are necessary, as the area of the annulus of planetary atmosphere is $\sim0.01 \times$ the area of the planet, which is again $\sim0.01 \times$ the area of the star.

In transmission, atomic metals have been observed in the UV and the optical, with previous Na and K detections \citep{Charb02,Redfield:2008aa,Sing:2011aa,Nikolov:2014aa,Sing:2015aa,Wyttenbach:2015} due to these species having particularly prominent features at 589 nm and 770 nm, respectively.  The presence of these strong, very pressure broadened optical features give rise to the very low geometric albedos measured and inferred for hot Jupiters, often below 0.1 \citep{Esteves:2015,Parmentier:2015}, which is much lower than Jupiter's~$\sim$0.5.  In the near-IR, \emph{Hubble}'s Wide-Field Camera 3 sits atop a water band from 1.1 to 1.7 $\mu$m, which has allowed for water vapor detections in many objects \citep{Deming:2013aa,Huitson:2013aa,Kreidberg14b,McCullough:2014aa,krei15,Line16L,Sing16} (see \citealp{Kreidberg17} for a recent review covering molecular abundance constraints) with the signal-to-noise ratio sometimes high enough to determine the water mixing ratio. \emph{Spitzer} allowed for photometry in the 3.6 $\mu$m and 4.5 $\mu$m bands, which has suggested roles for CO and CO$_2$ absorption in the 4.5 $\mu$m band \citep{desert09,Madhu11,Stevenson2016}.

Transmission spectroscopy at higher resolution has yielded information on atmospheric dynamics, as subtle Doppler shifts of absorption lines can be used to assess the velocity and direction of atmospheric flow \citep{Kempton12,Showman13,Flowers:2018aa}.  This has enabled planetary-scale wind determinations \citep{Snellen:2010}, differential wind measurements between leading and trailing hemispheres \citep[e.g.,][]{Louden15}, and constraints on the planetary rotation rate \citep{Brogi:2015}.  \cite{Ehrenreich:2020aa} found evidence for strong ($\approx 5.3~\mathrm{km}~\mathrm{s}^{-1}$) day-night winds in the ultra-hot Jupiter WASP-76b, along with evidence for condensation of iron on the planetary nightside due to a hemispheric asymmetry in its absorption signal. Recently, \cite{Tabernero:2020aa} confirmed this detection of day-night winds for WASP-76b over a broader range of atomic and molecular signatures, including ionized calcium, atomic manganese, magnesium, sodium, potassium, and lithium. These are just a handful of the broad range of atomic and ionized metallic species and molecules that have been detected in hot and ultra-hot Jupiter atmospheres with high-resolution transmission spectroscopy (e.g., \citealp{Birkby:2013aa,Brogi:2014aa,Birkby17,Nugroho:2017aa,Piskorz:2017aa,Wyttenbach:2017aa,Hoeijmakers:2018aa,Seidel:2019aa,Hoeijmakers:2020aa}, for a recent review see \citealp{Birkby18}).

\subsubsection{Transmission spectroscopy and atmospheric escape}
A special case of transmission spectroscopy, including studies at both low (via \emph{Hubble}) and high resolution (from the ground), is the study of atmospheric escape from hot Jupiters.  The high X-ray and UV fluxes are typically absorbed by hydrogen and metals, via photoionization, quite high in the planetary atmosphere (the thermosphere), where infrared opacities are low and cooling is inefficient, such that temperatures can reach $\sim 10^4$ K.  This can drive a hydrodynamic flow that can become unbound from the planet.  The amount of escape driven by this heating depends on how quickly the thermosphere cools through collisionally-excited atomic lines.  Mass loss is considerable, and hot Jupiters may typically lose $\sim$~1\% of their total mass, particularly at young ages when XUV fluxes are highest. 

After nearly 20 years of observations \citep{Vidal03,Vidal04,Ehrenreich15} and theory \citep{Yelle04,Murray08,Owen14} a considerable literature has developed.  A recent review can be found in \citet{Owen19}. Escaping hydrogen has been found from both Lyman-$\alpha$ observations \citep{Vidal03,Lecavelier10} and observations of the hydrogen Balmer series \citep{Jensen:2018aa,Cabot:2020aa,Wyttenbach:2020aa,Yan:2020aa}. Of particular recent interest has been the detection of escaping neutral helium \citep{Spake18,Ninan20} via the 1083 nm triplet.  This helium is presumably a part of the bulk flow from the planets \citep{Oklopcic18}. The combination of more detailed modeling studies \citep{Oklopcic20} and high-spectral resolution observations from ground-based telescopes that resolve velocity information for the He triplet \citep{Allart19} make this an important area for growth.

%cite brogi, flowers, using high-res to constrain rotation. Kempton & Rauscher, Showman using high-res to constrain jet speeds.
\subsubsection{Emission spectroscopy}
As the name implies, hot Jupiters are ``hot" and hence have substantial thermal emission.  For transiting planets, at the planetary occultation, or secondary eclipse, the planet passes behind the parent star.  The combined amount of star+planet flux decreases for the duration of the occultation, as the planet is blocked from view.  The power of emission spectroscopy, in principle, is that one could learn about atmospheric abundances (as in transmission spectroscopy) but also about the atmosphere's temperature structure, which is typically quite difficult to probe in transmission.

\begin{figure}
    \centering
    \includegraphics[width=0.8\textwidth]{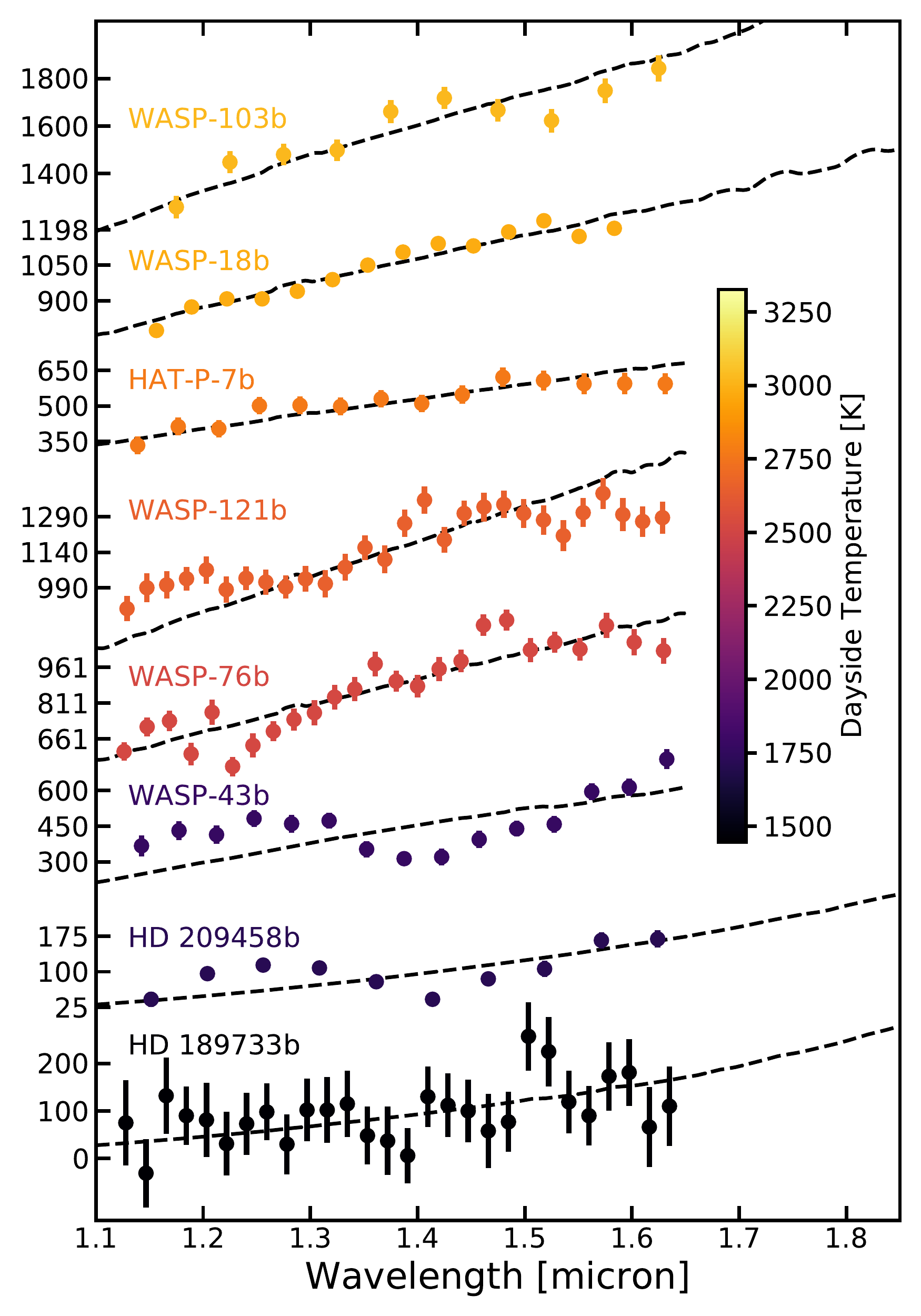}
    \caption{\emph{HST/WFC3} secondary eclipse spectra of hot and ultra-hot Jupiters observed in spatial scan mode as a function of dayside temperature. With increasing temperature, the 1.4$\mu$m water vapor absorption feature clearly seen for WASP-43b becomes diminished. In some cases (e.g., WASP-121b), this water feature can be seen in emission, indicative of an inverted temperature profile. Figure courtesy of Megan Mansfield.}
    \label{fig:sececlipse}
\end{figure}

The first emission detections were by \citet{Deming05a} and \citet{Charb05} for HD 209458b and TrES-1b, respectively.  Over the years, particularly in the Warm \emph{Spitzer} extended mission, large samples of 2-band emission photometry at 3.6 and 4.5 $\mu$m was obtained for dozens of planets.  These growing samples are now being assessed for trends that may suggest changes in atmospheric chemistry or temperature structure, as a function of the stellar incident flux \citep{Wallack19, Garhart20,Baxter:2020aa}, as shown in Figure \ref{fig:sececlipse}.  Furthermore, the idea of ``eclipse mapping'' uses differences in time-resolved ingress and egress lightcurves to create a brightness map, with latitude and longitude information, across the full day-side of the transiting planet \citep{Rauscher07b,Majeau12,Wit:2012}. 

Recently, it was discovered that the emission spectra of ultra-hot Jupiters are distinct from those of cooler hot Jupiters. \emph{HST/WFC3} emission spectra from a wide range of ultra-hot Jupiters (e.g., \citealp{Stevenson:2014aa,Haynes:2015,Beatty:2017aa,Evans:2017aa,Arcangeli:2018aa,Kreidberg:2018aa,Mansfield:2018aa,Baxter:2020aa,Fu:2020aa,Mikal-Evans:2020aa}) show black-body like, almost featureless spectra, with much weaker water features than are commonly found for cooler hot Jupiters (see Figure \ref{fig:sececlipse}).  It has been suggested that the almost featureless emission spectra of ultra-hot Jupiters is due to the dissociation of water molecules at the high temperatures on their daysides in concert with the formation of H$^{-}$, which acts as a strong continuum opacity source  \citep{Kitzmann:2018aa,Lothringer:2018aa,Parmentier:2018aa}. Additionally, \textit{Spitzer} photometry has shown that the ultra-hot Jupiters generally have stratospheres with inverted temperature-pressure profiles (potentially due to TiO/VO), while cooler hot Jupiters do not host inversions \citep{Garhart20,Baxter:2020aa}. These observed differences in the atmospheric composition and thermal structure between hot and ultra-hot Jupiters has led to the understanding that ultra-hot Jupiters represent a distinct class of exoplanet from cooler hot Jupiters.

\subsubsection{Phase curves}
Not long after the first detection of thermal emission from hot Jupiters, it was suggested that obtaining a phase curve of thermal emission over part of or all of a planetary orbit would be diagnostic of day-to-night temperature contrast and hot-spot offsets, first predicted in \citet{Showman02}.  \citet{Knutson_2007} published a half-orbit phase curve at 8 $\mu$m for planet HD 189733b, validating several aspects of these general circulation models.  This paper, and later \citet{Cowan2009}, described how phase curves could be inverted to understanding planetary brightness vs.~longitude. Phase curves for eccentric hot Jupiters, which are rare, are especially valuable as they show the planet's response near periapse, including the timescale of heating and subsequent cooling of the planetary atmosphere \citep{Laughlin11,Lewis13,Wit:2016aa}.

The first spectroscopic phase curve was of the planet WASP-43b \citep{Stevenson14} with \emph{Hubble} from 1.1 to 1.7 $\mu$m.  Importantly, it allowed one to probe atmospheric conditions and the hot-spot offset as a function of atmospheric depth, as observations within the water band at 1.4 $\mu$m probe less-deeply into the atmosphere than neighboring wavelengths with lower opacity.

\subsection{Temperature structure, clouds, and composition}
\label{sec:tempstructure}
The temperature structure of the planetary atmosphere is sensitive to many factors, including the absorption of stellar flux as a function of depth and wavelength, the emission of the absorbed flux and any intrinsic flux as a function of depth, and wavelength, and advection of energy to cooler parts of the atmosphere, including the night side.  This is a complex problem and a variety of 1D, 2D, and 3D models have been brought to bear.

\begin{figure}
    \centering
    \includegraphics[width=0.8\textwidth]{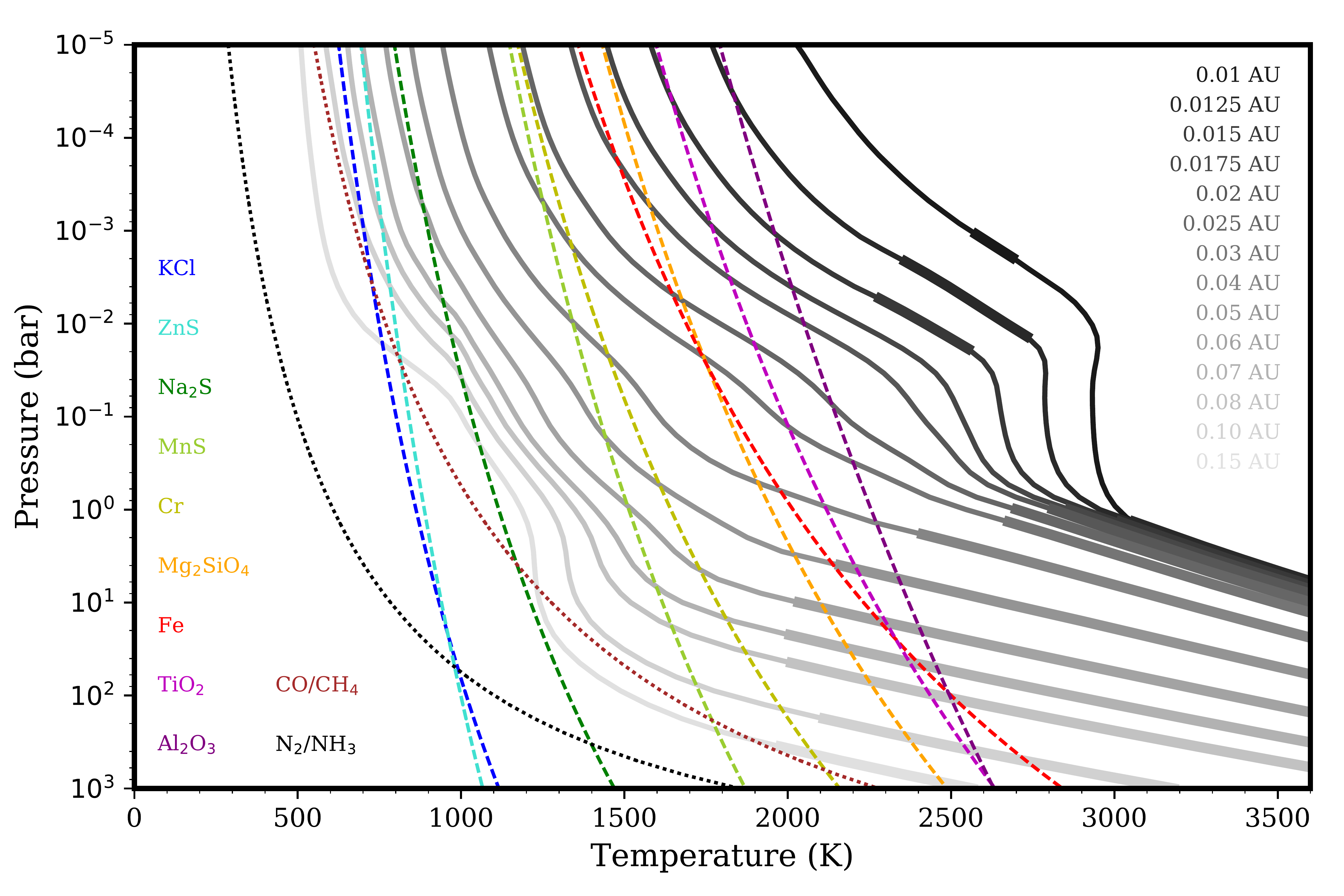}
    \caption{Atmospheric \emph{P--T} profiles as a function of distance from the Sun.  Thicker lines show convective regions and thinner lines show radiative regions.  Cloud condensation curves for relevant condensates are shown with colored dashed curves.  The equal-abundance curves of CO/CH$_4$ and N$_2$/NH$_3$ are shown with colored dotted curves.  Figure courtesy of Peter Gao.}
    \label{fig:gaoPT}
\end{figure}

Figure \ref{fig:gaoPT} shows a sampling of 1D atmospheric pressure-temperature (\emph{P--T}) profiles for metal-enriched ($10\times$ solar, or Saturn-like) atmospheres at small orbital distances.  The hottest models show a relatively shallow radiative-convective boundary (RCB) due to the high internal fluxes from whatever mechanism causes the radius anomaly \citep{Thorngren19}.  For slightly cooler planets the RCB may reach $\sim$~ 1 kbar.

The potential for gaseous TiO and VO, which dominate the optical spectra of M stars, to be found in the hottest hot Jupiters deserves a specific mention. \citet{Hubeny03} suggested that the high opacity of TiO and VO at low pressure, coupled with the lower opacity of coolants like water vapor, may drive a temperature inversion in these planets, which is not shown in Figure \ref{fig:gaoPT}.  \citet{Fortney08a} extended this work and suggest classifying planets in equilibrium temperature based on the presence or absence of TiO/VO.  The role of TiO/VO in these hottest atmospheres is still not clear, since these heavy molecules could be difficult to mix into the upper atmosphere, or could be lost into refractory clouds on the planetary night \citep{Spiegel:2009,parmentier_2013}. Significant additional work has gone to understand other aspects of how atmospheric chemistry and incident stellar flux changes the temperature structure, atmospheric abundances, and resulting spectra \citep[e.g.,][]{Madhu11b,Molliere15,Lothringer18}.

The equilibrium chemistry of these hot atmospheres has been explored by a number of authors, sometimes in conjunction with brown dwarf chemistry over this same \emph{P--T} space \citep{Lodders99,Lodders02,Visscher10,Madhu11b,Blecic16,Woitke18}.  Figure \ref{fig:gaoPT} shows the equal-abundance curves for CO/CH$_4$ (dashed red) and N$_2$/NH$_3$ (dashed black), showing the epectation that the dominant carbon carrier for hot Jupiters is CO, the dominant nitrogen carrier is N$_2$, and oxygen is mostly found in CO and H$_2$O.  A variety of atomic metals are found in the vapor phase, such as Na and K, which can dominated the optical opacity over a wide temperature range \citep{Sudar03}.  For the hottest profiles, abundances can be more akin to M-type (or even late K) stars, leading to even more atomic metals present in the ultra-hot Jupiters \citep{Kitzmann:2018aa}.

Studies of kinetic chemistry and photochemistry in hot Jupiters include \citet{Zahnle09,Moses11,Venot12,Tsai17,Hobbs19,Venot20}, and others.  Kinetic effects include the relevant chemical conversion and mixing timescales for these atmospheres.  \citet{Thorngren19} and \citet{Fortney20} have stressed that understanding the temperature structure of the (not visible) deep atmosphere, from thermal evolution models, is important for understanding these timescales and the associated atmospheric abundances.  Products that are typically enhanced compared to equilibrium chemistry, due to a variety of kinetic and photochemical effects, are molecules such as HCN, C$_2$H$_2$, and CO$_2$.  

Within the framework of chemistry calculations, a wide range of refractory clouds were long expected to be found in hot Jupiter atmospheres \citep{Marley99,Sudar00}, in particular Mg-bearing silicates and iron, as shown in Figure \ref{fig:gaoPT}.  The chemistry of this condensation sequence is now reasonably well understood from the atmospheres of brown dwarfs over this similar temperature range \citep{Ackerman2001,Helling14}.  One important aspect of the formation of condensates is that due to the loss of some elements from the gas phase, as they are sequestered in condensates, the atmospheric chemistry of remaining gases can be altered compared to simple expectations \citep{Lodders02,Helling19}.  Observationally, \citet{Sing16} demonstrated the near-ubiquity of clouds in affecting transmission spectra, which is well matched by models that include silicate clouds \citep{Gao:2020aa}.  However, the role of clouds in emission spectra is lacking in data, and the role in phase-dependent reflection spectra is still emerging \citep{Heng13,Esteves:2015,Hu:2015,Oreshenko:2016,Parmentier16,Roman:2017aa}.  Significant ongoing theoretical work is underway in combining 3D circulation models with cloud particle formation and transport, as discussed in Section \ref{tadnum}. 

\begin{figure}
    \centering
    \includegraphics[width=0.8\textwidth]{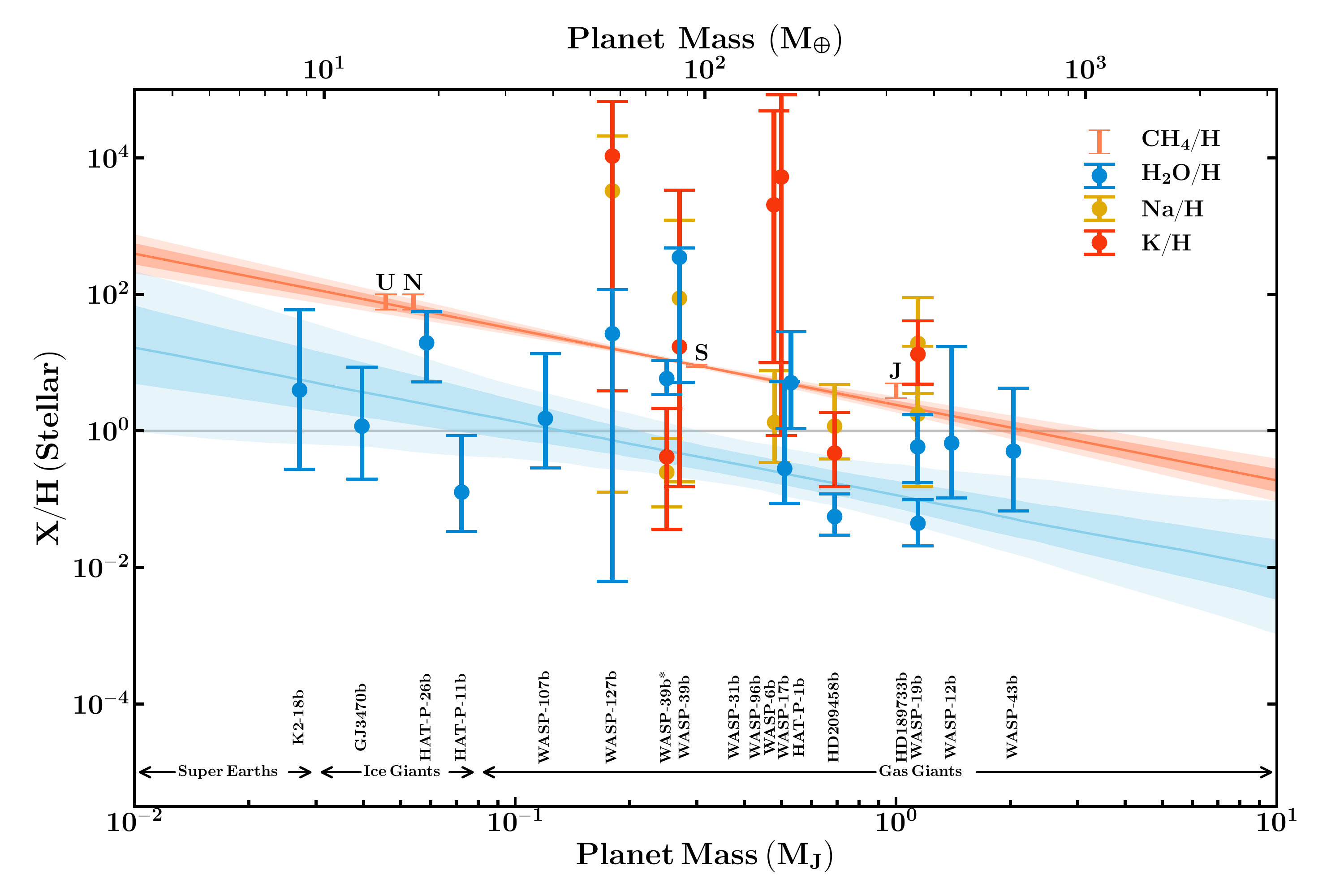}
    \caption{Abundances of atoms and molecules in the atmospheres of hot Jupiters (and hot Neptunes) from \citet{Welbanks19}.  The solar system mass-atmospheric metallicity trend, based on CH$_4$ is clear, shown in orange.  A similar trend \emph{may} be emerging for H$_2$O in hot Jupiters, but with a lower mixing ratio.}
    \label{welb}
\end{figure}

The high atmospheric temperatures of hot Jupiters, compared to our solar system's giant planets, make them fantastic laboratories for unveiling giant planet atmospheric abundances, as nearly all volatile species should be found in the vapor phase, rather than sequestered into clouds.  \citet{Oberg11} and \citet{Fortney13} have suggested links between atmospheric abundances and giant planet formation, the former in terms of carbon-to-oxygen (C/O) ratios, and the latter in terms of overall metallicity enhancements.  These ideas have been greatly developed by additional models, including \citet{Madhu14}, \citet{Mordasini16}, and \citet{Cridland19}.  The observational determination of C/O ratios have been hampered by the limited bandpass for \emph{Hubble} spectroscopy.  The determination of atmospheric metallicities, as a function of planet mass, began with \citet{Kreidberg14b} for the planet WASP-43b.  Error bars on abundance determinations with current instruments have been large, and there are not necessarily robust trends in metal-enrichment as a function of planet mass \citep{Fisher18,Welbanks19}.  A recent compilation of atmospheric abundances is shown in Figure \ref{welb} from \citet{Welbanks19}.

Several frameworks for simultaneously retrieving atmospheric abundances and temperature structure in a Bayesian framework, termed ``retrieval," are essential to this work, build on solar system heritage \citep{Irwin08,Line13}, and are reviewed in \citet{Madhu18} and \citet{Barstow20}.  More comprehensive reviews of hot Jupiter models and their comparison with data can be found in \citet{Madhu16}, \citet{Helling19}, and \citet{Madhu19}.

\subsection{Atmospheric circulation}
\label{sec:atmcirc}

\subsubsection{Numerical simulations}  \label{tadnum}
%current state of GCM simulations (model hierarchy encompassing primitive equations, correlated-k RT, etc.)
%where the field is moving toward: longer simulations, fullly equilibrated, non-hydrostatic, treating deep levels properly
\indent Over the past two decades, a wide range of general circulation models (GCMs) have been developed to simulate the atmospheric circulation of hot Jupiters. GCMs solve the dynamical equations of fluid motion coupled to a radiative transfer model that determines the radiative heating/cooling rate. The reader is referred to \cite{Showman_2009}, a detailed review of exoplanet atmospheric fluid dynamics, to \cite{Heng15} (see their Table 1) for a review of established hot Jupiter GCMs, to \cite{Showman:2020aa} for a review of recent advances in the study of hot gas giant exoplanets, and to \cite{Zhang:2020aa} for an expansive review of the atmospheric circulation regimes of exoplanets. Many hot Jupiter GCMs solve the primitive equations of meteorology (e.g. \citealp{Showman02,Menou:2009,Rauscher:2010,Heng:2011a,Lewis:2014,Polichtchouk:2014,Kataria:2014,Carone:2019aa}),
%\citealp{Showman02,Cooper:2005,Showman:2008,Menou:2009,Showman09,Rauscher:2010,Heng:2011a,Rauscher_2012,Liu:2013,parmentier_2013,Lewis:2014,Cho:2015,Kataria:2014,Carone:2019aa,Tan:2019aa})
which are a simplified form of the Navier-Stokes equations relevant to thin, locally hydrostatic atmospheres \citep{Vallis:2006aa,Holton:2013}. A subset of models solve the non-hydrostatic Euler or Navier-Stokes equation sets (e.g., \citealp{Dobbs-Dixon:2013,Mayne:2014,Mendonca:2016aa})
%,Drummond:2018ab}) 
that capture the propagation of sound waves, which is important because the winds in hot Jupiter atmospheres can approach or exceed the speed of sound.
%GCM simulations, flavors. Groups to cite: carone, Mayne, Heng, Rauscher, Menou, Showman/Kataria/Lewis/Parmentier, Dobbs-Dixon

%superrotation
%Large-scale flow manifests as superrotation (give background theory -- tsai, showman, hammond)
Though hot Jupiter GCMs have quantitative differences in atmospheric climate and wind speeds, they show qualitatively similar large-scale atmospheric circulation patterns (see Figure 9 of \citealp{Heng15}).
%, in agreement with that expected from our simple dynamical scaling in \Sec{sec:atmregimes}. 
%xxx Figure here comparing a subset of GCMs, like Figure 9 of Heng and Showman xxx.
%\begin{figure}
%    \centering
%    \includegraphics[width=0.6\textwidth]{showman_2013_schematic}
%    \caption{Schematic showing how large-scale atmospheric wave propagation drives superrotation in hot Jupiter atmospheres. From \cite{showman_2013_doppler}.}
%    \label{fig:showman2013schematic}
%\end{figure}
GCMs find that the atmospheric circulations of hot Jupiters are characterized by large dayside-to-nightside temperature contrasts. The large dayside-to-nightside temperature gradient forces fast winds which manifest as eastward (propagating in the direction of rotation) jets with maximum speeds in equatorial regions \citep{Showman_Polvani_2011,Tsai:2014,Hammond:2018aa}. These fast winds lead to a shift in the temperature maximum (or ``hot spot'') eastward (downwind) of the substellar point. The strong eastward flow near the equator of hot Jupiters is known as ``superrotation'' because the angular momentum of the atmospheric circulation exceeds the rotational angular momentum of the interior of the planet \citep{Vallis:2019aa}. 
%this paragraph below can be removed to save space/if it's too complicated. It's probably the most important theoretical result in the sub-field, but is hard to explain in one paragraph.

%Superrotation cannot be maintained by the longitudinally symmetric ``mean'' flow \citep{Hide:1969} --- instead, the existence of superrotation implies that flow features not associated with the mean flow (known as eddies) lead to up-gradient momentum transport toward the local maximum in angular momentum. The superrotation in hot Jupiter atmospheres is thought to be maintained by the differential wave propagation of Kelvin waves, which propagate eastward and are confined to near the equator, and Rossby waves, which propagate westward and extend over a broader range of latitudes \citep{Showman_Polvani_2011,Tsai:2014,Hammond:2018aa}. This differential wave propagation leads to the maintenance of northwest-southeast eddy phase tilts in the northern hemisphere and southwest-northeast phase tilts in the southern hemisphere (shown in \Fig{fig:showman2013schematic}), which causes the up-gradient momentum transport necessary to maintain the superrotating equatorial jet. 

%Complications: clouds (cite Lines, Lee trying to integrate in), MHD (rogers, batygin), non-hydrostatic (mayne, mendonca)/sub-grid instabilities/shocks (Fromang), integration time (mayne, mendonca)
Even though the overarching picture of hot Jupiter atmospheres has remained unchanged since \cite{Showman02}, there are many new frontiers that models are currently exploring.
%ing frontiers have shown that current understanding may be incomplete. 
Cloud formation and evolution can have important radiative impacts on climate and observable properties \citep{Helling:2016,Wakeford:2017,Powell:2018aa,Gao:2020aa}, which can feed back onto atmospheric dynamics. Recent modeling efforts (e.g., \citealp{Lee:2016,Lines:2018,lines:2019,Roman:2019aa,Parmentier:2020aa,Roman:2020aa}) have shown that cloud-radiative feedbacks can affect the climate and observable properties of hot Jupiters. Due to winds transporting species before they can relax to chemical equilibrium, the effects of chemical disequilibrium are expected to impact the atmospheric composition and observable properties of hot Jupiters \citep{Cooper:2006,Drummond:2018aa,Zhang18b,Mendonca:2018ab,Steinrueck:2019aa}. It is also expected that magneto-hydro-dynamic (MHD) effects play a role in the circulation of hot Jupiters.
%, as hot Jupiters likely have strong internal magnetic field strengths \citep{Yadav:2017,Cauley:2019aa} and hot enough atmospheres that species with low ionization potentials (e.g., Na, K) are partially ionized \citep{Perna_2010_1}. 
A wide range of efforts have included MHD effects in GCMs of hot Jupiter atmospheres (e.g., \citealp{Perna_2010_1,batygin_2013,Rauscher_2013,Rogers:2020}),
%,Rogers:2014,Rogers:2017})
and have shown that MHD can impact the strength and even direction of the equatorial jet. 
%However, to date no model has self-consistently coupled the magnetic induction equation to a GCM and included a realistic dayside-to-nightside conductivity contrast. 
Due to the strong spatial dependence of winds, it is expected that shear instabilities and shocks play a role in the dissipation of the circulation of hot Jupiters \citep{Li:2010,Heng:2012a,perna_2012,Fromang:2016}, but high-resolution non-hydrostatic global models are required to fully assess the impact of instabilities and shocks on the circulation. Additionally, long integration timescales are needed to properly model the impact of the deep atmospheric circulation of hot Jupiters on observable levels, as the timescale for radiative adjustment of the deep atmospheric layers is thousands of years at the $\sim 100~\mathrm{bar}$ pressures at the base of typical GCM models \citep{Arras06,Sainsbury-Martinez:2019aa}. Recent simulations that were run to an order of magnitude longer model time than was typical show a qualitatively different flow structure for both hot Jupiters and sub-Neptunes \citep{Mayne:2017,Mendonca:2020aa,Wang:2020aa}, pointing toward the need to characterize the deep flow of gaseous exoplanets.

\subsubsection{Interpreting observations with dynamical models}
%largely phase curves, also talk about high-eccentricity planets (Lewis and Kataria papers) as a unique regime to probe radiative timescale. Talk about fundamental challenge that you can't run too many GCMs, so currently can't do a retrieval using GCMs. This section could be fairly short. 
Numerical models of atmospheric circulation play a key role in understanding observations of exoplanet atmospheres. GCMs are uniquely suited to compare with phase curve observations, as understanding the longitudinal dependence of planetary flux requires a model for the atmospheric heat transport. GCMs have been directly compared to phase curve observations for a wide range of hot Jupiters (see \citealp{Parmentier:2017} for a recent review). As expected from GCM simulations, the majority of phase curves find large dayside-to-nightside temperature contrasts and eastward bright spot offsets.
%, starting with the \textit{Spitzer} observations of HD 189733b by \cite{Knutson_2007}. %\citep{Showman:2008,Showman09,Knutson:2012,Dobbs-Dixon:2013,Drummond:2018aa,Steinrueck:2019aa}, WASP-43b \citep{Kataria:2014,Mendonca:2018aa}, HD 209458b \citep{Zellem:2014,Amundsen:2016aa,lines:2019}, Kepler-7b \citep{Roman:2019aa}, HD 149026b \citep{Zhang:2017a}, WASP-14b \citep{Wong:2015}, WASP-19b \citep{Wong:2015a}, HAT-P-7b \citep{Wong:2015a}, WASP-18b \citep{Arcangeli:2019aa}, WASP-103b \citep{Kreidberg:2018aa}, WASP-33b \citep{Zhang:2017a}, KELT-9b \citep{Mansfield:2020aa} and the eccentric hot Jupiter HAT-P-2b \citep{Lewis:2014}. 
Additionally, GCMs have been shown to provide a good match to individual observed phase curves of hot Jupiters over a broad range of equilibrium temperature (e.g., \citealp{Showman09,Dobbs-Dixon:2013,Zellem:2014,Drummond:2018aa,Kreidberg:2018aa,Arcangeli:2019aa}).
%(e.g., metallicity, which sets the pressure level at which the planet emits to space). 
%, but due to computational limitations it is often challenging to provide a model that fits the phase curve in detail. 

Recent observations have found that the eastward hot spot offsets predicted by standard GCMs may not be ubiquitous. \cite{Dang:2018aa} found a westward phase offset in the \textit{Spitzer} observations of CoRoT-2b, and there is tentative evidence for westward infrared phase offsets in the atmospheres of HD 149026b \citep{Zhang:2017a} and HAT-P-7b \citep{Wong:2015a}. Additionally, time-variability in the phase offset has been detected on three planets to date: HAT-P-7b \citep{Mooij2016}, Kepler-76b \citep{Jackson:2019aa}, and WASP-12b \citep{Bell:2019aa}. A range of mechanisms have been proposed to explain westward phase offsets, including clouds \citep{Mooij2016,Parmentier16,Powell:2018aa},
%,Lines:2018,Powell:2018aa}
MHD effects \citep{Rogers:2017,Hindle:2019aa}, and non-synchronous rotation \citep{Rauscher:2014}. 
%Paragraph about full sample of planets -- perez-becker and showman, my, zhang, cowan work. Add discussion about UHJs and H dissociation here! Add Tan & Komacek figure?

Thanks to the sample of 26 exoplanets with observed \textit{Spitzer} phase curves \citep{Deming:2020aa}, some of which have additional \textit{Hubble} phase curves (e.g., \citealp{Stevenson:2014,Kreidberg:2018aa,Arcangeli:2019aa}), it is now possible to understand hot Jupiter atmospheric circulation through population-level studies. Both analytic theory \citep{Cowan:2011,Perez-Becker:2013fv,Komacek:2015,Zhang:2016} and GCMs \citep{Kataria2016,Komacek:2017,Parmentier:2017,Tan:2019aa} predict that for hot Jupiters with equilibrium temperatures below $2200~\mathrm{K}$, the phase curve amplitude increases and the phase curve offset decreases with increasing equilibrium temperature. This is because the large-scale wave motions that drive circulation and act to induce hot spot offsets and reduce day-to-night temperature contrasts become increasingly damped with increasing equilibrium temperature due to radiative cooling \citep{showman_2013_doppler,Perez-Becker:2013fv}. 

\begin{figure}[h]
    \centering
    \includegraphics[width=0.9\textwidth]{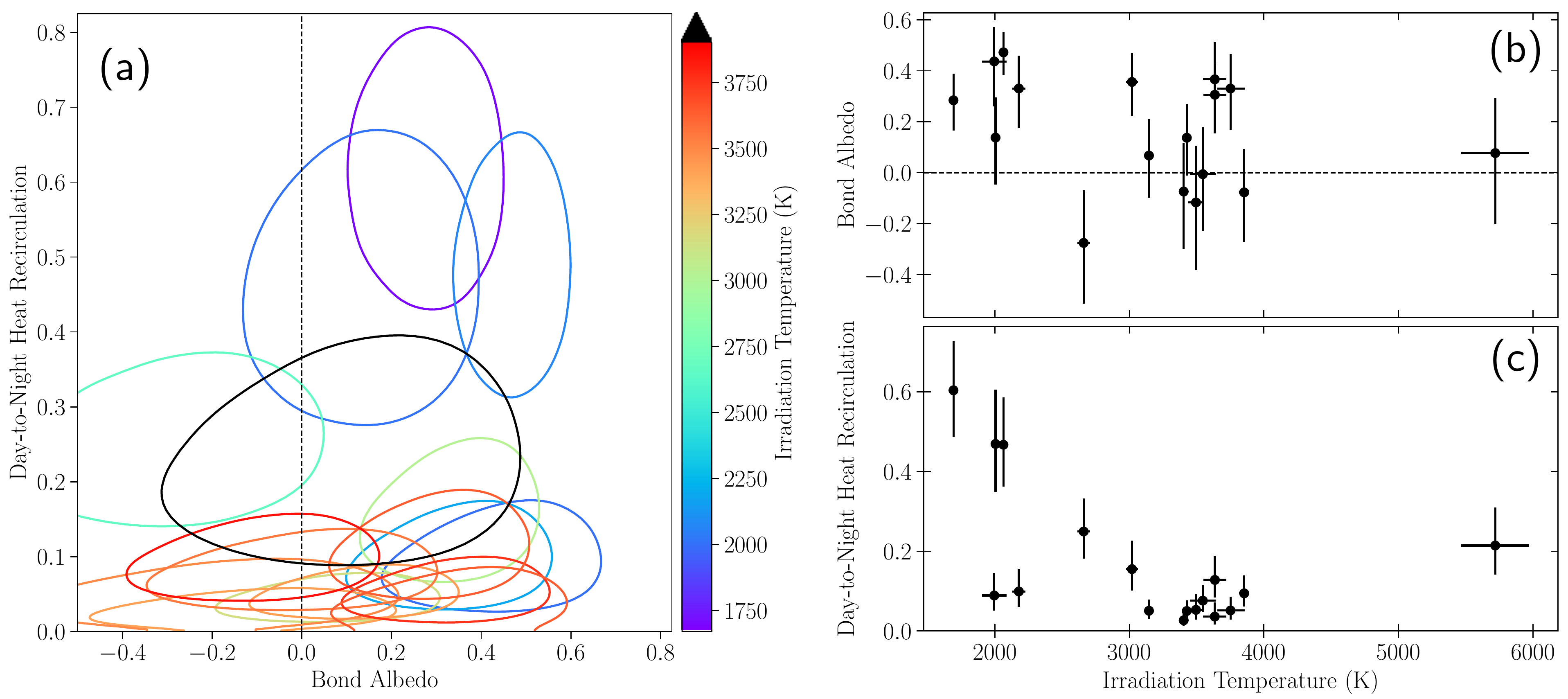}
    \caption{The atmospheric heat transport of hot Jupiters is strongly dependent on the incident stellar flux. Panel (a) shows joint constraints on day-night heat recirculation parameter (calculated as in \citealp{Cowan:2011}) and Bond albedo from literature Spitzer Channel 2 ($4.5~\mu\mathrm{m}$) phase curve observations, where the color of each contour refers to the planetary irradiation temperature. Panels (b) and (c) show the dependence of Bond albedo and heat recirculation on the irradiation temperature. Negative Bond albedos can occur when the region probed by the Spitzer Channel 2 is a hotter-than-typical region of the atmosphere. The dayside and nightside effective temperatures and their associated uncertainties were estimated using only the published $4.5~\mu\mathrm{m}$ brightness temperatures of \cite{Bell:2020aa}, combined with the Gaussian process regression technique of \cite{Pass:2019aa}. The inferred heat recirculation appears to decrease with increasing irradiation temperature until the regime of ultra-hot Jupiters, while there is no clear dependence of Bond albedo on incident stellar flux. Figure courtesy of Taylor Bell.}
    \label{fig:Bell}
\end{figure}

The basic expectations of the increase in the phase curve amplitude and decrease in phase offset with increasing equilibrium temperature below $2200~\mathrm{K}$ 
appear to be borne out by observations \citep{Cowan_2011,Perez-Becker:2013fv,Schwartz:2015,Schwartz:2017aa,Keating:2019aa}. Figure \ref{fig:Bell} shows joint constraints on the heat recirculation efficiency (which is inversely related to phase curve amplitude) and Bond albedo for the sample of observed hot Jupiters with \textit{Spitzer}, showing an apparent decrease in the heat recirculation with increasing planetary temperature as expected from previous analytic theory. For ultra-hot Jupiters with equilibrium temperatures in excess of $2200~\mathrm{K}$, it is expected that the hydrogen that comprises the bulk of the atmosphere will become partially thermally dissociated, with a resulting net cooling on the dayside and latent heating on the nightside \citep{Bell:2018aa}. Both theoretical models \citep{Bell:2018aa,Komacek:2018aa} and GCMs \citep{Tan:2019aa} predict that hydrogen dissociation and recombination leads to a reduction in the phase curve amplitude of ultra-hot Jupiters, with phase curve observations of the ultra-hot Jupiters WASP-103b \citep{Kreidberg:2018aa}, WASP-33b \citep{Zhang:2017a}, and KELT-9b \citep{Mansfield:2020aa} showing tentative evidence for this effect. Additionally, both hot and ultra-hot Jupiters generally have low Bond albedos relative to the value of $0.503 \pm 0.012$ measured for Jupiter \citep{Li:2018aa}. This may suggest that silicate and other condensate clouds do not dominate the dayside optical opacity of hot gas giants to the same extent that thick ammonia clouds do so on Jupiter.

%,Komacek:2017,
Determining if the trend in phase curve amplitude with equilibrium temperature is statistically significant requires further observations \citep{Parmentier:2017} and an improved understanding of detector systematics \citep{May:2020aa,Bell:2020aa}. Analysis of the growing sample of \textit{TESS} phase curves (e.g., \citealp{Daylan:2019aa,Shporer:2019aa,Wong:2019aa,Jansen:2020aa,Essen:2020aa}),
%Wong:2020ab,Wong:2020aa}
unpublished \textit{Spitzer} phase curves \citep{Deming:2020aa}, and future phase curve observations with \textit{ARIEL} \citep{Tinetti18,Zellem:2019aa} and \textit{JWST} \citep{Bean:2018aa} will provide a more complete understanding of how observable properties of hot Jupiter atmospheres depend on their planetary parameters.
%This has been done by studying the dependence of the amplitude and offset of the phase curve both on one another and on planetary parameters, including the equilibrium temperature  
%Talk about future with JWST/ARIEL (cite CASE paper, JWST ERS bean paper): secondary eclipse mapping (cite rauscher papers), spectroscopic phase curves. this can be a sentence or two at the end of paragraph above.

\section{Conclusions and Prospects for Future Characterization}
Hot Jupiters have been objects of intense study for 25 years, and vast strides have been made through the synergy between astronomical observations and astro- and geo-physical models.
There is evidence for at least two origins channels, with some hot Jupiters arriving through high eccentricity tidal migration and others through either disk migration or in situ formation. Future observations will help distinguish the importance of each channel. Better constraints on disk conditions and more detailed head-to-head comparisons on predicted occurrence rates and architectures will help distinguish whether in situ formation vs. disk migration is the predominant second channel.

The \emph{TESS} Mission is expected to provide longer period and/or smaller targets for ground-based spin-orbit alignment measurements that will clarify the interpretation of hot Jupiters' spin-orbit alignments. \emph{TESS} may also discover -- in the mission itself or through ground-based radial-velocity follow up -- more exceptional systems like WASP-47 featuring a hot Jupiter with nearby planets that likely originated through the second channel. The \emph{Gaia} Mission will enable astrometry of hot Jupiters' companions to measure the mutual inclination and further investigations of connections between hot Jupiter occurrence and host star properties. For example, \citet{Winter2020} recently found a correlation between hot Jupiters and present-day stellar clustering. If this present-day clustering reflects the early environment, the correlation suggests that dense stellar cluster environments facilitate processes that lead to hot Jupiters. Both missions will contribute to the goal of constructing large samples of hot Jupiters with a wide range of well-constrained host star ages to determine how often hot Jupiters have already formed or arrived during the gas disk stage vs. arriving via high-eccentricity migration throughout the star's lifetime. Further theoretical study is needed to understand whether hot Jupiters' atmospheric properties can feasibly trace formation location despite the diversity of disk conditions, enabling interpretation of \emph{JWST}'s measurements of atmospheric properties of a large sample of hot Jupiters.

In terms of planetary structure and evolution, the sheer number of planet detections means that we are firmly in the statistical era, where the large number of observed masses and radii strongly constrain evolution models.  Only radius inflation mechanisms that are closely tied to the magnitude of the stellar incident flux can explain the population.  Giant planet bulk metal-enrichment, as seen in the solar system, is the rule for giant exoplanets, and is itself a strongly inverse function of planet mass.  Looking ahead, more detailed multi-dimensional simulations of processes that may affect the radius anomaly are needed. Additionally, detecting still more planets is definitely needed, in particular in two areas. First, a focus on hot Jupiters orbiting stars evolving off the main sequence, where the stellar forcing changes on relatively fast timescales, will give new insight into how planetary structure responds.  Second, detections of planets $<1000 $ K around main sequence stars, below the radius anomaly threshold, is essential to understanding planetary bulk composition, to look for trends in planetary bulk composition with planetary and stellar properties \citep[e.g.,][]{Teske19}.

In the realm of hot Jupiter atmospheres, broad trends have emerged.  Albedos are typically low, due to alkali metal absorption and perhaps TiO vapor in the hottest objects.  Silicate clouds appear to be an important opacity source across a wide temperature range.  Trends in the day/night flux contrast are generally well-fit by models.  In the near term detailed spectroscopy across a wide wavelength range with \emph{JWST} and with large ground-based telescopes for benchmark objects is essential.  It is quite likely that high-quality spectra will yield surprises in terms of atmospheric abundances and trends with planet mass and temperature.  Entirely new paradigms may well emerge.  Later, the field will move to the statistical assessment of exoplanet atmospheres with the European Space Agency's \emph{ARIEL} mission.  This must be done in tandem with a hierarchy of modeling approaches, including those based on solar system and exoplanetary expertise.  This is because we have likely only glimpsed at the true 3D nature of these atmospheres, and future observations will push models toward even higher levels of complexity.  Very soon, \emph{JWST} and high-resolution ground-based observations of the lower and upper atmosphere in transmission and emission will also pay large dividends in understanding atmospheric composition, circulation, and escape.

\appendix 

\section{Glossary}
\noindent {\bf Astrometry:} The measurement of the positions, motions, and magnitudes of stars, which enables a measurement of their distance. \\ 
\noindent {\bf Atmospheric column mass:} Mass per unit area of the atmosphere, equal to $p/g$, where $p$ is the atmospheric pressure and $g$ is gravity. \\
\noindent {\bf Bayesian:} A statistical approach that uses Bayes' theorem to assign probabilities to events that are updated based on new data. \\
\noindent {\bf Bond albedo:} The fraction of the total incident electromagnetic radiation upon a planet that is scattered back to space. \\
\noindent {\bf Corotation period:} Orbital period of a planet at which its orbital period is equal to the stellar rotation period. \\ %(please check this one!)  
\noindent {\bf Double-diffusive convection:} A form of convection driven by two different density gradients with separate diffusivities. \\
\noindent {\bf Embryo:} A rocky building block of planets that is larger than $\sim 1000~\mathrm{km}$ in diameter. \\
\noindent {\bf Hydrostatic equilibrium:} An atmosphere in which the force due to the decrease in pressure with height balances gravity. \\ 
\noindent {\bf Ionization potential:} Energy necessary to remove an electron from a neutral atom. \\
\noindent {\bf Irradiation temperature:} The temperature that the planet would have on the star-facing side if there were no horizontal heat transport. The irradiation temperature $T_\mathrm{irr}$ is related to the zero-Bond albedo equilibrium temperature $T_\mathrm{eq}$ as $T_\mathrm{irr} = 4^{1/4}T_\mathrm{eq}$. \\
%\noindent {\bf Equatorial Kelvin wave:} An atmospheric wave whose restoring force is gravity and is trapped near the equator due to the change in the Coriolis (rotational) force with latitude. \\
\noindent {\bf Kozai-Lidov cycles:} An orbital cycle between the eccentricity and inclination that caused by the perturbation of a distant third body.  \\ 
\noindent {\bf M star:} The smallest and coolest type of star, with effective temperatures below 3900 K. \\ 
\noindent {\bf Magnetocavity:} The innermost region of the proto-planetary disk that is cleared out due to truncation by the stellar magnetic field. \\
\noindent {\bf Magneto-hydro-dynamics (MHD):} The study of the motions of electrically conducting fluids. \\ 
\noindent {\bf Metallicity (astrophysical):} Abundance of elements that are heavier than hydrogen and helium. \\
\noindent {\bf Migration:} The radial movement of the orbit of a planet due to interactions with a disk or other planets.  \\ 
\noindent {\bf Navier-Stokes equations:} A set of differential equations that can be applied to study the motions of geophysical fluids. \\ 
\noindent {\bf Ohmic dissipation:} Transfer of electrical energy to thermal energy when current flows through a resistance. \\
\noindent {\bf Opacity:} The amount of absorption and scattering of incident electromagnetic radiation by a material.  \\
\noindent {\bf Pebbles:} Centimeter to meter sized particles in proto-planetary disks that undergo radial drift due to gas drag. \\
\noindent {\bf Phase amplitude:} Difference between the maximum and minimum observed emergent planetary flux over a full orbital phase. \\
\noindent {\bf Phase offset:} Shift in orbital phase between secondary eclipse and the maximum observed emergent planetary flux. \\ 
\noindent {\bf Photoionization:} The loss of an electron from an atom or molecule due to absorption of electromagnetic radiation. \\
\noindent {\bf Planetesimal:} Solid body that forms during the era of planet formation that is larger than $\sim 1~\mathrm{km}$. \\
\noindent {\bf Proto-planetary disk:} A disk of gas and dust that surrounds a young star.  \\ 
\noindent {\bf Radial Velocity:} A method used to detect planets by observing the periodic effect that the gravity of a planet causes on the location of spectral lines of the host star. \\ 
\noindent {\bf Refractory:} A solid material that is resistant to becoming gaseous at high temperatures. \\ 
%\noindent {\bf Rossby wave:} An atmospheric wave whose restoring force is rotation. \\
\noindent {\bf Secular:} Taken or averaged over many orbits. \\
\noindent {\bf Shear instability:} A dynamical instability that occurs when there is sufficient velocity shear in a fluid. \\
\noindent {\bf Substellar point:} The point on a planet's surface where the host star is directly overhead and incident rays of light from the host star strike perpendicularly to the surface.  \\ 
\noindent {\bf Superrotation:} A phenomenon that occurs when a planetary atmosphere has greater angular momentum than the  solid-body rotation of the interior. \\
\noindent {\bf T Tauri star:} Pre-main-sequence stars that vary in their emitted light and are less than $\sim 10~\mathrm{Myr}$ old and still contracting. \\ 
\noindent {\bf Transit:} A method of detecting planets by observing the periodic decrease in stellar brightness due to the planet blocking light from the host star.  \\ 
\noindent {\bf XUV:} An acronym for Extreme UltraViolet radiation, which is light with wavelengths between $\sim 0.1 - 0.01 \mu\mathrm{m}$. \\
\noindent{\bf Zero Bond albedo equilibrium temperature:} The temperature a planet would have assuming that it does not reflect any incident stellar radiation and perfectly redistributed heat across its surface, equal to $\left[F_\star/(4\sigma)\right]^{1/4}$, where $F_\star$ is the incident stellar flux and $\sigma$ is the Stefan-Boltzmann constant.

\acknowledgments
%This document contains \total{citnum}\ references.

JJF acknowledges supports from grant 80NSSC19K0446 awarded by the NASA Exoplanets Research Program.  RID acknowledge support from grant NNX16AB50G awarded by the NASA Exoplanets Research Program and the Alfred P. Sloan Foundation's Sloan Research Fellowship. TDK acknowledges support from the 51 Pegasi b Fellowship in Planetary Astronomy sponsored by the Heising-Simons Foundation.

Data Availability Statement:  Data were not used, nor created for this research.

%% ------------------------------------------------------------------------ %%
%% References and Citations

%%%%%%%%%%%%%%%%%%%%%%%%%%%%%%%%%%%%%%%%%%%%%%%
%
% \bibliography{<name of your .bib file>} don't specify the file extension
%
% don't specify bibliographystyle
%%%%%%%%%%%%%%%%%%%%%%%%%%%%%%%%%%%%%%%%%%%%%%%

\bibliography{references}

%Reference citation instructions and examples:
%
% Please use ONLY \cite and \citeA for reference citations.
% \cite for parenthetical references
% ...as shown in recent studies (Simpson et al., 2019)
% \citeA for in-text citations
% ...Simpson et al. (2019) have shown...
%
%
%...as shown by \citeA{jskilby}.
%...as shown by \citeA{lewin76}, \citeA{carson86}, \citeA{bartoldy02}, and \citeA{rinaldi03}.
%...has been shown \cite{jskilbye}.
%...has been shown \cite{lewin76,carson86,bartoldy02,rinaldi03}.
%... \cite <i.e.>[]{lewin76,carson86,bartoldy02,rinaldi03}.
%...has been shown by \cite <e.g.,>[and others]{lewin76}.
%
% apacite uses < > for prenotes and [ ] for postnotes
% DO NOT use other cite commands (e.g., \citet, \citep, \citeyear, \nocite, \citealp, etc.).
%

\end{document}

%% file: bibsymlist.tex
\def\apjl{{ApJ}}		
\def\apjs{{ApJS}}

\def\aap{{A\&A}}

\newcommand{\Eq}[1]{Equation\,(\ref{#1})}

\newcommand{\Sec}[1]{Section~\ref{#1}}

\newcommand{\Fig}[1]{Figure~\ref{#1}}

\newcommand{\mj}{$M_{\mathrm{J}}$} 

\newcommand{\me}{$M_{\oplus}$}

%% file: revised.bbl
\begin{thebibliography}{}

\bibitem [\protect \citeauthoryear {%
Ackerman%
\ \BBA {} Marley%
}{%
Ackerman%
\ \BBA {} Marley%
}{%
{\protect \APACyear {2001}}%
}]{%
Ackerman2001}
\APACinsertmetastar {%
Ackerman2001}%
\begin{APACrefauthors}%
Ackerman, A\BPBI S.%
\BCBT {}\ \BBA {} Marley, M\BPBI S.%
\end{APACrefauthors}%
\unskip\
\newblock
\APACrefYearMonthDay{2001}{}{}.
\newblock
{\BBOQ}\APACrefatitle {Precipitating condensation clouds in substellar
  atmospheres} {Precipitating condensation clouds in substellar
  atmospheres}.{\BBCQ}
\newblock
\APACjournalVolNumPages{The Astrophysical Journal}{556}{}{872}.
\newblock
\begin{APACrefURL}
  \url{http://iopscience.iop.org/0004-637X/556/2/872/fulltext/{\%}5Cnhttp://arxiv.org/abs/astro-ph/0103423{\%}5Cnhttp://dx.doi.org/10.1086/321540}
  \end{APACrefURL}
\PrintBackRefs{\CurrentBib}

\bibitem [\protect \citeauthoryear {%
{Albrecht}%
\ \protect \BOthers {.}}{%
{Albrecht}%
\ \protect \BOthers {.}}{%
{\protect \APACyear {2012}}%
}]{%
albr12}
\APACinsertmetastar {%
albr12}%
\begin{APACrefauthors}%
{Albrecht}, S.%
, {Winn}, J\BPBI N.%
, {Johnson}, J\BPBI A.%
, {Howard}, A\BPBI W.%
, {Marcy}, G\BPBI W.%
, {Butler}, R\BPBI P.%
\BDBL {}{Hartman}, J\BPBI D.%
\end{APACrefauthors}%
\unskip\
\newblock
\APACrefYearMonthDay{2012}{{\APACmonth{09}}}{}.
\newblock
{\BBOQ}\APACrefatitle {{Obliquities of Hot Jupiter Host Stars: Evidence for
  Tidal Interactions and Primordial Misalignments}} {{Obliquities of Hot
  Jupiter Host Stars: Evidence for Tidal Interactions and Primordial
  Misalignments}}.{\BBCQ}
\newblock
\APACjournalVolNumPages{\apj}{757}{}{18}.
\newblock
\begin{APACrefDOI} \doi{10.1088/0004-637X/757/1/18} \end{APACrefDOI}
\PrintBackRefs{\CurrentBib}

\bibitem [\protect \citeauthoryear {%
{Allart}%
\ \protect \BOthers {.}}{%
{Allart}%
\ \protect \BOthers {.}}{%
{\protect \APACyear {2019}}%
}]{%
Allart19}
\APACinsertmetastar {%
Allart19}%
\begin{APACrefauthors}%
{Allart}, R.%
, {Bourrier}, V.%
, {Lovis}, C.%
, {Ehrenreich}, D.%
, {Aceituno}, J.%
, {Guijarro}, A.%
\BDBL {}{Wyttenbach}, A.%
\end{APACrefauthors}%
\unskip\
\newblock
\APACrefYearMonthDay{2019}{{\APACmonth{03}}}{}.
\newblock
{\BBOQ}\APACrefatitle {{High-resolution confirmation of an extended helium
  atmosphere around WASP-107b}} {{High-resolution confirmation of an extended
  helium atmosphere around WASP-107b}}.{\BBCQ}
\newblock
\APACjournalVolNumPages{\aap}{623}{}{A58}.
\newblock
\begin{APACrefDOI} \doi{10.1051/0004-6361/201834917} \end{APACrefDOI}
\PrintBackRefs{\CurrentBib}

\bibitem [\protect \citeauthoryear {%
Arcangeli%
, Desert%
, Line%
\BCBL {}\ \BBA {} et al.%
}{%
Arcangeli%
\ \protect \BOthers {.}}{%
{\protect \APACyear {2018}}%
}]{%
Arcangeli:2018aa}
\APACinsertmetastar {%
Arcangeli:2018aa}%
\begin{APACrefauthors}%
Arcangeli, J.%
, Desert, J.%
, Line, M.%
\BCBL {}\ \BBA {} et al.%
\end{APACrefauthors}%
\unskip\
\newblock
\APACrefYearMonthDay{2018}{}{}.
\newblock

\newblock
\APACjournalVolNumPages{The Astrophysical Journal Letters}{855}{}{L30}.
\PrintBackRefs{\CurrentBib}

\bibitem [\protect \citeauthoryear {%
Arcangeli%
, D\'{e}sert%
, Parmentier%
\BCBL {}\ \BBA {} {et al.}%
}{%
Arcangeli%
\ \protect \BOthers {.}}{%
{\protect \APACyear {2019}}%
}]{%
Arcangeli:2019aa}
\APACinsertmetastar {%
Arcangeli:2019aa}%
\begin{APACrefauthors}%
Arcangeli, J.%
, D\'{e}sert, J.%
, Parmentier, V.%
\BCBL {}\ \BBA {} {et al.}%
\end{APACrefauthors}%
\unskip\
\newblock
\APACrefYearMonthDay{2019}{}{}.
\newblock
{\BBOQ}\APACrefatitle {Climate of an ultra hot Jupiter. Spectroscopic phase
  curve of WASP-18b with HST/WFC3} {Climate of an ultra hot jupiter.
  spectroscopic phase curve of wasp-18b with hst/wfc3}.{\BBCQ}
\newblock
\APACjournalVolNumPages{Astronomy {\&} Astrophysics}{625}{}{A136}.
\PrintBackRefs{\CurrentBib}

\bibitem [\protect \citeauthoryear {%
Armstrong%
\ \protect \BOthers {.}}{%
Armstrong%
\ \protect \BOthers {.}}{%
{\protect \APACyear {2016}}%
}]{%
Mooij2016}
\APACinsertmetastar {%
Mooij2016}%
\begin{APACrefauthors}%
Armstrong, D.%
, de Mooij, E.%
, Barstow, J.%
, Osborn, H\BPBI P.%
, Blake, J.%
\BCBL {}\ \BBA {} Fereshteh~Saniee, N.%
\end{APACrefauthors}%
\unskip\
\newblock
\APACrefYearMonthDay{2016}{}{}.
\newblock
{\BBOQ}\APACrefatitle {{Variability in the atmosphere of the hot giant planet
  HAT-P-7 b}} {{Variability in the atmosphere of the hot giant planet HAT-P-7
  b}}.{\BBCQ}
\newblock
\APACjournalVolNumPages{Nature Astronomy}{1}{}{4}.
\PrintBackRefs{\CurrentBib}

\bibitem [\protect \citeauthoryear {%
{Arras}%
\ \BBA {} {Bildsten}%
}{%
{Arras}%
\ \BBA {} {Bildsten}%
}{%
{\protect \APACyear {2006}}%
}]{%
Arras06}
\APACinsertmetastar {%
Arras06}%
\begin{APACrefauthors}%
{Arras}, P.%
\BCBT {}\ \BBA {} {Bildsten}, L.%
\end{APACrefauthors}%
\unskip\
\newblock
\APACrefYearMonthDay{2006}{{\APACmonth{10}}}{}.
\newblock
{\BBOQ}\APACrefatitle {{Thermal Structure and Radius Evolution of Irradiated
  Gas Giant Planets}} {{Thermal Structure and Radius Evolution of Irradiated
  Gas Giant Planets}}.{\BBCQ}
\newblock
\APACjournalVolNumPages{\apj}{650}{}{394-407}.
\newblock
\begin{APACrefDOI} \doi{10.1086/506011} \end{APACrefDOI}
\PrintBackRefs{\CurrentBib}

\bibitem [\protect \citeauthoryear {%
{Arras}%
\ \BBA {} {Socrates}%
}{%
{Arras}%
\ \BBA {} {Socrates}%
}{%
{\protect \APACyear {2010}}%
}]{%
arra10}
\APACinsertmetastar {%
arra10}%
\begin{APACrefauthors}%
{Arras}, P.%
\BCBT {}\ \BBA {} {Socrates}, A.%
\end{APACrefauthors}%
\unskip\
\newblock
\APACrefYearMonthDay{2010}{{\APACmonth{05}}}{}.
\newblock
{\BBOQ}\APACrefatitle {{Thermal Tides in Fluid Extrasolar Planets}} {{Thermal
  Tides in Fluid Extrasolar Planets}}.{\BBCQ}
\newblock
\APACjournalVolNumPages{\apj}{714}{}{1-12}.
\newblock
\begin{APACrefDOI} \doi{10.1088/0004-637X/714/1/1} \end{APACrefDOI}
\PrintBackRefs{\CurrentBib}

\bibitem [\protect \citeauthoryear {%
Bakos%
, Noyes%
, Latham%
\BCBL {}\ \BBA {} {et al.}%
}{%
Bakos%
\ \protect \BOthers {.}}{%
{\protect \APACyear {2005}}%
}]{%
Bakos:2005aa}
\APACinsertmetastar {%
Bakos:2005aa}%
\begin{APACrefauthors}%
Bakos, G.%
, Noyes, R.%
, Latham, D.%
\BCBL {}\ \BBA {} {et al.}%
\end{APACrefauthors}%
\unskip\
\newblock
\APACrefYearMonthDay{2005}{}{}.
\newblock
{\BBOQ}\APACrefatitle {Extrasolar planet search with the HAT network}
  {Extrasolar planet search with the hat network}.{\BBCQ}
\newblock
\BIn{} L.~Arnold, F.~Bouchy\BCBL {}\ \BBA {} C.~Moutou\ (\BEDS), \APACrefbtitle
  {Tenth Anniversary of 51 Peg-b: Status of and prospects for hot Jupiter
  studies} {Tenth anniversary of 51 peg-b: Status of and prospects for hot
  jupiter studies}\ (\BPG~184).
\PrintBackRefs{\CurrentBib}

\bibitem [\protect \citeauthoryear {%
{Baraffe}%
, {Chabrier}%
\BCBL {}\ \BBA {} {Barman}%
}{%
{Baraffe}%
\ \protect \BOthers {.}}{%
{\protect \APACyear {2010}}%
}]{%
Baraffe10}
\APACinsertmetastar {%
Baraffe10}%
\begin{APACrefauthors}%
{Baraffe}, I.%
, {Chabrier}, G.%
\BCBL {}\ \BBA {} {Barman}, T.%
\end{APACrefauthors}%
\unskip\
\newblock
\APACrefYearMonthDay{2010}{{\APACmonth{01}}}{}.
\newblock
{\BBOQ}\APACrefatitle {{The physical properties of extra-solar planets}} {{The
  physical properties of extra-solar planets}}.{\BBCQ}
\newblock
\APACjournalVolNumPages{Reports on Progress in Physics}{73}{1}{016901-+}.
\newblock
\begin{APACrefDOI} \doi{10.1088/0034-4885/73/1/016901} \end{APACrefDOI}
\PrintBackRefs{\CurrentBib}

\bibitem [\protect \citeauthoryear {%
Baraffe%
, Chabrier%
, Fortney%
\BCBL {}\ \BBA {} Sotin%
}{%
Baraffe%
\ \protect \BOthers {.}}{%
{\protect \APACyear {2014}}%
}]{%
Baraffe:2014}
\APACinsertmetastar {%
Baraffe:2014}%
\begin{APACrefauthors}%
Baraffe, I.%
, Chabrier, G.%
, Fortney, J.%
\BCBL {}\ \BBA {} Sotin, C.%
\end{APACrefauthors}%
\unskip\
\newblock
\APACrefYearMonthDay{2014}{}{}.
\newblock
{\BBOQ}\APACrefatitle {Protostars and Planets VI} {Protostars and planets
  vi}.{\BBCQ}
\newblock
\BIn{} H.~Beuther, R.~Klessen, C.~Dullemond\BCBL {}\ \BBA {} T.~Henning\
  (\BEDS), (\BCHAP\ Planetary internal structures).
\newblock
\APACaddressPublisher{Tucson, AZ}{University of Arizona Press}.
\PrintBackRefs{\CurrentBib}

\bibitem [\protect \citeauthoryear {%
{Barstow}%
\ \BBA {} {Heng}%
}{%
{Barstow}%
\ \BBA {} {Heng}%
}{%
{\protect \APACyear {2020}}%
}]{%
Barstow20}
\APACinsertmetastar {%
Barstow20}%
\begin{APACrefauthors}%
{Barstow}, J\BPBI K.%
\BCBT {}\ \BBA {} {Heng}, K.%
\end{APACrefauthors}%
\unskip\
\newblock
\APACrefYearMonthDay{2020}{{\APACmonth{06}}}{}.
\newblock
{\BBOQ}\APACrefatitle {{Outstanding Challenges of Exoplanet Atmospheric
  Retrievals}} {{Outstanding Challenges of Exoplanet Atmospheric
  Retrievals}}.{\BBCQ}
\newblock
\APACjournalVolNumPages{\ssr}{216}{5}{82}.
\newblock
\begin{APACrefDOI} \doi{10.1007/s11214-020-00666-x} \end{APACrefDOI}
\PrintBackRefs{\CurrentBib}

\bibitem [\protect \citeauthoryear {%
{Baruteau}%
\ \protect \BOthers {.}}{%
{Baruteau}%
\ \protect \BOthers {.}}{%
{\protect \APACyear {2014}}%
}]{%
baru14}
\APACinsertmetastar {%
baru14}%
\begin{APACrefauthors}%
{Baruteau}, C.%
, {Crida}, A.%
, {Paardekooper}, S\BHBI J.%
, {Masset}, F.%
, {Guilet}, J.%
, {Bitsch}, B.%
\BDBL {}{Papaloizou}, J.%
\end{APACrefauthors}%
\unskip\
\newblock
\APACrefYearMonthDay{2014}{}{}.
\newblock
{\BBOQ}\APACrefatitle {{Planet-Disk Interactions and Early Evolution of
  Planetary Systems}} {{Planet-Disk Interactions and Early Evolution of
  Planetary Systems}}.{\BBCQ}
\newblock
\APACjournalVolNumPages{Protostars and Planets VI}{}{}{667-689}.
\newblock
\begin{APACrefDOI} \doi{10.2458/azu_uapress_9780816531240-ch029}
  \end{APACrefDOI}
\PrintBackRefs{\CurrentBib}

\bibitem [\protect \citeauthoryear {%
{Batygin}%
}{%
{Batygin}%
}{%
{\protect \APACyear {2012}}%
}]{%
baty12}
\APACinsertmetastar {%
baty12}%
\begin{APACrefauthors}%
{Batygin}, K.%
\end{APACrefauthors}%
\unskip\
\newblock
\APACrefYearMonthDay{2012}{{\APACmonth{11}}}{}.
\newblock
{\BBOQ}\APACrefatitle {{A primordial origin for misalignments between stellar
  spin axes and planetary orbits}} {{A primordial origin for misalignments
  between stellar spin axes and planetary orbits}}.{\BBCQ}
\newblock
\APACjournalVolNumPages{\nat}{491}{}{418-420}.
\newblock
\begin{APACrefDOI} \doi{10.1038/nature11560} \end{APACrefDOI}
\PrintBackRefs{\CurrentBib}

\bibitem [\protect \citeauthoryear {%
{Batygin}%
, {Bodenheimer}%
\BCBL {}\ \BBA {} {Laughlin}%
}{%
{Batygin}%
\ \protect \BOthers {.}}{%
{\protect \APACyear {2016}}%
}]{%
baty15}
\APACinsertmetastar {%
baty15}%
\begin{APACrefauthors}%
{Batygin}, K.%
, {Bodenheimer}, P\BPBI H.%
\BCBL {}\ \BBA {} {Laughlin}, G\BPBI P.%
\end{APACrefauthors}%
\unskip\
\newblock
\APACrefYearMonthDay{2016}{{\APACmonth{10}}}{}.
\newblock
{\BBOQ}\APACrefatitle {{In Situ Formation and Dynamical Evolution of Hot
  Jupiter Systems}} {{In Situ Formation and Dynamical Evolution of Hot Jupiter
  Systems}}.{\BBCQ}
\newblock
\APACjournalVolNumPages{\apj}{829}{}{114}.
\newblock
\begin{APACrefDOI} \doi{10.3847/0004-637X/829/2/114} \end{APACrefDOI}
\PrintBackRefs{\CurrentBib}

\bibitem [\protect \citeauthoryear {%
Batygin%
, Stanley%
\BCBL {}\ \BBA {} Stevenson%
}{%
Batygin%
\ \protect \BOthers {.}}{%
{\protect \APACyear {2013}}%
}]{%
batygin_2013}
\APACinsertmetastar {%
batygin_2013}%
\begin{APACrefauthors}%
Batygin, K.%
, Stanley, S.%
\BCBL {}\ \BBA {} Stevenson, D.%
\end{APACrefauthors}%
\unskip\
\newblock
\APACrefYearMonthDay{2013}{}{}.
\newblock
{\BBOQ}\APACrefatitle {Magnetically controlled circulation on hot extrasolar
  planets} {Magnetically controlled circulation on hot extrasolar
  planets}.{\BBCQ}
\newblock
\APACjournalVolNumPages{The Astrophysical Journal}{776}{}{53}.
\PrintBackRefs{\CurrentBib}

\bibitem [\protect \citeauthoryear {%
Batygin%
, Stevenson%
\BCBL {}\ \BBA {} Bodenheimer%
}{%
Batygin%
\ \protect \BOthers {.}}{%
{\protect \APACyear {2011}}%
}]{%
Batygin_2011}
\APACinsertmetastar {%
Batygin_2011}%
\begin{APACrefauthors}%
Batygin, K.%
, Stevenson, D.%
\BCBL {}\ \BBA {} Bodenheimer, P.%
\end{APACrefauthors}%
\unskip\
\newblock
\APACrefYearMonthDay{2011}{}{}.
\newblock
{\BBOQ}\APACrefatitle {Evolution of {O}hmically heated hot {J}upiters}
  {Evolution of {O}hmically heated hot {J}upiters}.{\BBCQ}
\newblock
\APACjournalVolNumPages{The Astrophysical Journal}{738}{}{1}.
\PrintBackRefs{\CurrentBib}

\bibitem [\protect \citeauthoryear {%
{Batygin}%
\ \BBA {} {Stevenson}%
}{%
{Batygin}%
\ \BBA {} {Stevenson}%
}{%
{\protect \APACyear {2010}}%
}]{%
baty10}
\APACinsertmetastar {%
baty10}%
\begin{APACrefauthors}%
{Batygin}, K.%
\BCBT {}\ \BBA {} {Stevenson}, D\BPBI J.%
\end{APACrefauthors}%
\unskip\
\newblock
\APACrefYearMonthDay{2010}{{\APACmonth{05}}}{}.
\newblock
{\BBOQ}\APACrefatitle {{Inflating Hot Jupiters with Ohmic Dissipation}}
  {{Inflating Hot Jupiters with Ohmic Dissipation}}.{\BBCQ}
\newblock
\APACjournalVolNumPages{\apjl}{714}{}{L238-L243}.
\newblock
\begin{APACrefDOI} \doi{10.1088/2041-8205/714/2/L238} \end{APACrefDOI}
\PrintBackRefs{\CurrentBib}

\bibitem [\protect \citeauthoryear {%
Baxter%
, D{\'{e}}sert%
, Parmentier%
\BCBL {}\ \BBA {} {et al.}%
}{%
Baxter%
\ \protect \BOthers {.}}{%
{\protect \APACyear {2020}}%
}]{%
Baxter:2020aa}
\APACinsertmetastar {%
Baxter:2020aa}%
\begin{APACrefauthors}%
Baxter, C.%
, D{\'{e}}sert, J.%
, Parmentier, V.%
\BCBL {}\ \BBA {} {et al.}%
\end{APACrefauthors}%
\unskip\
\newblock
\APACrefYearMonthDay{2020}{}{}.
\newblock
{\BBOQ}\APACrefatitle {A transition between the hot and the ultra-hot Jupiter
  atmospheres} {A transition between the hot and the ultra-hot jupiter
  atmospheres}.{\BBCQ}
\newblock
\APACjournalVolNumPages{ArXiv e-prints:2007.15287}{}{}{}.
\PrintBackRefs{\CurrentBib}

\bibitem [\protect \citeauthoryear {%
Bean%
, Stevenson%
, Batalha%
\BCBL {}\ \BBA {} {et al.}%
}{%
Bean%
\ \protect \BOthers {.}}{%
{\protect \APACyear {2018}}%
}]{%
Bean:2018aa}
\APACinsertmetastar {%
Bean:2018aa}%
\begin{APACrefauthors}%
Bean, J.%
, Stevenson, K.%
, Batalha, N.%
\BCBL {}\ \BBA {} {et al.}%
\end{APACrefauthors}%
\unskip\
\newblock
\APACrefYearMonthDay{2018}{}{}.
\newblock
{\BBOQ}\APACrefatitle {The Transiting Exoplanet Community Early Release Science
  Program for JWST} {The transiting exoplanet community early release science
  program for jwst}.{\BBCQ}
\newblock
\APACjournalVolNumPages{Publications of the Astronomical Society of the
  Pacific}{130}{}{114402}.
\PrintBackRefs{\CurrentBib}

\bibitem [\protect \citeauthoryear {%
Beatty%
, Madhusudhan%
, Tsiaras%
\BCBL {}\ \BBA {} et al.%
}{%
Beatty%
\ \protect \BOthers {.}}{%
{\protect \APACyear {2017}}%
}]{%
Beatty:2017aa}
\APACinsertmetastar {%
Beatty:2017aa}%
\begin{APACrefauthors}%
Beatty, T.%
, Madhusudhan, N.%
, Tsiaras, A.%
\BCBL {}\ \BBA {} et al.%
\end{APACrefauthors}%
\unskip\
\newblock
\APACrefYearMonthDay{2017}{}{}.
\newblock

\newblock
\APACjournalVolNumPages{The Astronomical Journal}{154}{}{158}.
\PrintBackRefs{\CurrentBib}

\bibitem [\protect \citeauthoryear {%
{Becker}%
, {Vanderburg}%
, {Adams}%
, {Rappaport}%
\BCBL {}\ \BBA {} {Schwengeler}%
}{%
{Becker}%
\ \protect \BOthers {.}}{%
{\protect \APACyear {2015}}%
}]{%
beck15}
\APACinsertmetastar {%
beck15}%
\begin{APACrefauthors}%
{Becker}, J\BPBI C.%
, {Vanderburg}, A.%
, {Adams}, F\BPBI C.%
, {Rappaport}, S\BPBI A.%
\BCBL {}\ \BBA {} {Schwengeler}, H\BPBI M.%
\end{APACrefauthors}%
\unskip\
\newblock
\APACrefYearMonthDay{2015}{{\APACmonth{10}}}{}.
\newblock
{\BBOQ}\APACrefatitle {{WASP-47: A Hot Jupiter System with Two Additional
  Planets Discovered by K2}} {{WASP-47: A Hot Jupiter System with Two
  Additional Planets Discovered by K2}}.{\BBCQ}
\newblock
\APACjournalVolNumPages{\apjl}{812}{}{L18}.
\newblock
\begin{APACrefDOI} \doi{10.1088/2041-8205/812/2/L18} \end{APACrefDOI}
\PrintBackRefs{\CurrentBib}

\bibitem [\protect \citeauthoryear {%
Bell%
\ \BBA {} Cowan%
}{%
Bell%
\ \BBA {} Cowan%
}{%
{\protect \APACyear {2018}}%
}]{%
Bell:2018aa}
\APACinsertmetastar {%
Bell:2018aa}%
\begin{APACrefauthors}%
Bell, T.%
\BCBT {}\ \BBA {} Cowan, N.%
\end{APACrefauthors}%
\unskip\
\newblock
\APACrefYearMonthDay{2018}{}{}.
\newblock
{\BBOQ}\APACrefatitle {Increased Heat Transport in Ultra-hot Jupiter
  Atmospheres through H2 Dissociation and Recombination} {Increased heat
  transport in ultra-hot jupiter atmospheres through h2 dissociation and
  recombination}.{\BBCQ}
\newblock
\APACjournalVolNumPages{The Astrophysical Journal Letters}{857}{}{L20}.
\PrintBackRefs{\CurrentBib}

\bibitem [\protect \citeauthoryear {%
Bell%
, Dang%
, Cowan%
\BCBL {}\ \BBA {} {et al.}%
}{%
Bell%
\ \protect \BOthers {.}}{%
{\protect \APACyear {2020}}%
}]{%
Bell:2020aa}
\APACinsertmetastar {%
Bell:2020aa}%
\begin{APACrefauthors}%
Bell, T.%
, Dang, L.%
, Cowan, N.%
\BCBL {}\ \BBA {} {et al.}%
\end{APACrefauthors}%
\unskip\
\newblock
\APACrefYearMonthDay{2020}{}{}.
\newblock

\newblock
\APACjournalVolNumPages{arXiv e-prints:2010.00687}{}{}{}.
\PrintBackRefs{\CurrentBib}

\bibitem [\protect \citeauthoryear {%
Bell%
\ \protect \BOthers {.}}{%
Bell%
\ \protect \BOthers {.}}{%
{\protect \APACyear {2019}}%
}]{%
Bell:2019aa}
\APACinsertmetastar {%
Bell:2019aa}%
\begin{APACrefauthors}%
Bell, T.%
, Zhang, M.%
, Cubillos, P.%
, Dang, L.%
, Fossati, L.%
, Todorov, K.%
\BDBL {}Line, M.%
\end{APACrefauthors}%
\unskip\
\newblock
\APACrefYearMonthDay{2019}{}{}.
\newblock
{\BBOQ}\APACrefatitle {Mass loss from the exoplanet WASP-12b inferred from
  Spitzer phase curves} {Mass loss from the exoplanet wasp-12b inferred from
  spitzer phase curves}.{\BBCQ}
\newblock
\APACjournalVolNumPages{Monthly Notices of the Royal Astronomical
  Society}{489}{}{1995}.
\PrintBackRefs{\CurrentBib}

\bibitem [\protect \citeauthoryear {%
Birkby%
, de Kok%
, Brogi%
\BCBL {}\ \BBA {} {et al.}%
}{%
Birkby%
\ \protect \BOthers {.}}{%
{\protect \APACyear {2013}}%
}]{%
Birkby:2013aa}
\APACinsertmetastar {%
Birkby:2013aa}%
\begin{APACrefauthors}%
Birkby, J.%
, de Kok, R.%
, Brogi, M.%
\BCBL {}\ \BBA {} {et al.}%
\end{APACrefauthors}%
\unskip\
\newblock
\APACrefYearMonthDay{2013}{}{}.
\newblock
{\BBOQ}\APACrefatitle {Detection of water absorption in the day side atmosphere
  of HD 189733 b using ground-based high-resolution spectroscopy at 3.2 $\mu$m}
  {Detection of water absorption in the day side atmosphere of hd 189733 b
  using ground-based high-resolution spectroscopy at 3.2 $\mu$m}.{\BBCQ}
\newblock
\APACjournalVolNumPages{Monthly Notices of the Royal Astronomical
  Society}{436}{}{L35}.
\PrintBackRefs{\CurrentBib}

\bibitem [\protect \citeauthoryear {%
{Birkby}%
}{%
{Birkby}%
}{%
{\protect \APACyear {2018}}%
}]{%
Birkby18}
\APACinsertmetastar {%
Birkby18}%
\begin{APACrefauthors}%
{Birkby}, J\BPBI L.%
\end{APACrefauthors}%
\unskip\
\newblock
\APACrefYearMonthDay{2018}{{\APACmonth{06}}}{}.
\newblock
{\BBOQ}\APACrefatitle {{Exoplanet Atmospheres at High Spectral Resolution}}
  {{Exoplanet Atmospheres at High Spectral Resolution}}.{\BBCQ}
\newblock
\APACjournalVolNumPages{arXiv e-prints}{}{}{arXiv:1806.04617}.
\PrintBackRefs{\CurrentBib}

\bibitem [\protect \citeauthoryear {%
{Birkby}%
, {de Kok}%
, {Brogi}%
, {Schwarz}%
\BCBL {}\ \BBA {} {Snellen}%
}{%
{Birkby}%
\ \protect \BOthers {.}}{%
{\protect \APACyear {2017}}%
}]{%
Birkby17}
\APACinsertmetastar {%
Birkby17}%
\begin{APACrefauthors}%
{Birkby}, J\BPBI L.%
, {de Kok}, R\BPBI J.%
, {Brogi}, M.%
, {Schwarz}, H.%
\BCBL {}\ \BBA {} {Snellen}, I\BPBI A\BPBI G.%
\end{APACrefauthors}%
\unskip\
\newblock
\APACrefYearMonthDay{2017}{{\APACmonth{03}}}{}.
\newblock
{\BBOQ}\APACrefatitle {{Discovery of Water at High Spectral Resolution in the
  Atmosphere of 51 Peg b}} {{Discovery of Water at High Spectral Resolution in
  the Atmosphere of 51 Peg b}}.{\BBCQ}
\newblock
\APACjournalVolNumPages{\aj}{153}{}{138}.
\newblock
\begin{APACrefDOI} \doi{10.3847/1538-3881/aa5c87} \end{APACrefDOI}
\PrintBackRefs{\CurrentBib}

\bibitem [\protect \citeauthoryear {%
{Blecic}%
, {Harrington}%
\BCBL {}\ \BBA {} {Bowman}%
}{%
{Blecic}%
\ \protect \BOthers {.}}{%
{\protect \APACyear {2016}}%
}]{%
Blecic16}
\APACinsertmetastar {%
Blecic16}%
\begin{APACrefauthors}%
{Blecic}, J.%
, {Harrington}, J.%
\BCBL {}\ \BBA {} {Bowman}, M\BPBI O.%
\end{APACrefauthors}%
\unskip\
\newblock
\APACrefYearMonthDay{2016}{{\APACmonth{07}}}{}.
\newblock
{\BBOQ}\APACrefatitle {{TEA: A Code Calculating Thermochemical Equilibrium
  Abundances}} {{TEA: A Code Calculating Thermochemical Equilibrium
  Abundances}}.{\BBCQ}
\newblock
\APACjournalVolNumPages{\apjs}{225}{1}{4}.
\newblock
\begin{APACrefDOI} \doi{10.3847/0067-0049/225/1/4} \end{APACrefDOI}
\PrintBackRefs{\CurrentBib}

\bibitem [\protect \citeauthoryear {%
Bodenheimer%
, Laughlin%
\BCBL {}\ \BBA {} Lin%
}{%
Bodenheimer%
\ \protect \BOthers {.}}{%
{\protect \APACyear {2003}}%
}]{%
Bodenheimer:2003}
\APACinsertmetastar {%
Bodenheimer:2003}%
\begin{APACrefauthors}%
Bodenheimer, P.%
, Laughlin, G.%
\BCBL {}\ \BBA {} Lin, D.%
\end{APACrefauthors}%
\unskip\
\newblock
\APACrefYearMonthDay{2003}{}{}.
\newblock
{\BBOQ}\APACrefatitle {On the radii of extrasolar giant planets} {On the radii
  of extrasolar giant planets}.{\BBCQ}
\newblock
\APACjournalVolNumPages{The Astrophysical Journal}{592}{}{555-563}.
\PrintBackRefs{\CurrentBib}

\bibitem [\protect \citeauthoryear {%
{Bodenheimer}%
, {Lin}%
\BCBL {}\ \BBA {} {Mardling}%
}{%
{Bodenheimer}%
\ \protect \BOthers {.}}{%
{\protect \APACyear {2001}}%
}]{%
Bodenheimer01}
\APACinsertmetastar {%
Bodenheimer01}%
\begin{APACrefauthors}%
{Bodenheimer}, P.%
, {Lin}, D\BPBI N\BPBI C.%
\BCBL {}\ \BBA {} {Mardling}, R\BPBI A.%
\end{APACrefauthors}%
\unskip\
\newblock
\APACrefYearMonthDay{2001}{{\APACmonth{02}}}{}.
\newblock
{\BBOQ}\APACrefatitle {{On the Tidal Inflation of Short-Period Extrasolar
  Planets}} {{On the Tidal Inflation of Short-Period Extrasolar
  Planets}}.{\BBCQ}
\newblock
\APACjournalVolNumPages{\apj}{548}{}{466-472}.
\PrintBackRefs{\CurrentBib}

\bibitem [\protect \citeauthoryear {%
{Boley}%
, {Granados Contreras}%
\BCBL {}\ \BBA {} {Gladman}%
}{%
{Boley}%
\ \protect \BOthers {.}}{%
{\protect \APACyear {2016}}%
}]{%
bole16}
\APACinsertmetastar {%
bole16}%
\begin{APACrefauthors}%
{Boley}, A\BPBI C.%
, {Granados Contreras}, A\BPBI P.%
\BCBL {}\ \BBA {} {Gladman}, B.%
\end{APACrefauthors}%
\unskip\
\newblock
\APACrefYearMonthDay{2016}{{\APACmonth{02}}}{}.
\newblock
{\BBOQ}\APACrefatitle {{The In Situ Formation of Giant Planets at Short Orbital
  Periods}} {{The In Situ Formation of Giant Planets at Short Orbital
  Periods}}.{\BBCQ}
\newblock
\APACjournalVolNumPages{\apjl}{817}{}{L17}.
\newblock
\begin{APACrefDOI} \doi{10.3847/2041-8205/817/2/L17} \end{APACrefDOI}
\PrintBackRefs{\CurrentBib}

\bibitem [\protect \citeauthoryear {%
{Boss}%
}{%
{Boss}%
}{%
{\protect \APACyear {1997}}%
}]{%
boss97}
\APACinsertmetastar {%
boss97}%
\begin{APACrefauthors}%
{Boss}, A\BPBI P.%
\end{APACrefauthors}%
\unskip\
\newblock
\APACrefYearMonthDay{1997}{}{}.
\newblock
{\BBOQ}\APACrefatitle {{Giant planet formation by gravitational instability.}}
  {{Giant planet formation by gravitational instability.}}{\BBCQ}
\newblock
\APACjournalVolNumPages{Science}{276}{}{1836-1839}.
\newblock
\begin{APACrefDOI} \doi{10.1126/science.276.5320.1836} \end{APACrefDOI}
\PrintBackRefs{\CurrentBib}

\bibitem [\protect \citeauthoryear {%
Brogi%
\ \protect \BOthers {.}}{%
Brogi%
\ \protect \BOthers {.}}{%
{\protect \APACyear {2016}}%
}]{%
Brogi:2015}
\APACinsertmetastar {%
Brogi:2015}%
\begin{APACrefauthors}%
Brogi, M.%
, de Kok, R.%
, Albrecht, S.%
, Snellen, I.%
, Birkby, J.%
\BCBL {}\ \BBA {} Schwarz, H.%
\end{APACrefauthors}%
\unskip\
\newblock
\APACrefYearMonthDay{2016}{}{}.
\newblock
{\BBOQ}\APACrefatitle {Rotation and winds of exoplanet {HD} 189733b measured
  with high-dispersion tramission spectrosocpy} {Rotation and winds of
  exoplanet {HD} 189733b measured with high-dispersion tramission
  spectrosocpy}.{\BBCQ}
\newblock
\APACjournalVolNumPages{The Astrophysical Journal}{817}{}{106}.
\PrintBackRefs{\CurrentBib}

\bibitem [\protect \citeauthoryear {%
Brogi%
, de Kok%
, Birkby%
\BCBL {}\ \BBA {} {et al.}%
}{%
Brogi%
\ \protect \BOthers {.}}{%
{\protect \APACyear {2014}}%
}]{%
Brogi:2014aa}
\APACinsertmetastar {%
Brogi:2014aa}%
\begin{APACrefauthors}%
Brogi, M.%
, de Kok, R.%
, Birkby, J.%
\BCBL {}\ \BBA {} {et al.}%
\end{APACrefauthors}%
\unskip\
\newblock
\APACrefYearMonthDay{2014}{}{}.
\newblock
{\BBOQ}\APACrefatitle {Carbon monoxide and water vapor in the atmosphere of the
  non-transiting exoplanet HD 179949 b} {Carbon monoxide and water vapor in the
  atmosphere of the non-transiting exoplanet hd 179949 b}.{\BBCQ}
\newblock
\APACjournalVolNumPages{Astronomy {\&} Astrophysics}{565}{}{A124}.
\PrintBackRefs{\CurrentBib}

\bibitem [\protect \citeauthoryear {%
{Brogi}%
\ \BBA {} {Line}%
}{%
{Brogi}%
\ \BBA {} {Line}%
}{%
{\protect \APACyear {2019}}%
}]{%
Brogi19}
\APACinsertmetastar {%
Brogi19}%
\begin{APACrefauthors}%
{Brogi}, M.%
\BCBT {}\ \BBA {} {Line}, M\BPBI R.%
\end{APACrefauthors}%
\unskip\
\newblock
\APACrefYearMonthDay{2019}{{\APACmonth{03}}}{}.
\newblock
{\BBOQ}\APACrefatitle {{Retrieving Temperatures and Abundances of Exoplanet
  Atmospheres with High-resolution Cross-correlation Spectroscopy}}
  {{Retrieving Temperatures and Abundances of Exoplanet Atmospheres with
  High-resolution Cross-correlation Spectroscopy}}.{\BBCQ}
\newblock
\APACjournalVolNumPages{\aj}{157}{3}{114}.
\newblock
\begin{APACrefDOI} \doi{10.3847/1538-3881/aaffd3} \end{APACrefDOI}
\PrintBackRefs{\CurrentBib}

\bibitem [\protect \citeauthoryear {%
{Brown}%
}{%
{Brown}%
}{%
{\protect \APACyear {2001}}%
}]{%
Brown01}
\APACinsertmetastar {%
Brown01}%
\begin{APACrefauthors}%
{Brown}, T\BPBI M.%
\end{APACrefauthors}%
\unskip\
\newblock
\APACrefYearMonthDay{2001}{{\APACmonth{06}}}{}.
\newblock
{\BBOQ}\APACrefatitle {{Transmission Spectra as Diagnostics of Extrasolar Giant
  Planet Atmospheres}} {{Transmission Spectra as Diagnostics of Extrasolar
  Giant Planet Atmospheres}}.{\BBCQ}
\newblock
\APACjournalVolNumPages{\apj}{553}{}{1006-1026}.
\newblock
\begin{APACrefURL}
  \url{http://adsabs.harvard.edu/cgi-bin/nph-bib_query?bibcode=2001ApJ...553.1006B&db_key=AST}
  \end{APACrefURL}
\PrintBackRefs{\CurrentBib}

\bibitem [\protect \citeauthoryear {%
{Bryan}%
\ \protect \BOthers {.}}{%
{Bryan}%
\ \protect \BOthers {.}}{%
{\protect \APACyear {2016}}%
}]{%
brya16}
\APACinsertmetastar {%
brya16}%
\begin{APACrefauthors}%
{Bryan}, M\BPBI L.%
, {Knutson}, H\BPBI A.%
, {Howard}, A\BPBI W.%
, {Ngo}, H.%
, {Batygin}, K.%
, {Crepp}, J\BPBI R.%
\BDBL {}{Wright}, J\BPBI T.%
\end{APACrefauthors}%
\unskip\
\newblock
\APACrefYearMonthDay{2016}{{\APACmonth{04}}}{}.
\newblock
{\BBOQ}\APACrefatitle {{Statistics of Long Period Gas Giant Planets in Known
  Planetary Systems}} {{Statistics of Long Period Gas Giant Planets in Known
  Planetary Systems}}.{\BBCQ}
\newblock
\APACjournalVolNumPages{\apj}{821}{}{89}.
\newblock
\begin{APACrefDOI} \doi{10.3847/0004-637X/821/2/89} \end{APACrefDOI}
\PrintBackRefs{\CurrentBib}

\bibitem [\protect \citeauthoryear {%
{Burrows}%
, {Hubeny}%
, {Budaj}%
\BCBL {}\ \BBA {} {Hubbard}%
}{%
{Burrows}%
\ \protect \BOthers {.}}{%
{\protect \APACyear {2007}}%
}]{%
Burrows07}
\APACinsertmetastar {%
Burrows07}%
\begin{APACrefauthors}%
{Burrows}, A.%
, {Hubeny}, I.%
, {Budaj}, J.%
\BCBL {}\ \BBA {} {Hubbard}, W\BPBI B.%
\end{APACrefauthors}%
\unskip\
\newblock
\APACrefYearMonthDay{2007}{{\APACmonth{05}}}{}.
\newblock
{\BBOQ}\APACrefatitle {{Possible Solutions to the Radius Anomalies of
  Transiting Giant Planets}} {{Possible Solutions to the Radius Anomalies of
  Transiting Giant Planets}}.{\BBCQ}
\newblock
\APACjournalVolNumPages{\apj}{661}{}{502-514}.
\newblock
\begin{APACrefDOI} \doi{10.1086/514326} \end{APACrefDOI}
\PrintBackRefs{\CurrentBib}

\bibitem [\protect \citeauthoryear {%
{Burrows}%
\ \BBA {} {Lunine}%
}{%
{Burrows}%
\ \BBA {} {Lunine}%
}{%
{\protect \APACyear {1995}}%
}]{%
Burrows95}
\APACinsertmetastar {%
Burrows95}%
\begin{APACrefauthors}%
{Burrows}, A.%
\BCBT {}\ \BBA {} {Lunine}, J.%
\end{APACrefauthors}%
\unskip\
\newblock
\APACrefYearMonthDay{1995}{{\APACmonth{11}}}{}.
\newblock
{\BBOQ}\APACrefatitle {{Astronomical questions of origin and survival}}
  {{Astronomical questions of origin and survival}}.{\BBCQ}
\newblock
\APACjournalVolNumPages{\nat}{378}{6555}{333}.
\newblock
\begin{APACrefDOI} \doi{10.1038/378333a0} \end{APACrefDOI}
\PrintBackRefs{\CurrentBib}

\bibitem [\protect \citeauthoryear {%
{Butler}%
, {Marcy}%
, {Williams}%
, {Hauser}%
\BCBL {}\ \BBA {} {Shirts}%
}{%
{Butler}%
\ \protect \BOthers {.}}{%
{\protect \APACyear {1997}}%
}]{%
Butler97}
\APACinsertmetastar {%
Butler97}%
\begin{APACrefauthors}%
{Butler}, R\BPBI P.%
, {Marcy}, G\BPBI W.%
, {Williams}, E.%
, {Hauser}, H.%
\BCBL {}\ \BBA {} {Shirts}, P.%
\end{APACrefauthors}%
\unskip\
\newblock
\APACrefYearMonthDay{1997}{{\APACmonth{01}}}{}.
\newblock
{\BBOQ}\APACrefatitle {{Three New ''51 Pegasi--Type'' Planets}} {{Three New
  ''51 Pegasi--Type'' Planets}}.{\BBCQ}
\newblock
\APACjournalVolNumPages{\apjl}{474}{}{L115+}.
\newblock
\begin{APACrefDOI} \doi{10.1086/310444} \end{APACrefDOI}
\PrintBackRefs{\CurrentBib}

\bibitem [\protect \citeauthoryear {%
Cabot%
, Madhusudhan%
, Welbanks%
\BCBL {}\ \BBA {} {et al.}%
}{%
Cabot%
\ \protect \BOthers {.}}{%
{\protect \APACyear {2020}}%
}]{%
Cabot:2020aa}
\APACinsertmetastar {%
Cabot:2020aa}%
\begin{APACrefauthors}%
Cabot, H.%
, Madhusudhan, N.%
, Welbanks, L.%
\BCBL {}\ \BBA {} {et al.}%
\end{APACrefauthors}%
\unskip\
\newblock
\APACrefYearMonthDay{2020}{}{}.
\newblock
{\BBOQ}\APACrefatitle {Detection of neutral atomic species in the ultra-hot
  Jupiter WASP-121b} {Detection of neutral atomic species in the ultra-hot
  jupiter wasp-121b}.{\BBCQ}
\newblock
\APACjournalVolNumPages{Monthly Notices of the Royal Astronomical
  Society}{494}{}{363}.
\PrintBackRefs{\CurrentBib}

\bibitem [\protect \citeauthoryear {%
Carone%
, Baeyens%
, Molli\'{e}re%
\BCBL {}\ \BBA {} {et al.}%
}{%
Carone%
\ \protect \BOthers {.}}{%
{\protect \APACyear {2019}}%
}]{%
Carone:2019aa}
\APACinsertmetastar {%
Carone:2019aa}%
\begin{APACrefauthors}%
Carone, L.%
, Baeyens, R.%
, Molli\'{e}re, P.%
\BCBL {}\ \BBA {} {et al.}%
\end{APACrefauthors}%
\unskip\
\newblock
\APACrefYearMonthDay{2019}{}{}.
\newblock
{\BBOQ}\APACrefatitle {Equatorial anti-rotating day side wind flow in WASP-43b
  elicited by deep wind jets?} {Equatorial anti-rotating day side wind flow in
  wasp-43b elicited by deep wind jets?}{\BBCQ}
\newblock
\APACjournalVolNumPages{arXiv e-prints:1904.13334}{}{}{}.
\PrintBackRefs{\CurrentBib}

\bibitem [\protect \citeauthoryear {%
Cauley%
, Shkolnik%
, Llama%
\BCBL {}\ \BBA {} Lanza%
}{%
Cauley%
\ \protect \BOthers {.}}{%
{\protect \APACyear {2019}}%
}]{%
Cauley:2019aa}
\APACinsertmetastar {%
Cauley:2019aa}%
\begin{APACrefauthors}%
Cauley, P.%
, Shkolnik, E.%
, Llama, J.%
\BCBL {}\ \BBA {} Lanza, A.%
\end{APACrefauthors}%
\unskip\
\newblock
\APACrefYearMonthDay{2019}{}{}.
\newblock
{\BBOQ}\APACrefatitle {Magnetic field strengths of hot Jupiters from signals of
  star-planet interactions} {Magnetic field strengths of hot jupiters from
  signals of star-planet interactions}.{\BBCQ}
\newblock
\APACjournalVolNumPages{Nature Astronomy}{3}{}{1128}.
\PrintBackRefs{\CurrentBib}

\bibitem [\protect \citeauthoryear {%
Chabrier%
\ \BBA {} Baraffe%
}{%
Chabrier%
\ \BBA {} Baraffe%
}{%
{\protect \APACyear {2007}}%
}]{%
Chabrier:2007}
\APACinsertmetastar {%
Chabrier:2007}%
\begin{APACrefauthors}%
Chabrier, G.%
\BCBT {}\ \BBA {} Baraffe, I.%
\end{APACrefauthors}%
\unskip\
\newblock
\APACrefYearMonthDay{2007}{}{}.
\newblock
{\BBOQ}\APACrefatitle {Heat transport in giant (exo)planets: a new perspective}
  {Heat transport in giant (exo)planets: a new perspective}.{\BBCQ}
\newblock
\APACjournalVolNumPages{The Astrophysical Journal Letters}{661}{}{81}.
\PrintBackRefs{\CurrentBib}

\bibitem [\protect \citeauthoryear {%
{Chabrier}%
, {Johansen}%
, {Janson}%
\BCBL {}\ \BBA {} {Rafikov}%
}{%
{Chabrier}%
\ \protect \BOthers {.}}{%
{\protect \APACyear {2014}}%
}]{%
chab14}
\APACinsertmetastar {%
chab14}%
\begin{APACrefauthors}%
{Chabrier}, G.%
, {Johansen}, A.%
, {Janson}, M.%
\BCBL {}\ \BBA {} {Rafikov}, R.%
\end{APACrefauthors}%
\unskip\
\newblock
\APACrefYearMonthDay{2014}{}{}.
\newblock
{\BBOQ}\APACrefatitle {{Giant Planet and Brown Dwarf Formation}} {{Giant Planet
  and Brown Dwarf Formation}}.{\BBCQ}
\newblock
\APACjournalVolNumPages{Protostars and Planets VI}{}{}{619-642}.
\newblock
\begin{APACrefDOI} \doi{10.2458/azu_uapress_9780816531240-ch027}
  \end{APACrefDOI}
\PrintBackRefs{\CurrentBib}

\bibitem [\protect \citeauthoryear {%
{Chang}%
, {Gu}%
\BCBL {}\ \BBA {} {Bodenheimer}%
}{%
{Chang}%
\ \protect \BOthers {.}}{%
{\protect \APACyear {2010}}%
}]{%
chan10}
\APACinsertmetastar {%
chan10}%
\begin{APACrefauthors}%
{Chang}, S\BHBI H.%
, {Gu}, P\BHBI G.%
\BCBL {}\ \BBA {} {Bodenheimer}, P\BPBI H.%
\end{APACrefauthors}%
\unskip\
\newblock
\APACrefYearMonthDay{2010}{{\APACmonth{01}}}{}.
\newblock
{\BBOQ}\APACrefatitle {{Tidal and Magnetic Interactions Between a Hot Jupiter
  and its Host Star in the Magnetospheric Cavity of a Protoplanetary Disk}}
  {{Tidal and Magnetic Interactions Between a Hot Jupiter and its Host Star in
  the Magnetospheric Cavity of a Protoplanetary Disk}}.{\BBCQ}
\newblock
\APACjournalVolNumPages{\apj}{708}{}{1692-1702}.
\newblock
\begin{APACrefDOI} \doi{10.1088/0004-637X/708/2/1692} \end{APACrefDOI}
\PrintBackRefs{\CurrentBib}

\bibitem [\protect \citeauthoryear {%
{Charbonneau}%
\ \protect \BOthers {.}}{%
{Charbonneau}%
\ \protect \BOthers {.}}{%
{\protect \APACyear {2005}}%
}]{%
Charb05}
\APACinsertmetastar {%
Charb05}%
\begin{APACrefauthors}%
{Charbonneau}, D.%
, {Allen}, L\BPBI E.%
, {Megeath}, S\BPBI T.%
, {Torres}, G.%
, {Alonso}, R.%
, {Brown}, T\BPBI M.%
\BDBL {}{Sozzetti}, A.%
\end{APACrefauthors}%
\unskip\
\newblock
\APACrefYearMonthDay{2005}{{\APACmonth{06}}}{}.
\newblock
{\BBOQ}\APACrefatitle {{Detection of Thermal Emission from an Extrasolar
  Planet}} {{Detection of Thermal Emission from an Extrasolar Planet}}.{\BBCQ}
\newblock
\APACjournalVolNumPages{ApJ}{626}{}{523-529}.
\PrintBackRefs{\CurrentBib}

\bibitem [\protect \citeauthoryear {%
{Charbonneau}%
, {Brown}%
, {Latham}%
\BCBL {}\ \BBA {} {Mayor}%
}{%
{Charbonneau}%
\ \protect \BOthers {.}}{%
{\protect \APACyear {2000}}%
}]{%
Charb00}
\APACinsertmetastar {%
Charb00}%
\begin{APACrefauthors}%
{Charbonneau}, D.%
, {Brown}, T\BPBI M.%
, {Latham}, D\BPBI W.%
\BCBL {}\ \BBA {} {Mayor}, M.%
\end{APACrefauthors}%
\unskip\
\newblock
\APACrefYearMonthDay{2000}{{\APACmonth{01}}}{}.
\newblock
{\BBOQ}\APACrefatitle {{Detection of Planetary Transits Across a Sun-like
  Star}} {{Detection of Planetary Transits Across a Sun-like Star}}.{\BBCQ}
\newblock
\APACjournalVolNumPages{\apjl}{529}{}{L45--L48}.
\PrintBackRefs{\CurrentBib}

\bibitem [\protect \citeauthoryear {%
{Charbonneau}%
, {Brown}%
, {Noyes}%
\BCBL {}\ \BBA {} {Gilliland}%
}{%
{Charbonneau}%
\ \protect \BOthers {.}}{%
{\protect \APACyear {2002}}%
}]{%
Charb02}
\APACinsertmetastar {%
Charb02}%
\begin{APACrefauthors}%
{Charbonneau}, D.%
, {Brown}, T\BPBI M.%
, {Noyes}, R\BPBI W.%
\BCBL {}\ \BBA {} {Gilliland}, R\BPBI L.%
\end{APACrefauthors}%
\unskip\
\newblock
\APACrefYearMonthDay{2002}{{\APACmonth{03}}}{}.
\newblock
{\BBOQ}\APACrefatitle {{Detection of an Extrasolar Planet Atmosphere}}
  {{Detection of an Extrasolar Planet Atmosphere}}.{\BBCQ}
\newblock
\APACjournalVolNumPages{\apj}{568}{}{377-384}.
\newblock
\begin{APACrefURL}
  \url{http://adsabs.harvard.edu/cgi-bin/nph-bib_query?bibcode=2002ApJ...568..377C&db_key=AST}
  \end{APACrefURL}
\PrintBackRefs{\CurrentBib}

\bibitem [\protect \citeauthoryear {%
{Chatterjee}%
, {Ford}%
, {Matsumura}%
\BCBL {}\ \BBA {} {Rasio}%
}{%
{Chatterjee}%
\ \protect \BOthers {.}}{%
{\protect \APACyear {2008}}%
}]{%
chat08}
\APACinsertmetastar {%
chat08}%
\begin{APACrefauthors}%
{Chatterjee}, S.%
, {Ford}, E\BPBI B.%
, {Matsumura}, S.%
\BCBL {}\ \BBA {} {Rasio}, F\BPBI A.%
\end{APACrefauthors}%
\unskip\
\newblock
\APACrefYearMonthDay{2008}{{\APACmonth{10}}}{}.
\newblock
{\BBOQ}\APACrefatitle {{Dynamical Outcomes of Planet-Planet Scattering}}
  {{Dynamical Outcomes of Planet-Planet Scattering}}.{\BBCQ}
\newblock
\APACjournalVolNumPages{\apj}{686}{}{580-602}.
\newblock
\begin{APACrefDOI} \doi{10.1086/590227} \end{APACrefDOI}
\PrintBackRefs{\CurrentBib}

\bibitem [\protect \citeauthoryear {%
{Coleman}%
\ \BBA {} {Nelson}%
}{%
{Coleman}%
\ \BBA {} {Nelson}%
}{%
{\protect \APACyear {2016}}%
}]{%
cole16}
\APACinsertmetastar {%
cole16}%
\begin{APACrefauthors}%
{Coleman}, G\BPBI A\BPBI L.%
\BCBT {}\ \BBA {} {Nelson}, R\BPBI P.%
\end{APACrefauthors}%
\unskip\
\newblock
\APACrefYearMonthDay{2016}{{\APACmonth{08}}}{}.
\newblock
{\BBOQ}\APACrefatitle {{Giant planet formation in radially structured
  protoplanetary discs}} {{Giant planet formation in radially structured
  protoplanetary discs}}.{\BBCQ}
\newblock
\APACjournalVolNumPages{\mnras}{460}{}{2779-2795}.
\newblock
\begin{APACrefDOI} \doi{10.1093/mnras/stw1177} \end{APACrefDOI}
\PrintBackRefs{\CurrentBib}

\bibitem [\protect \citeauthoryear {%
Cooper%
\ \BBA {} Showman%
}{%
Cooper%
\ \BBA {} Showman%
}{%
{\protect \APACyear {2006}}%
}]{%
Cooper:2006}
\APACinsertmetastar {%
Cooper:2006}%
\begin{APACrefauthors}%
Cooper, C.%
\BCBT {}\ \BBA {} Showman, A.%
\end{APACrefauthors}%
\unskip\
\newblock
\APACrefYearMonthDay{2006}{}{}.
\newblock
{\BBOQ}\APACrefatitle {Dynamics and disequilibrium carbon chemistry in hot
  {J}upiter atmospheres, with application to {HD} 209458b} {Dynamics and
  disequilibrium carbon chemistry in hot {J}upiter atmospheres, with
  application to {HD} 209458b}.{\BBCQ}
\newblock
\APACjournalVolNumPages{The Astrophysical Journal}{649}{}{1048}.
\PrintBackRefs{\CurrentBib}

\bibitem [\protect \citeauthoryear {%
Cowan%
\ \BBA {} Agol%
}{%
Cowan%
\ \BBA {} Agol%
}{%
{\protect \APACyear {2011}}%
{\protect \APACexlab {{\protect \BCnt {1}}}}}]{%
Cowan:2011}
\APACinsertmetastar {%
Cowan:2011}%
\begin{APACrefauthors}%
Cowan, N.%
\BCBT {}\ \BBA {} Agol, E.%
\end{APACrefauthors}%
\unskip\
\newblock
\APACrefYearMonthDay{2011{\protect \BCnt {1}}}{}{}.
\newblock
{\BBOQ}\APACrefatitle {A model for thermal phase variations of circular and
  eccentric exoplanets} {A model for thermal phase variations of circular and
  eccentric exoplanets}.{\BBCQ}
\newblock
\APACjournalVolNumPages{The Astrophysical Journal}{726}{}{82}.
\PrintBackRefs{\CurrentBib}

\bibitem [\protect \citeauthoryear {%
Cowan%
\ \BBA {} Agol%
}{%
Cowan%
\ \BBA {} Agol%
}{%
{\protect \APACyear {2011}}%
{\protect \APACexlab {{\protect \BCnt {2}}}}}]{%
Cowan_2011}
\APACinsertmetastar {%
Cowan_2011}%
\begin{APACrefauthors}%
Cowan, N.%
\BCBT {}\ \BBA {} Agol, E.%
\end{APACrefauthors}%
\unskip\
\newblock
\APACrefYearMonthDay{2011{\protect \BCnt {2}}}{}{}.
\newblock
{\BBOQ}\APACrefatitle {The statistics of albedo and heat recirculation on hot
  exoplanets} {The statistics of albedo and heat recirculation on hot
  exoplanets}.{\BBCQ}
\newblock
\APACjournalVolNumPages{The Astrophysical Journal}{759}{}{54}.
\PrintBackRefs{\CurrentBib}

\bibitem [\protect \citeauthoryear {%
Cowan%
, Agol%
, Meadows%
\BCBL {}\ \BBA {} {et al.}%
}{%
Cowan%
\ \protect \BOthers {.}}{%
{\protect \APACyear {2009}}%
}]{%
Cowan2009}
\APACinsertmetastar {%
Cowan2009}%
\begin{APACrefauthors}%
Cowan, N.%
, Agol, E.%
, Meadows, V.%
\BCBL {}\ \BBA {} {et al.}%
\end{APACrefauthors}%
\unskip\
\newblock
\APACrefYearMonthDay{2009}{}{}.
\newblock
{\BBOQ}\APACrefatitle {Alien maps of an ocean-bearing world} {Alien maps of an
  ocean-bearing world}.{\BBCQ}
\newblock
\APACjournalVolNumPages{The Astrophysical Journal}{700}{}{915}.
\newblock
\begin{APACrefURL}
  \url{http://stacks.iop.org/0004-637X/700/i=2/a=915?key=crossref.434ee2bff413de58edf46e4a744124dd}
  \end{APACrefURL}
\PrintBackRefs{\CurrentBib}

\bibitem [\protect \citeauthoryear {%
{Cridland}%
, {van Dishoeck}%
, {Alessi}%
\BCBL {}\ \BBA {} {Pudritz}%
}{%
{Cridland}%
\ \protect \BOthers {.}}{%
{\protect \APACyear {2019}}%
}]{%
Cridland19}
\APACinsertmetastar {%
Cridland19}%
\begin{APACrefauthors}%
{Cridland}, A\BPBI J.%
, {van Dishoeck}, E\BPBI F.%
, {Alessi}, M.%
\BCBL {}\ \BBA {} {Pudritz}, R\BPBI E.%
\end{APACrefauthors}%
\unskip\
\newblock
\APACrefYearMonthDay{2019}{{\APACmonth{12}}}{}.
\newblock
{\BBOQ}\APACrefatitle {{Connecting planet formation and astrochemistry. A main
  sequence for C/O in hot exoplanetary atmospheres}} {{Connecting planet
  formation and astrochemistry. A main sequence for C/O in hot exoplanetary
  atmospheres}}.{\BBCQ}
\newblock
\APACjournalVolNumPages{\aap}{632}{}{A63}.
\newblock
\begin{APACrefDOI} \doi{10.1051/0004-6361/201936105} \end{APACrefDOI}
\PrintBackRefs{\CurrentBib}

\bibitem [\protect \citeauthoryear {%
{Crossfield}%
}{%
{Crossfield}%
}{%
{\protect \APACyear {2015}}%
}]{%
Crossfield15}
\APACinsertmetastar {%
Crossfield15}%
\begin{APACrefauthors}%
{Crossfield}, I\BPBI J\BPBI M.%
\end{APACrefauthors}%
\unskip\
\newblock
\APACrefYearMonthDay{2015}{{\APACmonth{10}}}{}.
\newblock
{\BBOQ}\APACrefatitle {{Observations of Exoplanet Atmospheres}} {{Observations
  of Exoplanet Atmospheres}}.{\BBCQ}
\newblock
\APACjournalVolNumPages{\pasp}{127}{956}{941}.
\newblock
\begin{APACrefDOI} \doi{10.1086/683115} \end{APACrefDOI}
\PrintBackRefs{\CurrentBib}

\bibitem [\protect \citeauthoryear {%
{Cumming}%
\ \protect \BOthers {.}}{%
{Cumming}%
\ \protect \BOthers {.}}{%
{\protect \APACyear {2008}}%
}]{%
cumm08}
\APACinsertmetastar {%
cumm08}%
\begin{APACrefauthors}%
{Cumming}, A.%
, {Butler}, R\BPBI P.%
, {Marcy}, G\BPBI W.%
, {Vogt}, S\BPBI S.%
, {Wright}, J\BPBI T.%
\BCBL {}\ \BBA {} {Fischer}, D\BPBI A.%
\end{APACrefauthors}%
\unskip\
\newblock
\APACrefYearMonthDay{2008}{{\APACmonth{05}}}{}.
\newblock
{\BBOQ}\APACrefatitle {{The Keck Planet Search: Detectability and the Minimum
  Mass and Orbital Period Distribution of Extrasolar Planets}} {{The Keck
  Planet Search: Detectability and the Minimum Mass and Orbital Period
  Distribution of Extrasolar Planets}}.{\BBCQ}
\newblock
\APACjournalVolNumPages{\pasp}{120}{}{531}.
\newblock
\begin{APACrefDOI} \doi{10.1086/588487} \end{APACrefDOI}
\PrintBackRefs{\CurrentBib}

\bibitem [\protect \citeauthoryear {%
Dang%
, Cowan%
, Schwartz%
\BCBL {}\ \BBA {} {et al.}%
}{%
Dang%
\ \protect \BOthers {.}}{%
{\protect \APACyear {2018}}%
}]{%
Dang:2018aa}
\APACinsertmetastar {%
Dang:2018aa}%
\begin{APACrefauthors}%
Dang, L.%
, Cowan, N.%
, Schwartz, J.%
\BCBL {}\ \BBA {} {et al.}%
\end{APACrefauthors}%
\unskip\
\newblock
\APACrefYearMonthDay{2018}{}{}.
\newblock
{\BBOQ}\APACrefatitle {Detection of a westward hotspot offset in the atmosphere
  of hot gas giant CoRoT-2b} {Detection of a westward hotspot offset in the
  atmosphere of hot gas giant corot-2b}.{\BBCQ}
\newblock
\APACjournalVolNumPages{Nature Astronomy}{2}{}{220}.
\PrintBackRefs{\CurrentBib}

\bibitem [\protect \citeauthoryear {%
{Dawson}%
\ \BBA {} {Johnson}%
}{%
{Dawson}%
\ \BBA {} {Johnson}%
}{%
{\protect \APACyear {2018}}%
}]{%
daws18}
\APACinsertmetastar {%
daws18}%
\begin{APACrefauthors}%
{Dawson}, R\BPBI I.%
\BCBT {}\ \BBA {} {Johnson}, J\BPBI A.%
\end{APACrefauthors}%
\unskip\
\newblock
\APACrefYearMonthDay{2018}{{\APACmonth{09}}}{}.
\newblock
{\BBOQ}\APACrefatitle {{Origins of Hot Jupiters}} {{Origins of Hot
  Jupiters}}.{\BBCQ}
\newblock
\APACjournalVolNumPages{\araa}{56}{}{175-221}.
\newblock
\begin{APACrefDOI} \doi{10.1146/annurev-astro-081817-051853} \end{APACrefDOI}
\PrintBackRefs{\CurrentBib}

\bibitem [\protect \citeauthoryear {%
{Dawson}%
\ \BBA {} {Murray-Clay}%
}{%
{Dawson}%
\ \BBA {} {Murray-Clay}%
}{%
{\protect \APACyear {2013}}%
}]{%
daws13}
\APACinsertmetastar {%
daws13}%
\begin{APACrefauthors}%
{Dawson}, R\BPBI I.%
\BCBT {}\ \BBA {} {Murray-Clay}, R\BPBI A.%
\end{APACrefauthors}%
\unskip\
\newblock
\APACrefYearMonthDay{2013}{{\APACmonth{04}}}{}.
\newblock
{\BBOQ}\APACrefatitle {{Giant Planets Orbiting Metal-rich Stars Show Signatures
  of Planet-Planet Interactions}} {{Giant Planets Orbiting Metal-rich Stars
  Show Signatures of Planet-Planet Interactions}}.{\BBCQ}
\newblock
\APACjournalVolNumPages{\apjl}{767}{}{L24}.
\newblock
\begin{APACrefDOI} \doi{10.1088/2041-8205/767/2/L24} \end{APACrefDOI}
\PrintBackRefs{\CurrentBib}

\bibitem [\protect \citeauthoryear {%
Daylan%
, Gunther%
, Mikal-Evans%
\BCBL {}\ \BBA {} {et al.}%
}{%
Daylan%
\ \protect \BOthers {.}}{%
{\protect \APACyear {2019}}%
}]{%
Daylan:2019aa}
\APACinsertmetastar {%
Daylan:2019aa}%
\begin{APACrefauthors}%
Daylan, T.%
, Gunther, M.%
, Mikal-Evans, T.%
\BCBL {}\ \BBA {} {et al.}%
\end{APACrefauthors}%
\unskip\
\newblock
\APACrefYearMonthDay{2019}{}{}.
\newblock
{\BBOQ}\APACrefatitle {TESS observations of the WASP-121 b phase curve} {Tess
  observations of the wasp-121 b phase curve}.{\BBCQ}
\newblock
\APACjournalVolNumPages{arXiv e-prints:1909.03000}{}{}{}.
\PrintBackRefs{\CurrentBib}

\bibitem [\protect \citeauthoryear {%
{De Wit}%
, Gillon%
, Demory%
\BCBL {}\ \BBA {} Seager%
}{%
{De Wit}%
\ \protect \BOthers {.}}{%
{\protect \APACyear {2012}}%
}]{%
Wit:2012}
\APACinsertmetastar {%
Wit:2012}%
\begin{APACrefauthors}%
{De Wit}, J.%
, Gillon, M.%
, Demory, B.%
\BCBL {}\ \BBA {} Seager, S.%
\end{APACrefauthors}%
\unskip\
\newblock
\APACrefYearMonthDay{2012}{}{}.
\newblock
{\BBOQ}\APACrefatitle {Towards consistent mapping of distant worlds:
  secondary-eclipse scanning of the exoplanet {HD} 189733b} {Towards consistent
  mapping of distant worlds: secondary-eclipse scanning of the exoplanet {HD}
  189733b}.{\BBCQ}
\newblock
\APACjournalVolNumPages{Astronomy and Astrophysics}{548}{}{A128}.
\PrintBackRefs{\CurrentBib}

\bibitem [\protect \citeauthoryear {%
{Deming}%
, {Brown}%
, {Charbonneau}%
, {Harrington}%
\BCBL {}\ \BBA {} {Richardson}%
}{%
{Deming}%
\ \protect \BOthers {.}}{%
{\protect \APACyear {2005}}%
}]{%
Deming05a}
\APACinsertmetastar {%
Deming05a}%
\begin{APACrefauthors}%
{Deming}, D.%
, {Brown}, T\BPBI M.%
, {Charbonneau}, D.%
, {Harrington}, J.%
\BCBL {}\ \BBA {} {Richardson}, L\BPBI J.%
\end{APACrefauthors}%
\unskip\
\newblock
\APACrefYearMonthDay{2005}{{\APACmonth{04}}}{}.
\newblock
{\BBOQ}\APACrefatitle {{A New Search for Carbon Monoxide Absorption in the
  Transmission Spectrum of the Extrasolar Planet HD 209458b}} {{A New Search
  for Carbon Monoxide Absorption in the Transmission Spectrum of the Extrasolar
  Planet HD 209458b}}.{\BBCQ}
\newblock
\APACjournalVolNumPages{\apj}{622}{}{1149-1159}.
\PrintBackRefs{\CurrentBib}

\bibitem [\protect \citeauthoryear {%
Deming%
\ \BBA {} Knutson%
}{%
Deming%
\ \BBA {} Knutson%
}{%
{\protect \APACyear {2020}}%
}]{%
Deming:2020aa}
\APACinsertmetastar {%
Deming:2020aa}%
\begin{APACrefauthors}%
Deming, D.%
\BCBT {}\ \BBA {} Knutson, H.%
\end{APACrefauthors}%
\unskip\
\newblock
\APACrefYearMonthDay{2020}{}{}.
\newblock
{\BBOQ}\APACrefatitle {Highlights of exoplanetary science from Spitzer}
  {Highlights of exoplanetary science from spitzer}.{\BBCQ}
\newblock
\APACjournalVolNumPages{Nature Astronomy}{4}{}{453}.
\PrintBackRefs{\CurrentBib}

\bibitem [\protect \citeauthoryear {%
{Deming}%
, {Louie}%
\BCBL {}\ \BBA {} {Sheets}%
}{%
{Deming}%
\ \protect \BOthers {.}}{%
{\protect \APACyear {2019}}%
}]{%
Deming19}
\APACinsertmetastar {%
Deming19}%
\begin{APACrefauthors}%
{Deming}, D.%
, {Louie}, D.%
\BCBL {}\ \BBA {} {Sheets}, H.%
\end{APACrefauthors}%
\unskip\
\newblock
\APACrefYearMonthDay{2019}{{\APACmonth{01}}}{}.
\newblock
{\BBOQ}\APACrefatitle {{How to Characterize the Atmosphere of a Transiting
  Exoplanet}} {{How to Characterize the Atmosphere of a Transiting
  Exoplanet}}.{\BBCQ}
\newblock
\APACjournalVolNumPages{\pasp}{131}{995}{013001}.
\newblock
\begin{APACrefDOI} \doi{10.1088/1538-3873/aae5c5} \end{APACrefDOI}
\PrintBackRefs{\CurrentBib}

\bibitem [\protect \citeauthoryear {%
Deming%
, Wilkins%
, McCullough%
\BCBL {}\ \BBA {} {et al.}%
}{%
Deming%
\ \protect \BOthers {.}}{%
{\protect \APACyear {2013}}%
}]{%
Deming:2013aa}
\APACinsertmetastar {%
Deming:2013aa}%
\begin{APACrefauthors}%
Deming, D.%
, Wilkins, A.%
, McCullough, P.%
\BCBL {}\ \BBA {} {et al.}%
\end{APACrefauthors}%
\unskip\
\newblock
\APACrefYearMonthDay{2013}{}{}.
\newblock
{\BBOQ}\APACrefatitle {Infrared Transmission Spectroscopy of the Exoplanets HD
  209458b and XO-1b Using the Wide Field Camera-3 on the Hubble Space
  Telescope} {Infrared transmission spectroscopy of the exoplanets hd 209458b
  and xo-1b using the wide field camera-3 on the hubble space
  telescope}.{\BBCQ}
\newblock
\APACjournalVolNumPages{The Astrophysical Journal}{774}{}{95}.
\PrintBackRefs{\CurrentBib}

\bibitem [\protect \citeauthoryear {%
Demory%
\ \BBA {} Seager%
}{%
Demory%
\ \BBA {} Seager%
}{%
{\protect \APACyear {2011}}%
}]{%
Demory:2011}
\APACinsertmetastar {%
Demory:2011}%
\begin{APACrefauthors}%
Demory, B.%
\BCBT {}\ \BBA {} Seager, S.%
\end{APACrefauthors}%
\unskip\
\newblock
\APACrefYearMonthDay{2011}{}{}.
\newblock
{\BBOQ}\APACrefatitle {Lack of inflated radii for {K}epler giant planet
  candidates receiving modest stellar irradiation} {Lack of inflated radii for
  {K}epler giant planet candidates receiving modest stellar
  irradiation}.{\BBCQ}
\newblock
\APACjournalVolNumPages{The Astrophysical Journal Supplement
  Series}{197}{}{12}.
\PrintBackRefs{\CurrentBib}

\bibitem [\protect \citeauthoryear {%
{D{\'e}sert}%
\ \protect \BOthers {.}}{%
{D{\'e}sert}%
\ \protect \BOthers {.}}{%
{\protect \APACyear {2009}}%
}]{%
desert09}
\APACinsertmetastar {%
desert09}%
\begin{APACrefauthors}%
{D{\'e}sert}, J.%
, {Lecavelier des Etangs}, A.%
, {H{\'e}brard}, G.%
, {Sing}, D\BPBI K.%
, {Ehrenreich}, D.%
, {Ferlet}, R.%
\BCBL {}\ \BBA {} {Vidal-Madjar}, A.%
\end{APACrefauthors}%
\unskip\
\newblock
\APACrefYearMonthDay{2009}{{\APACmonth{07}}}{}.
\newblock
{\BBOQ}\APACrefatitle {{Search for Carbon Monoxide in the Atmosphere of the
  Transiting Exoplanet HD 189733b}} {{Search for Carbon Monoxide in the
  Atmosphere of the Transiting Exoplanet HD 189733b}}.{\BBCQ}
\newblock
\APACjournalVolNumPages{\apj}{699}{}{478-485}.
\newblock
\begin{APACrefDOI} \doi{10.1088/0004-637X/699/1/478} \end{APACrefDOI}
\PrintBackRefs{\CurrentBib}

\bibitem [\protect \citeauthoryear {%
de Wit%
\ \protect \BOthers {.}}{%
de Wit%
\ \protect \BOthers {.}}{%
{\protect \APACyear {2016}}%
}]{%
Wit:2016aa}
\APACinsertmetastar {%
Wit:2016aa}%
\begin{APACrefauthors}%
de Wit, J.%
, Lewis, N.%
, Langton, J.%
, Laughlin, G.%
, Deming, D.%
, Batygin, K.%
\BCBL {}\ \BBA {} Fortney, J.%
\end{APACrefauthors}%
\unskip\
\newblock
\APACrefYearMonthDay{2016}{}{}.
\newblock
{\BBOQ}\APACrefatitle {Direct Measure of Radiative and Dynamical Properties of
  an Exoplanet Atmosphere} {Direct measure of radiative and dynamical
  properties of an exoplanet atmosphere}.{\BBCQ}
\newblock
\APACjournalVolNumPages{The Astrophysical Journal Letters}{820}{}{L33}.
\PrintBackRefs{\CurrentBib}

\bibitem [\protect \citeauthoryear {%
Dobbs-Dixon%
\ \BBA {} Agol%
}{%
Dobbs-Dixon%
\ \BBA {} Agol%
}{%
{\protect \APACyear {2013}}%
}]{%
Dobbs-Dixon:2013}
\APACinsertmetastar {%
Dobbs-Dixon:2013}%
\begin{APACrefauthors}%
Dobbs-Dixon, I.%
\BCBT {}\ \BBA {} Agol, E.%
\end{APACrefauthors}%
\unskip\
\newblock
\APACrefYearMonthDay{2013}{}{}.
\newblock
{\BBOQ}\APACrefatitle {Three-dimensional radiative-hydrodynamical simulations
  of the highly irradiated short-period exoplanet {HD} 189733b}
  {Three-dimensional radiative-hydrodynamical simulations of the highly
  irradiated short-period exoplanet {HD} 189733b}.{\BBCQ}
\newblock
\APACjournalVolNumPages{Monthly Notices of the Royal Astronomical
  Society}{435}{}{3159}.
\PrintBackRefs{\CurrentBib}

\bibitem [\protect \citeauthoryear {%
{Donati}%
\ \protect \BOthers {.}}{%
{Donati}%
\ \protect \BOthers {.}}{%
{\protect \APACyear {2016}}%
}]{%
dona16}
\APACinsertmetastar {%
dona16}%
\begin{APACrefauthors}%
{Donati}, J\BPBI F.%
, {Moutou}, C.%
, {Malo}, L.%
, {Baruteau}, C.%
, {Yu}, L.%
, {H{\'e}brard}, E.%
\BDBL {}{Cameron}, A\BPBI C.%
\end{APACrefauthors}%
\unskip\
\newblock
\APACrefYearMonthDay{2016}{{\APACmonth{06}}}{}.
\newblock
{\BBOQ}\APACrefatitle {{A hot Jupiter orbiting a 2-million-year-old solar-mass
  T Tauri star}} {{A hot Jupiter orbiting a 2-million-year-old solar-mass T
  Tauri star}}.{\BBCQ}
\newblock
\APACjournalVolNumPages{\nat}{534}{}{662-666}.
\newblock
\begin{APACrefDOI} \doi{10.1038/nature18305} \end{APACrefDOI}
\PrintBackRefs{\CurrentBib}

\bibitem [\protect \citeauthoryear {%
Drummond%
, Mayne%
, Manners%
\BCBL {}\ \BBA {} {et al.}%
}{%
Drummond%
\ \protect \BOthers {.}}{%
{\protect \APACyear {2018}}%
}]{%
Drummond:2018aa}
\APACinsertmetastar {%
Drummond:2018aa}%
\begin{APACrefauthors}%
Drummond, B.%
, Mayne, N.%
, Manners, J.%
\BCBL {}\ \BBA {} {et al.}%
\end{APACrefauthors}%
\unskip\
\newblock
\APACrefYearMonthDay{2018}{}{}.
\newblock
{\BBOQ}\APACrefatitle {The 3D Thermal, Dynamical, and Chemical Structure of the
  Atmosphere of HD 189733b: Implications of Wind-driven Chemistry for the
  Emission Phase Curve} {The 3d thermal, dynamical, and chemical structure of
  the atmosphere of hd 189733b: Implications of wind-driven chemistry for the
  emission phase curve}.{\BBCQ}
\newblock
\APACjournalVolNumPages{The Astrophysical Journal}{869}{}{28}.
\PrintBackRefs{\CurrentBib}

\bibitem [\protect \citeauthoryear {%
{Duffell}%
\ \BBA {} {Chiang}%
}{%
{Duffell}%
\ \BBA {} {Chiang}%
}{%
{\protect \APACyear {2015}}%
}]{%
duff15}
\APACinsertmetastar {%
duff15}%
\begin{APACrefauthors}%
{Duffell}, P\BPBI C.%
\BCBT {}\ \BBA {} {Chiang}, E.%
\end{APACrefauthors}%
\unskip\
\newblock
\APACrefYearMonthDay{2015}{{\APACmonth{10}}}{}.
\newblock
{\BBOQ}\APACrefatitle {{Eccentric Jupiters via Disk-Planet Interactions}}
  {{Eccentric Jupiters via Disk-Planet Interactions}}.{\BBCQ}
\newblock
\APACjournalVolNumPages{\apj}{812}{}{94}.
\newblock
\begin{APACrefDOI} \doi{10.1088/0004-637X/812/2/94} \end{APACrefDOI}
\PrintBackRefs{\CurrentBib}

\bibitem [\protect \citeauthoryear {%
{Duffell}%
, {Haiman}%
, {MacFadyen}%
, {D'Orazio}%
\BCBL {}\ \BBA {} {Farris}%
}{%
{Duffell}%
\ \protect \BOthers {.}}{%
{\protect \APACyear {2014}}%
}]{%
duff14}
\APACinsertmetastar {%
duff14}%
\begin{APACrefauthors}%
{Duffell}, P\BPBI C.%
, {Haiman}, Z.%
, {MacFadyen}, A\BPBI I.%
, {D'Orazio}, D\BPBI J.%
\BCBL {}\ \BBA {} {Farris}, B\BPBI D.%
\end{APACrefauthors}%
\unskip\
\newblock
\APACrefYearMonthDay{2014}{{\APACmonth{09}}}{}.
\newblock
{\BBOQ}\APACrefatitle {{The Migration of Gap-opening Planets is Not Locked to
  Viscous Disk Evolution}} {{The Migration of Gap-opening Planets is Not Locked
  to Viscous Disk Evolution}}.{\BBCQ}
\newblock
\APACjournalVolNumPages{\apjl}{792}{}{L10}.
\newblock
\begin{APACrefDOI} \doi{10.1088/2041-8205/792/1/L10} \end{APACrefDOI}
\PrintBackRefs{\CurrentBib}

\bibitem [\protect \citeauthoryear {%
{Durisen}%
\ \protect \BOthers {.}}{%
{Durisen}%
\ \protect \BOthers {.}}{%
{\protect \APACyear {2007}}%
}]{%
duri07}
\APACinsertmetastar {%
duri07}%
\begin{APACrefauthors}%
{Durisen}, R\BPBI H.%
, {Boss}, A\BPBI P.%
, {Mayer}, L.%
, {Nelson}, A\BPBI F.%
, {Quinn}, T.%
\BCBL {}\ \BBA {} {Rice}, W\BPBI K\BPBI M.%
\end{APACrefauthors}%
\unskip\
\newblock
\APACrefYearMonthDay{2007}{}{}.
\newblock
{\BBOQ}\APACrefatitle {{Gravitational Instabilities in Gaseous Protoplanetary
  Disks and Implications for Giant Planet Formation}} {{Gravitational
  Instabilities in Gaseous Protoplanetary Disks and Implications for Giant
  Planet Formation}}.{\BBCQ}
\newblock
\APACjournalVolNumPages{Protostars and Planets V}{}{}{607-622}.
\PrintBackRefs{\CurrentBib}

\bibitem [\protect \citeauthoryear {%
{Eggleton}%
, {Kiseleva}%
\BCBL {}\ \BBA {} {Hut}%
}{%
{Eggleton}%
\ \protect \BOthers {.}}{%
{\protect \APACyear {1998}}%
}]{%
eggl98}
\APACinsertmetastar {%
eggl98}%
\begin{APACrefauthors}%
{Eggleton}, P\BPBI P.%
, {Kiseleva}, L\BPBI G.%
\BCBL {}\ \BBA {} {Hut}, P.%
\end{APACrefauthors}%
\unskip\
\newblock
\APACrefYearMonthDay{1998}{{\APACmonth{05}}}{}.
\newblock
{\BBOQ}\APACrefatitle {{The Equilibrium Tide Model for Tidal Friction}} {{The
  Equilibrium Tide Model for Tidal Friction}}.{\BBCQ}
\newblock
\APACjournalVolNumPages{\apj}{499}{}{853-870}.
\newblock
\begin{APACrefDOI} \doi{10.1086/305670} \end{APACrefDOI}
\PrintBackRefs{\CurrentBib}

\bibitem [\protect \citeauthoryear {%
{Ehrenreich}%
\ \protect \BOthers {.}}{%
{Ehrenreich}%
\ \protect \BOthers {.}}{%
{\protect \APACyear {2015}}%
}]{%
Ehrenreich15}
\APACinsertmetastar {%
Ehrenreich15}%
\begin{APACrefauthors}%
{Ehrenreich}, D.%
, {Bourrier}, V.%
, {Wheatley}, P\BPBI J.%
, {Lecavelier des Etangs}, A.%
, {H{\'e}brard}, G.%
, {Udry}, S.%
\BDBL {}{Vidal-Madjar}, A.%
\end{APACrefauthors}%
\unskip\
\newblock
\APACrefYearMonthDay{2015}{{\APACmonth{06}}}{}.
\newblock
{\BBOQ}\APACrefatitle {{A giant comet-like cloud of hydrogen escaping the warm
  Neptune-mass exoplanet GJ 436b}} {{A giant comet-like cloud of hydrogen
  escaping the warm Neptune-mass exoplanet GJ 436b}}.{\BBCQ}
\newblock
\APACjournalVolNumPages{\nat}{522}{7557}{459-461}.
\newblock
\begin{APACrefDOI} \doi{10.1038/nature14501} \end{APACrefDOI}
\PrintBackRefs{\CurrentBib}

\bibitem [\protect \citeauthoryear {%
Ehrenreich%
, Lovis%
, Allart%
\BCBL {}\ \BBA {} {et al.}%
}{%
Ehrenreich%
\ \protect \BOthers {.}}{%
{\protect \APACyear {2020}}%
}]{%
Ehrenreich:2020aa}
\APACinsertmetastar {%
Ehrenreich:2020aa}%
\begin{APACrefauthors}%
Ehrenreich, D.%
, Lovis, C.%
, Allart, R.%
\BCBL {}\ \BBA {} {et al.}%
\end{APACrefauthors}%
\unskip\
\newblock
\APACrefYearMonthDay{2020}{}{}.
\newblock
{\BBOQ}\APACrefatitle {Nightside condensation of iron in an ultrahot giant
  exoplanet.} {Nightside condensation of iron in an ultrahot giant
  exoplanet.}{\BBCQ}
\newblock
\APACjournalVolNumPages{Nature}{580}{}{597}.
\PrintBackRefs{\CurrentBib}

\bibitem [\protect \citeauthoryear {%
Esteves%
, de Mooij%
\BCBL {}\ \BBA {} Jayawardhana%
}{%
Esteves%
\ \protect \BOthers {.}}{%
{\protect \APACyear {2015}}%
}]{%
Esteves:2015}
\APACinsertmetastar {%
Esteves:2015}%
\begin{APACrefauthors}%
Esteves, L.%
, de Mooij, E.%
\BCBL {}\ \BBA {} Jayawardhana, R.%
\end{APACrefauthors}%
\unskip\
\newblock
\APACrefYearMonthDay{2015}{}{}.
\newblock
{\BBOQ}\APACrefatitle {Changing phases of ailen worlds: probing atmospheres of
  {K}epler planets with high-precision photometry} {Changing phases of ailen
  worlds: probing atmospheres of {K}epler planets with high-precision
  photometry}.{\BBCQ}
\newblock
\APACjournalVolNumPages{The Astrophysical Journal}{804}{}{150}.
\PrintBackRefs{\CurrentBib}

\bibitem [\protect \citeauthoryear {%
Evans%
, Sing%
, Kataria%
\BCBL {}\ \BBA {} et al.%
}{%
Evans%
\ \protect \BOthers {.}}{%
{\protect \APACyear {2017}}%
}]{%
Evans:2017aa}
\APACinsertmetastar {%
Evans:2017aa}%
\begin{APACrefauthors}%
Evans, T.%
, Sing, D.%
, Kataria, T.%
\BCBL {}\ \BBA {} et al.%
\end{APACrefauthors}%
\unskip\
\newblock
\APACrefYearMonthDay{2017}{}{}.
\newblock

\newblock
\APACjournalVolNumPages{Nature}{548}{}{58}.
\PrintBackRefs{\CurrentBib}

\bibitem [\protect \citeauthoryear {%
{Fabrycky}%
\ \BBA {} {Tremaine}%
}{%
{Fabrycky}%
\ \BBA {} {Tremaine}%
}{%
{\protect \APACyear {2007}}%
}]{%
fabr07}
\APACinsertmetastar {%
fabr07}%
\begin{APACrefauthors}%
{Fabrycky}, D.%
\BCBT {}\ \BBA {} {Tremaine}, S.%
\end{APACrefauthors}%
\unskip\
\newblock
\APACrefYearMonthDay{2007}{{\APACmonth{11}}}{}.
\newblock
{\BBOQ}\APACrefatitle {{Shrinking Binary and Planetary Orbits by Kozai Cycles
  with Tidal Friction}} {{Shrinking Binary and Planetary Orbits by Kozai Cycles
  with Tidal Friction}}.{\BBCQ}
\newblock
\APACjournalVolNumPages{\apj}{669}{}{1298-1315}.
\newblock
\begin{APACrefDOI} \doi{10.1086/521702} \end{APACrefDOI}
\PrintBackRefs{\CurrentBib}

\bibitem [\protect \citeauthoryear {%
{Fischer}%
\ \BBA {} {Valenti}%
}{%
{Fischer}%
\ \BBA {} {Valenti}%
}{%
{\protect \APACyear {2005}}%
}]{%
fisc05}
\APACinsertmetastar {%
fisc05}%
\begin{APACrefauthors}%
{Fischer}, D\BPBI A.%
\BCBT {}\ \BBA {} {Valenti}, J.%
\end{APACrefauthors}%
\unskip\
\newblock
\APACrefYearMonthDay{2005}{{\APACmonth{04}}}{}.
\newblock
{\BBOQ}\APACrefatitle {{The Planet-Metallicity Correlation}} {{The
  Planet-Metallicity Correlation}}.{\BBCQ}
\newblock
\APACjournalVolNumPages{\apj}{622}{}{1102-1117}.
\newblock
\begin{APACrefDOI} \doi{10.1086/428383} \end{APACrefDOI}
\PrintBackRefs{\CurrentBib}

\bibitem [\protect \citeauthoryear {%
{Fisher}%
\ \BBA {} {Heng}%
}{%
{Fisher}%
\ \BBA {} {Heng}%
}{%
{\protect \APACyear {2018}}%
}]{%
Fisher18}
\APACinsertmetastar {%
Fisher18}%
\begin{APACrefauthors}%
{Fisher}, C.%
\BCBT {}\ \BBA {} {Heng}, K.%
\end{APACrefauthors}%
\unskip\
\newblock
\APACrefYearMonthDay{2018}{{\APACmonth{12}}}{}.
\newblock
{\BBOQ}\APACrefatitle {{Retrieval analysis of 38 WFC3 transmission spectra and
  resolution of the normalization degeneracy}} {{Retrieval analysis of 38 WFC3
  transmission spectra and resolution of the normalization degeneracy}}.{\BBCQ}
\newblock
\APACjournalVolNumPages{\mnras}{481}{4}{4698-4727}.
\newblock
\begin{APACrefDOI} \doi{10.1093/mnras/sty2550} \end{APACrefDOI}
\PrintBackRefs{\CurrentBib}

\bibitem [\protect \citeauthoryear {%
Flowers%
, Brogi%
, Rauscher%
, Kempton%
\BCBL {}\ \BBA {} Chiavassa%
}{%
Flowers%
\ \protect \BOthers {.}}{%
{\protect \APACyear {2019}}%
}]{%
Flowers:2018aa}
\APACinsertmetastar {%
Flowers:2018aa}%
\begin{APACrefauthors}%
Flowers, E.%
, Brogi, M.%
, Rauscher, E.%
, Kempton, E.%
\BCBL {}\ \BBA {} Chiavassa, A.%
\end{APACrefauthors}%
\unskip\
\newblock
\APACrefYearMonthDay{2019}{}{}.
\newblock
{\BBOQ}\APACrefatitle {The High-Resolution Transmission Spectrum of HD 189733b
  Interpreted with Atmospheric Doppler Shifts from Three-Dimensional General
  Circulation Models} {The high-resolution transmission spectrum of hd 189733b
  interpreted with atmospheric doppler shifts from three-dimensional general
  circulation models}.{\BBCQ}
\newblock
\APACjournalVolNumPages{The Astronomical Journal}{157}{}{209}.
\PrintBackRefs{\CurrentBib}

\bibitem [\protect \citeauthoryear {%
Fortney%
, Baraffe%
\BCBL {}\ \BBA {} Militzer%
}{%
Fortney%
\ \protect \BOthers {.}}{%
{\protect \APACyear {2010}}%
}]{%
fortney_2009}
\APACinsertmetastar {%
fortney_2009}%
\begin{APACrefauthors}%
Fortney, J.%
, Baraffe, I.%
\BCBL {}\ \BBA {} Militzer, B.%
\end{APACrefauthors}%
\unskip\
\newblock
\APACrefYearMonthDay{2010}{}{}.
\newblock
{\BBOQ}\APACrefatitle {Exoplanets} {Exoplanets}.{\BBCQ}
\newblock
\BIn{} S.~Seager\ (\BED), (\BCHAP\ Giant Planet Interior Structure and Thermal
  Evolution).
\newblock
\APACaddressPublisher{Tucson, AZ}{University of Arizona Press}.
\PrintBackRefs{\CurrentBib}

\bibitem [\protect \citeauthoryear {%
{Fortney}%
, {Lodders}%
, {Marley}%
\BCBL {}\ \BBA {} {Freedman}%
}{%
{Fortney}%
\ \protect \BOthers {.}}{%
{\protect \APACyear {2008}}%
}]{%
Fortney08a}
\APACinsertmetastar {%
Fortney08a}%
\begin{APACrefauthors}%
{Fortney}, J\BPBI J.%
, {Lodders}, K.%
, {Marley}, M\BPBI S.%
\BCBL {}\ \BBA {} {Freedman}, R\BPBI S.%
\end{APACrefauthors}%
\unskip\
\newblock
\APACrefYearMonthDay{2008}{{\APACmonth{05}}}{}.
\newblock
{\BBOQ}\APACrefatitle {{A Unified Theory for the Atmospheres of the Hot and
  Very Hot Jupiters: Two Classes of Irradiated Atmospheres}} {{A Unified Theory
  for the Atmospheres of the Hot and Very Hot Jupiters: Two Classes of
  Irradiated Atmospheres}}.{\BBCQ}
\newblock
\APACjournalVolNumPages{\apj}{678}{}{1419-1435}.
\newblock
\begin{APACrefDOI} \doi{10.1086/528370} \end{APACrefDOI}
\PrintBackRefs{\CurrentBib}

\bibitem [\protect \citeauthoryear {%
{Fortney}%
, {Marley}%
\BCBL {}\ \BBA {} {Barnes}%
}{%
{Fortney}%
\ \protect \BOthers {.}}{%
{\protect \APACyear {2007}}%
}]{%
Fortney07a}
\APACinsertmetastar {%
Fortney07a}%
\begin{APACrefauthors}%
{Fortney}, J\BPBI J.%
, {Marley}, M\BPBI S.%
\BCBL {}\ \BBA {} {Barnes}, J\BPBI W.%
\end{APACrefauthors}%
\unskip\
\newblock
\APACrefYearMonthDay{2007}{{\APACmonth{04}}}{}.
\newblock
{\BBOQ}\APACrefatitle {{Planetary Radii across Five Orders of Magnitude in Mass
  and Stellar Insolation: Application to Transits}} {{Planetary Radii across
  Five Orders of Magnitude in Mass and Stellar Insolation: Application to
  Transits}}.{\BBCQ}
\newblock
\APACjournalVolNumPages{\apj}{659}{}{1661-1672}.
\newblock
\begin{APACrefDOI} \doi{10.1086/512120} \end{APACrefDOI}
\PrintBackRefs{\CurrentBib}

\bibitem [\protect \citeauthoryear {%
{Fortney}%
\ \protect \BOthers {.}}{%
{Fortney}%
\ \protect \BOthers {.}}{%
{\protect \APACyear {2013}}%
}]{%
Fortney13}
\APACinsertmetastar {%
Fortney13}%
\begin{APACrefauthors}%
{Fortney}, J\BPBI J.%
, {Mordasini}, C.%
, {Nettelmann}, N.%
, {Kempton}, E\BPBI M\BHBI R.%
, {Greene}, T\BPBI P.%
\BCBL {}\ \BBA {} {Zahnle}, K.%
\end{APACrefauthors}%
\unskip\
\newblock
\APACrefYearMonthDay{2013}{{\APACmonth{09}}}{}.
\newblock
{\BBOQ}\APACrefatitle {{A Framework for Characterizing the Atmospheres of
  Low-mass Low-density Transiting Planets}} {{A Framework for Characterizing
  the Atmospheres of Low-mass Low-density Transiting Planets}}.{\BBCQ}
\newblock
\APACjournalVolNumPages{\apj}{775}{}{80}.
\newblock
\begin{APACrefDOI} \doi{10.1088/0004-637X/775/1/80} \end{APACrefDOI}
\PrintBackRefs{\CurrentBib}

\bibitem [\protect \citeauthoryear {%
{Fortney}%
, {Saumon}%
, {Marley}%
, {Lodders}%
\BCBL {}\ \BBA {} {Freedman}%
}{%
{Fortney}%
\ \protect \BOthers {.}}{%
{\protect \APACyear {2006}}%
}]{%
Fortney06}
\APACinsertmetastar {%
Fortney06}%
\begin{APACrefauthors}%
{Fortney}, J\BPBI J.%
, {Saumon}, D.%
, {Marley}, M\BPBI S.%
, {Lodders}, K.%
\BCBL {}\ \BBA {} {Freedman}, R\BPBI S.%
\end{APACrefauthors}%
\unskip\
\newblock
\APACrefYearMonthDay{2006}{{\APACmonth{05}}}{}.
\newblock
{\BBOQ}\APACrefatitle {{Atmosphere, Interior, and Evolution of the Metal-rich
  Transiting Planet HD 149026B}} {{Atmosphere, Interior, and Evolution of the
  Metal-rich Transiting Planet HD 149026B}}.{\BBCQ}
\newblock
\APACjournalVolNumPages{\apj}{642}{}{495-504}.
\PrintBackRefs{\CurrentBib}

\bibitem [\protect \citeauthoryear {%
{Fortney}%
\ \protect \BOthers {.}}{%
{Fortney}%
\ \protect \BOthers {.}}{%
{\protect \APACyear {2020}}%
}]{%
Fortney20}
\APACinsertmetastar {%
Fortney20}%
\begin{APACrefauthors}%
{Fortney}, J\BPBI J.%
, {Visscher}, C.%
, {Marley}, M\BPBI S.%
, {Hood}, C\BPBI E.%
, {Line}, M\BPBI R.%
, {Thorngren}, D\BPBI P.%
\BDBL {}{Lupu}, R.%
\end{APACrefauthors}%
\unskip\
\newblock
\APACrefYearMonthDay{2020}{{\APACmonth{12}}}{}.
\newblock
{\BBOQ}\APACrefatitle {{Beyond Equilibrium Temperature: How the
  Atmosphere/Interior Connection Affects the Onset of Methane, Ammonia, and
  Clouds in Warm Transiting Giant Planets}} {{Beyond Equilibrium Temperature:
  How the Atmosphere/Interior Connection Affects the Onset of Methane, Ammonia,
  and Clouds in Warm Transiting Giant Planets}}.{\BBCQ}
\newblock
\APACjournalVolNumPages{\aj}{160}{6}{288}.
\newblock
\begin{APACrefDOI} \doi{10.3847/1538-3881/abc5bd} \end{APACrefDOI}
\PrintBackRefs{\CurrentBib}

\bibitem [\protect \citeauthoryear {%
Fromang%
, Leconte%
\BCBL {}\ \BBA {} Heng%
}{%
Fromang%
\ \protect \BOthers {.}}{%
{\protect \APACyear {2016}}%
}]{%
Fromang:2016}
\APACinsertmetastar {%
Fromang:2016}%
\begin{APACrefauthors}%
Fromang, S.%
, Leconte, J.%
\BCBL {}\ \BBA {} Heng, K.%
\end{APACrefauthors}%
\unskip\
\newblock
\APACrefYearMonthDay{2016}{}{}.
\newblock
{\BBOQ}\APACrefatitle {Shear-driven instabilities and shocks in the atmospheres
  of hot {J}upiters} {Shear-driven instabilities and shocks in the atmospheres
  of hot {J}upiters}.{\BBCQ}
\newblock
\APACjournalVolNumPages{Astronomy and Astrophysics}{591}{}{A144}.
\PrintBackRefs{\CurrentBib}

\bibitem [\protect \citeauthoryear {%
Fu%
, Deming%
, Lothringer%
\BCBL {}\ \BBA {} {et al.}%
}{%
Fu%
\ \protect \BOthers {.}}{%
{\protect \APACyear {2020}}%
}]{%
Fu:2020aa}
\APACinsertmetastar {%
Fu:2020aa}%
\begin{APACrefauthors}%
Fu, G.%
, Deming, D.%
, Lothringer, J.%
\BCBL {}\ \BBA {} {et al.}%
\end{APACrefauthors}%
\unskip\
\newblock
\APACrefYearMonthDay{2020}{}{}.
\newblock
{\BBOQ}\APACrefatitle {The Hubble PanCET program: Transit and Eclipse
  Spectroscopy of the Strongly Irradiated Giant Exoplanet WASP-76b} {The hubble
  pancet program: Transit and eclipse spectroscopy of the strongly irradiated
  giant exoplanet wasp-76b}.{\BBCQ}
\newblock
\APACjournalVolNumPages{ArXiv e-prints:2005.02568}{}{}{}.
\PrintBackRefs{\CurrentBib}

\bibitem [\protect \citeauthoryear {%
Gao%
, Thorngren%
, Lee%
\BCBL {}\ \BBA {} {et al.}%
}{%
Gao%
\ \protect \BOthers {.}}{%
{\protect \APACyear {2020}}%
}]{%
Gao:2020aa}
\APACinsertmetastar {%
Gao:2020aa}%
\begin{APACrefauthors}%
Gao, P.%
, Thorngren, D.%
, Lee, G.%
\BCBL {}\ \BBA {} {et al.}%
\end{APACrefauthors}%
\unskip\
\newblock
\APACrefYearMonthDay{2020}{}{}.
\newblock
{\BBOQ}\APACrefatitle {Aerosol composition of hot giant exoplanets dominated by
  silicates and hydrocarbon hazes Show affiliations} {Aerosol composition of
  hot giant exoplanets dominated by silicates and hydrocarbon hazes show
  affiliations}.{\BBCQ}
\newblock
\APACjournalVolNumPages{Nature Astronomy}{}{}{}.
\PrintBackRefs{\CurrentBib}

\bibitem [\protect \citeauthoryear {%
{Garhart}%
\ \protect \BOthers {.}}{%
{Garhart}%
\ \protect \BOthers {.}}{%
{\protect \APACyear {2020}}%
}]{%
Garhart20}
\APACinsertmetastar {%
Garhart20}%
\begin{APACrefauthors}%
{Garhart}, E.%
, {Deming}, D.%
, {Mandell}, A.%
, {Knutson}, H\BPBI A.%
, {Wallack}, N.%
, {Burrows}, A.%
\BDBL {}{Wakeford}, H.%
\end{APACrefauthors}%
\unskip\
\newblock
\APACrefYearMonthDay{2020}{{\APACmonth{04}}}{}.
\newblock
{\BBOQ}\APACrefatitle {{Statistical Characterization of Hot Jupiter Atmospheres
  Using Spitzer's Secondary Eclipses}} {{Statistical Characterization of Hot
  Jupiter Atmospheres Using Spitzer's Secondary Eclipses}}.{\BBCQ}
\newblock
\APACjournalVolNumPages{\aj}{159}{4}{137}.
\newblock
\begin{APACrefDOI} \doi{10.3847/1538-3881/ab6cff} \end{APACrefDOI}
\PrintBackRefs{\CurrentBib}

\bibitem [\protect \citeauthoryear {%
Ginzburg%
\ \BBA {} Sari%
}{%
Ginzburg%
\ \BBA {} Sari%
}{%
{\protect \APACyear {2015}}%
}]{%
Ginzburg:2015}
\APACinsertmetastar {%
Ginzburg:2015}%
\begin{APACrefauthors}%
Ginzburg, S.%
\BCBT {}\ \BBA {} Sari, R.%
\end{APACrefauthors}%
\unskip\
\newblock
\APACrefYearMonthDay{2015}{January}{}.
\newblock
{\BBOQ}\APACrefatitle {Hot-jupiter inflation due to deep energy deposition}
  {Hot-jupiter inflation due to deep energy deposition}.{\BBCQ}
\newblock
\APACjournalVolNumPages{The Astrophysical Journal}{803}{}{111}.
\PrintBackRefs{\CurrentBib}

\bibitem [\protect \citeauthoryear {%
{Ginzburg}%
\ \BBA {} {Sari}%
}{%
{Ginzburg}%
\ \BBA {} {Sari}%
}{%
{\protect \APACyear {2016}}%
}]{%
ginz16}
\APACinsertmetastar {%
ginz16}%
\begin{APACrefauthors}%
{Ginzburg}, S.%
\BCBT {}\ \BBA {} {Sari}, R.%
\end{APACrefauthors}%
\unskip\
\newblock
\APACrefYearMonthDay{2016}{{\APACmonth{03}}}{}.
\newblock
{\BBOQ}\APACrefatitle {{Extended Heat Deposition in Hot Jupiters: Application
  to Ohmic Heating}} {{Extended Heat Deposition in Hot Jupiters: Application to
  Ohmic Heating}}.{\BBCQ}
\newblock
\APACjournalVolNumPages{\apj}{819}{}{116}.
\newblock
\begin{APACrefDOI} \doi{10.3847/0004-637X/819/2/116} \end{APACrefDOI}
\PrintBackRefs{\CurrentBib}

\bibitem [\protect \citeauthoryear {%
Gold%
\ \BBA {} Soter%
}{%
Gold%
\ \BBA {} Soter%
}{%
{\protect \APACyear {1969}}%
}]{%
Gold:1969aa}
\APACinsertmetastar {%
Gold:1969aa}%
\begin{APACrefauthors}%
Gold, T.%
\BCBT {}\ \BBA {} Soter, S.%
\end{APACrefauthors}%
\unskip\
\newblock
\APACrefYearMonthDay{1969}{}{}.
\newblock
{\BBOQ}\APACrefatitle {Atmospheric Tides and the Resonant Rotation of Venus}
  {Atmospheric tides and the resonant rotation of venus}.{\BBCQ}
\newblock
\APACjournalVolNumPages{Icarus}{11}{}{356}.
\PrintBackRefs{\CurrentBib}

\bibitem [\protect \citeauthoryear {%
{Goldreich}%
\ \BBA {} {Tremaine}%
}{%
{Goldreich}%
\ \BBA {} {Tremaine}%
}{%
{\protect \APACyear {1980}}%
}]{%
gold80}
\APACinsertmetastar {%
gold80}%
\begin{APACrefauthors}%
{Goldreich}, P.%
\BCBT {}\ \BBA {} {Tremaine}, S.%
\end{APACrefauthors}%
\unskip\
\newblock
\APACrefYearMonthDay{1980}{{\APACmonth{10}}}{}.
\newblock
{\BBOQ}\APACrefatitle {{Disk-satellite interactions}} {{Disk-satellite
  interactions}}.{\BBCQ}
\newblock
\APACjournalVolNumPages{\apj}{241}{}{425-441}.
\newblock
\begin{APACrefDOI} \doi{10.1086/158356} \end{APACrefDOI}
\PrintBackRefs{\CurrentBib}

\bibitem [\protect \citeauthoryear {%
{Gonzalez}%
}{%
{Gonzalez}%
}{%
{\protect \APACyear {1997}}%
}]{%
gonz97}
\APACinsertmetastar {%
gonz97}%
\begin{APACrefauthors}%
{Gonzalez}, G.%
\end{APACrefauthors}%
\unskip\
\newblock
\APACrefYearMonthDay{1997}{{\APACmonth{02}}}{}.
\newblock
{\BBOQ}\APACrefatitle {{The stellar metallicity-giant planet connection}} {{The
  stellar metallicity-giant planet connection}}.{\BBCQ}
\newblock
\APACjournalVolNumPages{\mnras}{285}{}{403-412}.
\newblock
\begin{APACrefDOI} \doi{10.1093/mnras/285.2.403} \end{APACrefDOI}
\PrintBackRefs{\CurrentBib}

\bibitem [\protect \citeauthoryear {%
Grunblatt%
, Huber%
, Gaidos%
\BCBL {}\ \BBA {} {et al.}%
}{%
Grunblatt%
\ \protect \BOthers {.}}{%
{\protect \APACyear {2016}}%
}]{%
Grunblatt2016}
\APACinsertmetastar {%
Grunblatt2016}%
\begin{APACrefauthors}%
Grunblatt, S.%
, Huber, D.%
, Gaidos, E.%
\BCBL {}\ \BBA {} {et al.}%
\end{APACrefauthors}%
\unskip\
\newblock
\APACrefYearMonthDay{2016}{}{}.
\newblock
{\BBOQ}\APACrefatitle {K2-97b : A ( re- ? ) inflated planet orbiting a red
  giant star} {K2-97b : A ( re- ? ) inflated planet orbiting a red giant
  star}.{\BBCQ}
\newblock
\APACjournalVolNumPages{The Astronomical Journal}{152}{6}{1}.
\PrintBackRefs{\CurrentBib}

\bibitem [\protect \citeauthoryear {%
Grunblatt%
\ \protect \BOthers {.}}{%
Grunblatt%
\ \protect \BOthers {.}}{%
{\protect \APACyear {2019}}%
}]{%
Grunblatt:2019aa}
\APACinsertmetastar {%
Grunblatt:2019aa}%
\begin{APACrefauthors}%
Grunblatt, S.%
, Huber, D.%
, Gaidos, E.%
, Hon, M.%
, Zinn, J.%
\BCBL {}\ \BBA {} Stello, D.%
\end{APACrefauthors}%
\unskip\
\newblock
\APACrefYearMonthDay{2019}{}{}.
\newblock
{\BBOQ}\APACrefatitle {GIANT PLANET OCCURRENCE WITHIN 0.2 AU OF LOW-LUMINOSITY
  RED GIANT BRANCH STARS WITH K2} {Giant planet occurrence within 0.2 au of
  low-luminosity red giant branch stars with k2}.{\BBCQ}
\newblock
\APACjournalVolNumPages{The Astronomical Journal}{158}{}{227}.
\PrintBackRefs{\CurrentBib}

\bibitem [\protect \citeauthoryear {%
Grunblatt%
\ \protect \BOthers {.}}{%
Grunblatt%
\ \protect \BOthers {.}}{%
{\protect \APACyear {2017}}%
}]{%
Grunblatt:2017aa}
\APACinsertmetastar {%
Grunblatt:2017aa}%
\begin{APACrefauthors}%
Grunblatt, S.%
, Huber, D.%
, Gaidos, E.%
, Lopez, E.%
, Howard, A.%
, Isaacson, H.%
\BDBL {}Lin, D.%
\end{APACrefauthors}%
\unskip\
\newblock
\APACrefYearMonthDay{2017}{}{}.
\newblock
{\BBOQ}\APACrefatitle {Seeing Double with {K2}: Testing Re-inflation with Two
  Remarkably Similar Planets around Red Giant Branch Stars} {Seeing double with
  {K2}: Testing re-inflation with two remarkably similar planets around red
  giant branch stars}.{\BBCQ}
\newblock
\APACjournalVolNumPages{The Astronomical Journal}{154}{}{254}.
\PrintBackRefs{\CurrentBib}

\bibitem [\protect \citeauthoryear {%
Gu%
, Bodenheimer%
\BCBL {}\ \BBA {} Lin%
}{%
Gu%
\ \protect \BOthers {.}}{%
{\protect \APACyear {2004}}%
}]{%
Gu:2004aa}
\APACinsertmetastar {%
Gu:2004aa}%
\begin{APACrefauthors}%
Gu, P.%
, Bodenheimer, P.%
\BCBL {}\ \BBA {} Lin, D.%
\end{APACrefauthors}%
\unskip\
\newblock
\APACrefYearMonthDay{2004}{}{}.
\newblock
{\BBOQ}\APACrefatitle {The Internal Structural Adjustment Due to Tidal Heating
  of Short-Period Inflated Giant Planets} {The internal structural adjustment
  due to tidal heating of short-period inflated giant planets}.{\BBCQ}
\newblock
\APACjournalVolNumPages{The Astrophysical Journal}{608}{}{1076}.
\PrintBackRefs{\CurrentBib}

\bibitem [\protect \citeauthoryear {%
Gu%
, Lin%
\BCBL {}\ \BBA {} Bodenheimer%
}{%
Gu%
\ \protect \BOthers {.}}{%
{\protect \APACyear {2003}}%
}]{%
Gu:2003aa}
\APACinsertmetastar {%
Gu:2003aa}%
\begin{APACrefauthors}%
Gu, P.%
, Lin, D.%
\BCBL {}\ \BBA {} Bodenheimer, P.%
\end{APACrefauthors}%
\unskip\
\newblock
\APACrefYearMonthDay{2003}{}{}.
\newblock
{\BBOQ}\APACrefatitle {The Effect of Tidal Inflation Instability on the Mass
  and Dynamical Evolution of Extrasolar Planets with Ultrashort Periods} {The
  effect of tidal inflation instability on the mass and dynamical evolution of
  extrasolar planets with ultrashort periods}.{\BBCQ}
\newblock
\APACjournalVolNumPages{The Astrophysical Journal}{588}{}{509}.
\PrintBackRefs{\CurrentBib}

\bibitem [\protect \citeauthoryear {%
Gu%
, Peng%
\BCBL {}\ \BBA {} Yen%
}{%
Gu%
\ \protect \BOthers {.}}{%
{\protect \APACyear {2019}}%
}]{%
Gu:2019aa}
\APACinsertmetastar {%
Gu:2019aa}%
\begin{APACrefauthors}%
Gu, P.%
, Peng, D.%
\BCBL {}\ \BBA {} Yen, C.%
\end{APACrefauthors}%
\unskip\
\newblock
\APACrefYearMonthDay{2019}{}{}.
\newblock
{\BBOQ}\APACrefatitle {Modeling the Thermal Bulge of a Hot Jupiter with the
  Two-stream Approximation} {Modeling the thermal bulge of a hot jupiter with
  the two-stream approximation}.{\BBCQ}
\newblock
\APACjournalVolNumPages{The Astrophysical Journal}{887}{}{228}.
\PrintBackRefs{\CurrentBib}

\bibitem [\protect \citeauthoryear {%
{Guillot}%
, {Burrows}%
, {Hubbard}%
, {Lunine}%
\BCBL {}\ \BBA {} {Saumon}%
}{%
{Guillot}%
\ \protect \BOthers {.}}{%
{\protect \APACyear {1996}}%
}]{%
Guillot96}
\APACinsertmetastar {%
Guillot96}%
\begin{APACrefauthors}%
{Guillot}, T.%
, {Burrows}, A.%
, {Hubbard}, W\BPBI B.%
, {Lunine}, J\BPBI I.%
\BCBL {}\ \BBA {} {Saumon}, D.%
\end{APACrefauthors}%
\unskip\
\newblock
\APACrefYearMonthDay{1996}{{\APACmonth{03}}}{}.
\newblock
{\BBOQ}\APACrefatitle {{Giant Planets at Small Orbital Distances}} {{Giant
  Planets at Small Orbital Distances}}.{\BBCQ}
\newblock
\APACjournalVolNumPages{\apjl}{459}{}{L35-L38}.
\PrintBackRefs{\CurrentBib}

\bibitem [\protect \citeauthoryear {%
{Guillot}%
\ \protect \BOthers {.}}{%
{Guillot}%
\ \protect \BOthers {.}}{%
{\protect \APACyear {2006}}%
}]{%
Guillot06}
\APACinsertmetastar {%
Guillot06}%
\begin{APACrefauthors}%
{Guillot}, T.%
, {Santos}, N\BPBI C.%
, {Pont}, F.%
, {Iro}, N.%
, {Melo}, C.%
\BCBL {}\ \BBA {} {Ribas}, I.%
\end{APACrefauthors}%
\unskip\
\newblock
\APACrefYearMonthDay{2006}{{\APACmonth{07}}}{}.
\newblock
{\BBOQ}\APACrefatitle {{A correlation between the heavy element content of
  transiting extrasolar planets and the metallicity of their parent stars}} {{A
  correlation between the heavy element content of transiting extrasolar
  planets and the metallicity of their parent stars}}.{\BBCQ}
\newblock
\APACjournalVolNumPages{\aap}{453}{}{L21-L24}.
\newblock
\begin{APACrefDOI} \doi{10.1051/0004-6361:20065476} \end{APACrefDOI}
\PrintBackRefs{\CurrentBib}

\bibitem [\protect \citeauthoryear {%
Guillot%
\ \BBA {} Showman%
}{%
Guillot%
\ \BBA {} Showman%
}{%
{\protect \APACyear {2002}}%
}]{%
Guillot_2002}
\APACinsertmetastar {%
Guillot_2002}%
\begin{APACrefauthors}%
Guillot, T.%
\BCBT {}\ \BBA {} Showman, A.%
\end{APACrefauthors}%
\unskip\
\newblock
\APACrefYearMonthDay{2002}{}{}.
\newblock
{\BBOQ}\APACrefatitle {Evolution of ``51 {P}egasus b-like'' planets} {Evolution
  of ``51 {P}egasus b-like'' planets}.{\BBCQ}
\newblock
\APACjournalVolNumPages{Astronomy and Astrophysics}{385}{}{156}.
\PrintBackRefs{\CurrentBib}

\bibitem [\protect \citeauthoryear {%
{Guo}%
\ \protect \BOthers {.}}{%
{Guo}%
\ \protect \BOthers {.}}{%
{\protect \APACyear {2017}}%
}]{%
guo17}
\APACinsertmetastar {%
guo17}%
\begin{APACrefauthors}%
{Guo}, X.%
, {Johnson}, J\BPBI A.%
, {Mann}, A\BPBI W.%
, {Kraus}, A\BPBI L.%
, {Curtis}, J\BPBI L.%
\BCBL {}\ \BBA {} {Latham}, D\BPBI W.%
\end{APACrefauthors}%
\unskip\
\newblock
\APACrefYearMonthDay{2017}{{\APACmonth{03}}}{}.
\newblock
{\BBOQ}\APACrefatitle {{The Metallicity Distribution and Hot Jupiter Rate of
  the Kepler Field: Hectochelle High-resolution Spectroscopy for 776 Kepler
  Target Stars}} {{The Metallicity Distribution and Hot Jupiter Rate of the
  Kepler Field: Hectochelle High-resolution Spectroscopy for 776 Kepler Target
  Stars}}.{\BBCQ}
\newblock
\APACjournalVolNumPages{\apj}{838}{}{25}.
\newblock
\begin{APACrefDOI} \doi{10.3847/1538-4357/aa6004} \end{APACrefDOI}
\PrintBackRefs{\CurrentBib}

\bibitem [\protect \citeauthoryear {%
Hammond%
\ \BBA {} Pierrehumbert%
}{%
Hammond%
\ \BBA {} Pierrehumbert%
}{%
{\protect \APACyear {2018}}%
}]{%
Hammond:2018aa}
\APACinsertmetastar {%
Hammond:2018aa}%
\begin{APACrefauthors}%
Hammond, M.%
\BCBT {}\ \BBA {} Pierrehumbert, R.%
\end{APACrefauthors}%
\unskip\
\newblock
\APACrefYearMonthDay{2018}{}{}.
\newblock
{\BBOQ}\APACrefatitle {Wave-Mean Flow Interactions in the Atmospheric
  Circulation of Tidally Locked Planets} {Wave-mean flow interactions in the
  atmospheric circulation of tidally locked planets}.{\BBCQ}
\newblock
\APACjournalVolNumPages{The Astrophysical Journal}{869}{}{65}.
\PrintBackRefs{\CurrentBib}

\bibitem [\protect \citeauthoryear {%
{Hartman}%
\ \protect \BOthers {.}}{%
{Hartman}%
\ \protect \BOthers {.}}{%
{\protect \APACyear {2016}}%
}]{%
hart16}
\APACinsertmetastar {%
hart16}%
\begin{APACrefauthors}%
{Hartman}, J\BPBI D.%
, {Bakos}, G\BPBI {\'A}.%
, {Bhatti}, W.%
, {Penev}, K.%
, {Bieryla}, A.%
, {Latham}, D\BPBI W.%
\BDBL {}{S{\'a}ri}, P.%
\end{APACrefauthors}%
\unskip\
\newblock
\APACrefYearMonthDay{2016}{{\APACmonth{12}}}{}.
\newblock
{\BBOQ}\APACrefatitle {{HAT-P-65b and HAT-P-66b: Two Transiting Inflated Hot
  Jupiters and Observational Evidence for the Reinflation of Close-in Giant
  Planets}} {{HAT-P-65b and HAT-P-66b: Two Transiting Inflated Hot Jupiters and
  Observational Evidence for the Reinflation of Close-in Giant
  Planets}}.{\BBCQ}
\newblock
\APACjournalVolNumPages{\aj}{152}{}{182}.
\newblock
\begin{APACrefDOI} \doi{10.3847/0004-6256/152/6/182} \end{APACrefDOI}
\PrintBackRefs{\CurrentBib}

\bibitem [\protect \citeauthoryear {%
Haynes%
, Mandell%
, Madhusudhan%
, Deming%
\BCBL {}\ \BBA {} Knutson%
}{%
Haynes%
\ \protect \BOthers {.}}{%
{\protect \APACyear {2015}}%
}]{%
Haynes:2015}
\APACinsertmetastar {%
Haynes:2015}%
\begin{APACrefauthors}%
Haynes, K.%
, Mandell, A.%
, Madhusudhan, N.%
, Deming, D.%
\BCBL {}\ \BBA {} Knutson, H.%
\end{APACrefauthors}%
\unskip\
\newblock
\APACrefYearMonthDay{2015}{}{}.
\newblock
{\BBOQ}\APACrefatitle {Spectroscopic evidence for a temperature inversion in
  the dayside atmosphere of the hot {J}upiter {WASP}-33b} {Spectroscopic
  evidence for a temperature inversion in the dayside atmosphere of the hot
  {J}upiter {WASP}-33b}.{\BBCQ}
\newblock
\APACjournalVolNumPages{The Astrophysical Journal}{806}{}{146-158}.
\PrintBackRefs{\CurrentBib}

\bibitem [\protect \citeauthoryear {%
{Helling}%
}{%
{Helling}%
}{%
{\protect \APACyear {2019}}%
}]{%
Helling19}
\APACinsertmetastar {%
Helling19}%
\begin{APACrefauthors}%
{Helling}, C.%
\end{APACrefauthors}%
\unskip\
\newblock
\APACrefYearMonthDay{2019}{{\APACmonth{05}}}{}.
\newblock
{\BBOQ}\APACrefatitle {{Exoplanet Clouds}} {{Exoplanet Clouds}}.{\BBCQ}
\newblock
\APACjournalVolNumPages{Annual Review of Earth and Planetary
  Sciences}{47}{}{583-606}.
\newblock
\begin{APACrefDOI} \doi{10.1146/annurev-earth-053018-060401} \end{APACrefDOI}
\PrintBackRefs{\CurrentBib}

\bibitem [\protect \citeauthoryear {%
{Helling}%
\ \BBA {} {Casewell}%
}{%
{Helling}%
\ \BBA {} {Casewell}%
}{%
{\protect \APACyear {2014}}%
}]{%
Helling14}
\APACinsertmetastar {%
Helling14}%
\begin{APACrefauthors}%
{Helling}, C.%
\BCBT {}\ \BBA {} {Casewell}, S.%
\end{APACrefauthors}%
\unskip\
\newblock
\APACrefYearMonthDay{2014}{{\APACmonth{11}}}{}.
\newblock
{\BBOQ}\APACrefatitle {{Atmospheres of brown dwarfs}} {{Atmospheres of brown
  dwarfs}}.{\BBCQ}
\newblock
\APACjournalVolNumPages{\aapr}{22}{}{80}.
\newblock
\begin{APACrefDOI} \doi{10.1007/s00159-014-0080-0} \end{APACrefDOI}
\PrintBackRefs{\CurrentBib}

\bibitem [\protect \citeauthoryear {%
Helling%
\ \protect \BOthers {.}}{%
Helling%
\ \protect \BOthers {.}}{%
{\protect \APACyear {2016}}%
}]{%
Helling:2016}
\APACinsertmetastar {%
Helling:2016}%
\begin{APACrefauthors}%
Helling, C.%
, Lee, I., G.; Dobbs-Dixon%
, Mayne, N.%
, Amundsen, D\BPBI S.%
, Khaimova, J.%
, Unger, A\BPBI A.%
\BDBL {}Smith, C.%
\end{APACrefauthors}%
\unskip\
\newblock
\APACrefYearMonthDay{2016}{}{}.
\newblock
{\BBOQ}\APACrefatitle {The mineral clouds on HD 209458b and HD 189733b} {The
  mineral clouds on hd 209458b and hd 189733b}.{\BBCQ}
\newblock
\APACjournalVolNumPages{Monthly Notices of the Royal Astronomical
  Society}{460}{}{855}.
\PrintBackRefs{\CurrentBib}

\bibitem [\protect \citeauthoryear {%
Heng%
}{%
Heng%
}{%
{\protect \APACyear {2012}}%
}]{%
Heng:2012a}
\APACinsertmetastar {%
Heng:2012a}%
\begin{APACrefauthors}%
Heng, K.%
\end{APACrefauthors}%
\unskip\
\newblock
\APACrefYearMonthDay{2012}{}{}.
\newblock
{\BBOQ}\APACrefatitle {On the existence of shocks in irradiated planetary
  atmospheres} {On the existence of shocks in irradiated planetary
  atmospheres}.{\BBCQ}
\newblock
\APACjournalVolNumPages{The Astrophysical Journal Letters}{761}{}{L1}.
\PrintBackRefs{\CurrentBib}

\bibitem [\protect \citeauthoryear {%
{Heng}%
\ \BBA {} {Demory}%
}{%
{Heng}%
\ \BBA {} {Demory}%
}{%
{\protect \APACyear {2013}}%
}]{%
Heng13}
\APACinsertmetastar {%
Heng13}%
\begin{APACrefauthors}%
{Heng}, K.%
\BCBT {}\ \BBA {} {Demory}, B\BHBI O.%
\end{APACrefauthors}%
\unskip\
\newblock
\APACrefYearMonthDay{2013}{{\APACmonth{11}}}{}.
\newblock
{\BBOQ}\APACrefatitle {{Understanding Trends Associated with Clouds in
  Irradiated Exoplanets}} {{Understanding Trends Associated with Clouds in
  Irradiated Exoplanets}}.{\BBCQ}
\newblock
\APACjournalVolNumPages{\apj}{777}{2}{100}.
\newblock
\begin{APACrefDOI} \doi{10.1088/0004-637X/777/2/100} \end{APACrefDOI}
\PrintBackRefs{\CurrentBib}

\bibitem [\protect \citeauthoryear {%
Heng%
, Frierson%
\BCBL {}\ \BBA {} Phillips%
}{%
Heng%
\ \protect \BOthers {.}}{%
{\protect \APACyear {2011}}%
}]{%
Heng:2011a}
\APACinsertmetastar {%
Heng:2011a}%
\begin{APACrefauthors}%
Heng, K.%
, Frierson, D.%
\BCBL {}\ \BBA {} Phillips, P.%
\end{APACrefauthors}%
\unskip\
\newblock
\APACrefYearMonthDay{2011}{}{}.
\newblock
{\BBOQ}\APACrefatitle {Atmospheric circulation of tidally locked exoplanets:
  II. Dual-band radiative transfer and convective adjustment} {Atmospheric
  circulation of tidally locked exoplanets: Ii. dual-band radiative transfer
  and convective adjustment}.{\BBCQ}
\newblock
\APACjournalVolNumPages{Monthly Notices of the Royal Astronomical
  Society}{418}{}{2669}.
\PrintBackRefs{\CurrentBib}

\bibitem [\protect \citeauthoryear {%
{Heng}%
\ \BBA {} {Showman}%
}{%
{Heng}%
\ \BBA {} {Showman}%
}{%
{\protect \APACyear {2015}}%
}]{%
Heng15}
\APACinsertmetastar {%
Heng15}%
\begin{APACrefauthors}%
{Heng}, K.%
\BCBT {}\ \BBA {} {Showman}, A\BPBI P.%
\end{APACrefauthors}%
\unskip\
\newblock
\APACrefYearMonthDay{2015}{{\APACmonth{05}}}{}.
\newblock
{\BBOQ}\APACrefatitle {{Atmospheric Dynamics of Hot Exoplanets}} {{Atmospheric
  Dynamics of Hot Exoplanets}}.{\BBCQ}
\newblock
\APACjournalVolNumPages{Annual Review of Earth and Planetary
  Sciences}{43}{}{509-540}.
\newblock
\begin{APACrefDOI} \doi{10.1146/annurev-earth-060614-105146} \end{APACrefDOI}
\PrintBackRefs{\CurrentBib}

\bibitem [\protect \citeauthoryear {%
{Henry}%
, {Marcy}%
, {Butler}%
\BCBL {}\ \BBA {} {Vogt}%
}{%
{Henry}%
\ \protect \BOthers {.}}{%
{\protect \APACyear {2000}}%
}]{%
Henry00}
\APACinsertmetastar {%
Henry00}%
\begin{APACrefauthors}%
{Henry}, G\BPBI W.%
, {Marcy}, G\BPBI W.%
, {Butler}, R\BPBI P.%
\BCBL {}\ \BBA {} {Vogt}, S\BPBI S.%
\end{APACrefauthors}%
\unskip\
\newblock
\APACrefYearMonthDay{2000}{{\APACmonth{01}}}{}.
\newblock
{\BBOQ}\APACrefatitle {{A Transiting ``51 Peg-like'' Planet}} {{A Transiting
  ``51 Peg-like'' Planet}}.{\BBCQ}
\newblock
\APACjournalVolNumPages{\apjl}{529}{}{L41--L44}.
\PrintBackRefs{\CurrentBib}

\bibitem [\protect \citeauthoryear {%
Hindle%
, Bushby%
\BCBL {}\ \BBA {} Rogers%
}{%
Hindle%
\ \protect \BOthers {.}}{%
{\protect \APACyear {2019}}%
}]{%
Hindle:2019aa}
\APACinsertmetastar {%
Hindle:2019aa}%
\begin{APACrefauthors}%
Hindle, A.%
, Bushby, P.%
\BCBL {}\ \BBA {} Rogers, T.%
\end{APACrefauthors}%
\unskip\
\newblock
\APACrefYearMonthDay{2019}{}{}.
\newblock
{\BBOQ}\APACrefatitle {Shallow-water magnetohydrodynamics for westward hotspots
  on hot Jupiters} {Shallow-water magnetohydrodynamics for westward hotspots on
  hot jupiters}.{\BBCQ}
\newblock
\APACjournalVolNumPages{The Astrophysical Journal Letters}{872}{}{L27}.
\PrintBackRefs{\CurrentBib}

\bibitem [\protect \citeauthoryear {%
{Hobbs}%
, {Shorttle}%
, {Madhusudhan}%
\BCBL {}\ \BBA {} {Rimmer}%
}{%
{Hobbs}%
\ \protect \BOthers {.}}{%
{\protect \APACyear {2019}}%
}]{%
Hobbs19}
\APACinsertmetastar {%
Hobbs19}%
\begin{APACrefauthors}%
{Hobbs}, R.%
, {Shorttle}, O.%
, {Madhusudhan}, N.%
\BCBL {}\ \BBA {} {Rimmer}, P.%
\end{APACrefauthors}%
\unskip\
\newblock
\APACrefYearMonthDay{2019}{{\APACmonth{08}}}{}.
\newblock
{\BBOQ}\APACrefatitle {{A chemical kinetics code for modelling exoplanetary
  atmospheres}} {{A chemical kinetics code for modelling exoplanetary
  atmospheres}}.{\BBCQ}
\newblock
\APACjournalVolNumPages{\mnras}{487}{2}{2242-2261}.
\newblock
\begin{APACrefDOI} \doi{10.1093/mnras/stz1333} \end{APACrefDOI}
\PrintBackRefs{\CurrentBib}

\bibitem [\protect \citeauthoryear {%
Hoeijmakers%
, Ehrenreich%
, Heng%
\BCBL {}\ \BBA {} {et al.}%
}{%
Hoeijmakers%
\ \protect \BOthers {.}}{%
{\protect \APACyear {2018}}%
}]{%
Hoeijmakers:2018aa}
\APACinsertmetastar {%
Hoeijmakers:2018aa}%
\begin{APACrefauthors}%
Hoeijmakers, H.%
, Ehrenreich, D.%
, Heng, K.%
\BCBL {}\ \BBA {} {et al.}%
\end{APACrefauthors}%
\unskip\
\newblock
\APACrefYearMonthDay{2018}{}{}.
\newblock
{\BBOQ}\APACrefatitle {Atomic iron and titanium in the atmosphere of the
  exoplanet KELT-9b} {Atomic iron and titanium in the atmosphere of the
  exoplanet kelt-9b}.{\BBCQ}
\newblock
\APACjournalVolNumPages{Nature}{560}{}{453}.
\PrintBackRefs{\CurrentBib}

\bibitem [\protect \citeauthoryear {%
Hoeijmakers%
, Seidel%
, Pino%
\BCBL {}\ \BBA {} {et al.}%
}{%
Hoeijmakers%
\ \protect \BOthers {.}}{%
{\protect \APACyear {2020}}%
}]{%
Hoeijmakers:2020aa}
\APACinsertmetastar {%
Hoeijmakers:2020aa}%
\begin{APACrefauthors}%
Hoeijmakers, H.%
, Seidel, J.%
, Pino, L.%
\BCBL {}\ \BBA {} {et al.}%
\end{APACrefauthors}%
\unskip\
\newblock
\APACrefYearMonthDay{2020}{}{}.
\newblock

\newblock
\APACjournalVolNumPages{Astronomy {\&} Astrophysics}{641}{}{A123}.
\PrintBackRefs{\CurrentBib}

\bibitem [\protect \citeauthoryear {%
Holton%
\ \BBA {} Hakim%
}{%
Holton%
\ \BBA {} Hakim%
}{%
{\protect \APACyear {2013}}%
}]{%
Holton:2013}
\APACinsertmetastar {%
Holton:2013}%
\begin{APACrefauthors}%
Holton, J.%
\BCBT {}\ \BBA {} Hakim, G.%
\end{APACrefauthors}%
\unskip\
\newblock
\APACrefYear{2013}.
\newblock
\APACrefbtitle {An introduction to dynamic meteorology} {An introduction to
  dynamic meteorology}\ (\PrintOrdinal{5th}\ \BEd).
\newblock
\APACaddressPublisher{New York, NY}{Academic Press}.
\PrintBackRefs{\CurrentBib}

\bibitem [\protect \citeauthoryear {%
{Howard}%
\ \protect \BOthers {.}}{%
{Howard}%
\ \protect \BOthers {.}}{%
{\protect \APACyear {2012}}%
{\protect \APACexlab {{\protect \BCnt {1}}}}}]{%
Howard12}
\APACinsertmetastar {%
Howard12}%
\begin{APACrefauthors}%
{Howard}, A\BPBI W.%
, {Marcy}, G\BPBI W.%
, {Bryson}, S\BPBI T.%
, {Jenkins}, J\BPBI M.%
, {Rowe}, J\BPBI F.%
, {Batalha}, N\BPBI M.%
\BDBL {}{MacQueen}, P\BPBI J.%
\end{APACrefauthors}%
\unskip\
\newblock
\APACrefYearMonthDay{2012{\protect \BCnt {1}}}{{\APACmonth{08}}}{}.
\newblock
{\BBOQ}\APACrefatitle {{Planet Occurrence within 0.25 AU of Solar-type Stars
  from Kepler}} {{Planet Occurrence within 0.25 AU of Solar-type Stars from
  Kepler}}.{\BBCQ}
\newblock
\APACjournalVolNumPages{\apjs}{201}{}{15}.
\newblock
\begin{APACrefDOI} \doi{10.1088/0067-0049/201/2/15} \end{APACrefDOI}
\PrintBackRefs{\CurrentBib}

\bibitem [\protect \citeauthoryear {%
{Howard}%
\ \protect \BOthers {.}}{%
{Howard}%
\ \protect \BOthers {.}}{%
{\protect \APACyear {2012}}%
{\protect \APACexlab {{\protect \BCnt {2}}}}}]{%
howa12}
\APACinsertmetastar {%
howa12}%
\begin{APACrefauthors}%
{Howard}, A\BPBI W.%
, {Marcy}, G\BPBI W.%
, {Bryson}, S\BPBI T.%
, {Jenkins}, J\BPBI M.%
, {Rowe}, J\BPBI F.%
, {Batalha}, N\BPBI M.%
\BDBL {}{MacQueen}, P\BPBI J.%
\end{APACrefauthors}%
\unskip\
\newblock
\APACrefYearMonthDay{2012{\protect \BCnt {2}}}{{\APACmonth{08}}}{}.
\newblock
{\BBOQ}\APACrefatitle {{Planet Occurrence within 0.25 AU of Solar-type Stars
  from Kepler}} {{Planet Occurrence within 0.25 AU of Solar-type Stars from
  Kepler}}.{\BBCQ}
\newblock
\APACjournalVolNumPages{\apjs}{201}{}{15}.
\newblock
\begin{APACrefDOI} \doi{10.1088/0067-0049/201/2/15} \end{APACrefDOI}
\PrintBackRefs{\CurrentBib}

\bibitem [\protect \citeauthoryear {%
Hu%
, Demory%
, Seager%
, Lewis%
\BCBL {}\ \BBA {} Showman%
}{%
Hu%
\ \protect \BOthers {.}}{%
{\protect \APACyear {2015}}%
}]{%
Hu:2015}
\APACinsertmetastar {%
Hu:2015}%
\begin{APACrefauthors}%
Hu, R.%
, Demory, B.%
, Seager, S.%
, Lewis, N.%
\BCBL {}\ \BBA {} Showman, A.%
\end{APACrefauthors}%
\unskip\
\newblock
\APACrefYearMonthDay{2015}{}{}.
\newblock
{\BBOQ}\APACrefatitle {A semi-analytical model of visible-wavelength phase
  curves of exoplanets and applications to {K}epler-7b and {K}epler-10b} {A
  semi-analytical model of visible-wavelength phase curves of exoplanets and
  applications to {K}epler-7b and {K}epler-10b}.{\BBCQ}
\newblock
\APACjournalVolNumPages{The Astrophysical Journal}{802}{}{51}.
\PrintBackRefs{\CurrentBib}

\bibitem [\protect \citeauthoryear {%
Huang%
\ \BBA {} Cumming%
}{%
Huang%
\ \BBA {} Cumming%
}{%
{\protect \APACyear {2012}}%
}]{%
Huang_2012}
\APACinsertmetastar {%
Huang_2012}%
\begin{APACrefauthors}%
Huang, X.%
\BCBT {}\ \BBA {} Cumming, A.%
\end{APACrefauthors}%
\unskip\
\newblock
\APACrefYearMonthDay{2012}{}{}.
\newblock
{\BBOQ}\APACrefatitle {Ohmic dissipation in the interiors of hot {J}upiters}
  {Ohmic dissipation in the interiors of hot {J}upiters}.{\BBCQ}
\newblock
\APACjournalVolNumPages{The Astrophysical Journal}{757}{}{47}.
\PrintBackRefs{\CurrentBib}

\bibitem [\protect \citeauthoryear {%
{Hubbard}%
}{%
{Hubbard}%
}{%
{\protect \APACyear {1977}}%
}]{%
Hubbard77}
\APACinsertmetastar {%
Hubbard77}%
\begin{APACrefauthors}%
{Hubbard}, W\BPBI B.%
\end{APACrefauthors}%
\unskip\
\newblock
\APACrefYearMonthDay{1977}{{\APACmonth{02}}}{}.
\newblock
{\BBOQ}\APACrefatitle {{The Jovian surface condition and cooling rate}} {{The
  Jovian surface condition and cooling rate}}.{\BBCQ}
\newblock
\APACjournalVolNumPages{Icarus}{30}{}{305-310}.
\PrintBackRefs{\CurrentBib}

\bibitem [\protect \citeauthoryear {%
{Hubbard}%
\ \protect \BOthers {.}}{%
{Hubbard}%
\ \protect \BOthers {.}}{%
{\protect \APACyear {2001}}%
}]{%
Hubbard01}
\APACinsertmetastar {%
Hubbard01}%
\begin{APACrefauthors}%
{Hubbard}, W\BPBI B.%
, {Fortney}, J\BPBI J.%
, {Lunine}, J\BPBI I.%
, {Burrows}, A.%
, {Sudarsky}, D.%
\BCBL {}\ \BBA {} {Pinto}, P.%
\end{APACrefauthors}%
\unskip\
\newblock
\APACrefYearMonthDay{2001}{{\APACmonth{10}}}{}.
\newblock
{\BBOQ}\APACrefatitle {{Theory of Extrasolar Giant Planet Transits}} {{Theory
  of Extrasolar Giant Planet Transits}}.{\BBCQ}
\newblock
\APACjournalVolNumPages{\apj}{560}{}{413-419}.
\newblock
\begin{APACrefURL}
  \url{http://adsabs.harvard.edu/cgi-bin/nph-bib_query?bibcode=2001ApJ...56
  0..413H&db_key=AST} \end{APACrefURL}
\PrintBackRefs{\CurrentBib}

\bibitem [\protect \citeauthoryear {%
{Hubeny}%
, {Burrows}%
\BCBL {}\ \BBA {} {Sudarsky}%
}{%
{Hubeny}%
\ \protect \BOthers {.}}{%
{\protect \APACyear {2003}}%
}]{%
Hubeny03}
\APACinsertmetastar {%
Hubeny03}%
\begin{APACrefauthors}%
{Hubeny}, I.%
, {Burrows}, A.%
\BCBL {}\ \BBA {} {Sudarsky}, D.%
\end{APACrefauthors}%
\unskip\
\newblock
\APACrefYearMonthDay{2003}{{\APACmonth{09}}}{}.
\newblock
{\BBOQ}\APACrefatitle {{A Possible Bifurcation in Atmospheres of Strongly
  Irradiated Stars and Planets}} {{A Possible Bifurcation in Atmospheres of
  Strongly Irradiated Stars and Planets}}.{\BBCQ}
\newblock
\APACjournalVolNumPages{\apj}{594}{}{1011-1018}.
\PrintBackRefs{\CurrentBib}

\bibitem [\protect \citeauthoryear {%
Huitson%
, Sing%
, Pont%
\BCBL {}\ \BBA {} {et al.}%
}{%
Huitson%
\ \protect \BOthers {.}}{%
{\protect \APACyear {2013}}%
}]{%
Huitson:2013aa}
\APACinsertmetastar {%
Huitson:2013aa}%
\begin{APACrefauthors}%
Huitson, C.%
, Sing, D.%
, Pont, F.%
\BCBL {}\ \BBA {} {et al.}%
\end{APACrefauthors}%
\unskip\
\newblock
\APACrefYearMonthDay{2013}{}{}.
\newblock
{\BBOQ}\APACrefatitle {An HST optical-to-near-IR transmission spectrum of the
  hot Jupiter WASP-19b: detection of atmospheric water and likely absence of
  TiO} {An hst optical-to-near-ir transmission spectrum of the hot jupiter
  wasp-19b: detection of atmospheric water and likely absence of tio}.{\BBCQ}
\newblock
\APACjournalVolNumPages{Monthly Notices of the Royal Astronomical
  Society}{434}{}{3252}.
\PrintBackRefs{\CurrentBib}

\bibitem [\protect \citeauthoryear {%
{Hut}%
}{%
{Hut}%
}{%
{\protect \APACyear {1981}}%
}]{%
hut81}
\APACinsertmetastar {%
hut81}%
\begin{APACrefauthors}%
{Hut}, P.%
\end{APACrefauthors}%
\unskip\
\newblock
\APACrefYearMonthDay{1981}{{\APACmonth{06}}}{}.
\newblock
{\BBOQ}\APACrefatitle {{Tidal evolution in close binary systems}} {{Tidal
  evolution in close binary systems}}.{\BBCQ}
\newblock
\APACjournalVolNumPages{\aap}{99}{}{126-140}.
\PrintBackRefs{\CurrentBib}

\bibitem [\protect \citeauthoryear {%
Ibgui%
\ \BBA {} Burrows%
}{%
Ibgui%
\ \BBA {} Burrows%
}{%
{\protect \APACyear {2009}}%
}]{%
Ibgui:2009}
\APACinsertmetastar {%
Ibgui:2009}%
\begin{APACrefauthors}%
Ibgui, L.%
\BCBT {}\ \BBA {} Burrows, A.%
\end{APACrefauthors}%
\unskip\
\newblock
\APACrefYearMonthDay{2009}{}{}.
\newblock
{\BBOQ}\APACrefatitle {Coupled evolution with tides of the radius and orbit of
  transiting giant planets: general results} {Coupled evolution with tides of
  the radius and orbit of transiting giant planets: general results}.{\BBCQ}
\newblock
\APACjournalVolNumPages{The Astrophysical Journal}{700}{}{1921}.
\PrintBackRefs{\CurrentBib}

\bibitem [\protect \citeauthoryear {%
{Irwin}%
\ \protect \BOthers {.}}{%
{Irwin}%
\ \protect \BOthers {.}}{%
{\protect \APACyear {2008}}%
}]{%
Irwin08}
\APACinsertmetastar {%
Irwin08}%
\begin{APACrefauthors}%
{Irwin}, P\BPBI G\BPBI J.%
, {Teanby}, N\BPBI A.%
, {de Kok}, R.%
, {Fletcher}, L\BPBI N.%
, {Howett}, C\BPBI J\BPBI A.%
, {Tsang}, C\BPBI C\BPBI C.%
\BDBL {}{Parrish}, P\BPBI D.%
\end{APACrefauthors}%
\unskip\
\newblock
\APACrefYearMonthDay{2008}{{\APACmonth{04}}}{}.
\newblock
{\BBOQ}\APACrefatitle {{The NEMESIS planetary atmosphere radiative transfer and
  retrieval tool}} {{The NEMESIS planetary atmosphere radiative transfer and
  retrieval tool}}.{\BBCQ}
\newblock
\APACjournalVolNumPages{\jqsrt}{109}{}{1136-1150}.
\newblock
\begin{APACrefDOI} \doi{10.1016/j.jqsrt.2007.11.006} \end{APACrefDOI}
\PrintBackRefs{\CurrentBib}

\bibitem [\protect \citeauthoryear {%
Jackson%
, Adams%
, Sandidge%
, Kreyche%
\BCBL {}\ \BBA {} Briggs%
}{%
Jackson%
\ \protect \BOthers {.}}{%
{\protect \APACyear {2019}}%
}]{%
Jackson:2019aa}
\APACinsertmetastar {%
Jackson:2019aa}%
\begin{APACrefauthors}%
Jackson, B.%
, Adams, E.%
, Sandidge, W.%
, Kreyche, S.%
\BCBL {}\ \BBA {} Briggs, J.%
\end{APACrefauthors}%
\unskip\
\newblock
\APACrefYearMonthDay{2019}{}{}.
\newblock
{\BBOQ}\APACrefatitle {Variability in the atmosphere of the hot Jupiter
  Kepler-76b} {Variability in the atmosphere of the hot jupiter
  kepler-76b}.{\BBCQ}
\newblock
\APACjournalVolNumPages{The Astronomical Journal}{157}{}{239}.
\PrintBackRefs{\CurrentBib}

\bibitem [\protect \citeauthoryear {%
{Jackson}%
, {Greenberg}%
\BCBL {}\ \BBA {} {Barnes}%
}{%
{Jackson}%
\ \protect \BOthers {.}}{%
{\protect \APACyear {2008}}%
}]{%
Jackson08a}
\APACinsertmetastar {%
Jackson08a}%
\begin{APACrefauthors}%
{Jackson}, B.%
, {Greenberg}, R.%
\BCBL {}\ \BBA {} {Barnes}, R.%
\end{APACrefauthors}%
\unskip\
\newblock
\APACrefYearMonthDay{2008}{{\APACmonth{05}}}{}.
\newblock
{\BBOQ}\APACrefatitle {{Tidal Evolution of Close-in Extrasolar Planets}}
  {{Tidal Evolution of Close-in Extrasolar Planets}}.{\BBCQ}
\newblock
\APACjournalVolNumPages{\apj}{678}{}{1396-1406}.
\newblock
\begin{APACrefDOI} \doi{10.1086/529187} \end{APACrefDOI}
\PrintBackRefs{\CurrentBib}

\bibitem [\protect \citeauthoryear {%
Jackson%
, Greenberg%
\BCBL {}\ \BBA {} Barnes%
}{%
Jackson%
\ \protect \BOthers {.}}{%
{\protect \APACyear {2008}}%
}]{%
Jackson:681}
\APACinsertmetastar {%
Jackson:681}%
\begin{APACrefauthors}%
Jackson, B.%
, Greenberg, R.%
\BCBL {}\ \BBA {} Barnes, R.%
\end{APACrefauthors}%
\unskip\
\newblock
\APACrefYearMonthDay{2008}{}{}.
\newblock
{\BBOQ}\APACrefatitle {Tidal heating of extrasolar planets} {Tidal heating of
  extrasolar planets}.{\BBCQ}
\newblock
\APACjournalVolNumPages{The Astrophysical Journal}{681}{}{1631}.
\PrintBackRefs{\CurrentBib}

\bibitem [\protect \citeauthoryear {%
Jansen%
\ \BBA {} Kipping%
}{%
Jansen%
\ \BBA {} Kipping%
}{%
{\protect \APACyear {2020}}%
}]{%
Jansen:2020aa}
\APACinsertmetastar {%
Jansen:2020aa}%
\begin{APACrefauthors}%
Jansen, T.%
\BCBT {}\ \BBA {} Kipping, D.%
\end{APACrefauthors}%
\unskip\
\newblock
\APACrefYearMonthDay{2020}{}{}.
\newblock
{\BBOQ}\APACrefatitle {Detection of the phase curve and occultation of
  {WASP-100b} with TESS} {Detection of the phase curve and occultation of
  {WASP-100b} with tess}.{\BBCQ}
\newblock
\APACjournalVolNumPages{Monthly Notices of the Royal Astronomical
  Society}{494}{}{4077}.
\PrintBackRefs{\CurrentBib}

\bibitem [\protect \citeauthoryear {%
{Jenkins}%
\ \protect \BOthers {.}}{%
{Jenkins}%
\ \protect \BOthers {.}}{%
{\protect \APACyear {2017}}%
}]{%
jenk16}
\APACinsertmetastar {%
jenk16}%
\begin{APACrefauthors}%
{Jenkins}, J\BPBI S.%
, {Jones}, H\BPBI R\BPBI A.%
, {Tuomi}, M.%
, {D{\'{\i}}az}, M.%
, {Cordero}, J\BPBI P.%
, {Aguayo}, A.%
\BDBL {}{Minniti}, D.%
\end{APACrefauthors}%
\unskip\
\newblock
\APACrefYearMonthDay{2017}{{\APACmonth{04}}}{}.
\newblock
{\BBOQ}\APACrefatitle {{New planetary systems from the Calan-Hertfordshire
  Extrasolar Planet Search}} {{New planetary systems from the
  Calan-Hertfordshire Extrasolar Planet Search}}.{\BBCQ}
\newblock
\APACjournalVolNumPages{\mnras}{466}{}{443-473}.
\newblock
\begin{APACrefDOI} \doi{10.1093/mnras/stw2811} \end{APACrefDOI}
\PrintBackRefs{\CurrentBib}

\bibitem [\protect \citeauthoryear {%
Jensen%
, Cauley%
, Redfield%
\BCBL {}\ \BBA {} {et al.}%
}{%
Jensen%
\ \protect \BOthers {.}}{%
{\protect \APACyear {2018}}%
}]{%
Jensen:2018aa}
\APACinsertmetastar {%
Jensen:2018aa}%
\begin{APACrefauthors}%
Jensen, A.%
, Cauley, P.%
, Redfield, S.%
\BCBL {}\ \BBA {} {et al.}%
\end{APACrefauthors}%
\unskip\
\newblock
\APACrefYearMonthDay{2018}{}{}.
\newblock
{\BBOQ}\APACrefatitle {Hydrogen and Sodium Absorption in the Optical
  Transmission Spectrum of WASP-12b} {Hydrogen and sodium absorption in the
  optical transmission spectrum of wasp-12b}.{\BBCQ}
\newblock
\APACjournalVolNumPages{The Astronomical Journal}{156}{}{154}.
\PrintBackRefs{\CurrentBib}

\bibitem [\protect \citeauthoryear {%
{Johansen}%
\ \BBA {} {Lambrechts}%
}{%
{Johansen}%
\ \BBA {} {Lambrechts}%
}{%
{\protect \APACyear {2017}}%
}]{%
joha17}
\APACinsertmetastar {%
joha17}%
\begin{APACrefauthors}%
{Johansen}, A.%
\BCBT {}\ \BBA {} {Lambrechts}, M.%
\end{APACrefauthors}%
\unskip\
\newblock
\APACrefYearMonthDay{2017}{{\APACmonth{08}}}{}.
\newblock
{\BBOQ}\APACrefatitle {{Forming Planets via Pebble Accretion}} {{Forming
  Planets via Pebble Accretion}}.{\BBCQ}
\newblock
\APACjournalVolNumPages{Annual Review of Earth and Planetary
  Sciences}{45}{}{359-387}.
\newblock
\begin{APACrefDOI} \doi{10.1146/annurev-earth-063016-020226} \end{APACrefDOI}
\PrintBackRefs{\CurrentBib}

\bibitem [\protect \citeauthoryear {%
Kataria%
\ \protect \BOthers {.}}{%
Kataria%
\ \protect \BOthers {.}}{%
{\protect \APACyear {2015}}%
}]{%
Kataria:2014}
\APACinsertmetastar {%
Kataria:2014}%
\begin{APACrefauthors}%
Kataria, T.%
, Showman, A.%
, Fortney, J.%
, Stevenson, K.%
, Line, M.%
, Kriedberg, L.%
\BDBL {}Desert, J.%
\end{APACrefauthors}%
\unskip\
\newblock
\APACrefYearMonthDay{2015}{}{}.
\newblock
{\BBOQ}\APACrefatitle {The atmospheric circulation of the hot {J}upiter
  {WASP}-43b: Comparing three-dimensional models to spectrophotometric data}
  {The atmospheric circulation of the hot {J}upiter {WASP}-43b: Comparing
  three-dimensional models to spectrophotometric data}.{\BBCQ}
\newblock
\APACjournalVolNumPages{The Astrophysical Journal}{801}{}{86}.
\PrintBackRefs{\CurrentBib}

\bibitem [\protect \citeauthoryear {%
Kataria%
\ \protect \BOthers {.}}{%
Kataria%
\ \protect \BOthers {.}}{%
{\protect \APACyear {2016}}%
}]{%
Kataria2016}
\APACinsertmetastar {%
Kataria2016}%
\begin{APACrefauthors}%
Kataria, T.%
, Sing, D.%
, Lewis, N.%
, Visscher, C.%
, Showman, A.%
, Fortney, J.%
\BCBL {}\ \BBA {} Marley, M.%
\end{APACrefauthors}%
\unskip\
\newblock
\APACrefYearMonthDay{2016}{}{}.
\newblock
{\BBOQ}\APACrefatitle {The atmospheric circulation of a nine-hot-Jupiter
  sample: Probing circulation and chemistry over a wide phase space} {The
  atmospheric circulation of a nine-hot-jupiter sample: Probing circulation and
  chemistry over a wide phase space}.{\BBCQ}
\newblock
\APACjournalVolNumPages{The Astrophysical Journal}{821}{}{9}.
\PrintBackRefs{\CurrentBib}

\bibitem [\protect \citeauthoryear {%
Keating%
, Cowan%
\BCBL {}\ \BBA {} Dang%
}{%
Keating%
\ \protect \BOthers {.}}{%
{\protect \APACyear {2019}}%
}]{%
Keating:2019aa}
\APACinsertmetastar {%
Keating:2019aa}%
\begin{APACrefauthors}%
Keating, D.%
, Cowan, N.%
\BCBL {}\ \BBA {} Dang, L.%
\end{APACrefauthors}%
\unskip\
\newblock
\APACrefYearMonthDay{2019}{}{}.
\newblock
{\BBOQ}\APACrefatitle {Uniformly hot nightside temperatures on short-period gas
  giants} {Uniformly hot nightside temperatures on short-period gas
  giants}.{\BBCQ}
\newblock
\APACjournalVolNumPages{Nature Astronomy}{3}{}{1092}.
\PrintBackRefs{\CurrentBib}

\bibitem [\protect \citeauthoryear {%
Kitzmann%
, Heng%
, Rimmer%
\BCBL {}\ \BBA {} et al.%
}{%
Kitzmann%
\ \protect \BOthers {.}}{%
{\protect \APACyear {2018}}%
}]{%
Kitzmann:2018aa}
\APACinsertmetastar {%
Kitzmann:2018aa}%
\begin{APACrefauthors}%
Kitzmann, D.%
, Heng, K.%
, Rimmer, P.%
\BCBL {}\ \BBA {} et al.%
\end{APACrefauthors}%
\unskip\
\newblock
\APACrefYearMonthDay{2018}{}{}.
\newblock
{\BBOQ}\APACrefatitle {The pecuilar atmospheric chemistry of {KELT}-9b} {The
  pecuilar atmospheric chemistry of {KELT}-9b}.{\BBCQ}
\newblock
\APACjournalVolNumPages{The Astrophysical Journal}{863}{}{183}.
\PrintBackRefs{\CurrentBib}

\bibitem [\protect \citeauthoryear {%
Knutson%
\ \protect \BOthers {.}}{%
Knutson%
\ \protect \BOthers {.}}{%
{\protect \APACyear {2007}}%
}]{%
Knutson_2007}
\APACinsertmetastar {%
Knutson_2007}%
\begin{APACrefauthors}%
Knutson, H.%
, Charbonneau, D.%
, Allen, L.%
, Fortney, J.%
, Agol, E.%
, Cowan, N.%
\BDBL {}Megeath, S.%
\end{APACrefauthors}%
\unskip\
\newblock
\APACrefYearMonthDay{2007}{}{}.
\newblock
{\BBOQ}\APACrefatitle {A map of the day-night contrast of the extrasolar planet
  {HD} 189733b} {A map of the day-night contrast of the extrasolar planet {HD}
  189733b}.{\BBCQ}
\newblock
\APACjournalVolNumPages{Nature}{447}{}{183}.
\PrintBackRefs{\CurrentBib}

\bibitem [\protect \citeauthoryear {%
{Knutson}%
\ \protect \BOthers {.}}{%
{Knutson}%
\ \protect \BOthers {.}}{%
{\protect \APACyear {2014}}%
}]{%
knut14}
\APACinsertmetastar {%
knut14}%
\begin{APACrefauthors}%
{Knutson}, H\BPBI A.%
, {Fulton}, B\BPBI J.%
, {Montet}, B\BPBI T.%
, {Kao}, M.%
, {Ngo}, H.%
, {Howard}, A\BPBI W.%
\BDBL {}{Muirhead}, P\BPBI S.%
\end{APACrefauthors}%
\unskip\
\newblock
\APACrefYearMonthDay{2014}{{\APACmonth{04}}}{}.
\newblock
{\BBOQ}\APACrefatitle {{Friends of Hot Jupiters. I. A Radial Velocity Search
  for Massive, Long-period Companions to Close-in Gas Giant Planets}} {{Friends
  of Hot Jupiters. I. A Radial Velocity Search for Massive, Long-period
  Companions to Close-in Gas Giant Planets}}.{\BBCQ}
\newblock
\APACjournalVolNumPages{\apj}{785}{}{126}.
\newblock
\begin{APACrefDOI} \doi{10.1088/0004-637X/785/2/126} \end{APACrefDOI}
\PrintBackRefs{\CurrentBib}

\bibitem [\protect \citeauthoryear {%
Komacek%
\ \BBA {} Showman%
}{%
Komacek%
\ \BBA {} Showman%
}{%
{\protect \APACyear {2016}}%
}]{%
Komacek:2015}
\APACinsertmetastar {%
Komacek:2015}%
\begin{APACrefauthors}%
Komacek, T.%
\BCBT {}\ \BBA {} Showman, A.%
\end{APACrefauthors}%
\unskip\
\newblock
\APACrefYearMonthDay{2016}{}{}.
\newblock
{\BBOQ}\APACrefatitle {Atmospheric circulation of hot {J}upiters:
  Dayside-nightside temperature differences} {Atmospheric circulation of hot
  {J}upiters: Dayside-nightside temperature differences}.{\BBCQ}
\newblock
\APACjournalVolNumPages{The Astrophysical Journal}{821}{}{16}.
\PrintBackRefs{\CurrentBib}

\bibitem [\protect \citeauthoryear {%
Komacek%
, Showman%
\BCBL {}\ \BBA {} Tan%
}{%
Komacek%
\ \protect \BOthers {.}}{%
{\protect \APACyear {2017}}%
}]{%
Komacek:2017}
\APACinsertmetastar {%
Komacek:2017}%
\begin{APACrefauthors}%
Komacek, T.%
, Showman, A.%
\BCBL {}\ \BBA {} Tan, X.%
\end{APACrefauthors}%
\unskip\
\newblock
\APACrefYearMonthDay{2017}{}{}.
\newblock
{\BBOQ}\APACrefatitle {Atmospheric circulation of hot Jupiters:
  dayside-nightside temperature differences. II. Comparison with observations}
  {Atmospheric circulation of hot jupiters: dayside-nightside temperature
  differences. ii. comparison with observations}.{\BBCQ}
\newblock
\APACjournalVolNumPages{The Astrophysical Journal}{835}{}{198}.
\PrintBackRefs{\CurrentBib}

\bibitem [\protect \citeauthoryear {%
Komacek%
\ \BBA {} Tan%
}{%
Komacek%
\ \BBA {} Tan%
}{%
{\protect \APACyear {2018}}%
}]{%
Komacek:2018aa}
\APACinsertmetastar {%
Komacek:2018aa}%
\begin{APACrefauthors}%
Komacek, T.%
\BCBT {}\ \BBA {} Tan, X.%
\end{APACrefauthors}%
\unskip\
\newblock
\APACrefYearMonthDay{2018}{}{}.
\newblock
{\BBOQ}\APACrefatitle {Effects of Dissociation/Recombination on the Day-Night
  Temperature Contrasts of Ultra-hot Jupiters} {Effects of
  dissociation/recombination on the day-night temperature contrasts of
  ultra-hot jupiters}.{\BBCQ}
\newblock
\APACjournalVolNumPages{Research notes of the AAS}{2}{}{36}.
\PrintBackRefs{\CurrentBib}

\bibitem [\protect \citeauthoryear {%
Komacek%
, Thorngren%
, Lopez%
\BCBL {}\ \BBA {} Ginzburg%
}{%
Komacek%
\ \protect \BOthers {.}}{%
{\protect \APACyear {2020}}%
}]{%
Komacek:2020ab}
\APACinsertmetastar {%
Komacek:2020ab}%
\begin{APACrefauthors}%
Komacek, T.%
, Thorngren, D.%
, Lopez, E.%
\BCBL {}\ \BBA {} Ginzburg, S.%
\end{APACrefauthors}%
\unskip\
\newblock
\APACrefYearMonthDay{2020}{}{}.
\newblock
{\BBOQ}\APACrefatitle {Reinflation of Warm and Hot Jupiters} {Reinflation of
  warm and hot jupiters}.{\BBCQ}
\newblock
\APACjournalVolNumPages{The Astrophysical Journal}{893}{}{36}.
\PrintBackRefs{\CurrentBib}

\bibitem [\protect \citeauthoryear {%
Komacek%
\ \BBA {} Youdin%
}{%
Komacek%
\ \BBA {} Youdin%
}{%
{\protect \APACyear {2017}}%
}]{%
Komacek:2017a}
\APACinsertmetastar {%
Komacek:2017a}%
\begin{APACrefauthors}%
Komacek, T.%
\BCBT {}\ \BBA {} Youdin, A.%
\end{APACrefauthors}%
\unskip\
\newblock
\APACrefYearMonthDay{2017}{}{}.
\newblock
{\BBOQ}\APACrefatitle {Structure and evolution of internally heated hot
  Jupiters} {Structure and evolution of internally heated hot jupiters}.{\BBCQ}
\newblock
\APACjournalVolNumPages{The Astrophysical Journal}{844}{}{94}.
\PrintBackRefs{\CurrentBib}

\bibitem [\protect \citeauthoryear {%
{Kozai}%
}{%
{Kozai}%
}{%
{\protect \APACyear {1962}}%
}]{%
koza62}
\APACinsertmetastar {%
koza62}%
\begin{APACrefauthors}%
{Kozai}, Y.%
\end{APACrefauthors}%
\unskip\
\newblock
\APACrefYearMonthDay{1962}{{\APACmonth{11}}}{}.
\newblock
{\BBOQ}\APACrefatitle {{Secular perturbations of asteroids with high
  inclination and eccentricity}} {{Secular perturbations of asteroids with high
  inclination and eccentricity}}.{\BBCQ}
\newblock
\APACjournalVolNumPages{\aj}{67}{}{591}.
\newblock
\begin{APACrefDOI} \doi{10.1086/108790} \end{APACrefDOI}
\PrintBackRefs{\CurrentBib}

\bibitem [\protect \citeauthoryear {%
{Kreidberg}%
}{%
{Kreidberg}%
}{%
{\protect \APACyear {2017}}%
}]{%
Kreidberg17}
\APACinsertmetastar {%
Kreidberg17}%
\begin{APACrefauthors}%
{Kreidberg}, L.%
\end{APACrefauthors}%
\unskip\
\newblock
\APACrefYearMonthDay{2017}{}{}.
\newblock
{\BBOQ}\APACrefatitle {{Exoplanet Atmosphere Measurements from Transmission
  Spectroscopy and Other Planet Star Combined Light Observations}} {{Exoplanet
  Atmosphere Measurements from Transmission Spectroscopy and Other Planet Star
  Combined Light Observations}}.{\BBCQ}
\newblock
\BIn{} \APACrefbtitle {Handbook of Exoplanets, Edited by Hans J.~Deeg and Juan
  Antonio Belmonte.~Springer Living Reference Work, ISBN: 978-3-319-30648-3,
  2017, id.100} {Handbook of exoplanets, edited by hans j.~deeg and juan
  antonio belmonte.~springer living reference work, isbn: 978-3-319-30648-3,
  2017, id.100}\ (\BPG~100).
\newblock
\begin{APACrefDOI} \doi{10.1007/978-3-319-30648-3_100-1} \end{APACrefDOI}
\PrintBackRefs{\CurrentBib}

\bibitem [\protect \citeauthoryear {%
{Kreidberg}%
\ \protect \BOthers {.}}{%
{Kreidberg}%
\ \protect \BOthers {.}}{%
{\protect \APACyear {2014}}%
}]{%
Kreidberg14b}
\APACinsertmetastar {%
Kreidberg14b}%
\begin{APACrefauthors}%
{Kreidberg}, L.%
, {Bean}, J\BPBI L.%
, {D{\'e}sert}, J\BHBI M.%
, {Line}, M\BPBI R.%
, {Fortney}, J\BPBI J.%
, {Madhusudhan}, N.%
\BDBL {}{Homeier}, D.%
\end{APACrefauthors}%
\unskip\
\newblock
\APACrefYearMonthDay{2014}{{\APACmonth{10}}}{}.
\newblock
{\BBOQ}\APACrefatitle {{A Precise Water Abundance Measurement for the Hot
  Jupiter WASP-43b}} {{A Precise Water Abundance Measurement for the Hot
  Jupiter WASP-43b}}.{\BBCQ}
\newblock
\APACjournalVolNumPages{\apjl}{793}{}{L27}.
\newblock
\begin{APACrefDOI} \doi{10.1088/2041-8205/793/2/L27} \end{APACrefDOI}
\PrintBackRefs{\CurrentBib}

\bibitem [\protect \citeauthoryear {%
Kreidberg%
, Line%
, Parmentier%
\BCBL {}\ \BBA {} {et al.}%
}{%
Kreidberg%
\ \protect \BOthers {.}}{%
{\protect \APACyear {2018}}%
}]{%
Kreidberg:2018aa}
\APACinsertmetastar {%
Kreidberg:2018aa}%
\begin{APACrefauthors}%
Kreidberg, L.%
, Line, M.%
, Parmentier, V.%
\BCBL {}\ \BBA {} {et al.}%
\end{APACrefauthors}%
\unskip\
\newblock
\APACrefYearMonthDay{2018}{}{}.
\newblock
{\BBOQ}\APACrefatitle {Global Climate and Atmospheric Composition of the
  Ultra-hot Jupiter WASP-103b from HST and Spitzer Phase Curve Observations}
  {Global climate and atmospheric composition of the ultra-hot jupiter
  wasp-103b from hst and spitzer phase curve observations}.{\BBCQ}
\newblock
\APACjournalVolNumPages{The Astronomical Journal}{156}{}{17}.
\PrintBackRefs{\CurrentBib}

\bibitem [\protect \citeauthoryear {%
{Kreidberg}%
\ \protect \BOthers {.}}{%
{Kreidberg}%
\ \protect \BOthers {.}}{%
{\protect \APACyear {2015}}%
}]{%
krei15}
\APACinsertmetastar {%
krei15}%
\begin{APACrefauthors}%
{Kreidberg}, L.%
, {Line}, M\BPBI R.%
, {Bean}, J\BPBI L.%
, {Stevenson}, K\BPBI B.%
, {D{\'e}sert}, J\BHBI M.%
, {Madhusudhan}, N.%
\BDBL {}{Showman}, A\BPBI P.%
\end{APACrefauthors}%
\unskip\
\newblock
\APACrefYearMonthDay{2015}{{\APACmonth{11}}}{}.
\newblock
{\BBOQ}\APACrefatitle {{A Detection of Water in the Transmission Spectrum of
  the Hot Jupiter WASP-12b and Implications for Its Atmospheric Composition}}
  {{A Detection of Water in the Transmission Spectrum of the Hot Jupiter
  WASP-12b and Implications for Its Atmospheric Composition}}.{\BBCQ}
\newblock
\APACjournalVolNumPages{\apj}{814}{}{66}.
\newblock
\begin{APACrefDOI} \doi{10.1088/0004-637X/814/1/66} \end{APACrefDOI}
\PrintBackRefs{\CurrentBib}

\bibitem [\protect \citeauthoryear {%
Kurokawa%
\ \BBA {} Inutsuka%
}{%
Kurokawa%
\ \BBA {} Inutsuka%
}{%
{\protect \APACyear {2015}}%
}]{%
Kurokawa:2015aa}
\APACinsertmetastar {%
Kurokawa:2015aa}%
\begin{APACrefauthors}%
Kurokawa, H.%
\BCBT {}\ \BBA {} Inutsuka, S.%
\end{APACrefauthors}%
\unskip\
\newblock
\APACrefYearMonthDay{2015}{}{}.
\newblock
{\BBOQ}\APACrefatitle {On the Radius Anomaly of Hot Jupiters: Reexamination of
  the Possibility and Impact of Layered Convection} {On the radius anomaly of
  hot jupiters: Reexamination of the possibility and impact of layered
  convection}.{\BBCQ}
\newblock
\APACjournalVolNumPages{The Astrophysical Journal}{815}{}{78}.
\PrintBackRefs{\CurrentBib}

\bibitem [\protect \citeauthoryear {%
{Latham}%
\ \protect \BOthers {.}}{%
{Latham}%
\ \protect \BOthers {.}}{%
{\protect \APACyear {2011}}%
}]{%
lath11}
\APACinsertmetastar {%
lath11}%
\begin{APACrefauthors}%
{Latham}, D\BPBI W.%
, {Rowe}, J\BPBI F.%
, {Quinn}, S\BPBI N.%
, {Batalha}, N\BPBI M.%
, {Borucki}, W\BPBI J.%
, {Brown}, T\BPBI M.%
\BDBL {}{Wohler}, B.%
\end{APACrefauthors}%
\unskip\
\newblock
\APACrefYearMonthDay{2011}{{\APACmonth{05}}}{}.
\newblock
{\BBOQ}\APACrefatitle {{A First Comparison of Kepler Planet Candidates in
  Single and Multiple Systems}} {{A First Comparison of Kepler Planet
  Candidates in Single and Multiple Systems}}.{\BBCQ}
\newblock
\APACjournalVolNumPages{\apjl}{732}{}{L24}.
\newblock
\begin{APACrefDOI} \doi{10.1088/2041-8205/732/2/L24} \end{APACrefDOI}
\PrintBackRefs{\CurrentBib}

\bibitem [\protect \citeauthoryear {%
Laughlin%
}{%
Laughlin%
}{%
{\protect \APACyear {2018}}%
}]{%
Laughlin:2018aa}
\APACinsertmetastar {%
Laughlin:2018aa}%
\begin{APACrefauthors}%
Laughlin, G.%
\end{APACrefauthors}%
\unskip\
\newblock
\APACrefYearMonthDay{2018}{}{}.
\newblock
{\BBOQ}\APACrefatitle {Handbook of Exoplanets} {Handbook of exoplanets}.{\BBCQ}
\newblock
\BIn{} H.~Deeg\ \BBA {} J.~Belmonte\ (\BEDS), (\BCHAP\ Mass-Radius Relations of
  Giant Planets: The Radius Anomaly and Interior Models).
\newblock
\APACaddressPublisher{}{Springer, Cham}.
\PrintBackRefs{\CurrentBib}

\bibitem [\protect \citeauthoryear {%
Laughlin%
, Crismani%
\BCBL {}\ \BBA {} Adams%
}{%
Laughlin%
\ \protect \BOthers {.}}{%
{\protect \APACyear {2011}}%
}]{%
Laughlin_2011}
\APACinsertmetastar {%
Laughlin_2011}%
\begin{APACrefauthors}%
Laughlin, G.%
, Crismani, M.%
\BCBL {}\ \BBA {} Adams, F.%
\end{APACrefauthors}%
\unskip\
\newblock
\APACrefYearMonthDay{2011}{}{}.
\newblock
{\BBOQ}\APACrefatitle {On the anomalous radii of the transiting extrasolar
  planets} {On the anomalous radii of the transiting extrasolar
  planets}.{\BBCQ}
\newblock
\APACjournalVolNumPages{The Astrophysical Journal Letters}{729}{}{L7}.
\PrintBackRefs{\CurrentBib}

\bibitem [\protect \citeauthoryear {%
{Laughlin}%
, {Crismani}%
\BCBL {}\ \BBA {} {Adams}%
}{%
{Laughlin}%
\ \protect \BOthers {.}}{%
{\protect \APACyear {2011}}%
}]{%
Laughlin11}
\APACinsertmetastar {%
Laughlin11}%
\begin{APACrefauthors}%
{Laughlin}, G.%
, {Crismani}, M.%
\BCBL {}\ \BBA {} {Adams}, F\BPBI C.%
\end{APACrefauthors}%
\unskip\
\newblock
\APACrefYearMonthDay{2011}{{\APACmonth{03}}}{}.
\newblock
{\BBOQ}\APACrefatitle {{On the Anomalous Radii of the Transiting Extrasolar
  Planets}} {{On the Anomalous Radii of the Transiting Extrasolar
  Planets}}.{\BBCQ}
\newblock
\APACjournalVolNumPages{\apjl}{729}{}{L7+}.
\newblock
\begin{APACrefDOI} \doi{10.1088/2041-8205/729/1/L7} \end{APACrefDOI}
\PrintBackRefs{\CurrentBib}

\bibitem [\protect \citeauthoryear {%
Laughlin%
\ \BBA {} Lissauer%
}{%
Laughlin%
\ \BBA {} Lissauer%
}{%
{\protect \APACyear {2015}}%
}]{%
Laughlin:2015}
\APACinsertmetastar {%
Laughlin:2015}%
\begin{APACrefauthors}%
Laughlin, G.%
\BCBT {}\ \BBA {} Lissauer, J.%
\end{APACrefauthors}%
\unskip\
\newblock
\APACrefYearMonthDay{2015}{}{}.
\newblock
{\BBOQ}\APACrefatitle {Treatise on Geopysics} {Treatise on geopysics}.{\BBCQ}
\newblock
\BIn{} G.~Schubert\ (\BED), (\PrintOrdinal{2nd}\ \BEd, \BCHAP\ Exoplanetary
  Geophysics: An Emerging Discipline).
\newblock
\APACaddressPublisher{}{Elsevier}.
\PrintBackRefs{\CurrentBib}

\bibitem [\protect \citeauthoryear {%
{Lecavelier Des Etangs}%
\ \protect \BOthers {.}}{%
{Lecavelier Des Etangs}%
\ \protect \BOthers {.}}{%
{\protect \APACyear {2010}}%
}]{%
Lecavelier10}
\APACinsertmetastar {%
Lecavelier10}%
\begin{APACrefauthors}%
{Lecavelier Des Etangs}, A.%
, {Ehrenreich}, D.%
, {Vidal-Madjar}, A.%
, {Ballester}, G\BPBI E.%
, {D{\'e}sert}, J\BHBI M.%
, {Ferlet}, R.%
\BDBL {}{Udry}, S.%
\end{APACrefauthors}%
\unskip\
\newblock
\APACrefYearMonthDay{2010}{{\APACmonth{05}}}{}.
\newblock
{\BBOQ}\APACrefatitle {{Evaporation of the planet HD 189733b observed in H I
  Lyman-{$\alpha$}}} {{Evaporation of the planet HD 189733b observed in H I
  Lyman-{$\alpha$}}}.{\BBCQ}
\newblock
\APACjournalVolNumPages{\aap}{514}{}{A72}.
\newblock
\begin{APACrefDOI} \doi{10.1051/0004-6361/200913347} \end{APACrefDOI}
\PrintBackRefs{\CurrentBib}

\bibitem [\protect \citeauthoryear {%
{Leconte}%
\ \BBA {} {Chabrier}%
}{%
{Leconte}%
\ \BBA {} {Chabrier}%
}{%
{\protect \APACyear {2012}}%
}]{%
Leconte12}
\APACinsertmetastar {%
Leconte12}%
\begin{APACrefauthors}%
{Leconte}, J.%
\BCBT {}\ \BBA {} {Chabrier}, G.%
\end{APACrefauthors}%
\unskip\
\newblock
\APACrefYearMonthDay{2012}{{\APACmonth{04}}}{}.
\newblock
{\BBOQ}\APACrefatitle {{A new vision of giant planet interiors: Impact of
  double diffusive convection}} {{A new vision of giant planet interiors:
  Impact of double diffusive convection}}.{\BBCQ}
\newblock
\APACjournalVolNumPages{\aap}{540}{}{A20}.
\newblock
\begin{APACrefDOI} \doi{10.1051/0004-6361/201117595} \end{APACrefDOI}
\PrintBackRefs{\CurrentBib}

\bibitem [\protect \citeauthoryear {%
Leconte%
, Chabrier%
, Baraffe%
\BCBL {}\ \BBA {} Levrard%
}{%
Leconte%
\ \protect \BOthers {.}}{%
{\protect \APACyear {2010}}%
}]{%
Leconte:2010a}
\APACinsertmetastar {%
Leconte:2010a}%
\begin{APACrefauthors}%
Leconte, J.%
, Chabrier, G.%
, Baraffe, I.%
\BCBL {}\ \BBA {} Levrard, B.%
\end{APACrefauthors}%
\unskip\
\newblock
\APACrefYearMonthDay{2010}{}{}.
\newblock
{\BBOQ}\APACrefatitle {Is tidal heating sufficient to explain bloated
  exoplanets? Consistent calculations accounting for finite initial
  eccentricity} {Is tidal heating sufficient to explain bloated exoplanets?
  consistent calculations accounting for finite initial eccentricity}.{\BBCQ}
\newblock
\APACjournalVolNumPages{Astronomy and Astrophysics}{516}{}{A64}.
\PrintBackRefs{\CurrentBib}

\bibitem [\protect \citeauthoryear {%
{Lee}%
\ \BBA {} {Chiang}%
}{%
{Lee}%
\ \BBA {} {Chiang}%
}{%
{\protect \APACyear {2016}}%
}]{%
Lee16}
\APACinsertmetastar {%
Lee16}%
\begin{APACrefauthors}%
{Lee}, E\BPBI J.%
\BCBT {}\ \BBA {} {Chiang}, E.%
\end{APACrefauthors}%
\unskip\
\newblock
\APACrefYearMonthDay{2016}{{\APACmonth{02}}}{}.
\newblock
{\BBOQ}\APACrefatitle {{Breeding Super-Earths and Birthing Super-puffs in
  Transitional Disks}} {{Breeding Super-Earths and Birthing Super-puffs in
  Transitional Disks}}.{\BBCQ}
\newblock
\APACjournalVolNumPages{\apj}{817}{}{90}.
\newblock
\begin{APACrefDOI} \doi{10.3847/0004-637X/817/2/90} \end{APACrefDOI}
\PrintBackRefs{\CurrentBib}

\bibitem [\protect \citeauthoryear {%
{Lee}%
\ \BBA {} {Chiang}%
}{%
{Lee}%
\ \BBA {} {Chiang}%
}{%
{\protect \APACyear {2017}}%
}]{%
lee17}
\APACinsertmetastar {%
lee17}%
\begin{APACrefauthors}%
{Lee}, E\BPBI J.%
\BCBT {}\ \BBA {} {Chiang}, E.%
\end{APACrefauthors}%
\unskip\
\newblock
\APACrefYearMonthDay{2017}{{\APACmonth{06}}}{}.
\newblock
{\BBOQ}\APACrefatitle {{Magnetospheric Truncation, Tidal Inspiral, and the
  Creation of Short-period and Ultra-short-period Planets}} {{Magnetospheric
  Truncation, Tidal Inspiral, and the Creation of Short-period and
  Ultra-short-period Planets}}.{\BBCQ}
\newblock
\APACjournalVolNumPages{\apj}{842}{}{40}.
\newblock
\begin{APACrefDOI} \doi{10.3847/1538-4357/aa6fb3} \end{APACrefDOI}
\PrintBackRefs{\CurrentBib}

\bibitem [\protect \citeauthoryear {%
{Lee}%
, {Chiang}%
\BCBL {}\ \BBA {} {Ormel}%
}{%
{Lee}%
\ \protect \BOthers {.}}{%
{\protect \APACyear {2014}}%
}]{%
lee14}
\APACinsertmetastar {%
lee14}%
\begin{APACrefauthors}%
{Lee}, E\BPBI J.%
, {Chiang}, E.%
\BCBL {}\ \BBA {} {Ormel}, C\BPBI W.%
\end{APACrefauthors}%
\unskip\
\newblock
\APACrefYearMonthDay{2014}{{\APACmonth{12}}}{}.
\newblock
{\BBOQ}\APACrefatitle {{Make Super-Earths, Not Jupiters: Accreting Nebular Gas
  onto Solid Cores at 0.1 AU and Beyond}} {{Make Super-Earths, Not Jupiters:
  Accreting Nebular Gas onto Solid Cores at 0.1 AU and Beyond}}.{\BBCQ}
\newblock
\APACjournalVolNumPages{\apj}{797}{}{95}.
\newblock
\begin{APACrefDOI} \doi{10.1088/0004-637X/797/2/95} \end{APACrefDOI}
\PrintBackRefs{\CurrentBib}

\bibitem [\protect \citeauthoryear {%
Lee%
, Dobbs-Dixon%
, Helling%
, Bognar%
\BCBL {}\ \BBA {} Woitke%
}{%
Lee%
\ \protect \BOthers {.}}{%
{\protect \APACyear {2016}}%
}]{%
Lee:2016}
\APACinsertmetastar {%
Lee:2016}%
\begin{APACrefauthors}%
Lee, G.%
, Dobbs-Dixon, I.%
, Helling, C.%
, Bognar, K.%
\BCBL {}\ \BBA {} Woitke, P.%
\end{APACrefauthors}%
\unskip\
\newblock
\APACrefYearMonthDay{2016}{}{}.
\newblock
{\BBOQ}\APACrefatitle {Dynamic mineral clouds on HD 189733b I. 3D RHD with
  kinetic, non-equilibrium cloud formation} {Dynamic mineral clouds on hd
  189733b i. 3d rhd with kinetic, non-equilibrium cloud formation}.{\BBCQ}
\newblock
\APACjournalVolNumPages{Astronomy and Astrophysics}{594}{}{A48}.
\PrintBackRefs{\CurrentBib}

\bibitem [\protect \citeauthoryear {%
{Lee}%
\ \BBA {} {Peale}%
}{%
{Lee}%
\ \BBA {} {Peale}%
}{%
{\protect \APACyear {2002}}%
}]{%
lee02}
\APACinsertmetastar {%
lee02}%
\begin{APACrefauthors}%
{Lee}, M\BPBI H.%
\BCBT {}\ \BBA {} {Peale}, S\BPBI J.%
\end{APACrefauthors}%
\unskip\
\newblock
\APACrefYearMonthDay{2002}{{\APACmonth{03}}}{}.
\newblock
{\BBOQ}\APACrefatitle {{Dynamics and Origin of the 2:1 Orbital Resonances of
  the GJ 876 Planets}} {{Dynamics and Origin of the 2:1 Orbital Resonances of
  the GJ 876 Planets}}.{\BBCQ}
\newblock
\APACjournalVolNumPages{\apj}{567}{}{596-609}.
\newblock
\begin{APACrefDOI} \doi{10.1086/338504} \end{APACrefDOI}
\PrintBackRefs{\CurrentBib}

\bibitem [\protect \citeauthoryear {%
Lewis%
, Showman%
, Fortney%
, Knutson%
\BCBL {}\ \BBA {} Marley%
}{%
Lewis%
\ \protect \BOthers {.}}{%
{\protect \APACyear {2014}}%
}]{%
Lewis:2014}
\APACinsertmetastar {%
Lewis:2014}%
\begin{APACrefauthors}%
Lewis, N.%
, Showman, A.%
, Fortney, J.%
, Knutson, H.%
\BCBL {}\ \BBA {} Marley, M.%
\end{APACrefauthors}%
\unskip\
\newblock
\APACrefYearMonthDay{2014}{}{}.
\newblock
{\BBOQ}\APACrefatitle {Atmospheric circulation of the eccentric hot {J}upiter
  {HAT}-{P}-2b} {Atmospheric circulation of the eccentric hot {J}upiter
  {HAT}-{P}-2b}.{\BBCQ}
\newblock
\APACjournalVolNumPages{The Astrophysical Journal}{795}{}{150}.
\PrintBackRefs{\CurrentBib}

\bibitem [\protect \citeauthoryear {%
{Lewis}%
\ \protect \BOthers {.}}{%
{Lewis}%
\ \protect \BOthers {.}}{%
{\protect \APACyear {2013}}%
}]{%
Lewis13}
\APACinsertmetastar {%
Lewis13}%
\begin{APACrefauthors}%
{Lewis}, N\BPBI K.%
, {Knutson}, H\BPBI A.%
, {Showman}, A\BPBI P.%
, {Cowan}, N\BPBI B.%
, {Laughlin}, G.%
, {Burrows}, A.%
\BDBL {}{Marcy}, G\BPBI W.%
\end{APACrefauthors}%
\unskip\
\newblock
\APACrefYearMonthDay{2013}{{\APACmonth{04}}}{}.
\newblock
{\BBOQ}\APACrefatitle {{Orbital Phase Variations of the Eccentric Giant Planet
  HAT-P-2b}} {{Orbital Phase Variations of the Eccentric Giant Planet
  HAT-P-2b}}.{\BBCQ}
\newblock
\APACjournalVolNumPages{\apj}{766}{2}{95}.
\newblock
\begin{APACrefDOI} \doi{10.1088/0004-637X/766/2/95} \end{APACrefDOI}
\PrintBackRefs{\CurrentBib}

\bibitem [\protect \citeauthoryear {%
J.~Li%
\ \BBA {} Goodman%
}{%
J.~Li%
\ \BBA {} Goodman%
}{%
{\protect \APACyear {2010}}%
}]{%
Li:2010}
\APACinsertmetastar {%
Li:2010}%
\begin{APACrefauthors}%
Li, J.%
\BCBT {}\ \BBA {} Goodman, J.%
\end{APACrefauthors}%
\unskip\
\newblock
\APACrefYearMonthDay{2010}{}{}.
\newblock
{\BBOQ}\APACrefatitle {Circulation and dissipation on hot {J}upiters}
  {Circulation and dissipation on hot {J}upiters}.{\BBCQ}
\newblock
\APACjournalVolNumPages{The Astrophysical Journal}{725}{}{1146}.
\PrintBackRefs{\CurrentBib}

\bibitem [\protect \citeauthoryear {%
L.~Li%
, Jiang%
, West%
\BCBL {}\ \BBA {} {et al.}%
}{%
L.~Li%
\ \protect \BOthers {.}}{%
{\protect \APACyear {2018}}%
}]{%
Li:2018aa}
\APACinsertmetastar {%
Li:2018aa}%
\begin{APACrefauthors}%
Li, L.%
, Jiang, X.%
, West, R.%
\BCBL {}\ \BBA {} {et al.}%
\end{APACrefauthors}%
\unskip\
\newblock
\APACrefYearMonthDay{2018}{}{}.
\newblock
{\BBOQ}\APACrefatitle {Less absorbed solar energy and more internal heat for
  Jupiter} {Less absorbed solar energy and more internal heat for
  jupiter}.{\BBCQ}
\newblock
\APACjournalVolNumPages{Nature Communications}{9}{}{3709}.
\PrintBackRefs{\CurrentBib}

\bibitem [\protect \citeauthoryear {%
{Lidov}%
}{%
{Lidov}%
}{%
{\protect \APACyear {1962}}%
}]{%
lido62}
\APACinsertmetastar {%
lido62}%
\begin{APACrefauthors}%
{Lidov}, M\BPBI L.%
\end{APACrefauthors}%
\unskip\
\newblock
\APACrefYearMonthDay{1962}{{\APACmonth{10}}}{}.
\newblock
{\BBOQ}\APACrefatitle {{The evolution of orbits of artificial satellites of
  planets under the action of gravitational perturbations of external bodies}}
  {{The evolution of orbits of artificial satellites of planets under the
  action of gravitational perturbations of external bodies}}.{\BBCQ}
\newblock
\APACjournalVolNumPages{\planss}{9}{}{719-759}.
\newblock
\begin{APACrefDOI} \doi{10.1016/0032-0633(62)90129-0} \end{APACrefDOI}
\PrintBackRefs{\CurrentBib}

\bibitem [\protect \citeauthoryear {%
{Lin}%
, {Bodenheimer}%
\BCBL {}\ \BBA {} {Richardson}%
}{%
{Lin}%
\ \protect \BOthers {.}}{%
{\protect \APACyear {1996}}%
}]{%
Lin96}
\APACinsertmetastar {%
Lin96}%
\begin{APACrefauthors}%
{Lin}, D\BPBI N\BPBI C.%
, {Bodenheimer}, P.%
\BCBL {}\ \BBA {} {Richardson}, D\BPBI C.%
\end{APACrefauthors}%
\unskip\
\newblock
\APACrefYearMonthDay{1996}{{\APACmonth{04}}}{}.
\newblock
{\BBOQ}\APACrefatitle {{Orbital migration of the planetary companion of 51
  Pegasi to its present location}} {{Orbital migration of the planetary
  companion of 51 Pegasi to its present location}}.{\BBCQ}
\newblock
\APACjournalVolNumPages{\nat}{380}{}{606-607}.
\newblock
\begin{APACrefDOI} \doi{10.1038/380606a0} \end{APACrefDOI}
\PrintBackRefs{\CurrentBib}

\bibitem [\protect \citeauthoryear {%
{Lin}%
\ \BBA {} {Papaloizou}%
}{%
{Lin}%
\ \BBA {} {Papaloizou}%
}{%
{\protect \APACyear {1986}}%
}]{%
Lin86}
\APACinsertmetastar {%
Lin86}%
\begin{APACrefauthors}%
{Lin}, D\BPBI N\BPBI C.%
\BCBT {}\ \BBA {} {Papaloizou}, J.%
\end{APACrefauthors}%
\unskip\
\newblock
\APACrefYearMonthDay{1986}{{\APACmonth{10}}}{}.
\newblock
{\BBOQ}\APACrefatitle {{On the tidal interaction between protoplanets and the
  protoplanetary disk. III - Orbital migration of protoplanets}} {{On the tidal
  interaction between protoplanets and the protoplanetary disk. III - Orbital
  migration of protoplanets}}.{\BBCQ}
\newblock
\APACjournalVolNumPages{\apj}{309}{}{846-857}.
\newblock
\begin{APACrefDOI} \doi{10.1086/164653} \end{APACrefDOI}
\PrintBackRefs{\CurrentBib}

\bibitem [\protect \citeauthoryear {%
{Line}%
\ \protect \BOthers {.}}{%
{Line}%
\ \protect \BOthers {.}}{%
{\protect \APACyear {2016}}%
}]{%
Line16L}
\APACinsertmetastar {%
Line16L}%
\begin{APACrefauthors}%
{Line}, M\BPBI R.%
, {Stevenson}, K\BPBI B.%
, {Bean}, J.%
, {Desert}, J\BHBI M.%
, {Fortney}, J\BPBI J.%
, {Kreidberg}, L.%
\BDBL {}{Diamond-Lowe}, H.%
\end{APACrefauthors}%
\unskip\
\newblock
\APACrefYearMonthDay{2016}{{\APACmonth{12}}}{}.
\newblock
{\BBOQ}\APACrefatitle {{No Thermal Inversion and a Solar Water Abundance for
  the Hot Jupiter HD 209458b from HST/WFC3 Spectroscopy}} {{No Thermal
  Inversion and a Solar Water Abundance for the Hot Jupiter HD 209458b from
  HST/WFC3 Spectroscopy}}.{\BBCQ}
\newblock
\APACjournalVolNumPages{\aj}{152}{}{203}.
\newblock
\begin{APACrefDOI} \doi{10.3847/0004-6256/152/6/203} \end{APACrefDOI}
\PrintBackRefs{\CurrentBib}

\bibitem [\protect \citeauthoryear {%
{Line}%
\ \protect \BOthers {.}}{%
{Line}%
\ \protect \BOthers {.}}{%
{\protect \APACyear {2013}}%
}]{%
Line13}
\APACinsertmetastar {%
Line13}%
\begin{APACrefauthors}%
{Line}, M\BPBI R.%
, {Wolf}, A\BPBI S.%
, {Zhang}, X.%
, {Knutson}, H.%
, {Kammer}, J\BPBI A.%
, {Ellison}, E.%
\BDBL {}{Yung}, Y\BPBI L.%
\end{APACrefauthors}%
\unskip\
\newblock
\APACrefYearMonthDay{2013}{{\APACmonth{10}}}{}.
\newblock
{\BBOQ}\APACrefatitle {{A Systematic Retrieval Analysis of Secondary Eclipse
  Spectra. I. A Comparison of Atmospheric Retrieval Techniques}} {{A Systematic
  Retrieval Analysis of Secondary Eclipse Spectra. I. A Comparison of
  Atmospheric Retrieval Techniques}}.{\BBCQ}
\newblock
\APACjournalVolNumPages{\apj}{775}{}{137}.
\newblock
\begin{APACrefDOI} \doi{10.1088/0004-637X/775/2/137} \end{APACrefDOI}
\PrintBackRefs{\CurrentBib}

\bibitem [\protect \citeauthoryear {%
Lines%
\ \protect \BOthers {.}}{%
Lines%
\ \protect \BOthers {.}}{%
{\protect \APACyear {2018}}%
}]{%
Lines:2018}
\APACinsertmetastar {%
Lines:2018}%
\begin{APACrefauthors}%
Lines, S.%
, Mayne, N.%
, Boutle, I.%
, Manners, J.%
, Lee, G.%
, Helling, C.%
\BDBL {}Kerslake, M.%
\end{APACrefauthors}%
\unskip\
\newblock
\APACrefYearMonthDay{2018}{}{}.
\newblock
{\BBOQ}\APACrefatitle {Simulating the cloudy atmospheres of {HD 209458b} and
  {HD 189733b} with the {3D} Met Office Unifed Model} {Simulating the cloudy
  atmospheres of {HD 209458b} and {HD 189733b} with the {3D} met office unifed
  model}.{\BBCQ}
\newblock
\APACjournalVolNumPages{Astronomy {\&} Astrophysics}{615}{}{A97}.
\PrintBackRefs{\CurrentBib}

\bibitem [\protect \citeauthoryear {%
Lines%
\ \protect \BOthers {.}}{%
Lines%
\ \protect \BOthers {.}}{%
{\protect \APACyear {2019}}%
}]{%
lines:2019}
\APACinsertmetastar {%
lines:2019}%
\begin{APACrefauthors}%
Lines, S.%
, Mayne, N.%
, Manners, J.%
, Boutle, I.%
, Drummond, B.%
, Mikal-Evans, T.%
\BDBL {}Sing, D.%
\end{APACrefauthors}%
\unskip\
\newblock
\APACrefYearMonthDay{2019}{}{}.
\newblock
{\BBOQ}\APACrefatitle {Overcast on Osiris: 3D radiative--hydrodynamical
  simulations of a cloudy hot Jupiter using the parameterised,
  phase--equilibrium cloud formation code EddySed} {Overcast on osiris: 3d
  radiative--hydrodynamical simulations of a cloudy hot jupiter using the
  parameterised, phase--equilibrium cloud formation code eddysed}.{\BBCQ}
\newblock
\APACjournalVolNumPages{Monthly Notices of the Royal Astronomical
  Society}{488}{}{1332}.
\PrintBackRefs{\CurrentBib}

\bibitem [\protect \citeauthoryear {%
{Lodders}%
}{%
{Lodders}%
}{%
{\protect \APACyear {1999}}%
}]{%
Lodders99}
\APACinsertmetastar {%
Lodders99}%
\begin{APACrefauthors}%
{Lodders}, K.%
\end{APACrefauthors}%
\unskip\
\newblock
\APACrefYearMonthDay{1999}{{\APACmonth{07}}}{}.
\newblock
{\BBOQ}\APACrefatitle {{Alkali Element Chemistry in Cool Dwarf Atmospheres}}
  {{Alkali Element Chemistry in Cool Dwarf Atmospheres}}.{\BBCQ}
\newblock
\APACjournalVolNumPages{\apj}{519}{}{793-801}.
\PrintBackRefs{\CurrentBib}

\bibitem [\protect \citeauthoryear {%
{Lodders}%
\ \BBA {} {Fegley}%
}{%
{Lodders}%
\ \BBA {} {Fegley}%
}{%
{\protect \APACyear {2002}}%
}]{%
Lodders02}
\APACinsertmetastar {%
Lodders02}%
\begin{APACrefauthors}%
{Lodders}, K.%
\BCBT {}\ \BBA {} {Fegley}, B.%
\end{APACrefauthors}%
\unskip\
\newblock
\APACrefYearMonthDay{2002}{{\APACmonth{02}}}{}.
\newblock
{\BBOQ}\APACrefatitle {{Atmospheric Chemistry in Giant Planets, Brown Dwarfs,
  and Low-Mass Dwarf Stars. I. Carbon, Nitrogen, and Oxygen}} {{Atmospheric
  Chemistry in Giant Planets, Brown Dwarfs, and Low-Mass Dwarf Stars. I.
  Carbon, Nitrogen, and Oxygen}}.{\BBCQ}
\newblock
\APACjournalVolNumPages{Icarus}{155}{}{393-424}.
\PrintBackRefs{\CurrentBib}

\bibitem [\protect \citeauthoryear {%
Lopez%
\ \BBA {} Fortney%
}{%
Lopez%
\ \BBA {} Fortney%
}{%
{\protect \APACyear {2016}}%
}]{%
Lopez:2015}
\APACinsertmetastar {%
Lopez:2015}%
\begin{APACrefauthors}%
Lopez, E.%
\BCBT {}\ \BBA {} Fortney, J.%
\end{APACrefauthors}%
\unskip\
\newblock
\APACrefYearMonthDay{2016}{}{}.
\newblock
{\BBOQ}\APACrefatitle {Re-inflated warm {J}upiters around gas giants}
  {Re-inflated warm {J}upiters around gas giants}.{\BBCQ}
\newblock
\APACjournalVolNumPages{The Astrophysical Journal}{818}{}{4}.
\PrintBackRefs{\CurrentBib}

\bibitem [\protect \citeauthoryear {%
Lothringer%
, Barman%
\BCBL {}\ \BBA {} Koskinen%
}{%
Lothringer%
\ \protect \BOthers {.}}{%
{\protect \APACyear {2018}}%
}]{%
Lothringer:2018aa}
\APACinsertmetastar {%
Lothringer:2018aa}%
\begin{APACrefauthors}%
Lothringer, J.%
, Barman, T.%
\BCBL {}\ \BBA {} Koskinen, T.%
\end{APACrefauthors}%
\unskip\
\newblock
\APACrefYearMonthDay{2018}{}{}.
\newblock
{\BBOQ}\APACrefatitle {Extremely Irradiated Hot Jupiters: Non-Oxide Inversions,
  H- Opacity, and Thermal Dissociation of Molecules} {Extremely irradiated hot
  jupiters: Non-oxide inversions, h- opacity, and thermal dissociation of
  molecules}.{\BBCQ}
\newblock
\APACjournalVolNumPages{The Astrophysical Journal}{866}{}{27}.
\PrintBackRefs{\CurrentBib}

\bibitem [\protect \citeauthoryear {%
{Lothringer}%
, {Barman}%
\BCBL {}\ \BBA {} {Koskinen}%
}{%
{Lothringer}%
\ \protect \BOthers {.}}{%
{\protect \APACyear {2018}}%
}]{%
Lothringer18}
\APACinsertmetastar {%
Lothringer18}%
\begin{APACrefauthors}%
{Lothringer}, J\BPBI D.%
, {Barman}, T.%
\BCBL {}\ \BBA {} {Koskinen}, T.%
\end{APACrefauthors}%
\unskip\
\newblock
\APACrefYearMonthDay{2018}{{\APACmonth{10}}}{}.
\newblock
{\BBOQ}\APACrefatitle {{Extremely Irradiated Hot Jupiters: Non-oxide
  Inversions, H$^{-}$ Opacity, and Thermal Dissociation of Molecules}}
  {{Extremely Irradiated Hot Jupiters: Non-oxide Inversions, H$^{-}$ Opacity,
  and Thermal Dissociation of Molecules}}.{\BBCQ}
\newblock
\APACjournalVolNumPages{\apj}{866}{1}{27}.
\newblock
\begin{APACrefDOI} \doi{10.3847/1538-4357/aadd9e} \end{APACrefDOI}
\PrintBackRefs{\CurrentBib}

\bibitem [\protect \citeauthoryear {%
{Louden}%
\ \BBA {} {Wheatley}%
}{%
{Louden}%
\ \BBA {} {Wheatley}%
}{%
{\protect \APACyear {2015}}%
}]{%
Louden15}
\APACinsertmetastar {%
Louden15}%
\begin{APACrefauthors}%
{Louden}, T.%
\BCBT {}\ \BBA {} {Wheatley}, P\BPBI J.%
\end{APACrefauthors}%
\unskip\
\newblock
\APACrefYearMonthDay{2015}{{\APACmonth{12}}}{}.
\newblock
{\BBOQ}\APACrefatitle {{Spatially Resolved Eastward Winds and Rotation of HD
  189733b}} {{Spatially Resolved Eastward Winds and Rotation of HD
  189733b}}.{\BBCQ}
\newblock
\APACjournalVolNumPages{\apjl}{814}{2}{L24}.
\newblock
\begin{APACrefDOI} \doi{10.1088/2041-8205/814/2/L24} \end{APACrefDOI}
\PrintBackRefs{\CurrentBib}

\bibitem [\protect \citeauthoryear {%
{Madhusudhan}%
}{%
{Madhusudhan}%
}{%
{\protect \APACyear {2018}}%
}]{%
Madhu18}
\APACinsertmetastar {%
Madhu18}%
\begin{APACrefauthors}%
{Madhusudhan}, N.%
\end{APACrefauthors}%
\unskip\
\newblock
\APACrefYearMonthDay{2018}{}{}.
\newblock
{\BBOQ}\APACrefatitle {{Atmospheric Retrieval of Exoplanets}} {{Atmospheric
  Retrieval of Exoplanets}}.{\BBCQ}
\newblock
\BIn{} \APACrefbtitle {Handbook of Exoplanets} {Handbook of exoplanets}\
  (\BPG~104).
\newblock
\begin{APACrefDOI} \doi{10.1007/978-3-319-55333-7_104} \end{APACrefDOI}
\PrintBackRefs{\CurrentBib}

\bibitem [\protect \citeauthoryear {%
{Madhusudhan}%
}{%
{Madhusudhan}%
}{%
{\protect \APACyear {2019}}%
}]{%
Madhu19}
\APACinsertmetastar {%
Madhu19}%
\begin{APACrefauthors}%
{Madhusudhan}, N.%
\end{APACrefauthors}%
\unskip\
\newblock
\APACrefYearMonthDay{2019}{{\APACmonth{08}}}{}.
\newblock
{\BBOQ}\APACrefatitle {{Exoplanetary Atmospheres: Key Insights, Challenges, and
  Prospects}} {{Exoplanetary Atmospheres: Key Insights, Challenges, and
  Prospects}}.{\BBCQ}
\newblock
\APACjournalVolNumPages{\araa}{57}{}{617-663}.
\newblock
\begin{APACrefDOI} \doi{10.1146/annurev-astro-081817-051846} \end{APACrefDOI}
\PrintBackRefs{\CurrentBib}

\bibitem [\protect \citeauthoryear {%
{Madhusudhan}%
, {Ag{\'u}ndez}%
, {Moses}%
\BCBL {}\ \BBA {} {Hu}%
}{%
{Madhusudhan}%
\ \protect \BOthers {.}}{%
{\protect \APACyear {2016}}%
}]{%
Madhu16}
\APACinsertmetastar {%
Madhu16}%
\begin{APACrefauthors}%
{Madhusudhan}, N.%
, {Ag{\'u}ndez}, M.%
, {Moses}, J\BPBI I.%
\BCBL {}\ \BBA {} {Hu}, Y.%
\end{APACrefauthors}%
\unskip\
\newblock
\APACrefYearMonthDay{2016}{{\APACmonth{12}}}{}.
\newblock
{\BBOQ}\APACrefatitle {{Exoplanetary Atmospheres{\textemdash}Chemistry,
  Formation Conditions, and Habitability}} {{Exoplanetary
  Atmospheres{\textemdash}Chemistry, Formation Conditions, and
  Habitability}}.{\BBCQ}
\newblock
\APACjournalVolNumPages{\ssr}{205}{1-4}{285-348}.
\newblock
\begin{APACrefDOI} \doi{10.1007/s11214-016-0254-3} \end{APACrefDOI}
\PrintBackRefs{\CurrentBib}

\bibitem [\protect \citeauthoryear {%
{Madhusudhan}%
, {Amin}%
\BCBL {}\ \BBA {} {Kennedy}%
}{%
{Madhusudhan}%
\ \protect \BOthers {.}}{%
{\protect \APACyear {2014}}%
}]{%
Madhu14}
\APACinsertmetastar {%
Madhu14}%
\begin{APACrefauthors}%
{Madhusudhan}, N.%
, {Amin}, M\BPBI A.%
\BCBL {}\ \BBA {} {Kennedy}, G\BPBI M.%
\end{APACrefauthors}%
\unskip\
\newblock
\APACrefYearMonthDay{2014}{{\APACmonth{10}}}{}.
\newblock
{\BBOQ}\APACrefatitle {{Toward Chemical Constraints on Hot Jupiter Migration}}
  {{Toward Chemical Constraints on Hot Jupiter Migration}}.{\BBCQ}
\newblock
\APACjournalVolNumPages{\apjl}{794}{}{L12}.
\newblock
\begin{APACrefDOI} \doi{10.1088/2041-8205/794/1/L12} \end{APACrefDOI}
\PrintBackRefs{\CurrentBib}

\bibitem [\protect \citeauthoryear {%
{Madhusudhan}%
, {Harrington}%
\BCBL {}\ \protect \BOthers {.}}{%
{Madhusudhan}%
, {Harrington}%
\BCBL {}\ \protect \BOthers {.}}{%
{\protect \APACyear {2011}}%
}]{%
Madhu11}
\APACinsertmetastar {%
Madhu11}%
\begin{APACrefauthors}%
{Madhusudhan}, N.%
, {Harrington}, J.%
, {Stevenson}, K\BPBI B.%
, {Nymeyer}, S.%
, {Campo}, C\BPBI J.%
, {Wheatley}, P\BPBI J.%
\BDBL {}{West}, R\BPBI G.%
\end{APACrefauthors}%
\unskip\
\newblock
\APACrefYearMonthDay{2011}{{\APACmonth{01}}}{}.
\newblock
{\BBOQ}\APACrefatitle {{A high C/O ratio and weak thermal inversion in the
  atmosphere of exoplanet WASP-12b}} {{A high C/O ratio and weak thermal
  inversion in the atmosphere of exoplanet WASP-12b}}.{\BBCQ}
\newblock
\APACjournalVolNumPages{\nat}{469}{}{64-67}.
\newblock
\begin{APACrefDOI} \doi{10.1038/nature09602} \end{APACrefDOI}
\PrintBackRefs{\CurrentBib}

\bibitem [\protect \citeauthoryear {%
{Madhusudhan}%
, {Mousis}%
, {Johnson}%
\BCBL {}\ \BBA {} {Lunine}%
}{%
{Madhusudhan}%
, {Mousis}%
\BCBL {}\ \protect \BOthers {.}}{%
{\protect \APACyear {2011}}%
}]{%
Madhu11b}
\APACinsertmetastar {%
Madhu11b}%
\begin{APACrefauthors}%
{Madhusudhan}, N.%
, {Mousis}, O.%
, {Johnson}, T\BPBI V.%
\BCBL {}\ \BBA {} {Lunine}, J\BPBI I.%
\end{APACrefauthors}%
\unskip\
\newblock
\APACrefYearMonthDay{2011}{{\APACmonth{12}}}{}.
\newblock
{\BBOQ}\APACrefatitle {{Carbon-rich Giant Planets: Atmospheric Chemistry,
  Thermal Inversions, Spectra, and Formation Conditions}} {{Carbon-rich Giant
  Planets: Atmospheric Chemistry, Thermal Inversions, Spectra, and Formation
  Conditions}}.{\BBCQ}
\newblock
\APACjournalVolNumPages{\apj}{743}{}{191}.
\newblock
\begin{APACrefDOI} \doi{10.1088/0004-637X/743/2/191} \end{APACrefDOI}
\PrintBackRefs{\CurrentBib}

\bibitem [\protect \citeauthoryear {%
{Majeau}%
, {Agol}%
\BCBL {}\ \BBA {} {Cowan}%
}{%
{Majeau}%
\ \protect \BOthers {.}}{%
{\protect \APACyear {2012}}%
}]{%
Majeau12}
\APACinsertmetastar {%
Majeau12}%
\begin{APACrefauthors}%
{Majeau}, C.%
, {Agol}, E.%
\BCBL {}\ \BBA {} {Cowan}, N\BPBI B.%
\end{APACrefauthors}%
\unskip\
\newblock
\APACrefYearMonthDay{2012}{{\APACmonth{03}}}{}.
\newblock
{\BBOQ}\APACrefatitle {{A Two-dimensional Infrared Map of the Extrasolar Planet
  HD 189733b}} {{A Two-dimensional Infrared Map of the Extrasolar Planet HD
  189733b}}.{\BBCQ}
\newblock
\APACjournalVolNumPages{\apjl}{747}{2}{L20}.
\newblock
\begin{APACrefDOI} \doi{10.1088/2041-8205/747/2/L20} \end{APACrefDOI}
\PrintBackRefs{\CurrentBib}

\bibitem [\protect \citeauthoryear {%
{Malhotra}%
}{%
{Malhotra}%
}{%
{\protect \APACyear {1993}}%
}]{%
malh93}
\APACinsertmetastar {%
malh93}%
\begin{APACrefauthors}%
{Malhotra}, R.%
\end{APACrefauthors}%
\unskip\
\newblock
\APACrefYearMonthDay{1993}{{\APACmonth{10}}}{}.
\newblock
{\BBOQ}\APACrefatitle {{The origin of Pluto's peculiar orbit}} {{The origin of
  Pluto's peculiar orbit}}.{\BBCQ}
\newblock
\APACjournalVolNumPages{\nat}{365}{}{819-821}.
\newblock
\begin{APACrefDOI} \doi{10.1038/365819a0} \end{APACrefDOI}
\PrintBackRefs{\CurrentBib}

\bibitem [\protect \citeauthoryear {%
{Mandushev}%
\ \protect \BOthers {.}}{%
{Mandushev}%
\ \protect \BOthers {.}}{%
{\protect \APACyear {2005}}%
}]{%
Mandushev05}
\APACinsertmetastar {%
Mandushev05}%
\begin{APACrefauthors}%
{Mandushev}, G.%
, {Torres}, G.%
, {Latham}, D\BPBI W.%
, {Charbonneau}, D.%
, {Alonso}, R.%
, {White}, R\BPBI J.%
\BDBL {}{O'Donovan}, F\BPBI T.%
\end{APACrefauthors}%
\unskip\
\newblock
\APACrefYearMonthDay{2005}{{\APACmonth{03}}}{}.
\newblock
{\BBOQ}\APACrefatitle {{The Challenge of Wide-Field Transit Surveys: The Case
  of GSC 01944-02289}} {{The Challenge of Wide-Field Transit Surveys: The Case
  of GSC 01944-02289}}.{\BBCQ}
\newblock
\APACjournalVolNumPages{\apj}{621}{2}{1061-1071}.
\newblock
\begin{APACrefDOI} \doi{10.1086/427727} \end{APACrefDOI}
\PrintBackRefs{\CurrentBib}

\bibitem [\protect \citeauthoryear {%
Mankovich%
, Fortney%
\BCBL {}\ \BBA {} Moore%
}{%
Mankovich%
\ \protect \BOthers {.}}{%
{\protect \APACyear {2016}}%
}]{%
Mankovich:2016aa}
\APACinsertmetastar {%
Mankovich:2016aa}%
\begin{APACrefauthors}%
Mankovich, C.%
, Fortney, J.%
\BCBL {}\ \BBA {} Moore, K.%
\end{APACrefauthors}%
\unskip\
\newblock
\APACrefYearMonthDay{2016}{}{}.
\newblock
{\BBOQ}\APACrefatitle {Bayesian Evolution Models for Jupiter with Helium Rain
  and Double-diffusive Convection Show affiliations} {Bayesian evolution models
  for jupiter with helium rain and double-diffusive convection show
  affiliations}.{\BBCQ}
\newblock
\APACjournalVolNumPages{The Astrophysical Journal}{832}{}{113}.
\PrintBackRefs{\CurrentBib}

\bibitem [\protect \citeauthoryear {%
Mansfield%
, Bean%
, Line%
\BCBL {}\ \BBA {} et al.%
}{%
Mansfield%
\ \protect \BOthers {.}}{%
{\protect \APACyear {2018}}%
}]{%
Mansfield:2018aa}
\APACinsertmetastar {%
Mansfield:2018aa}%
\begin{APACrefauthors}%
Mansfield, M.%
, Bean, J.%
, Line, M.%
\BCBL {}\ \BBA {} et al.%
\end{APACrefauthors}%
\unskip\
\newblock
\APACrefYearMonthDay{2018}{}{}.
\newblock

\newblock
\APACjournalVolNumPages{The Astronomical Journal}{156}{}{10}.
\PrintBackRefs{\CurrentBib}

\bibitem [\protect \citeauthoryear {%
Mansfield%
, Bean%
, Stevenson%
\BCBL {}\ \BBA {} {et al.}%
}{%
Mansfield%
\ \protect \BOthers {.}}{%
{\protect \APACyear {2020}}%
}]{%
Mansfield:2020aa}
\APACinsertmetastar {%
Mansfield:2020aa}%
\begin{APACrefauthors}%
Mansfield, M.%
, Bean, J.%
, Stevenson, K.%
\BCBL {}\ \BBA {} {et al.}%
\end{APACrefauthors}%
\unskip\
\newblock
\APACrefYearMonthDay{2020}{}{}.
\newblock
{\BBOQ}\APACrefatitle {Evidence for {H2} Dissociation and Recombination Heat
  Transport in the Atmosphere of {KELT-9b}} {Evidence for {H2} dissociation and
  recombination heat transport in the atmosphere of {KELT-9b}}.{\BBCQ}
\newblock
\APACjournalVolNumPages{The Astrophysical Journal Letters}{888}{}{L15}.
\PrintBackRefs{\CurrentBib}

\bibitem [\protect \citeauthoryear {%
{Marley}%
, {Gelino}%
, {Stephens}%
, {Lunine}%
\BCBL {}\ \BBA {} {Freedman}%
}{%
{Marley}%
\ \protect \BOthers {.}}{%
{\protect \APACyear {1999}}%
}]{%
Marley99}
\APACinsertmetastar {%
Marley99}%
\begin{APACrefauthors}%
{Marley}, M\BPBI S.%
, {Gelino}, C.%
, {Stephens}, D.%
, {Lunine}, J\BPBI I.%
\BCBL {}\ \BBA {} {Freedman}, R.%
\end{APACrefauthors}%
\unskip\
\newblock
\APACrefYearMonthDay{1999}{{\APACmonth{03}}}{}.
\newblock
{\BBOQ}\APACrefatitle {{Reflected Spectra and Albedos of Extrasolar Giant
  Planets. I. Clear and Cloudy Atmospheres}} {{Reflected Spectra and Albedos of
  Extrasolar Giant Planets. I. Clear and Cloudy Atmospheres}}.{\BBCQ}
\newblock
\APACjournalVolNumPages{\apj}{513}{}{879-893}.
\PrintBackRefs{\CurrentBib}

\bibitem [\protect \citeauthoryear {%
May%
\ \BBA {} Stevenson%
}{%
May%
\ \BBA {} Stevenson%
}{%
{\protect \APACyear {2020}}%
}]{%
May:2020aa}
\APACinsertmetastar {%
May:2020aa}%
\begin{APACrefauthors}%
May, E.%
\BCBT {}\ \BBA {} Stevenson, K.%
\end{APACrefauthors}%
\unskip\
\newblock
\APACrefYearMonthDay{2020}{}{}.
\newblock
{\BBOQ}\APACrefatitle {Introducing a New Spitzer Master BLISS Map to Remove the
  Instrument-Systematic -- Phase-Curve-Parameter Degeneracy, as Demonstrated by
  a Reanalysis of the 4.5$\mu m$ WASP-43b Phase Curve} {Introducing a new
  spitzer master bliss map to remove the instrument-systematic --
  phase-curve-parameter degeneracy, as demonstrated by a reanalysis of the
  4.5$\mu m$ wasp-43b phase curve}.{\BBCQ}
\newblock
\APACjournalVolNumPages{arXiv e-prints:2007.06618}{}{}{}.
\PrintBackRefs{\CurrentBib}

\bibitem [\protect \citeauthoryear {%
Mayne%
\ \protect \BOthers {.}}{%
Mayne%
\ \protect \BOthers {.}}{%
{\protect \APACyear {2014}}%
}]{%
Mayne:2014}
\APACinsertmetastar {%
Mayne:2014}%
\begin{APACrefauthors}%
Mayne, N.%
, Baraffe, I.%
, Acreman, D.%
, Smith, C.%
, Browning, M.%
, Amundsen, D.%
\BDBL {}Jackson, D.%
\end{APACrefauthors}%
\unskip\
\newblock
\APACrefYearMonthDay{2014}{}{}.
\newblock
{\BBOQ}\APACrefatitle {The unified model, a fully-compressible,
  non-hydrostatic, deep atmosphere global circulation model, applied to hot
  {J}upiters: {ENDG}ame for a {HD} 209458b test case} {The unified model, a
  fully-compressible, non-hydrostatic, deep atmosphere global circulation
  model, applied to hot {J}upiters: {ENDG}ame for a {HD} 209458b test
  case}.{\BBCQ}
\newblock
\APACjournalVolNumPages{Astronomy and Astrophysics}{561}{}{A1}.
\PrintBackRefs{\CurrentBib}

\bibitem [\protect \citeauthoryear {%
Mayne%
\ \protect \BOthers {.}}{%
Mayne%
\ \protect \BOthers {.}}{%
{\protect \APACyear {2017}}%
}]{%
Mayne:2017}
\APACinsertmetastar {%
Mayne:2017}%
\begin{APACrefauthors}%
Mayne, N.%
, Debras, F.%
, Baraffe, I.%
, Thuburn, J.%
, Amundsen, D.%
, Acreman, D.%
\BDBL {}Wood, N.%
\end{APACrefauthors}%
\unskip\
\newblock
\APACrefYearMonthDay{2017}{}{}.
\newblock
{\BBOQ}\APACrefatitle {Results from a set of three-dimensional numerical
  experiments of a hot Jupiter atmosphere} {Results from a set of
  three-dimensional numerical experiments of a hot jupiter atmosphere}.{\BBCQ}
\newblock
\APACjournalVolNumPages{Astronomy {\&} Astrophysics}{604}{}{A79}.
\PrintBackRefs{\CurrentBib}

\bibitem [\protect \citeauthoryear {%
{Mayor}%
\ \BBA {} {Queloz}%
}{%
{Mayor}%
\ \BBA {} {Queloz}%
}{%
{\protect \APACyear {1995}}%
}]{%
Mayor95}
\APACinsertmetastar {%
Mayor95}%
\begin{APACrefauthors}%
{Mayor}, M.%
\BCBT {}\ \BBA {} {Queloz}, D.%
\end{APACrefauthors}%
\unskip\
\newblock
\APACrefYearMonthDay{1995}{{\APACmonth{11}}}{}.
\newblock
{\BBOQ}\APACrefatitle {{A Jupiter-Mass Companion to a Solar-Type Star}} {{A
  Jupiter-Mass Companion to a Solar-Type Star}}.{\BBCQ}
\newblock
\APACjournalVolNumPages{\nat}{378}{}{355}.
\PrintBackRefs{\CurrentBib}

\bibitem [\protect \citeauthoryear {%
McCullough%
, Crouzet%
, Deming%
\BCBL {}\ \BBA {} {et al.}%
}{%
McCullough%
\ \protect \BOthers {.}}{%
{\protect \APACyear {2014}}%
}]{%
McCullough:2014aa}
\APACinsertmetastar {%
McCullough:2014aa}%
\begin{APACrefauthors}%
McCullough, P.%
, Crouzet, N.%
, Deming, D.%
\BCBL {}\ \BBA {} {et al.}%
\end{APACrefauthors}%
\unskip\
\newblock
\APACrefYearMonthDay{2014}{}{}.
\newblock
{\BBOQ}\APACrefatitle {Water Vapor in the Spectrum of the Extrasolar Planet HD
  189733b. I. The Transit} {Water vapor in the spectrum of the extrasolar
  planet hd 189733b. i. the transit}.{\BBCQ}
\newblock
\APACjournalVolNumPages{The Astrophysical Journal}{791}{}{55}.
\PrintBackRefs{\CurrentBib}

\bibitem [\protect \citeauthoryear {%
Mendon\c{c}a%
}{%
Mendon\c{c}a%
}{%
{\protect \APACyear {2020}}%
}]{%
Mendonca:2020aa}
\APACinsertmetastar {%
Mendonca:2020aa}%
\begin{APACrefauthors}%
Mendon\c{c}a, J.%
\end{APACrefauthors}%
\unskip\
\newblock
\APACrefYearMonthDay{2020}{}{}.
\newblock
{\BBOQ}\APACrefatitle {Angular momentum and heat transport on tidally locked
  hot Jupiter planets} {Angular momentum and heat transport on tidally locked
  hot jupiter planets}.{\BBCQ}
\newblock
\APACjournalVolNumPages{Monthly Notices of the Royal Astronomical
  Society}{491}{}{1456}.
\PrintBackRefs{\CurrentBib}

\bibitem [\protect \citeauthoryear {%
Mendon\c{c}a%
, Grimm%
, Grosheintz%
\BCBL {}\ \BBA {} Heng%
}{%
Mendon\c{c}a%
\ \protect \BOthers {.}}{%
{\protect \APACyear {2016}}%
}]{%
Mendonca:2016aa}
\APACinsertmetastar {%
Mendonca:2016aa}%
\begin{APACrefauthors}%
Mendon\c{c}a, J.%
, Grimm, S.%
, Grosheintz, L.%
\BCBL {}\ \BBA {} Heng, K.%
\end{APACrefauthors}%
\unskip\
\newblock
\APACrefYearMonthDay{2016}{}{}.
\newblock
{\BBOQ}\APACrefatitle {THOR: A New and Flexible Global Circulation Model to
  Explore Planetary Atmospheres} {Thor: A new and flexible global circulation
  model to explore planetary atmospheres}.{\BBCQ}
\newblock
\APACjournalVolNumPages{The Astrophysical Journal}{829}{}{115}.
\PrintBackRefs{\CurrentBib}

\bibitem [\protect \citeauthoryear {%
Mendon\c{c}a%
, Tsai%
, Malik%
, Grimm%
\BCBL {}\ \BBA {} Heng%
}{%
Mendon\c{c}a%
\ \protect \BOthers {.}}{%
{\protect \APACyear {2018}}%
}]{%
Mendonca:2018ab}
\APACinsertmetastar {%
Mendonca:2018ab}%
\begin{APACrefauthors}%
Mendon\c{c}a, J.%
, Tsai, S.%
, Malik, M.%
, Grimm, S.%
\BCBL {}\ \BBA {} Heng, K.%
\end{APACrefauthors}%
\unskip\
\newblock
\APACrefYearMonthDay{2018}{}{}.
\newblock
{\BBOQ}\APACrefatitle {Three-dimensional Circulation Driving Chemical
  Disequilibrium in WASP-43b} {Three-dimensional circulation driving chemical
  disequilibrium in wasp-43b}.{\BBCQ}
\newblock
\APACjournalVolNumPages{The Astrophysical Journal}{869}{}{107}.
\PrintBackRefs{\CurrentBib}

\bibitem [\protect \citeauthoryear {%
Menou%
}{%
Menou%
}{%
{\protect \APACyear {2012}}%
}]{%
Menou:2012fu}
\APACinsertmetastar {%
Menou:2012fu}%
\begin{APACrefauthors}%
Menou, K.%
\end{APACrefauthors}%
\unskip\
\newblock
\APACrefYearMonthDay{2012}{}{}.
\newblock
{\BBOQ}\APACrefatitle {Magnetic scaling laws for the atmospheres of hot giant
  exoplanets} {Magnetic scaling laws for the atmospheres of hot giant
  exoplanets}.{\BBCQ}
\newblock
\APACjournalVolNumPages{The Astrophysical Journal}{745}{}{138}.
\PrintBackRefs{\CurrentBib}

\bibitem [\protect \citeauthoryear {%
Menou%
\ \BBA {} Rauscher%
}{%
Menou%
\ \BBA {} Rauscher%
}{%
{\protect \APACyear {2009}}%
}]{%
Menou:2009}
\APACinsertmetastar {%
Menou:2009}%
\begin{APACrefauthors}%
Menou, K.%
\BCBT {}\ \BBA {} Rauscher, E.%
\end{APACrefauthors}%
\unskip\
\newblock
\APACrefYearMonthDay{2009}{}{}.
\newblock
{\BBOQ}\APACrefatitle {Atmospheric circulation of hot {J}upiters: a shallow
  three-dimensional model} {Atmospheric circulation of hot {J}upiters: a
  shallow three-dimensional model}.{\BBCQ}
\newblock
\APACjournalVolNumPages{The Astrophysical Journal}{700}{}{887}.
\PrintBackRefs{\CurrentBib}

\bibitem [\protect \citeauthoryear {%
Mikal-Evans%
, Sing%
, Kataria%
\BCBL {}\ \BBA {} {et al.}%
}{%
Mikal-Evans%
\ \protect \BOthers {.}}{%
{\protect \APACyear {2020}}%
}]{%
Mikal-Evans:2020aa}
\APACinsertmetastar {%
Mikal-Evans:2020aa}%
\begin{APACrefauthors}%
Mikal-Evans, T.%
, Sing, D.%
, Kataria, T.%
\BCBL {}\ \BBA {} {et al.}%
\end{APACrefauthors}%
\unskip\
\newblock
\APACrefYearMonthDay{2020}{}{}.
\newblock
{\BBOQ}\APACrefatitle {Confirmation of water emission in the dayside spectrum
  of the ultrahot Jupiter WASP-121b} {Confirmation of water emission in the
  dayside spectrum of the ultrahot jupiter wasp-121b}.{\BBCQ}
\newblock
\APACjournalVolNumPages{Monthly Notices of the Royal Astronomical
  Society}{496}{}{1638}.
\PrintBackRefs{\CurrentBib}

\bibitem [\protect \citeauthoryear {%
Miller%
\ \BBA {} Fortney%
}{%
Miller%
\ \BBA {} Fortney%
}{%
{\protect \APACyear {2011}}%
}]{%
Miller:2011}
\APACinsertmetastar {%
Miller:2011}%
\begin{APACrefauthors}%
Miller, N.%
\BCBT {}\ \BBA {} Fortney, J.%
\end{APACrefauthors}%
\unskip\
\newblock
\APACrefYearMonthDay{2011}{}{}.
\newblock
{\BBOQ}\APACrefatitle {The heavy-element masses of extrasolar giant planets,
  revealed} {The heavy-element masses of extrasolar giant planets,
  revealed}.{\BBCQ}
\newblock
\APACjournalVolNumPages{The Astrophysical Journal Letters}{736}{}{L29}.
\PrintBackRefs{\CurrentBib}

\bibitem [\protect \citeauthoryear {%
Miller%
, Fortney%
\BCBL {}\ \BBA {} Jackson%
}{%
Miller%
\ \protect \BOthers {.}}{%
{\protect \APACyear {2009}}%
}]{%
Miller:2009}
\APACinsertmetastar {%
Miller:2009}%
\begin{APACrefauthors}%
Miller, N.%
, Fortney, J.%
\BCBL {}\ \BBA {} Jackson, B.%
\end{APACrefauthors}%
\unskip\
\newblock
\APACrefYearMonthDay{2009}{}{}.
\newblock
{\BBOQ}\APACrefatitle {Inflating and deflating hot {J}upiters: Coupled tidal
  and thermal evolution of known transiting planets} {Inflating and deflating
  hot {J}upiters: Coupled tidal and thermal evolution of known transiting
  planets}.{\BBCQ}
\newblock
\APACjournalVolNumPages{The Astrophysical Journal}{702}{}{1413}.
\PrintBackRefs{\CurrentBib}

\bibitem [\protect \citeauthoryear {%
{Miller}%
\ \BBA {} {Fortney}%
}{%
{Miller}%
\ \BBA {} {Fortney}%
}{%
{\protect \APACyear {2011}}%
}]{%
Miller11}
\APACinsertmetastar {%
Miller11}%
\begin{APACrefauthors}%
{Miller}, N.%
\BCBT {}\ \BBA {} {Fortney}, J\BPBI J.%
\end{APACrefauthors}%
\unskip\
\newblock
\APACrefYearMonthDay{2011}{{\APACmonth{08}}}{}.
\newblock
{\BBOQ}\APACrefatitle {{The Heavy-element Masses of Extrasolar Giant Planets,
  Revealed}} {{The Heavy-element Masses of Extrasolar Giant Planets,
  Revealed}}.{\BBCQ}
\newblock
\APACjournalVolNumPages{\apjl}{736}{}{L29}.
\newblock
\begin{APACrefDOI} \doi{10.1088/2041-8205/736/2/L29} \end{APACrefDOI}
\PrintBackRefs{\CurrentBib}

\bibitem [\protect \citeauthoryear {%
{Miller-Ricci Kempton}%
, {Zahnle}%
\BCBL {}\ \BBA {} {Fortney}%
}{%
{Miller-Ricci Kempton}%
\ \protect \BOthers {.}}{%
{\protect \APACyear {2012}}%
}]{%
Kempton12}
\APACinsertmetastar {%
Kempton12}%
\begin{APACrefauthors}%
{Miller-Ricci Kempton}, E.%
, {Zahnle}, K.%
\BCBL {}\ \BBA {} {Fortney}, J\BPBI J.%
\end{APACrefauthors}%
\unskip\
\newblock
\APACrefYearMonthDay{2012}{{\APACmonth{01}}}{}.
\newblock
{\BBOQ}\APACrefatitle {{The Atmospheric Chemistry of GJ 1214b: Photochemistry
  and Clouds}} {{The Atmospheric Chemistry of GJ 1214b: Photochemistry and
  Clouds}}.{\BBCQ}
\newblock
\APACjournalVolNumPages{\apj}{745}{}{3}.
\newblock
\begin{APACrefDOI} \doi{10.1088/0004-637X/745/1/3} \end{APACrefDOI}
\PrintBackRefs{\CurrentBib}

\bibitem [\protect \citeauthoryear {%
{Moe}%
\ \BBA {} {Kratter}%
}{%
{Moe}%
\ \BBA {} {Kratter}%
}{%
{\protect \APACyear {2019}}%
}]{%
moe19}
\APACinsertmetastar {%
moe19}%
\begin{APACrefauthors}%
{Moe}, M.%
\BCBT {}\ \BBA {} {Kratter}, K\BPBI M.%
\end{APACrefauthors}%
\unskip\
\newblock
\APACrefYearMonthDay{2019}{{\APACmonth{12}}}{}.
\newblock
{\BBOQ}\APACrefatitle {{Impact of Binary Stars on Planet Statistics -- I.
  Planet Occurrence Rates, Trends with Stellar Mass, and Wide Companions to Hot
  Jupiter Hosts}} {{Impact of Binary Stars on Planet Statistics -- I. Planet
  Occurrence Rates, Trends with Stellar Mass, and Wide Companions to Hot
  Jupiter Hosts}}.{\BBCQ}
\newblock
\APACjournalVolNumPages{arXiv e-prints}{}{}{arXiv:1912.01699}.
\PrintBackRefs{\CurrentBib}

\bibitem [\protect \citeauthoryear {%
{Molli{\`e}re}%
, {van Boekel}%
, {Dullemond}%
, {Henning}%
\BCBL {}\ \BBA {} {Mordasini}%
}{%
{Molli{\`e}re}%
\ \protect \BOthers {.}}{%
{\protect \APACyear {2015}}%
}]{%
Molliere15}
\APACinsertmetastar {%
Molliere15}%
\begin{APACrefauthors}%
{Molli{\`e}re}, P.%
, {van Boekel}, R.%
, {Dullemond}, C.%
, {Henning}, T.%
\BCBL {}\ \BBA {} {Mordasini}, C.%
\end{APACrefauthors}%
\unskip\
\newblock
\APACrefYearMonthDay{2015}{{\APACmonth{11}}}{}.
\newblock
{\BBOQ}\APACrefatitle {{Model Atmospheres of Irradiated Exoplanets: The
  Influence of Stellar Parameters, Metallicity, and the C/O Ratio}} {{Model
  Atmospheres of Irradiated Exoplanets: The Influence of Stellar Parameters,
  Metallicity, and the C/O Ratio}}.{\BBCQ}
\newblock
\APACjournalVolNumPages{\apj}{813}{}{47}.
\newblock
\begin{APACrefDOI} \doi{10.1088/0004-637X/813/1/47} \end{APACrefDOI}
\PrintBackRefs{\CurrentBib}

\bibitem [\protect \citeauthoryear {%
{Mordasini}%
, {Klahr}%
, {Alibert}%
, {Miller}%
\BCBL {}\ \BBA {} {Henning}%
}{%
{Mordasini}%
\ \protect \BOthers {.}}{%
{\protect \APACyear {2014}}%
}]{%
Mordasini14}
\APACinsertmetastar {%
Mordasini14}%
\begin{APACrefauthors}%
{Mordasini}, C.%
, {Klahr}, H.%
, {Alibert}, Y.%
, {Miller}, N.%
\BCBL {}\ \BBA {} {Henning}, T.%
\end{APACrefauthors}%
\unskip\
\newblock
\APACrefYearMonthDay{2014}{{\APACmonth{03}}}{}.
\newblock
{\BBOQ}\APACrefatitle {{Grain opacity and the bulk composition of extrasolar
  planets. I. Results from scaling the ISM opacity}} {{Grain opacity and the
  bulk composition of extrasolar planets. I. Results from scaling the ISM
  opacity}}.{\BBCQ}
\newblock
\APACjournalVolNumPages{ArXiv:1403.5272}{}{}{}.
\PrintBackRefs{\CurrentBib}

\bibitem [\protect \citeauthoryear {%
{Mordasini}%
, {van Boekel}%
, {Molli{\`e}re}%
, {Henning}%
\BCBL {}\ \BBA {} {Benneke}%
}{%
{Mordasini}%
\ \protect \BOthers {.}}{%
{\protect \APACyear {2016}}%
}]{%
Mordasini16}
\APACinsertmetastar {%
Mordasini16}%
\begin{APACrefauthors}%
{Mordasini}, C.%
, {van Boekel}, R.%
, {Molli{\`e}re}, P.%
, {Henning}, T.%
\BCBL {}\ \BBA {} {Benneke}, B.%
\end{APACrefauthors}%
\unskip\
\newblock
\APACrefYearMonthDay{2016}{{\APACmonth{11}}}{}.
\newblock
{\BBOQ}\APACrefatitle {{The Imprint of Exoplanet Formation History on
  Observable Present-day Spectra of Hot Jupiters}} {{The Imprint of Exoplanet
  Formation History on Observable Present-day Spectra of Hot Jupiters}}.{\BBCQ}
\newblock
\APACjournalVolNumPages{\apj}{832}{}{41}.
\newblock
\begin{APACrefDOI} \doi{10.3847/0004-637X/832/1/41} \end{APACrefDOI}
\PrintBackRefs{\CurrentBib}

\bibitem [\protect \citeauthoryear {%
{Moses}%
\ \protect \BOthers {.}}{%
{Moses}%
\ \protect \BOthers {.}}{%
{\protect \APACyear {2011}}%
}]{%
Moses11}
\APACinsertmetastar {%
Moses11}%
\begin{APACrefauthors}%
{Moses}, J\BPBI I.%
, {Visscher}, C.%
, {Fortney}, J\BPBI J.%
, {Showman}, A\BPBI P.%
, {Lewis}, N\BPBI K.%
, {Griffith}, C\BPBI A.%
\BDBL {}{Freedman}, R\BPBI S.%
\end{APACrefauthors}%
\unskip\
\newblock
\APACrefYearMonthDay{2011}{{\APACmonth{08}}}{}.
\newblock
{\BBOQ}\APACrefatitle {{Disequilibrium Carbon, Oxygen, and Nitrogen Chemistry
  in the Atmospheres of HD 189733b and HD 209458b}} {{Disequilibrium Carbon,
  Oxygen, and Nitrogen Chemistry in the Atmospheres of HD 189733b and HD
  209458b}}.{\BBCQ}
\newblock
\APACjournalVolNumPages{\apj}{737}{1}{15}.
\newblock
\begin{APACrefDOI} \doi{10.1088/0004-637X/737/1/15} \end{APACrefDOI}
\PrintBackRefs{\CurrentBib}

\bibitem [\protect \citeauthoryear {%
{Mu{\~n}oz}%
, {Lai}%
\BCBL {}\ \BBA {} {Liu}%
}{%
{Mu{\~n}oz}%
\ \protect \BOthers {.}}{%
{\protect \APACyear {2016}}%
}]{%
muno16}
\APACinsertmetastar {%
muno16}%
\begin{APACrefauthors}%
{Mu{\~n}oz}, D\BPBI J.%
, {Lai}, D.%
\BCBL {}\ \BBA {} {Liu}, B.%
\end{APACrefauthors}%
\unskip\
\newblock
\APACrefYearMonthDay{2016}{{\APACmonth{07}}}{}.
\newblock
{\BBOQ}\APACrefatitle {{The formation efficiency of close-in planets via
  Lidov-Kozai migration: analytic calculations}} {{The formation efficiency of
  close-in planets via Lidov-Kozai migration: analytic calculations}}.{\BBCQ}
\newblock
\APACjournalVolNumPages{\mnras}{460}{}{1086-1093}.
\newblock
\begin{APACrefDOI} \doi{10.1093/mnras/stw983} \end{APACrefDOI}
\PrintBackRefs{\CurrentBib}

\bibitem [\protect \citeauthoryear {%
{Murray-Clay}%
, {Chiang}%
\BCBL {}\ \BBA {} {Murray}%
}{%
{Murray-Clay}%
\ \protect \BOthers {.}}{%
{\protect \APACyear {2008}}%
}]{%
Murray08}
\APACinsertmetastar {%
Murray08}%
\begin{APACrefauthors}%
{Murray-Clay}, R.%
, {Chiang}, E.%
\BCBL {}\ \BBA {} {Murray}, N.%
\end{APACrefauthors}%
\unskip\
\newblock
\APACrefYearMonthDay{2008}{{\APACmonth{11}}}{}.
\newblock
{\BBOQ}\APACrefatitle {{Atmospheric Escape from Hot Jupiters}} {{Atmospheric
  Escape from Hot Jupiters}}.{\BBCQ}
\newblock
\APACjournalVolNumPages{ArXiv e-prints/0811.0006}{}{}{}.
\PrintBackRefs{\CurrentBib}

\bibitem [\protect \citeauthoryear {%
{Mustill}%
, {Davies}%
\BCBL {}\ \BBA {} {Johansen}%
}{%
{Mustill}%
\ \protect \BOthers {.}}{%
{\protect \APACyear {2015}}%
}]{%
must15}
\APACinsertmetastar {%
must15}%
\begin{APACrefauthors}%
{Mustill}, A\BPBI J.%
, {Davies}, M\BPBI B.%
\BCBL {}\ \BBA {} {Johansen}, A.%
\end{APACrefauthors}%
\unskip\
\newblock
\APACrefYearMonthDay{2015}{{\APACmonth{07}}}{}.
\newblock
{\BBOQ}\APACrefatitle {{The Destruction of Inner Planetary Systems during
  High-eccentricity Migration of Gas Giants}} {{The Destruction of Inner
  Planetary Systems during High-eccentricity Migration of Gas Giants}}.{\BBCQ}
\newblock
\APACjournalVolNumPages{\apj}{808}{}{14}.
\newblock
\begin{APACrefDOI} \doi{10.1088/0004-637X/808/1/14} \end{APACrefDOI}
\PrintBackRefs{\CurrentBib}

\bibitem [\protect \citeauthoryear {%
{Naoz}%
}{%
{Naoz}%
}{%
{\protect \APACyear {2016}}%
}]{%
naoz16}
\APACinsertmetastar {%
naoz16}%
\begin{APACrefauthors}%
{Naoz}, S.%
\end{APACrefauthors}%
\unskip\
\newblock
\APACrefYearMonthDay{2016}{{\APACmonth{09}}}{}.
\newblock
{\BBOQ}\APACrefatitle {{The Eccentric Kozai-Lidov Effect and Its Applications}}
  {{The Eccentric Kozai-Lidov Effect and Its Applications}}.{\BBCQ}
\newblock
\APACjournalVolNumPages{\araa}{54}{}{441-489}.
\newblock
\begin{APACrefDOI} \doi{10.1146/annurev-astro-081915-023315} \end{APACrefDOI}
\PrintBackRefs{\CurrentBib}

\bibitem [\protect \citeauthoryear {%
{Nettelmann}%
, {Fortney}%
, {Moore}%
\BCBL {}\ \BBA {} {Mankovich}%
}{%
{Nettelmann}%
\ \protect \BOthers {.}}{%
{\protect \APACyear {2015}}%
}]{%
Nettelmann15}
\APACinsertmetastar {%
Nettelmann15}%
\begin{APACrefauthors}%
{Nettelmann}, N.%
, {Fortney}, J\BPBI J.%
, {Moore}, K.%
\BCBL {}\ \BBA {} {Mankovich}, C.%
\end{APACrefauthors}%
\unskip\
\newblock
\APACrefYearMonthDay{2015}{{\APACmonth{03}}}{}.
\newblock
{\BBOQ}\APACrefatitle {{An exploration of double diffusive convection in
  Jupiter as a result of hydrogen-helium phase separation}} {{An exploration of
  double diffusive convection in Jupiter as a result of hydrogen-helium phase
  separation}}.{\BBCQ}
\newblock
\APACjournalVolNumPages{\mnras}{447}{}{3422-3441}.
\newblock
\begin{APACrefDOI} \doi{10.1093/mnras/stu2634} \end{APACrefDOI}
\PrintBackRefs{\CurrentBib}

\bibitem [\protect \citeauthoryear {%
{Ngo}%
\ \protect \BOthers {.}}{%
{Ngo}%
\ \protect \BOthers {.}}{%
{\protect \APACyear {2016}}%
}]{%
ngo16}
\APACinsertmetastar {%
ngo16}%
\begin{APACrefauthors}%
{Ngo}, H.%
, {Knutson}, H\BPBI A.%
, {Hinkley}, S.%
, {Bryan}, M.%
, {Crepp}, J\BPBI R.%
, {Batygin}, K.%
\BDBL {}{Wang}, J.%
\end{APACrefauthors}%
\unskip\
\newblock
\APACrefYearMonthDay{2016}{{\APACmonth{08}}}{}.
\newblock
{\BBOQ}\APACrefatitle {{Friends of Hot Jupiters. IV. Stellar Companions Beyond
  50 au Might Facilitate Giant Planet Formation, but Most are Unlikely to Cause
  Kozai-Lidov Migration}} {{Friends of Hot Jupiters. IV. Stellar Companions
  Beyond 50 au Might Facilitate Giant Planet Formation, but Most are Unlikely
  to Cause Kozai-Lidov Migration}}.{\BBCQ}
\newblock
\APACjournalVolNumPages{\apj}{827}{}{8}.
\newblock
\begin{APACrefDOI} \doi{10.3847/0004-637X/827/1/8} \end{APACrefDOI}
\PrintBackRefs{\CurrentBib}

\bibitem [\protect \citeauthoryear {%
Nikolov%
, Sing%
, Pont%
\BCBL {}\ \BBA {} {et al.}%
}{%
Nikolov%
\ \protect \BOthers {.}}{%
{\protect \APACyear {2014}}%
}]{%
Nikolov:2014aa}
\APACinsertmetastar {%
Nikolov:2014aa}%
\begin{APACrefauthors}%
Nikolov, N.%
, Sing, D.%
, Pont, F.%
\BCBL {}\ \BBA {} {et al.}%
\end{APACrefauthors}%
\unskip\
\newblock
\APACrefYearMonthDay{2014}{}{}.
\newblock

\newblock
\APACjournalVolNumPages{Monthly Notices of the Royal Astronomical
  Society}{437}{}{46}.
\PrintBackRefs{\CurrentBib}

\bibitem [\protect \citeauthoryear {%
{Ninan}%
\ \protect \BOthers {.}}{%
{Ninan}%
\ \protect \BOthers {.}}{%
{\protect \APACyear {2020}}%
}]{%
Ninan20}
\APACinsertmetastar {%
Ninan20}%
\begin{APACrefauthors}%
{Ninan}, J\BPBI P.%
, {Stefansson}, G.%
, {Mahadevan}, S.%
, {Bender}, C.%
, {Robertson}, P.%
, {Ramsey}, L.%
\BDBL {}{Schwab}, C.%
\end{APACrefauthors}%
\unskip\
\newblock
\APACrefYearMonthDay{2020}{{\APACmonth{05}}}{}.
\newblock
{\BBOQ}\APACrefatitle {{Evidence for He I 10830 {\r{A}} Absorption during the
  Transit of a Warm Neptune around the M-dwarf GJ 3470 with the Habitable-zone
  Planet Finder}} {{Evidence for He I 10830 {\r{A}} Absorption during the
  Transit of a Warm Neptune around the M-dwarf GJ 3470 with the Habitable-zone
  Planet Finder}}.{\BBCQ}
\newblock
\APACjournalVolNumPages{\apj}{894}{2}{97}.
\newblock
\begin{APACrefDOI} \doi{10.3847/1538-4357/ab8559} \end{APACrefDOI}
\PrintBackRefs{\CurrentBib}

\bibitem [\protect \citeauthoryear {%
Nugroho%
, Kawahara%
, Masuda%
\BCBL {}\ \BBA {} {et al.}%
}{%
Nugroho%
\ \protect \BOthers {.}}{%
{\protect \APACyear {2017}}%
}]{%
Nugroho:2017aa}
\APACinsertmetastar {%
Nugroho:2017aa}%
\begin{APACrefauthors}%
Nugroho, S.%
, Kawahara, H.%
, Masuda, K.%
\BCBL {}\ \BBA {} {et al.}%
\end{APACrefauthors}%
\unskip\
\newblock
\APACrefYearMonthDay{2017}{}{}.
\newblock
{\BBOQ}\APACrefatitle {The American Astronomical Society, find out more The
  Institute of Physics, find out more High-resolution Spectroscopic Detection
  of TiO and a Stratosphere in the Day-side of WASP-33b} {The american
  astronomical society, find out more the institute of physics, find out more
  high-resolution spectroscopic detection of tio and a stratosphere in the
  day-side of wasp-33b}.{\BBCQ}
\newblock
\APACjournalVolNumPages{The Astronomical Journal}{156}{}{221}.
\PrintBackRefs{\CurrentBib}

\bibitem [\protect \citeauthoryear {%
{{\"O}berg}%
, {Murray-Clay}%
\BCBL {}\ \BBA {} {Bergin}%
}{%
{{\"O}berg}%
\ \protect \BOthers {.}}{%
{\protect \APACyear {2011}}%
}]{%
Oberg11}
\APACinsertmetastar {%
Oberg11}%
\begin{APACrefauthors}%
{{\"O}berg}, K\BPBI I.%
, {Murray-Clay}, R.%
\BCBL {}\ \BBA {} {Bergin}, E\BPBI A.%
\end{APACrefauthors}%
\unskip\
\newblock
\APACrefYearMonthDay{2011}{{\APACmonth{12}}}{}.
\newblock
{\BBOQ}\APACrefatitle {{The Effects of Snowlines on C/O in Planetary
  Atmospheres}} {{The Effects of Snowlines on C/O in Planetary
  Atmospheres}}.{\BBCQ}
\newblock
\APACjournalVolNumPages{\apjl}{743}{}{L16}.
\newblock
\begin{APACrefDOI} \doi{10.1088/2041-8205/743/1/L16} \end{APACrefDOI}
\PrintBackRefs{\CurrentBib}

\bibitem [\protect \citeauthoryear {%
{Oklop{\v{c}}i{\'c}}%
\ \BBA {} {Hirata}%
}{%
{Oklop{\v{c}}i{\'c}}%
\ \BBA {} {Hirata}%
}{%
{\protect \APACyear {2018}}%
}]{%
Oklopcic18}
\APACinsertmetastar {%
Oklopcic18}%
\begin{APACrefauthors}%
{Oklop{\v{c}}i{\'c}}, A.%
\BCBT {}\ \BBA {} {Hirata}, C\BPBI M.%
\end{APACrefauthors}%
\unskip\
\newblock
\APACrefYearMonthDay{2018}{{\APACmonth{03}}}{}.
\newblock
{\BBOQ}\APACrefatitle {{A New Window into Escaping Exoplanet Atmospheres: 10830
  {\r{A}} Line of Helium}} {{A New Window into Escaping Exoplanet Atmospheres:
  10830 {\r{A}} Line of Helium}}.{\BBCQ}
\newblock
\APACjournalVolNumPages{\apjl}{855}{1}{L11}.
\newblock
\begin{APACrefDOI} \doi{10.3847/2041-8213/aaada9} \end{APACrefDOI}
\PrintBackRefs{\CurrentBib}

\bibitem [\protect \citeauthoryear {%
{Oklop{\v{c}}i{\'c}}%
, {Silva}%
, {Montero-Camacho}%
\BCBL {}\ \BBA {} {Hirata}%
}{%
{Oklop{\v{c}}i{\'c}}%
\ \protect \BOthers {.}}{%
{\protect \APACyear {2020}}%
}]{%
Oklopcic20}
\APACinsertmetastar {%
Oklopcic20}%
\begin{APACrefauthors}%
{Oklop{\v{c}}i{\'c}}, A.%
, {Silva}, M.%
, {Montero-Camacho}, P.%
\BCBL {}\ \BBA {} {Hirata}, C\BPBI M.%
\end{APACrefauthors}%
\unskip\
\newblock
\APACrefYearMonthDay{2020}{{\APACmonth{02}}}{}.
\newblock
{\BBOQ}\APACrefatitle {{Detecting Magnetic Fields in Exoplanets with
  Spectropolarimetry of the Helium Line at 1083 nm}} {{Detecting Magnetic
  Fields in Exoplanets with Spectropolarimetry of the Helium Line at 1083
  nm}}.{\BBCQ}
\newblock
\APACjournalVolNumPages{\apj}{890}{1}{88}.
\newblock
\begin{APACrefDOI} \doi{10.3847/1538-4357/ab67c6} \end{APACrefDOI}
\PrintBackRefs{\CurrentBib}

\bibitem [\protect \citeauthoryear {%
Oreshenko%
, Heng%
\BCBL {}\ \BBA {} Demory%
}{%
Oreshenko%
\ \protect \BOthers {.}}{%
{\protect \APACyear {2016}}%
}]{%
Oreshenko:2016}
\APACinsertmetastar {%
Oreshenko:2016}%
\begin{APACrefauthors}%
Oreshenko, M.%
, Heng, K.%
\BCBL {}\ \BBA {} Demory, B.%
\end{APACrefauthors}%
\unskip\
\newblock
\APACrefYearMonthDay{2016}{}{}.
\newblock
{\BBOQ}\APACrefatitle {Optical phase curves as diagnostics for aerosol
  composition in exoplanetary atmospheres} {Optical phase curves as diagnostics
  for aerosol composition in exoplanetary atmospheres}.{\BBCQ}
\newblock
\APACjournalVolNumPages{Monthly Notices of the Royal Astronomical
  Society}{457}{}{3420}.
\PrintBackRefs{\CurrentBib}

\bibitem [\protect \citeauthoryear {%
{Owen}%
}{%
{Owen}%
}{%
{\protect \APACyear {2019}}%
}]{%
Owen19}
\APACinsertmetastar {%
Owen19}%
\begin{APACrefauthors}%
{Owen}, J\BPBI E.%
\end{APACrefauthors}%
\unskip\
\newblock
\APACrefYearMonthDay{2019}{{\APACmonth{05}}}{}.
\newblock
{\BBOQ}\APACrefatitle {{Atmospheric Escape and the Evolution of Close-In
  Exoplanets}} {{Atmospheric Escape and the Evolution of Close-In
  Exoplanets}}.{\BBCQ}
\newblock
\APACjournalVolNumPages{Annual Review of Earth and Planetary
  Sciences}{47}{}{67-90}.
\newblock
\begin{APACrefDOI} \doi{10.1146/annurev-earth-053018-060246} \end{APACrefDOI}
\PrintBackRefs{\CurrentBib}

\bibitem [\protect \citeauthoryear {%
{Owen}%
\ \BBA {} {Adams}%
}{%
{Owen}%
\ \BBA {} {Adams}%
}{%
{\protect \APACyear {2014}}%
}]{%
Owen14}
\APACinsertmetastar {%
Owen14}%
\begin{APACrefauthors}%
{Owen}, J\BPBI E.%
\BCBT {}\ \BBA {} {Adams}, F\BPBI C.%
\end{APACrefauthors}%
\unskip\
\newblock
\APACrefYearMonthDay{2014}{{\APACmonth{11}}}{}.
\newblock
{\BBOQ}\APACrefatitle {{Magnetically controlled mass-loss from extrasolar
  planets in close orbits}} {{Magnetically controlled mass-loss from extrasolar
  planets in close orbits}}.{\BBCQ}
\newblock
\APACjournalVolNumPages{\mnras}{444}{4}{3761-3779}.
\newblock
\begin{APACrefDOI} \doi{10.1093/mnras/stu1684} \end{APACrefDOI}
\PrintBackRefs{\CurrentBib}

\bibitem [\protect \citeauthoryear {%
{Paardekooper}%
\ \BBA {} {Mellema}%
}{%
{Paardekooper}%
\ \BBA {} {Mellema}%
}{%
{\protect \APACyear {2006}}%
}]{%
paar06}
\APACinsertmetastar {%
paar06}%
\begin{APACrefauthors}%
{Paardekooper}, S\BHBI J.%
\BCBT {}\ \BBA {} {Mellema}, G.%
\end{APACrefauthors}%
\unskip\
\newblock
\APACrefYearMonthDay{2006}{{\APACmonth{11}}}{}.
\newblock
{\BBOQ}\APACrefatitle {{Halting type I planet migration in non-isothermal
  disks}} {{Halting type I planet migration in non-isothermal disks}}.{\BBCQ}
\newblock
\APACjournalVolNumPages{\aap}{459}{}{L17-L20}.
\newblock
\begin{APACrefDOI} \doi{10.1051/0004-6361:20066304} \end{APACrefDOI}
\PrintBackRefs{\CurrentBib}

\bibitem [\protect \citeauthoryear {%
Parmentier%
\ \BBA {} Crossfield%
}{%
Parmentier%
\ \BBA {} Crossfield%
}{%
{\protect \APACyear {2018}}%
}]{%
Parmentier:2017}
\APACinsertmetastar {%
Parmentier:2017}%
\begin{APACrefauthors}%
Parmentier, V.%
\BCBT {}\ \BBA {} Crossfield, I.%
\end{APACrefauthors}%
\unskip\
\newblock
\APACrefYearMonthDay{2018}{}{}.
\newblock
{\BBOQ}\APACrefatitle {The Exoplanet Handbook} {The exoplanet handbook}.{\BBCQ}
\newblock
\BIn{} (\BCHAP\ Exoplanet phase curves: observations and theory).
\newblock
\APACaddressPublisher{}{Springer}.
\PrintBackRefs{\CurrentBib}

\bibitem [\protect \citeauthoryear {%
Parmentier%
, Fortney%
, Showman%
, Morley%
\BCBL {}\ \BBA {} Marley%
}{%
Parmentier%
\ \protect \BOthers {.}}{%
{\protect \APACyear {2016}}%
}]{%
Parmentier:2015}
\APACinsertmetastar {%
Parmentier:2015}%
\begin{APACrefauthors}%
Parmentier, V.%
, Fortney, J.%
, Showman, A.%
, Morley, C.%
\BCBL {}\ \BBA {} Marley, M.%
\end{APACrefauthors}%
\unskip\
\newblock
\APACrefYearMonthDay{2016}{}{}.
\newblock
{\BBOQ}\APACrefatitle {A transition in the composition of clouds in hot
  Jupiters} {A transition in the composition of clouds in hot jupiters}.{\BBCQ}
\newblock
\APACjournalVolNumPages{The Astrophysical Journal}{828}{}{22}.
\PrintBackRefs{\CurrentBib}

\bibitem [\protect \citeauthoryear {%
{Parmentier}%
, {Fortney}%
, {Showman}%
, {Morley}%
\BCBL {}\ \BBA {} {Marley}%
}{%
{Parmentier}%
\ \protect \BOthers {.}}{%
{\protect \APACyear {2016}}%
}]{%
Parmentier16}
\APACinsertmetastar {%
Parmentier16}%
\begin{APACrefauthors}%
{Parmentier}, V.%
, {Fortney}, J\BPBI J.%
, {Showman}, A\BPBI P.%
, {Morley}, C.%
\BCBL {}\ \BBA {} {Marley}, M\BPBI S.%
\end{APACrefauthors}%
\unskip\
\newblock
\APACrefYearMonthDay{2016}{{\APACmonth{09}}}{}.
\newblock
{\BBOQ}\APACrefatitle {{Transitions in the Cloud Composition of Hot Jupiters}}
  {{Transitions in the Cloud Composition of Hot Jupiters}}.{\BBCQ}
\newblock
\APACjournalVolNumPages{\apj}{828}{}{22}.
\newblock
\begin{APACrefDOI} \doi{10.3847/0004-637X/828/1/22} \end{APACrefDOI}
\PrintBackRefs{\CurrentBib}

\bibitem [\protect \citeauthoryear {%
Parmentier%
, Line%
, Bean%
\BCBL {}\ \BBA {} et al.%
}{%
Parmentier%
\ \protect \BOthers {.}}{%
{\protect \APACyear {2018}}%
}]{%
Parmentier:2018aa}
\APACinsertmetastar {%
Parmentier:2018aa}%
\begin{APACrefauthors}%
Parmentier, V.%
, Line, M.%
, Bean, J.%
\BCBL {}\ \BBA {} et al.%
\end{APACrefauthors}%
\unskip\
\newblock
\APACrefYearMonthDay{2018}{}{}.
\newblock

\newblock
\APACjournalVolNumPages{Astronomy {\&} Astrophysics}{617}{}{A110}.
\PrintBackRefs{\CurrentBib}

\bibitem [\protect \citeauthoryear {%
Parmentier%
, Showman%
\BCBL {}\ \BBA {} Fortney%
}{%
Parmentier%
\ \protect \BOthers {.}}{%
{\protect \APACyear {2020}}%
}]{%
Parmentier:2020aa}
\APACinsertmetastar {%
Parmentier:2020aa}%
\begin{APACrefauthors}%
Parmentier, V.%
, Showman, A.%
\BCBL {}\ \BBA {} Fortney, J.%
\end{APACrefauthors}%
\unskip\
\newblock
\APACrefYearMonthDay{2020}{}{}.
\newblock
{\BBOQ}\APACrefatitle {The cloudy shape of hot Jupiter thermal phase curves}
  {The cloudy shape of hot jupiter thermal phase curves}.{\BBCQ}
\newblock
\APACjournalVolNumPages{Monthly Notices of the Royal Astronomical
  Society}{}{}{}.
\PrintBackRefs{\CurrentBib}

\bibitem [\protect \citeauthoryear {%
Parmentier%
, Showman%
\BCBL {}\ \BBA {} Lian%
}{%
Parmentier%
\ \protect \BOthers {.}}{%
{\protect \APACyear {2013}}%
}]{%
parmentier_2013}
\APACinsertmetastar {%
parmentier_2013}%
\begin{APACrefauthors}%
Parmentier, V.%
, Showman, A.%
\BCBL {}\ \BBA {} Lian, Y.%
\end{APACrefauthors}%
\unskip\
\newblock
\APACrefYearMonthDay{2013}{}{}.
\newblock
{\BBOQ}\APACrefatitle {3{D} mixing in hot {J}upiters atmospheres} {3{D} mixing
  in hot {J}upiters atmospheres}.{\BBCQ}
\newblock
\APACjournalVolNumPages{Astronomy and Astrophysics}{558}{}{A91}.
\PrintBackRefs{\CurrentBib}

\bibitem [\protect \citeauthoryear {%
Pass%
, Cowan%
, Cubillos%
\BCBL {}\ \BBA {} Sklar%
}{%
Pass%
\ \protect \BOthers {.}}{%
{\protect \APACyear {2019}}%
}]{%
Pass:2019aa}
\APACinsertmetastar {%
Pass:2019aa}%
\begin{APACrefauthors}%
Pass, E.%
, Cowan, N.%
, Cubillos, P.%
\BCBL {}\ \BBA {} Sklar, J.%
\end{APACrefauthors}%
\unskip\
\newblock
\APACrefYearMonthDay{2019}{}{}.
\newblock
{\BBOQ}\APACrefatitle {Estimating dayside effective temperatures of hot
  Jupiters and associated uncertainties through Gaussian process regression}
  {Estimating dayside effective temperatures of hot jupiters and associated
  uncertainties through gaussian process regression}.{\BBCQ}
\newblock
\APACjournalVolNumPages{Monthly Notices of the Royal Astronomical
  Society}{489}{}{941}.
\PrintBackRefs{\CurrentBib}

\bibitem [\protect \citeauthoryear {%
Perez-Becker%
\ \BBA {} Showman%
}{%
Perez-Becker%
\ \BBA {} Showman%
}{%
{\protect \APACyear {2013}}%
}]{%
Perez-Becker:2013fv}
\APACinsertmetastar {%
Perez-Becker:2013fv}%
\begin{APACrefauthors}%
Perez-Becker, D.%
\BCBT {}\ \BBA {} Showman, A.%
\end{APACrefauthors}%
\unskip\
\newblock
\APACrefYearMonthDay{2013}{}{}.
\newblock
{\BBOQ}\APACrefatitle {Atmospheric heat redistribution on hot {J}upiters}
  {Atmospheric heat redistribution on hot {J}upiters}.{\BBCQ}
\newblock
\APACjournalVolNumPages{The Astrophysical Journal}{776}{}{134}.
\PrintBackRefs{\CurrentBib}

\bibitem [\protect \citeauthoryear {%
Perna%
, Heng%
\BCBL {}\ \BBA {} Pont%
}{%
Perna%
\ \protect \BOthers {.}}{%
{\protect \APACyear {2012}}%
}]{%
perna_2012}
\APACinsertmetastar {%
perna_2012}%
\begin{APACrefauthors}%
Perna, R.%
, Heng, K.%
\BCBL {}\ \BBA {} Pont, F.%
\end{APACrefauthors}%
\unskip\
\newblock
\APACrefYearMonthDay{2012}{}{}.
\newblock
{\BBOQ}\APACrefatitle {The effects of irradiation on hot {J}ovian atmospheres:
  Heat redistribution and energy dissipation} {The effects of irradiation on
  hot {J}ovian atmospheres: Heat redistribution and energy dissipation}.{\BBCQ}
\newblock
\APACjournalVolNumPages{The Astrophysical Journal}{751}{}{59}.
\PrintBackRefs{\CurrentBib}

\bibitem [\protect \citeauthoryear {%
Perna%
, Menou%
\BCBL {}\ \BBA {} Rauscher%
}{%
Perna%
\ \protect \BOthers {.}}{%
{\protect \APACyear {2010}}%
{\protect \APACexlab {{\protect \BCnt {1}}}}}]{%
Perna_2010_1}
\APACinsertmetastar {%
Perna_2010_1}%
\begin{APACrefauthors}%
Perna, R.%
, Menou, K.%
\BCBL {}\ \BBA {} Rauscher, E.%
\end{APACrefauthors}%
\unskip\
\newblock
\APACrefYearMonthDay{2010{\protect \BCnt {1}}}{}{}.
\newblock
{\BBOQ}\APACrefatitle {Magnetic drag on hot {J}upiter atmospheric winds}
  {Magnetic drag on hot {J}upiter atmospheric winds}.{\BBCQ}
\newblock
\APACjournalVolNumPages{The Astrophysical Journal}{719}{}{1421}.
\PrintBackRefs{\CurrentBib}

\bibitem [\protect \citeauthoryear {%
Perna%
, Menou%
\BCBL {}\ \BBA {} Rauscher%
}{%
Perna%
\ \protect \BOthers {.}}{%
{\protect \APACyear {2010}}%
{\protect \APACexlab {{\protect \BCnt {2}}}}}]{%
Perna_2010_2}
\APACinsertmetastar {%
Perna_2010_2}%
\begin{APACrefauthors}%
Perna, R.%
, Menou, K.%
\BCBL {}\ \BBA {} Rauscher, E.%
\end{APACrefauthors}%
\unskip\
\newblock
\APACrefYearMonthDay{2010{\protect \BCnt {2}}}{}{}.
\newblock
{\BBOQ}\APACrefatitle {Ohmic dissipation in the atmospheres of hot {J}upiters}
  {Ohmic dissipation in the atmospheres of hot {J}upiters}.{\BBCQ}
\newblock
\APACjournalVolNumPages{The Astrophysical Journal}{724}{}{313}.
\PrintBackRefs{\CurrentBib}

\bibitem [\protect \citeauthoryear {%
{Perri}%
\ \BBA {} {Cameron}%
}{%
{Perri}%
\ \BBA {} {Cameron}%
}{%
{\protect \APACyear {1974}}%
}]{%
peri74}
\APACinsertmetastar {%
peri74}%
\begin{APACrefauthors}%
{Perri}, F.%
\BCBT {}\ \BBA {} {Cameron}, A\BPBI G\BPBI W.%
\end{APACrefauthors}%
\unskip\
\newblock
\APACrefYearMonthDay{1974}{{\APACmonth{08}}}{}.
\newblock
{\BBOQ}\APACrefatitle {{Hydrodynamic instability of the solar nebula in the
  presence of a planetary core}} {{Hydrodynamic instability of the solar nebula
  in the presence of a planetary core}}.{\BBCQ}
\newblock
\APACjournalVolNumPages{\icarus}{22}{}{416-425}.
\newblock
\begin{APACrefDOI} \doi{10.1016/0019-1035(74)90074-8} \end{APACrefDOI}
\PrintBackRefs{\CurrentBib}

\bibitem [\protect \citeauthoryear {%
{Petrovich}%
, {Tremaine}%
\BCBL {}\ \BBA {} {Rafikov}%
}{%
{Petrovich}%
\ \protect \BOthers {.}}{%
{\protect \APACyear {2014}}%
}]{%
petr14}
\APACinsertmetastar {%
petr14}%
\begin{APACrefauthors}%
{Petrovich}, C.%
, {Tremaine}, S.%
\BCBL {}\ \BBA {} {Rafikov}, R.%
\end{APACrefauthors}%
\unskip\
\newblock
\APACrefYearMonthDay{2014}{{\APACmonth{05}}}{}.
\newblock
{\BBOQ}\APACrefatitle {{Scattering Outcomes of Close-in Planets: Constraints on
  Planet Migration}} {{Scattering Outcomes of Close-in Planets: Constraints on
  Planet Migration}}.{\BBCQ}
\newblock
\APACjournalVolNumPages{\apj}{786}{}{101}.
\newblock
\begin{APACrefDOI} \doi{10.1088/0004-637X/786/2/101} \end{APACrefDOI}
\PrintBackRefs{\CurrentBib}

\bibitem [\protect \citeauthoryear {%
Piskorz%
, Benneke%
, Crockett%
\BCBL {}\ \BBA {} {et al.}%
}{%
Piskorz%
\ \protect \BOthers {.}}{%
{\protect \APACyear {2017}}%
}]{%
Piskorz:2017aa}
\APACinsertmetastar {%
Piskorz:2017aa}%
\begin{APACrefauthors}%
Piskorz, D.%
, Benneke, B.%
, Crockett, N.%
\BCBL {}\ \BBA {} {et al.}%
\end{APACrefauthors}%
\unskip\
\newblock
\APACrefYearMonthDay{2017}{}{}.
\newblock
{\BBOQ}\APACrefatitle {Detection of Water Vapor in the Thermal Spectrum of the
  Non-transiting Hot Jupiter Upsilon Andromedae b} {Detection of water vapor in
  the thermal spectrum of the non-transiting hot jupiter upsilon andromedae
  b}.{\BBCQ}
\newblock
\APACjournalVolNumPages{The Astronomical Journal}{154}{}{78}.
\PrintBackRefs{\CurrentBib}

\bibitem [\protect \citeauthoryear {%
Polichtchouk%
\ \protect \BOthers {.}}{%
Polichtchouk%
\ \protect \BOthers {.}}{%
{\protect \APACyear {2014}}%
}]{%
Polichtchouk:2014}
\APACinsertmetastar {%
Polichtchouk:2014}%
\begin{APACrefauthors}%
Polichtchouk, I.%
, Cho, J.%
, Watkins, C.%
, Thrastarson, H.%
, Umurhan, O.%
\BCBL {}\ \BBA {} de~la Torre~Juarez, M.%
\end{APACrefauthors}%
\unskip\
\newblock
\APACrefYearMonthDay{2014}{}{}.
\newblock
{\BBOQ}\APACrefatitle {Intercomparison of general circulation models for hot
  extrasolar planets} {Intercomparison of general circulation models for hot
  extrasolar planets}.{\BBCQ}
\newblock
\APACjournalVolNumPages{Icarus}{229}{}{355-377}.
\PrintBackRefs{\CurrentBib}

\bibitem [\protect \citeauthoryear {%
{Pollack}%
\ \protect \BOthers {.}}{%
{Pollack}%
\ \protect \BOthers {.}}{%
{\protect \APACyear {1996}}%
}]{%
poll96}
\APACinsertmetastar {%
poll96}%
\begin{APACrefauthors}%
{Pollack}, J\BPBI B.%
, {Hubickyj}, O.%
, {Bodenheimer}, P.%
, {Lissauer}, J\BPBI J.%
, {Podolak}, M.%
\BCBL {}\ \BBA {} {Greenzweig}, Y.%
\end{APACrefauthors}%
\unskip\
\newblock
\APACrefYearMonthDay{1996}{{\APACmonth{11}}}{}.
\newblock
{\BBOQ}\APACrefatitle {{Formation of the Giant Planets by Concurrent Accretion
  of Solids and Gas}} {{Formation of the Giant Planets by Concurrent Accretion
  of Solids and Gas}}.{\BBCQ}
\newblock
\APACjournalVolNumPages{\icarus}{124}{}{62-85}.
\newblock
\begin{APACrefDOI} \doi{10.1006/icar.1996.0190} \end{APACrefDOI}
\PrintBackRefs{\CurrentBib}

\bibitem [\protect \citeauthoryear {%
Powell%
, Zhang%
, Gao%
\BCBL {}\ \BBA {} Parmentier%
}{%
Powell%
\ \protect \BOthers {.}}{%
{\protect \APACyear {2018}}%
}]{%
Powell:2018aa}
\APACinsertmetastar {%
Powell:2018aa}%
\begin{APACrefauthors}%
Powell, D.%
, Zhang, X.%
, Gao, P.%
\BCBL {}\ \BBA {} Parmentier, V.%
\end{APACrefauthors}%
\unskip\
\newblock
\APACrefYearMonthDay{2018}{}{}.
\newblock
{\BBOQ}\APACrefatitle {Formation of silicate and titanium clouds on hot
  Jupiters} {Formation of silicate and titanium clouds on hot jupiters}.{\BBCQ}
\newblock
\APACjournalVolNumPages{The Astrophysical Journal}{860}{}{18}.
\PrintBackRefs{\CurrentBib}

\bibitem [\protect \citeauthoryear {%
{Rafikov}%
}{%
{Rafikov}%
}{%
{\protect \APACyear {2005}}%
}]{%
rafi05}
\APACinsertmetastar {%
rafi05}%
\begin{APACrefauthors}%
{Rafikov}, R\BPBI R.%
\end{APACrefauthors}%
\unskip\
\newblock
\APACrefYearMonthDay{2005}{{\APACmonth{03}}}{}.
\newblock
{\BBOQ}\APACrefatitle {{Can Giant Planets Form by Direct Gravitational
  Instability?}} {{Can Giant Planets Form by Direct Gravitational
  Instability?}}{\BBCQ}
\newblock
\APACjournalVolNumPages{\apjl}{621}{}{L69-L72}.
\newblock
\begin{APACrefDOI} \doi{10.1086/428899} \end{APACrefDOI}
\PrintBackRefs{\CurrentBib}

\bibitem [\protect \citeauthoryear {%
{Rasio}%
\ \BBA {} {Ford}%
}{%
{Rasio}%
\ \BBA {} {Ford}%
}{%
{\protect \APACyear {1996}}%
}]{%
rasi96}
\APACinsertmetastar {%
rasi96}%
\begin{APACrefauthors}%
{Rasio}, F\BPBI A.%
\BCBT {}\ \BBA {} {Ford}, E\BPBI B.%
\end{APACrefauthors}%
\unskip\
\newblock
\APACrefYearMonthDay{1996}{{\APACmonth{11}}}{}.
\newblock
{\BBOQ}\APACrefatitle {{Dynamical instabilities and the formation of extrasolar
  planetary systems}} {{Dynamical instabilities and the formation of extrasolar
  planetary systems}}.{\BBCQ}
\newblock
\APACjournalVolNumPages{Science}{274}{}{954-956}.
\newblock
\begin{APACrefDOI} \doi{10.1126/science.274.5289.954} \end{APACrefDOI}
\PrintBackRefs{\CurrentBib}

\bibitem [\protect \citeauthoryear {%
Rauscher%
\ \BBA {} Kempton%
}{%
Rauscher%
\ \BBA {} Kempton%
}{%
{\protect \APACyear {2014}}%
}]{%
Rauscher:2014}
\APACinsertmetastar {%
Rauscher:2014}%
\begin{APACrefauthors}%
Rauscher, E.%
\BCBT {}\ \BBA {} Kempton, E.%
\end{APACrefauthors}%
\unskip\
\newblock
\APACrefYearMonthDay{2014}{}{}.
\newblock
{\BBOQ}\APACrefatitle {The atmospheric circulation and observable properties of
  non-synchronously rotating hot {J}upiters} {The atmospheric circulation and
  observable properties of non-synchronously rotating hot {J}upiters}.{\BBCQ}
\newblock
\APACjournalVolNumPages{arXiv}{}{}{}.
\PrintBackRefs{\CurrentBib}

\bibitem [\protect \citeauthoryear {%
Rauscher%
\ \BBA {} Menou%
}{%
Rauscher%
\ \BBA {} Menou%
}{%
{\protect \APACyear {2010}}%
}]{%
Rauscher:2010}
\APACinsertmetastar {%
Rauscher:2010}%
\begin{APACrefauthors}%
Rauscher, E.%
\BCBT {}\ \BBA {} Menou, K.%
\end{APACrefauthors}%
\unskip\
\newblock
\APACrefYearMonthDay{2010}{}{}.
\newblock
{\BBOQ}\APACrefatitle {Three-dimensional modeling of hot {J}upiter atmospheric
  flows} {Three-dimensional modeling of hot {J}upiter atmospheric
  flows}.{\BBCQ}
\newblock
\APACjournalVolNumPages{The Astrophysical Journal}{714}{}{1334}.
\PrintBackRefs{\CurrentBib}

\bibitem [\protect \citeauthoryear {%
Rauscher%
\ \BBA {} Menou%
}{%
Rauscher%
\ \BBA {} Menou%
}{%
{\protect \APACyear {2013}}%
}]{%
Rauscher_2013}
\APACinsertmetastar {%
Rauscher_2013}%
\begin{APACrefauthors}%
Rauscher, E.%
\BCBT {}\ \BBA {} Menou, K.%
\end{APACrefauthors}%
\unskip\
\newblock
\APACrefYearMonthDay{2013}{}{}.
\newblock
{\BBOQ}\APACrefatitle {Three-dimensional atmospheric circulation models of {HD}
  189733b and {HD} 209458b with consistent magnetic drag and {O}hmic
  dissipation} {Three-dimensional atmospheric circulation models of {HD}
  189733b and {HD} 209458b with consistent magnetic drag and {O}hmic
  dissipation}.{\BBCQ}
\newblock
\APACjournalVolNumPages{The Astrophysical Journal}{764}{}{103}.
\PrintBackRefs{\CurrentBib}

\bibitem [\protect \citeauthoryear {%
{Rauscher}%
\ \protect \BOthers {.}}{%
{Rauscher}%
\ \protect \BOthers {.}}{%
{\protect \APACyear {2007}}%
}]{%
Rauscher07b}
\APACinsertmetastar {%
Rauscher07b}%
\begin{APACrefauthors}%
{Rauscher}, E.%
, {Menou}, K.%
, {Seager}, S.%
, {Deming}, D.%
, {Cho}, J\BPBI Y\BHBI K.%
\BCBL {}\ \BBA {} {Hansen}, B\BPBI M\BPBI S.%
\end{APACrefauthors}%
\unskip\
\newblock
\APACrefYearMonthDay{2007}{{\APACmonth{08}}}{}.
\newblock
{\BBOQ}\APACrefatitle {{Toward Eclipse Mapping of Hot Jupiters}} {{Toward
  Eclipse Mapping of Hot Jupiters}}.{\BBCQ}
\newblock
\APACjournalVolNumPages{\apj}{664}{}{1199-1209}.
\newblock
\begin{APACrefDOI} \doi{10.1086/519213} \end{APACrefDOI}
\PrintBackRefs{\CurrentBib}

\bibitem [\protect \citeauthoryear {%
Redfield%
, Endl%
, Cochran%
\BCBL {}\ \BBA {} Koesterke%
}{%
Redfield%
\ \protect \BOthers {.}}{%
{\protect \APACyear {2008}}%
}]{%
Redfield:2008aa}
\APACinsertmetastar {%
Redfield:2008aa}%
\begin{APACrefauthors}%
Redfield, S.%
, Endl, M.%
, Cochran, W.%
\BCBL {}\ \BBA {} Koesterke, L.%
\end{APACrefauthors}%
\unskip\
\newblock
\APACrefYearMonthDay{2008}{}{}.
\newblock

\newblock
\APACjournalVolNumPages{The Astrophysical Journal Letters}{673}{}{L87}.
\PrintBackRefs{\CurrentBib}

\bibitem [\protect \citeauthoryear {%
{Rice}%
, {Armitage}%
\BCBL {}\ \BBA {} {Hogg}%
}{%
{Rice}%
\ \protect \BOthers {.}}{%
{\protect \APACyear {2008}}%
}]{%
rice08}
\APACinsertmetastar {%
rice08}%
\begin{APACrefauthors}%
{Rice}, W\BPBI K\BPBI M.%
, {Armitage}, P\BPBI J.%
\BCBL {}\ \BBA {} {Hogg}, D\BPBI F.%
\end{APACrefauthors}%
\unskip\
\newblock
\APACrefYearMonthDay{2008}{{\APACmonth{03}}}{}.
\newblock
{\BBOQ}\APACrefatitle {{Why are there so few hot Jupiters?}} {{Why are there so
  few hot Jupiters?}}{\BBCQ}
\newblock
\APACjournalVolNumPages{\mnras}{384}{}{1242-1248}.
\newblock
\begin{APACrefDOI} \doi{10.1111/j.1365-2966.2007.12817.x} \end{APACrefDOI}
\PrintBackRefs{\CurrentBib}

\bibitem [\protect \citeauthoryear {%
Rogers%
}{%
Rogers%
}{%
{\protect \APACyear {2017}}%
}]{%
Rogers:2017}
\APACinsertmetastar {%
Rogers:2017}%
\begin{APACrefauthors}%
Rogers, T.%
\end{APACrefauthors}%
\unskip\
\newblock
\APACrefYearMonthDay{2017}{}{}.
\newblock
{\BBOQ}\APACrefatitle {Constraints on the magnetic field strength of {HAT-P-7b}
  and other hot giant exoplanets} {Constraints on the magnetic field strength
  of {HAT-P-7b} and other hot giant exoplanets}.{\BBCQ}
\newblock
\APACjournalVolNumPages{Nature Astronomy}{1}{}{131}.
\PrintBackRefs{\CurrentBib}

\bibitem [\protect \citeauthoryear {%
Rogers%
\ \BBA {} Komacek%
}{%
Rogers%
\ \BBA {} Komacek%
}{%
{\protect \APACyear {2014}}%
}]{%
Rogers:2014}
\APACinsertmetastar {%
Rogers:2014}%
\begin{APACrefauthors}%
Rogers, T.%
\BCBT {}\ \BBA {} Komacek, T.%
\end{APACrefauthors}%
\unskip\
\newblock
\APACrefYearMonthDay{2014}{}{}.
\newblock
{\BBOQ}\APACrefatitle {Magnetic effects in hot {J}upiter atmospheres} {Magnetic
  effects in hot {J}upiter atmospheres}.{\BBCQ}
\newblock
\APACjournalVolNumPages{The Astrophysical Journal}{794}{}{132}.
\PrintBackRefs{\CurrentBib}

\bibitem [\protect \citeauthoryear {%
Rogers%
\ \BBA {} Showman%
}{%
Rogers%
\ \BBA {} Showman%
}{%
{\protect \APACyear {2014}}%
}]{%
Rogers:2020}
\APACinsertmetastar {%
Rogers:2020}%
\begin{APACrefauthors}%
Rogers, T.%
\BCBT {}\ \BBA {} Showman, A.%
\end{APACrefauthors}%
\unskip\
\newblock
\APACrefYearMonthDay{2014}{}{}.
\newblock
{\BBOQ}\APACrefatitle {Magnetohydrodynamic simulations of the atmosphere of
  {HD}209458b} {Magnetohydrodynamic simulations of the atmosphere of
  {HD}209458b}.{\BBCQ}
\newblock
\APACjournalVolNumPages{The Astrophysical Journal Letters}{782}{}{L4}.
\PrintBackRefs{\CurrentBib}

\bibitem [\protect \citeauthoryear {%
{Rogers}%
, {Lin}%
\BCBL {}\ \BBA {} {Lau}%
}{%
{Rogers}%
\ \protect \BOthers {.}}{%
{\protect \APACyear {2012}}%
}]{%
roge12}
\APACinsertmetastar {%
roge12}%
\begin{APACrefauthors}%
{Rogers}, T\BPBI M.%
, {Lin}, D\BPBI N\BPBI C.%
\BCBL {}\ \BBA {} {Lau}, H\BPBI H\BPBI B.%
\end{APACrefauthors}%
\unskip\
\newblock
\APACrefYearMonthDay{2012}{{\APACmonth{10}}}{}.
\newblock
{\BBOQ}\APACrefatitle {{Internal Gravity Waves Modulate the Apparent
  Misalignment of Exoplanets around Hot Stars}} {{Internal Gravity Waves
  Modulate the Apparent Misalignment of Exoplanets around Hot Stars}}.{\BBCQ}
\newblock
\APACjournalVolNumPages{\apjl}{758}{}{L6}.
\newblock
\begin{APACrefDOI} \doi{10.1088/2041-8205/758/1/L6} \end{APACrefDOI}
\PrintBackRefs{\CurrentBib}

\bibitem [\protect \citeauthoryear {%
Roman%
, Kempton%
, Rauscher%
\BCBL {}\ \BBA {} {et al.}%
}{%
Roman%
\ \protect \BOthers {.}}{%
{\protect \APACyear {2020}}%
}]{%
Roman:2020aa}
\APACinsertmetastar {%
Roman:2020aa}%
\begin{APACrefauthors}%
Roman, M.%
, Kempton, E.%
, Rauscher, E.%
\BCBL {}\ \BBA {} {et al.}%
\end{APACrefauthors}%
\unskip\
\newblock
\APACrefYearMonthDay{2020}{}{}.
\newblock
{\BBOQ}\APACrefatitle {Clouds in Three-Dimensional Models of Hot Jupiters Over
  a Wide Range of Temperatures I: Thermal Structures and Broadband Phase Curve
  Predictions} {Clouds in three-dimensional models of hot jupiters over a wide
  range of temperatures i: Thermal structures and broadband phase curve
  predictions}.{\BBCQ}
\newblock
\APACjournalVolNumPages{ArXiv e-prints:2010.06936}{}{}{}.
\PrintBackRefs{\CurrentBib}

\bibitem [\protect \citeauthoryear {%
Roman%
\ \BBA {} Rauscher%
}{%
Roman%
\ \BBA {} Rauscher%
}{%
{\protect \APACyear {2017}}%
}]{%
Roman:2017aa}
\APACinsertmetastar {%
Roman:2017aa}%
\begin{APACrefauthors}%
Roman, M.%
\BCBT {}\ \BBA {} Rauscher, E.%
\end{APACrefauthors}%
\unskip\
\newblock
\APACrefYearMonthDay{2017}{}{}.
\newblock
{\BBOQ}\APACrefatitle {Modeling the Effects of Inhomogeneous Aerosols on the
  Hot Jupiter Kepler-7b's Atmospheric Circuluation} {Modeling the effects of
  inhomogeneous aerosols on the hot jupiter kepler-7b's atmospheric
  circuluation}.{\BBCQ}
\newblock
\APACjournalVolNumPages{The Astrophysical Journal}{850}{}{17}.
\PrintBackRefs{\CurrentBib}

\bibitem [\protect \citeauthoryear {%
Roman%
\ \BBA {} Rauscher%
}{%
Roman%
\ \BBA {} Rauscher%
}{%
{\protect \APACyear {2019}}%
}]{%
Roman:2019aa}
\APACinsertmetastar {%
Roman:2019aa}%
\begin{APACrefauthors}%
Roman, M.%
\BCBT {}\ \BBA {} Rauscher, E.%
\end{APACrefauthors}%
\unskip\
\newblock
\APACrefYearMonthDay{2019}{}{}.
\newblock
{\BBOQ}\APACrefatitle {Modeled Temperature-dependent Clouds with Radiative
  Feedback in Hot Jupiter Atmospheres} {Modeled temperature-dependent clouds
  with radiative feedback in hot jupiter atmospheres}.{\BBCQ}
\newblock
\APACjournalVolNumPages{The Astrophysical Journal}{872}{}{1}.
\PrintBackRefs{\CurrentBib}

\bibitem [\protect \citeauthoryear {%
Sainsbury-Martinez%
, Wang%
, Fromang%
\BCBL {}\ \BBA {} {et al.}%
}{%
Sainsbury-Martinez%
\ \protect \BOthers {.}}{%
{\protect \APACyear {2019}}%
}]{%
Sainsbury-Martinez:2019aa}
\APACinsertmetastar {%
Sainsbury-Martinez:2019aa}%
\begin{APACrefauthors}%
Sainsbury-Martinez, F.%
, Wang, P.%
, Fromang, S.%
\BCBL {}\ \BBA {} {et al.}%
\end{APACrefauthors}%
\unskip\
\newblock
\APACrefYearMonthDay{2019}{}{}.
\newblock
{\BBOQ}\APACrefatitle {Idealised simulations of the deep atmosphere of hot
  Jupiters: Deep, hot adiabats as a robust solution to the radius inflation
  problem} {Idealised simulations of the deep atmosphere of hot jupiters: Deep,
  hot adiabats as a robust solution to the radius inflation problem}.{\BBCQ}
\newblock
\APACjournalVolNumPages{Astronomy {\&} Astrophysics}{632}{}{A114}.
\PrintBackRefs{\CurrentBib}

\bibitem [\protect \citeauthoryear {%
{Santos}%
, {Israelian}%
\BCBL {}\ \BBA {} {Mayor}%
}{%
{Santos}%
\ \protect \BOthers {.}}{%
{\protect \APACyear {2001}}%
}]{%
sant01}
\APACinsertmetastar {%
sant01}%
\begin{APACrefauthors}%
{Santos}, N\BPBI C.%
, {Israelian}, G.%
\BCBL {}\ \BBA {} {Mayor}, M.%
\end{APACrefauthors}%
\unskip\
\newblock
\APACrefYearMonthDay{2001}{{\APACmonth{07}}}{}.
\newblock
{\BBOQ}\APACrefatitle {{The metal-rich nature of stars with planets}} {{The
  metal-rich nature of stars with planets}}.{\BBCQ}
\newblock
\APACjournalVolNumPages{\aap}{373}{}{1019-1031}.
\newblock
\begin{APACrefDOI} \doi{10.1051/0004-6361:20010648} \end{APACrefDOI}
\PrintBackRefs{\CurrentBib}

\bibitem [\protect \citeauthoryear {%
{Sarkis}%
, {Mordasini}%
, {Henning}%
, {Marleau}%
\BCBL {}\ \BBA {} {Molli{\`e}re}%
}{%
{Sarkis}%
\ \protect \BOthers {.}}{%
{\protect \APACyear {2020}}%
}]{%
Sarkis20}
\APACinsertmetastar {%
Sarkis20}%
\begin{APACrefauthors}%
{Sarkis}, P.%
, {Mordasini}, C.%
, {Henning}, T.%
, {Marleau}, G\BPBI D.%
\BCBL {}\ \BBA {} {Molli{\`e}re}, P.%
\end{APACrefauthors}%
\unskip\
\newblock
\APACrefYearMonthDay{2020}{{\APACmonth{09}}}{}.
\newblock
{\BBOQ}\APACrefatitle {{Evidence of Three Mechanisms Explaining the Radius
  Anomaly of Hot Jupiters}} {{Evidence of Three Mechanisms Explaining the
  Radius Anomaly of Hot Jupiters}}.{\BBCQ}
\newblock
\APACjournalVolNumPages{arXiv e-prints}{}{}{arXiv:2009.04291}.
\PrintBackRefs{\CurrentBib}

\bibitem [\protect \citeauthoryear {%
{Sato}%
\ \protect \BOthers {.}}{%
{Sato}%
\ \protect \BOthers {.}}{%
{\protect \APACyear {2005}}%
}]{%
Sato05}
\APACinsertmetastar {%
Sato05}%
\begin{APACrefauthors}%
{Sato}, B.%
, {Fischer}, D\BPBI A.%
, {Henry}, G\BPBI W.%
, {Laughlin}, G.%
, {Butler}, R\BPBI P.%
, {Marcy}, G\BPBI W.%
\BDBL {}{Minniti}, D.%
\end{APACrefauthors}%
\unskip\
\newblock
\APACrefYearMonthDay{2005}{{\APACmonth{11}}}{}.
\newblock
{\BBOQ}\APACrefatitle {{The N2K Consortium. II. A Transiting Hot Saturn around
  HD 149026 with a Large Dense Core}} {{The N2K Consortium. II. A Transiting
  Hot Saturn around HD 149026 with a Large Dense Core}}.{\BBCQ}
\newblock
\APACjournalVolNumPages{\apj}{633}{}{465-473}.
\newblock
\begin{APACrefDOI} \doi{10.1086/449306} \end{APACrefDOI}
\PrintBackRefs{\CurrentBib}

\bibitem [\protect \citeauthoryear {%
{Schlichting}%
}{%
{Schlichting}%
}{%
{\protect \APACyear {2014}}%
}]{%
schl14}
\APACinsertmetastar {%
schl14}%
\begin{APACrefauthors}%
{Schlichting}, H\BPBI E.%
\end{APACrefauthors}%
\unskip\
\newblock
\APACrefYearMonthDay{2014}{{\APACmonth{11}}}{}.
\newblock
{\BBOQ}\APACrefatitle {{Formation of Close in Super-Earths and Mini-Neptunes:
  Required Disk Masses and their Implications}} {{Formation of Close in
  Super-Earths and Mini-Neptunes: Required Disk Masses and their
  Implications}}.{\BBCQ}
\newblock
\APACjournalVolNumPages{\apjl}{795}{}{L15}.
\newblock
\begin{APACrefDOI} \doi{10.1088/2041-8205/795/1/L15} \end{APACrefDOI}
\PrintBackRefs{\CurrentBib}

\bibitem [\protect \citeauthoryear {%
Schwartz%
\ \BBA {} Cowan%
}{%
Schwartz%
\ \BBA {} Cowan%
}{%
{\protect \APACyear {2015}}%
}]{%
Schwartz:2015}
\APACinsertmetastar {%
Schwartz:2015}%
\begin{APACrefauthors}%
Schwartz, J.%
\BCBT {}\ \BBA {} Cowan, N.%
\end{APACrefauthors}%
\unskip\
\newblock
\APACrefYearMonthDay{2015}{}{}.
\newblock
{\BBOQ}\APACrefatitle {Balancing the energy budget of short-period giant
  planets: evidence for reflective clouds and optical absorbers} {Balancing the
  energy budget of short-period giant planets: evidence for reflective clouds
  and optical absorbers}.{\BBCQ}
\newblock
\APACjournalVolNumPages{Monthly Notices of the Royal Astronomical
  Society}{449}{}{4192}.
\PrintBackRefs{\CurrentBib}

\bibitem [\protect \citeauthoryear {%
Schwartz%
, Kashner%
, Jovmir%
\BCBL {}\ \BBA {} Cowan%
}{%
Schwartz%
\ \protect \BOthers {.}}{%
{\protect \APACyear {2017}}%
}]{%
Schwartz:2017aa}
\APACinsertmetastar {%
Schwartz:2017aa}%
\begin{APACrefauthors}%
Schwartz, J.%
, Kashner, Z.%
, Jovmir, D.%
\BCBL {}\ \BBA {} Cowan, N.%
\end{APACrefauthors}%
\unskip\
\newblock
\APACrefYearMonthDay{2017}{}{}.
\newblock
{\BBOQ}\APACrefatitle {Phase Offsets and the Energy Budgets of Hot Jupiters}
  {Phase offsets and the energy budgets of hot jupiters}.{\BBCQ}
\newblock
\APACjournalVolNumPages{The Astrophysical Journal}{850}{}{154}.
\PrintBackRefs{\CurrentBib}

\bibitem [\protect \citeauthoryear {%
{Seager}%
\ \BBA {} {Sasselov}%
}{%
{Seager}%
\ \BBA {} {Sasselov}%
}{%
{\protect \APACyear {1998}}%
}]{%
SS98}
\APACinsertmetastar {%
SS98}%
\begin{APACrefauthors}%
{Seager}, S.%
\BCBT {}\ \BBA {} {Sasselov}, D\BPBI D.%
\end{APACrefauthors}%
\unskip\
\newblock
\APACrefYearMonthDay{1998}{{\APACmonth{08}}}{}.
\newblock
{\BBOQ}\APACrefatitle {{Extrasolar Giant Planets under Strong Stellar
  Irradiation}} {{Extrasolar Giant Planets under Strong Stellar
  Irradiation}}.{\BBCQ}
\newblock
\APACjournalVolNumPages{\apjl}{502}{}{L157}.
\newblock
\begin{APACrefURL}
  \url{http://adsabs.harvard.edu/cgi-bin/nph-bib_query?bibcode=1998ApJ...502L.157S&db_key=AST}
  \end{APACrefURL}
\PrintBackRefs{\CurrentBib}

\bibitem [\protect \citeauthoryear {%
{Seager}%
\ \BBA {} {Sasselov}%
}{%
{Seager}%
\ \BBA {} {Sasselov}%
}{%
{\protect \APACyear {2000}}%
}]{%
SS00}
\APACinsertmetastar {%
SS00}%
\begin{APACrefauthors}%
{Seager}, S.%
\BCBT {}\ \BBA {} {Sasselov}, D\BPBI D.%
\end{APACrefauthors}%
\unskip\
\newblock
\APACrefYearMonthDay{2000}{{\APACmonth{07}}}{}.
\newblock
{\BBOQ}\APACrefatitle {{Theoretical Transmission Spectra during Extrasolar
  Giant Planet Transits}} {{Theoretical Transmission Spectra during Extrasolar
  Giant Planet Transits}}.{\BBCQ}
\newblock
\APACjournalVolNumPages{\apj}{537}{}{916-921}.
\newblock
\begin{APACrefURL}
  \url{http://adsabs.harvard.edu/cgi-bin/nph-bib_query?bibcode=2000ApJ...537..916S&db_key=AST}
  \end{APACrefURL}
\PrintBackRefs{\CurrentBib}

\bibitem [\protect \citeauthoryear {%
Seidel%
, Ehrenreich%
, Wyttenbach%
\BCBL {}\ \BBA {} {et al.}%
}{%
Seidel%
\ \protect \BOthers {.}}{%
{\protect \APACyear {2019}}%
}]{%
Seidel:2019aa}
\APACinsertmetastar {%
Seidel:2019aa}%
\begin{APACrefauthors}%
Seidel, J.%
, Ehrenreich, D.%
, Wyttenbach, A.%
\BCBL {}\ \BBA {} {et al.}%
\end{APACrefauthors}%
\unskip\
\newblock
\APACrefYearMonthDay{2019}{}{}.
\newblock
{\BBOQ}\APACrefatitle {HEARTS II. A broadened sodium feature on the ultra-hot
  giant WASP-76b} {Hearts ii. a broadened sodium feature on the ultra-hot giant
  wasp-76b}.{\BBCQ}
\newblock
\APACjournalVolNumPages{Astronomy {\&} Astrophysics}{623}{}{A166}.
\PrintBackRefs{\CurrentBib}

\bibitem [\protect \citeauthoryear {%
Showman%
, Cho%
\BCBL {}\ \BBA {} Menou%
}{%
Showman%
\ \protect \BOthers {.}}{%
{\protect \APACyear {2010}}%
}]{%
Showman_2009}
\APACinsertmetastar {%
Showman_2009}%
\begin{APACrefauthors}%
Showman, A.%
, Cho, J.%
\BCBL {}\ \BBA {} Menou, K.%
\end{APACrefauthors}%
\unskip\
\newblock
\APACrefYearMonthDay{2010}{}{}.
\newblock
{\BBOQ}\APACrefatitle {Exoplanets} {Exoplanets}.{\BBCQ}
\newblock
\BIn{} S.~Seager\ (\BED), (\BCHAP\ Atmospheric Circulation of Exoplanets).
\newblock
\APACaddressPublisher{Tucson, AZ}{University of Arizona Press}.
\PrintBackRefs{\CurrentBib}

\bibitem [\protect \citeauthoryear {%
Showman%
, Fortney%
, Lewis%
\BCBL {}\ \BBA {} Shabram%
}{%
Showman%
\ \protect \BOthers {.}}{%
{\protect \APACyear {2013}}%
}]{%
showman_2013_doppler}
\APACinsertmetastar {%
showman_2013_doppler}%
\begin{APACrefauthors}%
Showman, A.%
, Fortney, J.%
, Lewis, N.%
\BCBL {}\ \BBA {} Shabram, M.%
\end{APACrefauthors}%
\unskip\
\newblock
\APACrefYearMonthDay{2013}{}{}.
\newblock
{\BBOQ}\APACrefatitle {Doppler signatures of the atmospheric circulation on hot
  {J}upiters} {Doppler signatures of the atmospheric circulation on hot
  {J}upiters}.{\BBCQ}
\newblock
\APACjournalVolNumPages{The Astrophysical Journal}{762}{}{24}.
\PrintBackRefs{\CurrentBib}

\bibitem [\protect \citeauthoryear {%
Showman%
\ \BBA {} Polvani%
}{%
Showman%
\ \BBA {} Polvani%
}{%
{\protect \APACyear {2011}}%
}]{%
Showman_Polvani_2011}
\APACinsertmetastar {%
Showman_Polvani_2011}%
\begin{APACrefauthors}%
Showman, A.%
\BCBT {}\ \BBA {} Polvani, L.%
\end{APACrefauthors}%
\unskip\
\newblock
\APACrefYearMonthDay{2011}{}{}.
\newblock
{\BBOQ}\APACrefatitle {Equatorial superrotation on tidally locked exoplanets}
  {Equatorial superrotation on tidally locked exoplanets}.{\BBCQ}
\newblock
\APACjournalVolNumPages{The Astrophysical Journal}{738}{}{71}.
\PrintBackRefs{\CurrentBib}

\bibitem [\protect \citeauthoryear {%
Showman%
, Tan%
\BCBL {}\ \BBA {} Parmentier%
}{%
Showman%
\ \protect \BOthers {.}}{%
{\protect \APACyear {2020}}%
}]{%
Showman:2020aa}
\APACinsertmetastar {%
Showman:2020aa}%
\begin{APACrefauthors}%
Showman, A.%
, Tan, X.%
\BCBL {}\ \BBA {} Parmentier, V.%
\end{APACrefauthors}%
\unskip\
\newblock
\APACrefYearMonthDay{2020}{}{}.
\newblock
{\BBOQ}\APACrefatitle {Atmospheric Dynamics of Hot Giant Planets and Brown
  Dwarfs} {Atmospheric dynamics of hot giant planets and brown dwarfs}.{\BBCQ}
\newblock
\APACjournalVolNumPages{arXiv e-prints:2007.15363}{}{}{}.
\PrintBackRefs{\CurrentBib}

\bibitem [\protect \citeauthoryear {%
{Showman}%
\ \protect \BOthers {.}}{%
{Showman}%
\ \protect \BOthers {.}}{%
{\protect \APACyear {2009}}%
}]{%
Showman09}
\APACinsertmetastar {%
Showman09}%
\begin{APACrefauthors}%
{Showman}, A\BPBI P.%
, {Fortney}, J\BPBI J.%
, {Lian}, Y.%
, {Marley}, M\BPBI S.%
, {Freedman}, R\BPBI S.%
, {Knutson}, H\BPBI A.%
\BCBL {}\ \BBA {} {Charbonneau}, D.%
\end{APACrefauthors}%
\unskip\
\newblock
\APACrefYearMonthDay{2009}{{\APACmonth{07}}}{}.
\newblock
{\BBOQ}\APACrefatitle {{Atmospheric Circulation of Hot Jupiters: Coupled
  Radiative-Dynamical General Circulation Model Simulations of HD 189733b and
  HD 209458b}} {{Atmospheric Circulation of Hot Jupiters: Coupled
  Radiative-Dynamical General Circulation Model Simulations of HD 189733b and
  HD 209458b}}.{\BBCQ}
\newblock
\APACjournalVolNumPages{\apj}{699}{}{564-584}.
\newblock
\begin{APACrefDOI} \doi{10.1088/0004-637X/699/1/564} \end{APACrefDOI}
\PrintBackRefs{\CurrentBib}

\bibitem [\protect \citeauthoryear {%
{Showman}%
\ \BBA {} {Guillot}%
}{%
{Showman}%
\ \BBA {} {Guillot}%
}{%
{\protect \APACyear {2002}}%
}]{%
Showman02}
\APACinsertmetastar {%
Showman02}%
\begin{APACrefauthors}%
{Showman}, A\BPBI P.%
\BCBT {}\ \BBA {} {Guillot}, T.%
\end{APACrefauthors}%
\unskip\
\newblock
\APACrefYearMonthDay{2002}{{\APACmonth{04}}}{}.
\newblock
{\BBOQ}\APACrefatitle {{Atmospheric circulation and tides of ``51 Pegasus
  b-like'' planets}} {{Atmospheric circulation and tides of ``51 Pegasus
  b-like'' planets}}.{\BBCQ}
\newblock
\APACjournalVolNumPages{\aap}{385}{}{166-180}.
\newblock
\begin{APACrefURL}
  \url{http://adsabs.harvard.edu/cgi-bin/nph-bib_query?bibcode=2002A%26A...385..166S&db_key=AST}
  \end{APACrefURL}
\PrintBackRefs{\CurrentBib}

\bibitem [\protect \citeauthoryear {%
{Showman}%
, {Wordsworth}%
, {Merlis}%
\BCBL {}\ \BBA {} {Kaspi}%
}{%
{Showman}%
\ \protect \BOthers {.}}{%
{\protect \APACyear {2013}}%
}]{%
Showman13}
\APACinsertmetastar {%
Showman13}%
\begin{APACrefauthors}%
{Showman}, A\BPBI P.%
, {Wordsworth}, R\BPBI D.%
, {Merlis}, T\BPBI M.%
\BCBL {}\ \BBA {} {Kaspi}, Y.%
\end{APACrefauthors}%
\unskip\
\newblock
\APACrefYearMonthDay{2013}{}{}.
\newblock
{\BBOQ}\APACrefatitle {{Atmospheric Circulation of Terrestrial Exoplanets}}
  {{Atmospheric Circulation of Terrestrial Exoplanets}}.{\BBCQ}
\newblock
\BIn{} S\BPBI J.~{Mackwell}, A\BPBI A.~{Simon-Miller}, J\BPBI W.~{Harder}\BCBL
  {}\ \BBA {} M\BPBI A.~{Bullock}\ (\BEDS), \APACrefbtitle {Comparative
  Climatology of Terrestrial Planets} {Comparative climatology of terrestrial
  planets}\ (\BPG~277-326).
\newblock
\begin{APACrefDOI} \doi{10.2458/azu_uapress_9780816530595-ch12}
  \end{APACrefDOI}
\PrintBackRefs{\CurrentBib}

\bibitem [\protect \citeauthoryear {%
Shporer%
, Wong%
, Huang%
\BCBL {}\ \BBA {} {et al.}%
}{%
Shporer%
\ \protect \BOthers {.}}{%
{\protect \APACyear {2019}}%
}]{%
Shporer:2019aa}
\APACinsertmetastar {%
Shporer:2019aa}%
\begin{APACrefauthors}%
Shporer, A.%
, Wong, I.%
, Huang, C.%
\BCBL {}\ \BBA {} {et al.}%
\end{APACrefauthors}%
\unskip\
\newblock
\APACrefYearMonthDay{2019}{}{}.
\newblock
{\BBOQ}\APACrefatitle {TESS Full Orbital Phase Curve of the WASP-18b System}
  {Tess full orbital phase curve of the wasp-18b system}.{\BBCQ}
\newblock
\APACjournalVolNumPages{The Astronomical Journal}{157}{}{178}.
\PrintBackRefs{\CurrentBib}

\bibitem [\protect \citeauthoryear {%
Sing%
, D{\'{e}}sert%
, Fortney%
\BCBL {}\ \BBA {} {et al.}%
}{%
Sing%
\ \protect \BOthers {.}}{%
{\protect \APACyear {2011}}%
}]{%
Sing:2011aa}
\APACinsertmetastar {%
Sing:2011aa}%
\begin{APACrefauthors}%
Sing, D.%
, D{\'{e}}sert, J.%
, Fortney, J.%
\BCBL {}\ \BBA {} {et al.}%
\end{APACrefauthors}%
\unskip\
\newblock
\APACrefYearMonthDay{2011}{}{}.
\newblock

\newblock
\APACjournalVolNumPages{Astronomy {\&} Astrophysics}{527}{}{A73}.
\PrintBackRefs{\CurrentBib}

\bibitem [\protect \citeauthoryear {%
Sing%
, Wakeford%
, Showman%
\BCBL {}\ \BBA {} {et al.}%
}{%
Sing%
\ \protect \BOthers {.}}{%
{\protect \APACyear {2015}}%
}]{%
Sing:2015aa}
\APACinsertmetastar {%
Sing:2015aa}%
\begin{APACrefauthors}%
Sing, D.%
, Wakeford, H.%
, Showman, A.%
\BCBL {}\ \BBA {} {et al.}%
\end{APACrefauthors}%
\unskip\
\newblock
\APACrefYearMonthDay{2015}{}{}.
\newblock

\newblock
\APACjournalVolNumPages{Monthly Notices of the Royal Astronomical
  Society}{446}{}{2428}.
\PrintBackRefs{\CurrentBib}

\bibitem [\protect \citeauthoryear {%
{Sing}%
}{%
{Sing}%
}{%
{\protect \APACyear {2018}}%
}]{%
Sing18}
\APACinsertmetastar {%
Sing18}%
\begin{APACrefauthors}%
{Sing}, D\BPBI K.%
\end{APACrefauthors}%
\unskip\
\newblock
\APACrefYearMonthDay{2018}{{\APACmonth{04}}}{}.
\newblock
{\BBOQ}\APACrefatitle {{Observational Techniques With Transiting Exoplanetary
  Atmospheres}} {{Observational Techniques With Transiting Exoplanetary
  Atmospheres}}.{\BBCQ}
\newblock
\APACjournalVolNumPages{arXiv e-prints}{}{}{arXiv:1804.07357}.
\PrintBackRefs{\CurrentBib}

\bibitem [\protect \citeauthoryear {%
{Sing}%
\ \protect \BOthers {.}}{%
{Sing}%
\ \protect \BOthers {.}}{%
{\protect \APACyear {2016}}%
}]{%
Sing16}
\APACinsertmetastar {%
Sing16}%
\begin{APACrefauthors}%
{Sing}, D\BPBI K.%
, {Fortney}, J\BPBI J.%
, {Nikolov}, N.%
, {Wakeford}, H\BPBI R.%
, {Kataria}, T.%
, {Evans}, T\BPBI M.%
\BDBL {}{Wilson}, P\BPBI A.%
\end{APACrefauthors}%
\unskip\
\newblock
\APACrefYearMonthDay{2016}{{\APACmonth{01}}}{}.
\newblock
{\BBOQ}\APACrefatitle {{A continuum from clear to cloudy hot-Jupiter exoplanets
  without primordial water depletion}} {{A continuum from clear to cloudy
  hot-Jupiter exoplanets without primordial water depletion}}.{\BBCQ}
\newblock
\APACjournalVolNumPages{\nat}{529}{}{59-62}.
\newblock
\begin{APACrefDOI} \doi{10.1038/nature16068} \end{APACrefDOI}
\PrintBackRefs{\CurrentBib}

\bibitem [\protect \citeauthoryear {%
Snellen%
, de Kok%
, de Mooij%
\BCBL {}\ \BBA {} Albrecht%
}{%
Snellen%
\ \protect \BOthers {.}}{%
{\protect \APACyear {2010}}%
}]{%
Snellen:2010}
\APACinsertmetastar {%
Snellen:2010}%
\begin{APACrefauthors}%
Snellen, I.%
, de Kok, R.%
, de Mooij, E.%
\BCBL {}\ \BBA {} Albrecht, S.%
\end{APACrefauthors}%
\unskip\
\newblock
\APACrefYearMonthDay{2010}{}{}.
\newblock
{\BBOQ}\APACrefatitle {The orbital motion, absolute mass and high-altitude
  winds of exoplanet {HD} 209458b} {The orbital motion, absolute mass and
  high-altitude winds of exoplanet {HD} 209458b}.{\BBCQ}
\newblock
\APACjournalVolNumPages{Nature}{465}{}{1049}.
\PrintBackRefs{\CurrentBib}

\bibitem [\protect \citeauthoryear {%
{Socrates}%
}{%
{Socrates}%
}{%
{\protect \APACyear {2013}}%
}]{%
socr13}
\APACinsertmetastar {%
socr13}%
\begin{APACrefauthors}%
{Socrates}, A.%
\end{APACrefauthors}%
\unskip\
\newblock
\APACrefYearMonthDay{2013}{{\APACmonth{04}}}{}.
\newblock
{\BBOQ}\APACrefatitle {{Relationship Between Thermal Tides and Radius Excess}}
  {{Relationship Between Thermal Tides and Radius Excess}}.{\BBCQ}
\newblock
\APACjournalVolNumPages{ArXiv e-prints}{}{}{}.
\PrintBackRefs{\CurrentBib}

\bibitem [\protect \citeauthoryear {%
{Spake}%
\ \protect \BOthers {.}}{%
{Spake}%
\ \protect \BOthers {.}}{%
{\protect \APACyear {2018}}%
}]{%
Spake18}
\APACinsertmetastar {%
Spake18}%
\begin{APACrefauthors}%
{Spake}, J\BPBI J.%
, {Sing}, D\BPBI K.%
, {Evans}, T\BPBI M.%
, {Oklop{\v{c}}i{\'c}}%
, {}, A.%
, {Bourrier}, V.%
\BDBL {}{Madhusudhan}, N.%
\end{APACrefauthors}%
\unskip\
\newblock
\APACrefYearMonthDay{2018}{{\APACmonth{05}}}{}.
\newblock
{\BBOQ}\APACrefatitle {{Helium in the eroding atmosphere of an exoplanet}}
  {{Helium in the eroding atmosphere of an exoplanet}}.{\BBCQ}
\newblock
\APACjournalVolNumPages{\nat}{557}{7703}{68-70}.
\newblock
\begin{APACrefDOI} \doi{10.1038/s41586-018-0067-5} \end{APACrefDOI}
\PrintBackRefs{\CurrentBib}

\bibitem [\protect \citeauthoryear {%
Spiegel%
\ \BBA {} Burrows%
}{%
Spiegel%
\ \BBA {} Burrows%
}{%
{\protect \APACyear {2013}}%
}]{%
Spiegel:2013}
\APACinsertmetastar {%
Spiegel:2013}%
\begin{APACrefauthors}%
Spiegel, D.%
\BCBT {}\ \BBA {} Burrows, A.%
\end{APACrefauthors}%
\unskip\
\newblock
\APACrefYearMonthDay{2013}{}{}.
\newblock
{\BBOQ}\APACrefatitle {Thermal processes governing hot-{J}upiter radii}
  {Thermal processes governing hot-{J}upiter radii}.{\BBCQ}
\newblock
\APACjournalVolNumPages{The Astrophysical Journal}{772}{}{76}.
\PrintBackRefs{\CurrentBib}

\bibitem [\protect \citeauthoryear {%
Spiegel%
, Silverio%
\BCBL {}\ \BBA {} Burrows%
}{%
Spiegel%
\ \protect \BOthers {.}}{%
{\protect \APACyear {2009}}%
}]{%
Spiegel:2009}
\APACinsertmetastar {%
Spiegel:2009}%
\begin{APACrefauthors}%
Spiegel, D.%
, Silverio, K.%
\BCBL {}\ \BBA {} Burrows, A.%
\end{APACrefauthors}%
\unskip\
\newblock
\APACrefYearMonthDay{2009}{}{}.
\newblock
{\BBOQ}\APACrefatitle {Can Ti{O} explain thermal inversions in the upper
  atmospheres of irradiated giant planets?} {Can ti{O} explain thermal
  inversions in the upper atmospheres of irradiated giant planets?}{\BBCQ}
\newblock
\APACjournalVolNumPages{The Astrophysical Journal}{699}{}{1487}.
\PrintBackRefs{\CurrentBib}

\bibitem [\protect \citeauthoryear {%
{Steffen}%
\ \protect \BOthers {.}}{%
{Steffen}%
\ \protect \BOthers {.}}{%
{\protect \APACyear {2012}}%
}]{%
stef12}
\APACinsertmetastar {%
stef12}%
\begin{APACrefauthors}%
{Steffen}, J\BPBI H.%
, {Ragozzine}, D.%
, {Fabrycky}, D\BPBI C.%
, {Carter}, J\BPBI A.%
, {Ford}, E\BPBI B.%
, {Holman}, M\BPBI J.%
\BDBL {}{Quinn}, S\BPBI N.%
\end{APACrefauthors}%
\unskip\
\newblock
\APACrefYearMonthDay{2012}{{\APACmonth{05}}}{}.
\newblock
{\BBOQ}\APACrefatitle {{Kepler constraints on planets near hot Jupiters}}
  {{Kepler constraints on planets near hot Jupiters}}.{\BBCQ}
\newblock
\APACjournalVolNumPages{Proceedings of the National Academy of
  Science}{109}{}{7982-7987}.
\newblock
\begin{APACrefDOI} \doi{10.1073/pnas.1120970109} \end{APACrefDOI}
\PrintBackRefs{\CurrentBib}

\bibitem [\protect \citeauthoryear {%
Steinrueck%
, Parmentier%
, Showman%
\BCBL {}\ \BBA {} {et al.}%
}{%
Steinrueck%
\ \protect \BOthers {.}}{%
{\protect \APACyear {2019}}%
}]{%
Steinrueck:2019aa}
\APACinsertmetastar {%
Steinrueck:2019aa}%
\begin{APACrefauthors}%
Steinrueck, M.%
, Parmentier, V.%
, Showman, A.%
\BCBL {}\ \BBA {} {et al.}%
\end{APACrefauthors}%
\unskip\
\newblock
\APACrefYearMonthDay{2019}{}{}.
\newblock
{\BBOQ}\APACrefatitle {The Effect of 3D Transport-induced Disequilibrium Carbon
  Chemistry on the Atmospheric Structure, Phase Curves, and Emission Spectra of
  Hot Jupiter HD 189733b} {The effect of 3d transport-induced disequilibrium
  carbon chemistry on the atmospheric structure, phase curves, and emission
  spectra of hot jupiter hd 189733b}.{\BBCQ}
\newblock
\APACjournalVolNumPages{The Astrophysical Journal}{880}{}{14}.
\PrintBackRefs{\CurrentBib}

\bibitem [\protect \citeauthoryear {%
Stevenson%
, Bean%
, Madhusudhan%
\BCBL {}\ \BBA {} {et al.}%
}{%
Stevenson%
, Bean%
\BCBL {}\ \protect \BOthers {.}}{%
{\protect \APACyear {2014}}%
}]{%
Stevenson:2014aa}
\APACinsertmetastar {%
Stevenson:2014aa}%
\begin{APACrefauthors}%
Stevenson, K.%
, Bean, J.%
, Madhusudhan, N.%
\BCBL {}\ \BBA {} {et al.}%
\end{APACrefauthors}%
\unskip\
\newblock
\APACrefYearMonthDay{2014}{}{}.
\newblock

\newblock
\APACjournalVolNumPages{The Astrophysical Journal}{791}{}{36}.
\PrintBackRefs{\CurrentBib}

\bibitem [\protect \citeauthoryear {%
Stevenson%
, Desert%
\BCBL {}\ \protect \BOthers {.}}{%
Stevenson%
, Desert%
\BCBL {}\ \protect \BOthers {.}}{%
{\protect \APACyear {2014}}%
}]{%
Stevenson:2014}
\APACinsertmetastar {%
Stevenson:2014}%
\begin{APACrefauthors}%
Stevenson, K.%
, Desert, J.%
, Line, M.%
, Bean, J.%
, Fortney, J.%
, Showman, A.%
\BDBL {}Homeier, D.%
\end{APACrefauthors}%
\unskip\
\newblock
\APACrefYearMonthDay{2014}{}{}.
\newblock
{\BBOQ}\APACrefatitle {Thermal structure of an exoplanet atmosphere from
  phase-resolved emission spectroscopy} {Thermal structure of an exoplanet
  atmosphere from phase-resolved emission spectroscopy}.{\BBCQ}
\newblock
\APACjournalVolNumPages{Science}{346}{}{838}.
\PrintBackRefs{\CurrentBib}

\bibitem [\protect \citeauthoryear {%
{Stevenson}%
\ \protect \BOthers {.}}{%
{Stevenson}%
\ \protect \BOthers {.}}{%
{\protect \APACyear {2014}}%
}]{%
Stevenson14}
\APACinsertmetastar {%
Stevenson14}%
\begin{APACrefauthors}%
{Stevenson}, K\BPBI B.%
, {D{\'e}sert}, J\BHBI M.%
, {Line}, M\BPBI R.%
, {Bean}, J\BPBI L.%
, {Fortney}, J\BPBI J.%
, {Showman}, A\BPBI P.%
\BDBL {}{Homeier}, D.%
\end{APACrefauthors}%
\unskip\
\newblock
\APACrefYearMonthDay{2014}{{\APACmonth{11}}}{}.
\newblock
{\BBOQ}\APACrefatitle {{Thermal structure of an exoplanet atmosphere from
  phase-resolved emission spectroscopy}} {{Thermal structure of an exoplanet
  atmosphere from phase-resolved emission spectroscopy}}.{\BBCQ}
\newblock
\APACjournalVolNumPages{Science}{346}{}{838-841}.
\newblock
\begin{APACrefDOI} \doi{10.1126/science.1256758} \end{APACrefDOI}
\PrintBackRefs{\CurrentBib}

\bibitem [\protect \citeauthoryear {%
Stevenson%
\ \protect \BOthers {.}}{%
Stevenson%
\ \protect \BOthers {.}}{%
{\protect \APACyear {2017}}%
}]{%
Stevenson2016}
\APACinsertmetastar {%
Stevenson2016}%
\begin{APACrefauthors}%
Stevenson, K\BPBI B.%
, Line, M\BPBI R.%
, Bean, J\BPBI L.%
, Desert, J\BHBI M.%
, Fortney, J\BPBI J.%
, Showman, A\BPBI P.%
\BDBL {}Feng, Y\BPBI K.%
\end{APACrefauthors}%
\unskip\
\newblock
\APACrefYearMonthDay{2017}{}{}.
\newblock
{\BBOQ}\APACrefatitle {Spitzer phase curve constraints for WASP-43b at 3.6 and
  4.5 microns} {Spitzer phase curve constraints for wasp-43b at 3.6 and 4.5
  microns}.{\BBCQ}
\newblock
\APACjournalVolNumPages{The Astronomical Journal}{153}{}{68}.
\newblock
\begin{APACrefURL} \url{http://arxiv.org/abs/1608.00056} \end{APACrefURL}
\PrintBackRefs{\CurrentBib}

\bibitem [\protect \citeauthoryear {%
{Storch}%
, {Anderson}%
\BCBL {}\ \BBA {} {Lai}%
}{%
{Storch}%
\ \protect \BOthers {.}}{%
{\protect \APACyear {2014}}%
}]{%
stor14a}
\APACinsertmetastar {%
stor14a}%
\begin{APACrefauthors}%
{Storch}, N\BPBI I.%
, {Anderson}, K\BPBI R.%
\BCBL {}\ \BBA {} {Lai}, D.%
\end{APACrefauthors}%
\unskip\
\newblock
\APACrefYearMonthDay{2014}{{\APACmonth{09}}}{}.
\newblock
{\BBOQ}\APACrefatitle {{Chaotic dynamics of stellar spin in binaries and the
  production of misaligned hot Jupiters}} {{Chaotic dynamics of stellar spin in
  binaries and the production of misaligned hot Jupiters}}.{\BBCQ}
\newblock
\APACjournalVolNumPages{Science}{345}{}{1317-1321}.
\newblock
\begin{APACrefDOI} \doi{10.1126/science.1254358} \end{APACrefDOI}
\PrintBackRefs{\CurrentBib}

\bibitem [\protect \citeauthoryear {%
{Sudarsky}%
, {Burrows}%
\BCBL {}\ \BBA {} {Hubeny}%
}{%
{Sudarsky}%
\ \protect \BOthers {.}}{%
{\protect \APACyear {2003}}%
}]{%
Sudar03}
\APACinsertmetastar {%
Sudar03}%
\begin{APACrefauthors}%
{Sudarsky}, D.%
, {Burrows}, A.%
\BCBL {}\ \BBA {} {Hubeny}, I.%
\end{APACrefauthors}%
\unskip\
\newblock
\APACrefYearMonthDay{2003}{{\APACmonth{05}}}{}.
\newblock
{\BBOQ}\APACrefatitle {{Theoretical Spectra and Atmospheres of Extrasolar Giant
  Planets}} {{Theoretical Spectra and Atmospheres of Extrasolar Giant
  Planets}}.{\BBCQ}
\newblock
\APACjournalVolNumPages{\apj}{588}{}{1121-1148}.
\PrintBackRefs{\CurrentBib}

\bibitem [\protect \citeauthoryear {%
{Sudarsky}%
, {Burrows}%
\BCBL {}\ \BBA {} {Pinto}%
}{%
{Sudarsky}%
\ \protect \BOthers {.}}{%
{\protect \APACyear {2000}}%
}]{%
Sudar00}
\APACinsertmetastar {%
Sudar00}%
\begin{APACrefauthors}%
{Sudarsky}, D.%
, {Burrows}, A.%
\BCBL {}\ \BBA {} {Pinto}, P.%
\end{APACrefauthors}%
\unskip\
\newblock
\APACrefYearMonthDay{2000}{{\APACmonth{08}}}{}.
\newblock
{\BBOQ}\APACrefatitle {{Albedo and Reflection Spectra of Extrasolar Giant
  Planets}} {{Albedo and Reflection Spectra of Extrasolar Giant
  Planets}}.{\BBCQ}
\newblock
\APACjournalVolNumPages{\apj}{538}{}{885-903}.
\newblock
\begin{APACrefURL}
  \url{http://adsabs.harvard.edu/cgi-bin/nph-bib_query?bibcode=2000ApJ...538..885S&db_key=AST}
  \end{APACrefURL}
\PrintBackRefs{\CurrentBib}

\bibitem [\protect \citeauthoryear {%
Tabernero%
, Osorio%
, Allart%
\BCBL {}\ \BBA {} {et al.}%
}{%
Tabernero%
\ \protect \BOthers {.}}{%
{\protect \APACyear {2020}}%
}]{%
Tabernero:2020aa}
\APACinsertmetastar {%
Tabernero:2020aa}%
\begin{APACrefauthors}%
Tabernero, H.%
, Osorio, M\BPBI Z.%
, Allart, R.%
\BCBL {}\ \BBA {} {et al.}%
\end{APACrefauthors}%
\unskip\
\newblock
\APACrefYearMonthDay{2020}{}{}.
\newblock
{\BBOQ}\APACrefatitle {ESPRESSO high resolution transmission spectroscopy of
  WASP-76b} {Espresso high resolution transmission spectroscopy of
  wasp-76b}.{\BBCQ}
\newblock
\APACjournalVolNumPages{arXiv e-prints:2011.12197}{}{}{}.
\PrintBackRefs{\CurrentBib}

\bibitem [\protect \citeauthoryear {%
Tan%
\ \BBA {} Komacek%
}{%
Tan%
\ \BBA {} Komacek%
}{%
{\protect \APACyear {2019}}%
}]{%
Tan:2019aa}
\APACinsertmetastar {%
Tan:2019aa}%
\begin{APACrefauthors}%
Tan, X.%
\BCBT {}\ \BBA {} Komacek, T.%
\end{APACrefauthors}%
\unskip\
\newblock
\APACrefYearMonthDay{2019}{}{}.
\newblock
{\BBOQ}\APACrefatitle {The Atmospheric Circulation of Ultra-hot Jupiters} {The
  atmospheric circulation of ultra-hot jupiters}.{\BBCQ}
\newblock
\APACjournalVolNumPages{The Astrophysical Journal}{886}{}{26}.
\PrintBackRefs{\CurrentBib}

\bibitem [\protect \citeauthoryear {%
{Teske}%
, {Thorngren}%
, {Fortney}%
, {Hinkel}%
\BCBL {}\ \BBA {} {Brewer}%
}{%
{Teske}%
\ \protect \BOthers {.}}{%
{\protect \APACyear {2019}}%
}]{%
Teske19}
\APACinsertmetastar {%
Teske19}%
\begin{APACrefauthors}%
{Teske}, J\BPBI K.%
, {Thorngren}, D\BPBI P.%
, {Fortney}, J\BPBI J.%
, {Hinkel}, N.%
\BCBL {}\ \BBA {} {Brewer}, J\BPBI M.%
\end{APACrefauthors}%
\unskip\
\newblock
\APACrefYearMonthDay{2019}{{\APACmonth{12}}}{}.
\newblock
{\BBOQ}\APACrefatitle {{Do Metal-rich Stars Make Metal-rich Planets? New
  Insights on Giant Planet Formation from Host Star Abundances}} {{Do
  Metal-rich Stars Make Metal-rich Planets? New Insights on Giant Planet
  Formation from Host Star Abundances}}.{\BBCQ}
\newblock
\APACjournalVolNumPages{\aj}{158}{6}{239}.
\newblock
\begin{APACrefDOI} \doi{10.3847/1538-3881/ab4f79} \end{APACrefDOI}
\PrintBackRefs{\CurrentBib}

\bibitem [\protect \citeauthoryear {%
{Thommes}%
, {Duncan}%
\BCBL {}\ \BBA {} {Levison}%
}{%
{Thommes}%
\ \protect \BOthers {.}}{%
{\protect \APACyear {1999}}%
}]{%
thom99}
\APACinsertmetastar {%
thom99}%
\begin{APACrefauthors}%
{Thommes}, E\BPBI W.%
, {Duncan}, M\BPBI J.%
\BCBL {}\ \BBA {} {Levison}, H\BPBI F.%
\end{APACrefauthors}%
\unskip\
\newblock
\APACrefYearMonthDay{1999}{{\APACmonth{12}}}{}.
\newblock
{\BBOQ}\APACrefatitle {{The formation of Uranus and Neptune in the
  Jupiter-Saturn region of the Solar System}} {{The formation of Uranus and
  Neptune in the Jupiter-Saturn region of the Solar System}}.{\BBCQ}
\newblock
\APACjournalVolNumPages{\nat}{402}{}{635-638}.
\newblock
\begin{APACrefDOI} \doi{10.1038/45185} \end{APACrefDOI}
\PrintBackRefs{\CurrentBib}

\bibitem [\protect \citeauthoryear {%
Thompson%
}{%
Thompson%
}{%
{\protect \APACyear {1882}}%
}]{%
Thompson:1882aa}
\APACinsertmetastar {%
Thompson:1882aa}%
\begin{APACrefauthors}%
Thompson, W.%
\end{APACrefauthors}%
\unskip\
\newblock
\APACrefYearMonthDay{1882}{}{}.
\newblock
{\BBOQ}\APACrefatitle {On the thermodynamic acceleration of the Earth's
  rotation} {On the thermodynamic acceleration of the earth's rotation}.{\BBCQ}
\newblock
\APACjournalVolNumPages{Proceedings of the Royal Society of
  Edinburgh}{11}{}{396}.
\PrintBackRefs{\CurrentBib}

\bibitem [\protect \citeauthoryear {%
Thorngren%
\ \BBA {} Fortney%
}{%
Thorngren%
\ \BBA {} Fortney%
}{%
{\protect \APACyear {2018}}%
}]{%
Thorngren:2017}
\APACinsertmetastar {%
Thorngren:2017}%
\begin{APACrefauthors}%
Thorngren, D.%
\BCBT {}\ \BBA {} Fortney, J.%
\end{APACrefauthors}%
\unskip\
\newblock
\APACrefYearMonthDay{2018}{}{}.
\newblock
{\BBOQ}\APACrefatitle {Bayesian analysis of hot Jupiter radius anomalies:
  Evidence for Ohmic dissipation?} {Bayesian analysis of hot jupiter radius
  anomalies: Evidence for ohmic dissipation?}{\BBCQ}
\newblock
\APACjournalVolNumPages{The Astronomical Journal}{155}{}{214}.
\PrintBackRefs{\CurrentBib}

\bibitem [\protect \citeauthoryear {%
{Thorngren}%
, {Fortney}%
, {Murray-Clay}%
\BCBL {}\ \BBA {} {Lopez}%
}{%
{Thorngren}%
\ \protect \BOthers {.}}{%
{\protect \APACyear {2016}}%
}]{%
Thorngren16}
\APACinsertmetastar {%
Thorngren16}%
\begin{APACrefauthors}%
{Thorngren}, D\BPBI P.%
, {Fortney}, J\BPBI J.%
, {Murray-Clay}, R\BPBI A.%
\BCBL {}\ \BBA {} {Lopez}, E\BPBI D.%
\end{APACrefauthors}%
\unskip\
\newblock
\APACrefYearMonthDay{2016}{{\APACmonth{11}}}{}.
\newblock
{\BBOQ}\APACrefatitle {{The Mass-Metallicity Relation for Giant Planets}} {{The
  Mass-Metallicity Relation for Giant Planets}}.{\BBCQ}
\newblock
\APACjournalVolNumPages{\apj}{831}{}{64}.
\newblock
\begin{APACrefDOI} \doi{10.3847/0004-637X/831/1/64} \end{APACrefDOI}
\PrintBackRefs{\CurrentBib}

\bibitem [\protect \citeauthoryear {%
{Thorngren}%
, {Gao}%
\BCBL {}\ \BBA {} {Fortney}%
}{%
{Thorngren}%
\ \protect \BOthers {.}}{%
{\protect \APACyear {2019}}%
}]{%
Thorngren19}
\APACinsertmetastar {%
Thorngren19}%
\begin{APACrefauthors}%
{Thorngren}, D\BPBI P.%
, {Gao}, P.%
\BCBL {}\ \BBA {} {Fortney}, J\BPBI J.%
\end{APACrefauthors}%
\unskip\
\newblock
\APACrefYearMonthDay{2019}{{\APACmonth{10}}}{}.
\newblock
{\BBOQ}\APACrefatitle {{The Intrinsic Temperature and Radiative-Convective
  Boundary Depth in the Atmospheres of Hot Jupiters}} {{The Intrinsic
  Temperature and Radiative-Convective Boundary Depth in the Atmospheres of Hot
  Jupiters}}.{\BBCQ}
\newblock
\APACjournalVolNumPages{\apjl}{884}{1}{L6}.
\newblock
\begin{APACrefDOI} \doi{10.3847/2041-8213/ab43d0} \end{APACrefDOI}
\PrintBackRefs{\CurrentBib}

\bibitem [\protect \citeauthoryear {%
{Tinetti}%
\ \protect \BOthers {.}}{%
{Tinetti}%
\ \protect \BOthers {.}}{%
{\protect \APACyear {2018}}%
}]{%
Tinetti18}
\APACinsertmetastar {%
Tinetti18}%
\begin{APACrefauthors}%
{Tinetti}, G.%
, {Drossart}, P.%
, {Eccleston}, P.%
, {Hartogh}, P.%
, {Heske}, A.%
, {Leconte}, J.%
\BDBL {}{Zwart}, F.%
\end{APACrefauthors}%
\unskip\
\newblock
\APACrefYearMonthDay{2018}{{\APACmonth{11}}}{}.
\newblock
{\BBOQ}\APACrefatitle {{A chemical survey of exoplanets with ARIEL}} {{A
  chemical survey of exoplanets with ARIEL}}.{\BBCQ}
\newblock
\APACjournalVolNumPages{Experimental Astronomy}{46}{1}{135-209}.
\newblock
\begin{APACrefDOI} \doi{10.1007/s10686-018-9598-x} \end{APACrefDOI}
\PrintBackRefs{\CurrentBib}

\bibitem [\protect \citeauthoryear {%
Tremblin%
\ \protect \BOthers {.}}{%
Tremblin%
\ \protect \BOthers {.}}{%
{\protect \APACyear {2017}}%
}]{%
Tremblin:2017}
\APACinsertmetastar {%
Tremblin:2017}%
\begin{APACrefauthors}%
Tremblin, P.%
, Chabrier, G.%
, Mayne, N.%
, Amundsen, D.%
, Baraffe, I.%
, Debras, F.%
\BDBL {}Fromang, S.%
\end{APACrefauthors}%
\unskip\
\newblock
\APACrefYearMonthDay{2017}{}{}.
\newblock
{\BBOQ}\APACrefatitle {Advection of potential temperature in the atmosphere of
  irradiated exoplanets: A robust mechanism to explain radius inflation}
  {Advection of potential temperature in the atmosphere of irradiated
  exoplanets: A robust mechanism to explain radius inflation}.{\BBCQ}
\newblock
\APACjournalVolNumPages{The Astrophysical Journal}{843}{}{30}.
\PrintBackRefs{\CurrentBib}

\bibitem [\protect \citeauthoryear {%
{Trilling}%
\ \protect \BOthers {.}}{%
{Trilling}%
\ \protect \BOthers {.}}{%
{\protect \APACyear {1998}}%
}]{%
tril98}
\APACinsertmetastar {%
tril98}%
\begin{APACrefauthors}%
{Trilling}, D\BPBI E.%
, {Benz}, W.%
, {Guillot}, T.%
, {Lunine}, J\BPBI I.%
, {Hubbard}, W\BPBI B.%
\BCBL {}\ \BBA {} {Burrows}, A.%
\end{APACrefauthors}%
\unskip\
\newblock
\APACrefYearMonthDay{1998}{{\APACmonth{06}}}{}.
\newblock
{\BBOQ}\APACrefatitle {{Orbital Evolution and Migration of Giant Planets:
  Modeling Extrasolar Planets}} {{Orbital Evolution and Migration of Giant
  Planets: Modeling Extrasolar Planets}}.{\BBCQ}
\newblock
\APACjournalVolNumPages{\apj}{500}{}{428-439}.
\newblock
\begin{APACrefDOI} \doi{10.1086/305711} \end{APACrefDOI}
\PrintBackRefs{\CurrentBib}

\bibitem [\protect \citeauthoryear {%
Tsai%
, Dobbs-Dixon%
\BCBL {}\ \BBA {} Gu%
}{%
Tsai%
\ \protect \BOthers {.}}{%
{\protect \APACyear {2014}}%
}]{%
Tsai:2014}
\APACinsertmetastar {%
Tsai:2014}%
\begin{APACrefauthors}%
Tsai, S.%
, Dobbs-Dixon, I.%
\BCBL {}\ \BBA {} Gu, P.%
\end{APACrefauthors}%
\unskip\
\newblock
\APACrefYearMonthDay{2014}{}{}.
\newblock
{\BBOQ}\APACrefatitle {3{D} structures of equatorial waves and the resulting
  superrotation in the atmosphere of a tidally locked hot {J}upiter} {3{D}
  structures of equatorial waves and the resulting superrotation in the
  atmosphere of a tidally locked hot {J}upiter}.{\BBCQ}
\newblock
\APACjournalVolNumPages{The Astrophysical Journal}{793}{}{141}.
\PrintBackRefs{\CurrentBib}

\bibitem [\protect \citeauthoryear {%
{Tsai}%
\ \protect \BOthers {.}}{%
{Tsai}%
\ \protect \BOthers {.}}{%
{\protect \APACyear {2017}}%
}]{%
Tsai17}
\APACinsertmetastar {%
Tsai17}%
\begin{APACrefauthors}%
{Tsai}, S\BHBI M.%
, {Lyons}, J\BPBI R.%
, {Grosheintz}, L.%
, {Rimmer}, P\BPBI B.%
, {Kitzmann}, D.%
\BCBL {}\ \BBA {} {Heng}, K.%
\end{APACrefauthors}%
\unskip\
\newblock
\APACrefYearMonthDay{2017}{{\APACmonth{02}}}{}.
\newblock
{\BBOQ}\APACrefatitle {{VULCAN: An Open-source, Validated Chemical Kinetics
  Python Code for Exoplanetary Atmospheres}} {{VULCAN: An Open-source,
  Validated Chemical Kinetics Python Code for Exoplanetary
  Atmospheres}}.{\BBCQ}
\newblock
\APACjournalVolNumPages{\apjs}{228}{2}{20}.
\newblock
\begin{APACrefDOI} \doi{10.3847/1538-4365/228/2/20} \end{APACrefDOI}
\PrintBackRefs{\CurrentBib}

\bibitem [\protect \citeauthoryear {%
{Udalski}%
\ \protect \BOthers {.}}{%
{Udalski}%
\ \protect \BOthers {.}}{%
{\protect \APACyear {2002}}%
}]{%
Udalski02}
\APACinsertmetastar {%
Udalski02}%
\begin{APACrefauthors}%
{Udalski}, A.%
, {Szewczyk}, O.%
, {Zebrun}, K.%
, {Pietrzynski}, G.%
, {Szymanski}, M.%
, {Kubiak}, M.%
\BDBL {}{Wyrzykowski}, L.%
\end{APACrefauthors}%
\unskip\
\newblock
\APACrefYearMonthDay{2002}{{\APACmonth{12}}}{}.
\newblock
{\BBOQ}\APACrefatitle {{The Optical Gravitational Lensing Experiment. Planetary
  and Low-Luminosity Object Transits in the Carina Fields of the Galactic
  Disk}} {{The Optical Gravitational Lensing Experiment. Planetary and
  Low-Luminosity Object Transits in the Carina Fields of the Galactic
  Disk}}.{\BBCQ}
\newblock
\APACjournalVolNumPages{\actaa}{52}{}{317-359}.
\PrintBackRefs{\CurrentBib}

\bibitem [\protect \citeauthoryear {%
Vallis%
}{%
Vallis%
}{%
{\protect \APACyear {2006}}%
}]{%
Vallis:2006aa}
\APACinsertmetastar {%
Vallis:2006aa}%
\begin{APACrefauthors}%
Vallis, G.%
\end{APACrefauthors}%
\unskip\
\newblock
\APACrefYear{2006}.
\newblock
\APACrefbtitle {Atmospheric and Oceanic Fluid Dynamics: Fundamentals and
  Large-Scale Circulation} {Atmospheric and oceanic fluid dynamics:
  Fundamentals and large-scale circulation}.
\newblock
\APACaddressPublisher{Cambridge}{Cambridge University Press}.
\PrintBackRefs{\CurrentBib}

\bibitem [\protect \citeauthoryear {%
Vallis%
}{%
Vallis%
}{%
{\protect \APACyear {2019}}%
}]{%
Vallis:2019aa}
\APACinsertmetastar {%
Vallis:2019aa}%
\begin{APACrefauthors}%
Vallis, G.%
\end{APACrefauthors}%
\unskip\
\newblock
\APACrefYear{2019}.
\newblock
\APACrefbtitle {Essentials of Atmospheric and Oceanic Dynamics} {Essentials of
  atmospheric and oceanic dynamics}.
\newblock
\APACaddressPublisher{Cambridge, UK}{Cambdridge University Press}.
\PrintBackRefs{\CurrentBib}

\bibitem [\protect \citeauthoryear {%
{Valsecchi}%
, {Rappaport}%
, {Rasio}%
, {Marchant}%
\BCBL {}\ \BBA {} {Rogers}%
}{%
{Valsecchi}%
\ \protect \BOthers {.}}{%
{\protect \APACyear {2015}}%
}]{%
vals15}
\APACinsertmetastar {%
vals15}%
\begin{APACrefauthors}%
{Valsecchi}, F.%
, {Rappaport}, S.%
, {Rasio}, F\BPBI A.%
, {Marchant}, P.%
\BCBL {}\ \BBA {} {Rogers}, L\BPBI A.%
\end{APACrefauthors}%
\unskip\
\newblock
\APACrefYearMonthDay{2015}{{\APACmonth{11}}}{}.
\newblock
{\BBOQ}\APACrefatitle {{Tidally-driven Roche-lobe Overflow of Hot Jupiters with
  MESA}} {{Tidally-driven Roche-lobe Overflow of Hot Jupiters with
  MESA}}.{\BBCQ}
\newblock
\APACjournalVolNumPages{\apj}{813}{}{101}.
\newblock
\begin{APACrefDOI} \doi{10.1088/0004-637X/813/2/101} \end{APACrefDOI}
\PrintBackRefs{\CurrentBib}

\bibitem [\protect \citeauthoryear {%
{Valsecchi}%
, {Rasio}%
\BCBL {}\ \BBA {} {Steffen}%
}{%
{Valsecchi}%
\ \protect \BOthers {.}}{%
{\protect \APACyear {2014}}%
}]{%
vals14}
\APACinsertmetastar {%
vals14}%
\begin{APACrefauthors}%
{Valsecchi}, F.%
, {Rasio}, F\BPBI A.%
\BCBL {}\ \BBA {} {Steffen}, J\BPBI H.%
\end{APACrefauthors}%
\unskip\
\newblock
\APACrefYearMonthDay{2014}{{\APACmonth{09}}}{}.
\newblock
{\BBOQ}\APACrefatitle {{From Hot Jupiters to Super-Earths via Roche Lobe
  Overflow}} {{From Hot Jupiters to Super-Earths via Roche Lobe
  Overflow}}.{\BBCQ}
\newblock
\APACjournalVolNumPages{\apjl}{793}{}{L3}.
\newblock
\begin{APACrefDOI} \doi{10.1088/2041-8205/793/1/L3} \end{APACrefDOI}
\PrintBackRefs{\CurrentBib}

\bibitem [\protect \citeauthoryear {%
Vazan%
, Helled%
\BCBL {}\ \BBA {} Guillot%
}{%
Vazan%
\ \protect \BOthers {.}}{%
{\protect \APACyear {2018}}%
}]{%
Vazan:2018aa}
\APACinsertmetastar {%
Vazan:2018aa}%
\begin{APACrefauthors}%
Vazan, A.%
, Helled, R.%
\BCBL {}\ \BBA {} Guillot, T.%
\end{APACrefauthors}%
\unskip\
\newblock
\APACrefYearMonthDay{2018}{}{}.
\newblock
{\BBOQ}\APACrefatitle {Jupiter's Evolution with Primordial Composition
  Gradients} {Jupiter's evolution with primordial composition
  gradients}.{\BBCQ}
\newblock
\APACjournalVolNumPages{Astronomy {\&} Astrophysics}{610}{}{L14}.
\PrintBackRefs{\CurrentBib}

\bibitem [\protect \citeauthoryear {%
{Venot}%
\ \protect \BOthers {.}}{%
{Venot}%
\ \protect \BOthers {.}}{%
{\protect \APACyear {2020}}%
}]{%
Venot20}
\APACinsertmetastar {%
Venot20}%
\begin{APACrefauthors}%
{Venot}, O.%
, {Cavali{\'e}}, T.%
, {Bounaceur}, R.%
, {Tremblin}, P.%
, {Brouillard}, L.%
\BCBL {}\ \BBA {} {Lhoussaine Ben Brahim}, R.%
\end{APACrefauthors}%
\unskip\
\newblock
\APACrefYearMonthDay{2020}{{\APACmonth{02}}}{}.
\newblock
{\BBOQ}\APACrefatitle {{New chemical scheme for giant planet thermochemistry.
  Update of the methanol chemistry and new reduced chemical scheme}} {{New
  chemical scheme for giant planet thermochemistry. Update of the methanol
  chemistry and new reduced chemical scheme}}.{\BBCQ}
\newblock
\APACjournalVolNumPages{\aap}{634}{}{A78}.
\newblock
\begin{APACrefDOI} \doi{10.1051/0004-6361/201936697} \end{APACrefDOI}
\PrintBackRefs{\CurrentBib}

\bibitem [\protect \citeauthoryear {%
{Venot}%
\ \protect \BOthers {.}}{%
{Venot}%
\ \protect \BOthers {.}}{%
{\protect \APACyear {2012}}%
}]{%
Venot12}
\APACinsertmetastar {%
Venot12}%
\begin{APACrefauthors}%
{Venot}, O.%
, {H{\'e}brard}, E.%
, {Ag{\'u}ndez}, M.%
, {Dobrijevic}, M.%
, {Selsis}, F.%
, {Hersant}, F.%
\BDBL {}{Bounaceur}, R.%
\end{APACrefauthors}%
\unskip\
\newblock
\APACrefYearMonthDay{2012}{{\APACmonth{10}}}{}.
\newblock
{\BBOQ}\APACrefatitle {{A chemical model for the atmosphere of hot Jupiters}}
  {{A chemical model for the atmosphere of hot Jupiters}}.{\BBCQ}
\newblock
\APACjournalVolNumPages{\aap}{546}{}{A43}.
\newblock
\begin{APACrefDOI} \doi{10.1051/0004-6361/201219310} \end{APACrefDOI}
\PrintBackRefs{\CurrentBib}

\bibitem [\protect \citeauthoryear {%
{Vidal-Madjar}%
\ \protect \BOthers {.}}{%
{Vidal-Madjar}%
\ \protect \BOthers {.}}{%
{\protect \APACyear {2004}}%
}]{%
Vidal04}
\APACinsertmetastar {%
Vidal04}%
\begin{APACrefauthors}%
{Vidal-Madjar}, A.%
, {D{\'e}sert}, J\BHBI M.%
, {Lecavelier des Etangs}, A.%
, {H{\'e}brard}, G.%
, {Ballester}, G\BPBI E.%
, {Ehrenreich}, D.%
\BDBL {}{Parkinson}, C\BPBI D.%
\end{APACrefauthors}%
\unskip\
\newblock
\APACrefYearMonthDay{2004}{{\APACmonth{03}}}{}.
\newblock
{\BBOQ}\APACrefatitle {{Detection of Oxygen and Carbon in the Hydrodynamically
  Escaping Atmosphere of the Extrasolar Planet HD 209458b}} {{Detection of
  Oxygen and Carbon in the Hydrodynamically Escaping Atmosphere of the
  Extrasolar Planet HD 209458b}}.{\BBCQ}
\newblock
\APACjournalVolNumPages{\apjl}{604}{}{L69-L72}.
\PrintBackRefs{\CurrentBib}

\bibitem [\protect \citeauthoryear {%
{Vidal-Madjar}%
\ \protect \BOthers {.}}{%
{Vidal-Madjar}%
\ \protect \BOthers {.}}{%
{\protect \APACyear {2003}}%
}]{%
Vidal03}
\APACinsertmetastar {%
Vidal03}%
\begin{APACrefauthors}%
{Vidal-Madjar}, A.%
, {Lecavelier des Etangs}, A.%
, {D{\'e}sert}, J\BHBI M.%
, {Ballester}, G\BPBI E.%
, {Ferlet}, R.%
, {H{\'e}brard}, G.%
\BCBL {}\ \BBA {} {Mayor}, M.%
\end{APACrefauthors}%
\unskip\
\newblock
\APACrefYearMonthDay{2003}{{\APACmonth{03}}}{}.
\newblock
{\BBOQ}\APACrefatitle {{An extended upper atmosphere around the extrasolar
  planet HD209458b}} {{An extended upper atmosphere around the extrasolar
  planet HD209458b}}.{\BBCQ}
\newblock
\APACjournalVolNumPages{\nat}{422}{}{143-146}.
\PrintBackRefs{\CurrentBib}

\bibitem [\protect \citeauthoryear {%
{Visscher}%
, {Lodders}%
\BCBL {}\ \BBA {} {Fegley}%
}{%
{Visscher}%
\ \protect \BOthers {.}}{%
{\protect \APACyear {2010}}%
}]{%
Visscher10}
\APACinsertmetastar {%
Visscher10}%
\begin{APACrefauthors}%
{Visscher}, C.%
, {Lodders}, K.%
\BCBL {}\ \BBA {} {Fegley}, B., Jr.%
\end{APACrefauthors}%
\unskip\
\newblock
\APACrefYearMonthDay{2010}{{\APACmonth{06}}}{}.
\newblock
{\BBOQ}\APACrefatitle {{Atmospheric Chemistry in Giant Planets, Brown Dwarfs,
  and Low-mass Dwarf Stars. III. Iron, Magnesium, and Silicon}} {{Atmospheric
  Chemistry in Giant Planets, Brown Dwarfs, and Low-mass Dwarf Stars. III.
  Iron, Magnesium, and Silicon}}.{\BBCQ}
\newblock
\APACjournalVolNumPages{\apj}{716}{}{1060-1075}.
\newblock
\begin{APACrefDOI} \doi{10.1088/0004-637X/716/2/1060} \end{APACrefDOI}
\PrintBackRefs{\CurrentBib}

\bibitem [\protect \citeauthoryear {%
von Essen%
, Mallonn%
, Borre%
\BCBL {}\ \BBA {} {et al.}%
}{%
von Essen%
\ \protect \BOthers {.}}{%
{\protect \APACyear {2020}}%
}]{%
Essen:2020aa}
\APACinsertmetastar {%
Essen:2020aa}%
\begin{APACrefauthors}%
von Essen, C.%
, Mallonn, M.%
, Borre, C.%
\BCBL {}\ \BBA {} {et al.}%
\end{APACrefauthors}%
\unskip\
\newblock
\APACrefYearMonthDay{2020}{}{}.
\newblock
{\BBOQ}\APACrefatitle {TESS unveils the phase curve of WASP-33b.
  Characterization of the planetary atmosphere and the pulsations from the
  star} {Tess unveils the phase curve of wasp-33b. characterization of the
  planetary atmosphere and the pulsations from the star}.{\BBCQ}
\newblock
\APACjournalVolNumPages{arXiv e-prints:2004.10767}{}{}{}.
\PrintBackRefs{\CurrentBib}

\bibitem [\protect \citeauthoryear {%
Wakeford%
\ \protect \BOthers {.}}{%
Wakeford%
\ \protect \BOthers {.}}{%
{\protect \APACyear {2017}}%
}]{%
Wakeford:2017}
\APACinsertmetastar {%
Wakeford:2017}%
\begin{APACrefauthors}%
Wakeford, H\BPBI R.%
, Visscher, C.%
, Lewis, N\BPBI K.%
, Kataria, T.%
, Marley, M\BPBI S.%
, Fortney, J\BPBI J.%
\BCBL {}\ \BBA {} Mandell, A\BPBI M.%
\end{APACrefauthors}%
\unskip\
\newblock
\APACrefYearMonthDay{2017}{}{}.
\newblock
{\BBOQ}\APACrefatitle {High-temperature condensate clouds in super-hot Jupiter
  atmospheres} {High-temperature condensate clouds in super-hot jupiter
  atmospheres}.{\BBCQ}
\newblock
\APACjournalVolNumPages{Monthly Notices of the Royal Astronomical
  Society}{464}{}{4247}.
\PrintBackRefs{\CurrentBib}

\bibitem [\protect \citeauthoryear {%
{Wallack}%
\ \protect \BOthers {.}}{%
{Wallack}%
\ \protect \BOthers {.}}{%
{\protect \APACyear {2019}}%
}]{%
Wallack19}
\APACinsertmetastar {%
Wallack19}%
\begin{APACrefauthors}%
{Wallack}, N\BPBI L.%
, {Knutson}, H\BPBI A.%
, {Morley}, C\BPBI V.%
, {Moses}, J\BPBI I.%
, {Thomas}, N\BPBI H.%
, {Thorngren}, D\BPBI P.%
\BDBL {}{Kammer}, J\BPBI A.%
\end{APACrefauthors}%
\unskip\
\newblock
\APACrefYearMonthDay{2019}{{\APACmonth{12}}}{}.
\newblock
{\BBOQ}\APACrefatitle {{Investigating Trends in Atmospheric Compositions of
  Cool Gas Giant Planets Using Spitzer Secondary Eclipses}} {{Investigating
  Trends in Atmospheric Compositions of Cool Gas Giant Planets Using Spitzer
  Secondary Eclipses}}.{\BBCQ}
\newblock
\APACjournalVolNumPages{\aj}{158}{6}{217}.
\newblock
\begin{APACrefDOI} \doi{10.3847/1538-3881/ab2a05} \end{APACrefDOI}
\PrintBackRefs{\CurrentBib}

\bibitem [\protect \citeauthoryear {%
{Walsh}%
, {Morbidelli}%
, {Raymond}%
, {O'Brien}%
\BCBL {}\ \BBA {} {Mandell}%
}{%
{Walsh}%
\ \protect \BOthers {.}}{%
{\protect \APACyear {2011}}%
}]{%
Walsh11}
\APACinsertmetastar {%
Walsh11}%
\begin{APACrefauthors}%
{Walsh}, K\BPBI J.%
, {Morbidelli}, A.%
, {Raymond}, S\BPBI N.%
, {O'Brien}, D\BPBI P.%
\BCBL {}\ \BBA {} {Mandell}, A\BPBI M.%
\end{APACrefauthors}%
\unskip\
\newblock
\APACrefYearMonthDay{2011}{{\APACmonth{07}}}{}.
\newblock
{\BBOQ}\APACrefatitle {{A low mass for Mars from Jupiter's early gas-driven
  migration}} {{A low mass for Mars from Jupiter's early gas-driven
  migration}}.{\BBCQ}
\newblock
\APACjournalVolNumPages{\nat}{475}{}{206-209}.
\newblock
\begin{APACrefDOI} \doi{10.1038/nature10201} \end{APACrefDOI}
\PrintBackRefs{\CurrentBib}

\bibitem [\protect \citeauthoryear {%
Wang%
\ \BBA {} Wordsworth%
}{%
Wang%
\ \BBA {} Wordsworth%
}{%
{\protect \APACyear {2020}}%
}]{%
Wang:2020aa}
\APACinsertmetastar {%
Wang:2020aa}%
\begin{APACrefauthors}%
Wang, H.%
\BCBT {}\ \BBA {} Wordsworth, R.%
\end{APACrefauthors}%
\unskip\
\newblock
\APACrefYearMonthDay{2020}{}{}.
\newblock
{\BBOQ}\APACrefatitle {Extremely Long Convergence Times in a 3D GCM Simulation
  of the Sub-Neptune Gliese 1214b} {Extremely long convergence times in a 3d
  gcm simulation of the sub-neptune gliese 1214b}.{\BBCQ}
\newblock
\APACjournalVolNumPages{The Astrophysical Journal}{891}{}{7}.
\PrintBackRefs{\CurrentBib}

\bibitem [\protect \citeauthoryear {%
{Weidenschilling}%
\ \BBA {} {Marzari}%
}{%
{Weidenschilling}%
\ \BBA {} {Marzari}%
}{%
{\protect \APACyear {1996}}%
}]{%
weid96}
\APACinsertmetastar {%
weid96}%
\begin{APACrefauthors}%
{Weidenschilling}, S\BPBI J.%
\BCBT {}\ \BBA {} {Marzari}, F.%
\end{APACrefauthors}%
\unskip\
\newblock
\APACrefYearMonthDay{1996}{{\APACmonth{12}}}{}.
\newblock
{\BBOQ}\APACrefatitle {{Gravitational scattering as a possible origin for giant
  planets at small stellar distances}} {{Gravitational scattering as a possible
  origin for giant planets at small stellar distances}}.{\BBCQ}
\newblock
\APACjournalVolNumPages{\nat}{384}{}{619-621}.
\newblock
\begin{APACrefDOI} \doi{10.1038/384619a0} \end{APACrefDOI}
\PrintBackRefs{\CurrentBib}

\bibitem [\protect \citeauthoryear {%
{Weiss}%
\ \protect \BOthers {.}}{%
{Weiss}%
\ \protect \BOthers {.}}{%
{\protect \APACyear {2013}}%
}]{%
weis13}
\APACinsertmetastar {%
weis13}%
\begin{APACrefauthors}%
{Weiss}, L\BPBI M.%
, {Marcy}, G\BPBI W.%
, {Rowe}, J\BPBI F.%
, {Howard}, A\BPBI W.%
, {Isaacson}, H.%
, {Fortney}, J\BPBI J.%
\BDBL {}{Seager}, S.%
\end{APACrefauthors}%
\unskip\
\newblock
\APACrefYearMonthDay{2013}{{\APACmonth{05}}}{}.
\newblock
{\BBOQ}\APACrefatitle {{The Mass of KOI-94d and a Relation for Planet Radius,
  Mass, and Incident Flux}} {{The Mass of KOI-94d and a Relation for Planet
  Radius, Mass, and Incident Flux}}.{\BBCQ}
\newblock
\APACjournalVolNumPages{\apj}{768}{}{14}.
\newblock
\begin{APACrefDOI} \doi{10.1088/0004-637X/768/1/14} \end{APACrefDOI}
\PrintBackRefs{\CurrentBib}

\bibitem [\protect \citeauthoryear {%
{Welbanks}%
\ \protect \BOthers {.}}{%
{Welbanks}%
\ \protect \BOthers {.}}{%
{\protect \APACyear {2019}}%
}]{%
Welbanks19}
\APACinsertmetastar {%
Welbanks19}%
\begin{APACrefauthors}%
{Welbanks}, L.%
, {Madhusudhan}, N.%
, {Allard}, N\BPBI F.%
, {Hubeny}, I.%
, {Spiegelman}, F.%
\BCBL {}\ \BBA {} {Leininger}, T.%
\end{APACrefauthors}%
\unskip\
\newblock
\APACrefYearMonthDay{2019}{{\APACmonth{12}}}{}.
\newblock
{\BBOQ}\APACrefatitle {{Mass-Metallicity Trends in Transiting Exoplanets from
  Atmospheric Abundances of H$_{2}$O, Na, and K}} {{Mass-Metallicity Trends in
  Transiting Exoplanets from Atmospheric Abundances of H$_{2}$O, Na, and
  K}}.{\BBCQ}
\newblock
\APACjournalVolNumPages{\apjl}{887}{1}{L20}.
\newblock
\begin{APACrefDOI} \doi{10.3847/2041-8213/ab5a89} \end{APACrefDOI}
\PrintBackRefs{\CurrentBib}

\bibitem [\protect \citeauthoryear {%
{Winn}%
}{%
{Winn}%
}{%
{\protect \APACyear {2010}}%
{\protect \APACexlab {{\protect \BCnt {1}}}}}]{%
winn10}
\APACinsertmetastar {%
winn10}%
\begin{APACrefauthors}%
{Winn}, J\BPBI N.%
\end{APACrefauthors}%
\unskip\
\newblock
\APACrefYearMonthDay{2010{\protect \BCnt {1}}}{}{}.
\newblock
{\BBOQ}\APACrefatitle {{Exoplanet Transits and Occultations}} {{Exoplanet
  Transits and Occultations}}.{\BBCQ}
\newblock
\BIn{} S.~{Seager}\ (\BED), \APACrefbtitle {Exoplanets} {Exoplanets}\
  (\BPG~55-77).
\PrintBackRefs{\CurrentBib}

\bibitem [\protect \citeauthoryear {%
{Winn}%
}{%
{Winn}%
}{%
{\protect \APACyear {2010}}%
{\protect \APACexlab {{\protect \BCnt {2}}}}}]{%
Winn10book}
\APACinsertmetastar {%
Winn10book}%
\begin{APACrefauthors}%
{Winn}, J\BPBI N.%
\end{APACrefauthors}%
\unskip\
\newblock
\APACrefYearMonthDay{2010{\protect \BCnt {2}}}{{\APACmonth{12}}}{}.
\newblock
{\BBOQ}\APACrefatitle {{Exoplanet Transits and Occultations}} {{Exoplanet
  Transits and Occultations}}.{\BBCQ}
\newblock
\BIn{} S.~{Seager}\ (\BED), \APACrefbtitle {Exoplanets} {Exoplanets}\
  (\BPG~55-77).
\newblock
\APACaddressPublisher{}{University of Arizona Press}.
\PrintBackRefs{\CurrentBib}

\bibitem [\protect \citeauthoryear {%
{Winter}%
, {Kruijssen}%
, {Longmore}%
\BCBL {}\ \BBA {} {Chevance}%
}{%
{Winter}%
\ \protect \BOthers {.}}{%
{\protect \APACyear {2020}}%
}]{%
Winter2020}
\APACinsertmetastar {%
Winter2020}%
\begin{APACrefauthors}%
{Winter}, A\BPBI J.%
, {Kruijssen}, J\BPBI M\BPBI D.%
, {Longmore}, S\BPBI N.%
\BCBL {}\ \BBA {} {Chevance}, M.%
\end{APACrefauthors}%
\unskip\
\newblock
\APACrefYearMonthDay{2020}{{\APACmonth{10}}}{}.
\newblock
{\BBOQ}\APACrefatitle {{Stellar clustering shapes the architecture of planetary
  systems}} {{Stellar clustering shapes the architecture of planetary
  systems}}.{\BBCQ}
\newblock
\APACjournalVolNumPages{\nat}{586}{7830}{528-532}.
\newblock
\begin{APACrefDOI} \doi{10.1038/s41586-020-2800-0} \end{APACrefDOI}
\PrintBackRefs{\CurrentBib}

\bibitem [\protect \citeauthoryear {%
{Woitke}%
\ \protect \BOthers {.}}{%
{Woitke}%
\ \protect \BOthers {.}}{%
{\protect \APACyear {2018}}%
}]{%
Woitke18}
\APACinsertmetastar {%
Woitke18}%
\begin{APACrefauthors}%
{Woitke}, P.%
, {Helling}, C.%
, {Hunter}, G\BPBI H.%
, {Millard}, J\BPBI D.%
, {Turner}, G\BPBI E.%
, {Worters}, M.%
\BDBL {}{Stock}, J\BPBI W.%
\end{APACrefauthors}%
\unskip\
\newblock
\APACrefYearMonthDay{2018}{{\APACmonth{06}}}{}.
\newblock
{\BBOQ}\APACrefatitle {{Equilibrium chemistry down to 100 K. Impact of
  silicates and phyllosilicates on the carbon to oxygen ratio}} {{Equilibrium
  chemistry down to 100 K. Impact of silicates and phyllosilicates on the
  carbon to oxygen ratio}}.{\BBCQ}
\newblock
\APACjournalVolNumPages{\aap}{614}{}{A1}.
\newblock
\begin{APACrefDOI} \doi{10.1051/0004-6361/201732193} \end{APACrefDOI}
\PrintBackRefs{\CurrentBib}

\bibitem [\protect \citeauthoryear {%
Wong%
\ \protect \BOthers {.}}{%
Wong%
\ \protect \BOthers {.}}{%
{\protect \APACyear {2016}}%
}]{%
Wong:2015a}
\APACinsertmetastar {%
Wong:2015a}%
\begin{APACrefauthors}%
Wong, I.%
, Knutson, H.%
, Lewis, N.%
, Kataria, T.%
, Burrows, A.%
, Fortney, J.%
\BDBL {}Todorov, K.%
\end{APACrefauthors}%
\unskip\
\newblock
\APACrefYearMonthDay{2016}{}{}.
\newblock
{\BBOQ}\APACrefatitle {3.6 and 4.5 $\mu$m phase curves of the highly-irradiated
  hot {J}upiters {WASP}-19b and {HAT}-{P}-7b} {3.6 and 4.5 $\mu$m phase curves
  of the highly-irradiated hot {J}upiters {WASP}-19b and {HAT}-{P}-7b}.{\BBCQ}
\newblock
\APACjournalVolNumPages{The Astrophysical Journal}{823}{}{122}.
\PrintBackRefs{\CurrentBib}

\bibitem [\protect \citeauthoryear {%
Wong%
, Shporer%
, Morris%
\BCBL {}\ \BBA {} {et al.}%
}{%
Wong%
\ \protect \BOthers {.}}{%
{\protect \APACyear {2019}}%
}]{%
Wong:2019aa}
\APACinsertmetastar {%
Wong:2019aa}%
\begin{APACrefauthors}%
Wong, I.%
, Shporer, A.%
, Morris, B.%
\BCBL {}\ \BBA {} {et al.}%
\end{APACrefauthors}%
\unskip\
\newblock
\APACrefYearMonthDay{2019}{}{}.
\newblock
{\BBOQ}\APACrefatitle {Exploring the atmospheric dynamics of the extreme
  ultra-hot Jupiter {KELT-9b} using {TESS} photometry} {Exploring the
  atmospheric dynamics of the extreme ultra-hot jupiter {KELT-9b} using {TESS}
  photometry}.{\BBCQ}
\newblock
\APACjournalVolNumPages{arXiv e-prints:1910.01607}{}{}{}.
\PrintBackRefs{\CurrentBib}

\bibitem [\protect \citeauthoryear {%
{Wright}%
\ \protect \BOthers {.}}{%
{Wright}%
\ \protect \BOthers {.}}{%
{\protect \APACyear {2012}}%
}]{%
wrig12}
\APACinsertmetastar {%
wrig12}%
\begin{APACrefauthors}%
{Wright}, J\BPBI T.%
, {Marcy}, G\BPBI W.%
, {Howard}, A\BPBI W.%
, {Johnson}, J\BPBI A.%
, {Morton}, T\BPBI D.%
\BCBL {}\ \BBA {} {Fischer}, D\BPBI A.%
\end{APACrefauthors}%
\unskip\
\newblock
\APACrefYearMonthDay{2012}{{\APACmonth{07}}}{}.
\newblock
{\BBOQ}\APACrefatitle {{The Frequency of Hot Jupiters Orbiting nearby
  Solar-type Stars}} {{The Frequency of Hot Jupiters Orbiting nearby Solar-type
  Stars}}.{\BBCQ}
\newblock
\APACjournalVolNumPages{\apj}{753}{}{160}.
\newblock
\begin{APACrefDOI} \doi{10.1088/0004-637X/753/2/160} \end{APACrefDOI}
\PrintBackRefs{\CurrentBib}

\bibitem [\protect \citeauthoryear {%
{Wu}%
\ \BBA {} {Lithwick}%
}{%
{Wu}%
\ \BBA {} {Lithwick}%
}{%
{\protect \APACyear {2011}}%
}]{%
wu11}
\APACinsertmetastar {%
wu11}%
\begin{APACrefauthors}%
{Wu}, Y.%
\BCBT {}\ \BBA {} {Lithwick}, Y.%
\end{APACrefauthors}%
\unskip\
\newblock
\APACrefYearMonthDay{2011}{{\APACmonth{07}}}{}.
\newblock
{\BBOQ}\APACrefatitle {{Secular Chaos and the Production of Hot Jupiters}}
  {{Secular Chaos and the Production of Hot Jupiters}}.{\BBCQ}
\newblock
\APACjournalVolNumPages{\apj}{735}{}{109}.
\newblock
\begin{APACrefDOI} \doi{10.1088/0004-637X/735/2/109} \end{APACrefDOI}
\PrintBackRefs{\CurrentBib}

\bibitem [\protect \citeauthoryear {%
Wu%
\ \BBA {} Lithwick%
}{%
Wu%
\ \BBA {} Lithwick%
}{%
{\protect \APACyear {2013}}%
}]{%
Wu:2013}
\APACinsertmetastar {%
Wu:2013}%
\begin{APACrefauthors}%
Wu, Y.%
\BCBT {}\ \BBA {} Lithwick, Y.%
\end{APACrefauthors}%
\unskip\
\newblock
\APACrefYearMonthDay{2013}{}{}.
\newblock
{\BBOQ}\APACrefatitle {{O}hmic heating suspends, not reverses, the cooling
  contraction of hot {J}upiters} {{O}hmic heating suspends, not reverses, the
  cooling contraction of hot {J}upiters}.{\BBCQ}
\newblock
\APACjournalVolNumPages{The Astrophysical Journal}{763}{}{13}.
\PrintBackRefs{\CurrentBib}

\bibitem [\protect \citeauthoryear {%
{Wu}%
\ \BBA {} {Murray}%
}{%
{Wu}%
\ \BBA {} {Murray}%
}{%
{\protect \APACyear {2003}}%
}]{%
wu03}
\APACinsertmetastar {%
wu03}%
\begin{APACrefauthors}%
{Wu}, Y.%
\BCBT {}\ \BBA {} {Murray}, N.%
\end{APACrefauthors}%
\unskip\
\newblock
\APACrefYearMonthDay{2003}{{\APACmonth{05}}}{}.
\newblock
{\BBOQ}\APACrefatitle {{Planet Migration and Binary Companions: The Case of HD
  80606b}} {{Planet Migration and Binary Companions: The Case of HD
  80606b}}.{\BBCQ}
\newblock
\APACjournalVolNumPages{\apj}{589}{}{605-614}.
\newblock
\begin{APACrefDOI} \doi{10.1086/374598} \end{APACrefDOI}
\PrintBackRefs{\CurrentBib}

\bibitem [\protect \citeauthoryear {%
Wyttenbach%
, Ehrenreich%
, Lovis%
, Udry%
\BCBL {}\ \BBA {} Pepe%
}{%
Wyttenbach%
\ \protect \BOthers {.}}{%
{\protect \APACyear {2015}}%
}]{%
Wyttenbach:2015}
\APACinsertmetastar {%
Wyttenbach:2015}%
\begin{APACrefauthors}%
Wyttenbach, A.%
, Ehrenreich, D.%
, Lovis, C.%
, Udry, S.%
\BCBL {}\ \BBA {} Pepe, F.%
\end{APACrefauthors}%
\unskip\
\newblock
\APACrefYearMonthDay{2015}{}{}.
\newblock
{\BBOQ}\APACrefatitle {Spectrally resolved detection of sodium in the
  atmosphere of HD 189733b with the HARPS spectrograph} {Spectrally resolved
  detection of sodium in the atmosphere of hd 189733b with the harps
  spectrograph}.{\BBCQ}
\newblock
\APACjournalVolNumPages{Astronomy {\&} Astrophysics}{577}{}{A62}.
\PrintBackRefs{\CurrentBib}

\bibitem [\protect \citeauthoryear {%
Wyttenbach%
, Lovis%
, Ehrenreich%
\BCBL {}\ \BBA {} {et al.}%
}{%
Wyttenbach%
\ \protect \BOthers {.}}{%
{\protect \APACyear {2017}}%
}]{%
Wyttenbach:2017aa}
\APACinsertmetastar {%
Wyttenbach:2017aa}%
\begin{APACrefauthors}%
Wyttenbach, A.%
, Lovis, C.%
, Ehrenreich, D.%
\BCBL {}\ \BBA {} {et al.}%
\end{APACrefauthors}%
\unskip\
\newblock
\APACrefYearMonthDay{2017}{}{}.
\newblock
{\BBOQ}\APACrefatitle {HEARTS I. Detection of hot neutral sodium at high
  altitudes on WASP-49b} {Hearts i. detection of hot neutral sodium at high
  altitudes on wasp-49b}.{\BBCQ}
\newblock
\APACjournalVolNumPages{Astronomy {\&} Astrophysics}{602}{}{A36}.
\PrintBackRefs{\CurrentBib}

\bibitem [\protect \citeauthoryear {%
Wyttenbach%
, Molli\'{e}re%
, Ehrenreich%
\BCBL {}\ \BBA {} {et al.}%
}{%
Wyttenbach%
\ \protect \BOthers {.}}{%
{\protect \APACyear {2020}}%
}]{%
Wyttenbach:2020aa}
\APACinsertmetastar {%
Wyttenbach:2020aa}%
\begin{APACrefauthors}%
Wyttenbach, A.%
, Molli\'{e}re, P.%
, Ehrenreich, D.%
\BCBL {}\ \BBA {} {et al.}%
\end{APACrefauthors}%
\unskip\
\newblock
\APACrefYearMonthDay{2020}{}{}.
\newblock
{\BBOQ}\APACrefatitle {Mass-loss rate and local thermodynamic state of the
  KELT-9 b thermosphere from the hydrogen Balmer series} {Mass-loss rate and
  local thermodynamic state of the kelt-9 b thermosphere from the hydrogen
  balmer series}.{\BBCQ}
\newblock
\APACjournalVolNumPages{Astronomy {\&} Astrophysics}{638}{}{A87}.
\PrintBackRefs{\CurrentBib}

\bibitem [\protect \citeauthoryear {%
Yadav%
\ \BBA {} Thorngren%
}{%
Yadav%
\ \BBA {} Thorngren%
}{%
{\protect \APACyear {2017}}%
}]{%
Yadav:2017}
\APACinsertmetastar {%
Yadav:2017}%
\begin{APACrefauthors}%
Yadav, R.%
\BCBT {}\ \BBA {} Thorngren, D.%
\end{APACrefauthors}%
\unskip\
\newblock
\APACrefYearMonthDay{2017}{}{}.
\newblock
{\BBOQ}\APACrefatitle {Estimating the magnetic field strength in hot Jupiters}
  {Estimating the magnetic field strength in hot jupiters}.{\BBCQ}
\newblock
\APACjournalVolNumPages{The Astrophysical Journal Letters}{849}{}{L12}.
\PrintBackRefs{\CurrentBib}

\bibitem [\protect \citeauthoryear {%
Yan%
, Wyttenbach%
, Casasayas-Barris%
\BCBL {}\ \BBA {} {et al.}%
}{%
Yan%
\ \protect \BOthers {.}}{%
{\protect \APACyear {2020}}%
}]{%
Yan:2020aa}
\APACinsertmetastar {%
Yan:2020aa}%
\begin{APACrefauthors}%
Yan, F.%
, Wyttenbach, A.%
, Casasayas-Barris, N.%
\BCBL {}\ \BBA {} {et al.}%
\end{APACrefauthors}%
\unskip\
\newblock
\APACrefYearMonthDay{2020}{}{}.
\newblock
{\BBOQ}\APACrefatitle {Detection of the hydrogen Balmer lines in the ultra-hot
  Jupiter WASP-33b} {Detection of the hydrogen balmer lines in the ultra-hot
  jupiter wasp-33b}.{\BBCQ}
\newblock
\APACjournalVolNumPages{arXiv e-prints:2011.07888}{}{}{}.
\PrintBackRefs{\CurrentBib}

\bibitem [\protect \citeauthoryear {%
{Yelle}%
}{%
{Yelle}%
}{%
{\protect \APACyear {2004}}%
}]{%
Yelle04}
\APACinsertmetastar {%
Yelle04}%
\begin{APACrefauthors}%
{Yelle}, R\BPBI V.%
\end{APACrefauthors}%
\unskip\
\newblock
\APACrefYearMonthDay{2004}{{\APACmonth{07}}}{}.
\newblock
{\BBOQ}\APACrefatitle {{Aeronomy of extra-solar giant planets at small orbital
  distances}} {{Aeronomy of extra-solar giant planets at small orbital
  distances}}.{\BBCQ}
\newblock
\APACjournalVolNumPages{Icarus}{170}{}{167-179}.
\PrintBackRefs{\CurrentBib}

\bibitem [\protect \citeauthoryear {%
{Youdin}%
\ \BBA {} {Mitchell}%
}{%
{Youdin}%
\ \BBA {} {Mitchell}%
}{%
{\protect \APACyear {2010}}%
}]{%
youd10}
\APACinsertmetastar {%
youd10}%
\begin{APACrefauthors}%
{Youdin}, A\BPBI N.%
\BCBT {}\ \BBA {} {Mitchell}, J\BPBI L.%
\end{APACrefauthors}%
\unskip\
\newblock
\APACrefYearMonthDay{2010}{{\APACmonth{10}}}{}.
\newblock
{\BBOQ}\APACrefatitle {{The Mechanical Greenhouse: Burial of Heat by Turbulence
  in Hot Jupiter Atmospheres}} {{The Mechanical Greenhouse: Burial of Heat by
  Turbulence in Hot Jupiter Atmospheres}}.{\BBCQ}
\newblock
\APACjournalVolNumPages{\apj}{721}{}{1113-1126}.
\newblock
\begin{APACrefDOI} \doi{10.1088/0004-637X/721/2/1113} \end{APACrefDOI}
\PrintBackRefs{\CurrentBib}

\bibitem [\protect \citeauthoryear {%
{Yu}%
\ \protect \BOthers {.}}{%
{Yu}%
\ \protect \BOthers {.}}{%
{\protect \APACyear {2017}}%
}]{%
yu17}
\APACinsertmetastar {%
yu17}%
\begin{APACrefauthors}%
{Yu}, L.%
, {Donati}, J\BHBI F.%
, {H{\'e}brard}, E\BPBI M.%
, {Moutou}, C.%
, {Malo}, L.%
, {Grankin}, K.%
\BDBL {}{Matysse Collaboration}%
\end{APACrefauthors}%
\unskip\
\newblock
\APACrefYearMonthDay{2017}{{\APACmonth{05}}}{}.
\newblock
{\BBOQ}\APACrefatitle {{A hot Jupiter around the very active weak-line T Tauri
  star TAP 26}} {{A hot Jupiter around the very active weak-line T Tauri star
  TAP 26}}.{\BBCQ}
\newblock
\APACjournalVolNumPages{\mnras}{467}{}{1342-1359}.
\newblock
\begin{APACrefDOI} \doi{10.1093/mnras/stx009} \end{APACrefDOI}
\PrintBackRefs{\CurrentBib}

\bibitem [\protect \citeauthoryear {%
{Zahnle}%
, {Marley}%
, {Freedman}%
, {Lodders}%
\BCBL {}\ \BBA {} {Fortney}%
}{%
{Zahnle}%
\ \protect \BOthers {.}}{%
{\protect \APACyear {2009}}%
}]{%
Zahnle09}
\APACinsertmetastar {%
Zahnle09}%
\begin{APACrefauthors}%
{Zahnle}, K.%
, {Marley}, M\BPBI S.%
, {Freedman}, R\BPBI S.%
, {Lodders}, K.%
\BCBL {}\ \BBA {} {Fortney}, J\BPBI J.%
\end{APACrefauthors}%
\unskip\
\newblock
\APACrefYearMonthDay{2009}{{\APACmonth{08}}}{}.
\newblock
{\BBOQ}\APACrefatitle {{Atmospheric Sulfur Photochemistry on Hot Jupiters}}
  {{Atmospheric Sulfur Photochemistry on Hot Jupiters}}.{\BBCQ}
\newblock
\APACjournalVolNumPages{\apjl}{701}{}{L20-L24}.
\newblock
\begin{APACrefDOI} \doi{10.1088/0004-637X/701/1/L20} \end{APACrefDOI}
\PrintBackRefs{\CurrentBib}

\bibitem [\protect \citeauthoryear {%
Zellem%
\ \protect \BOthers {.}}{%
Zellem%
\ \protect \BOthers {.}}{%
{\protect \APACyear {2014}}%
}]{%
Zellem:2014}
\APACinsertmetastar {%
Zellem:2014}%
\begin{APACrefauthors}%
Zellem, R.%
, Lewis, N.%
, Knutson, H.%
, Griffith, C.%
, Showman, A.%
, Fortney, J.%
\BDBL {}Langton, J.%
\end{APACrefauthors}%
\unskip\
\newblock
\APACrefYearMonthDay{2014}{}{}.
\newblock
{\BBOQ}\APACrefatitle {The 4.5 $\mu$m full-orbit phase curve of the hot
  {J}upiter {HD} 209458b} {The 4.5 $\mu$m full-orbit phase curve of the hot
  {J}upiter {HD} 209458b}.{\BBCQ}
\newblock
\APACjournalVolNumPages{The Astrophysical Journal}{790}{}{53}.
\PrintBackRefs{\CurrentBib}

\bibitem [\protect \citeauthoryear {%
Zellem%
, Swain%
, Cowan%
\BCBL {}\ \BBA {} {et al.}%
}{%
Zellem%
\ \protect \BOthers {.}}{%
{\protect \APACyear {2019}}%
}]{%
Zellem:2019aa}
\APACinsertmetastar {%
Zellem:2019aa}%
\begin{APACrefauthors}%
Zellem, R.%
, Swain, M.%
, Cowan, N.%
\BCBL {}\ \BBA {} {et al.}%
\end{APACrefauthors}%
\unskip\
\newblock
\APACrefYearMonthDay{2019}{}{}.
\newblock
{\BBOQ}\APACrefatitle {Constraining Exoplanet Metallicities and Aerosols with
  the Contribution to ARIEL Spectroscopy of Exoplanets (CASE)} {Constraining
  exoplanet metallicities and aerosols with the contribution to ariel
  spectroscopy of exoplanets (case)}.{\BBCQ}
\newblock
\APACjournalVolNumPages{Publications of the Astronomical Society of the
  Pacific}{131}{}{094401}.
\PrintBackRefs{\CurrentBib}

\bibitem [\protect \citeauthoryear {%
M.~Zhang%
, Knutson%
, Kataria%
\BCBL {}\ \BBA {} {et al.}%
}{%
M.~Zhang%
\ \protect \BOthers {.}}{%
{\protect \APACyear {2018}}%
}]{%
Zhang:2017a}
\APACinsertmetastar {%
Zhang:2017a}%
\begin{APACrefauthors}%
Zhang, M.%
, Knutson, H.%
, Kataria, T.%
\BCBL {}\ \BBA {} {et al.}%
\end{APACrefauthors}%
\unskip\
\newblock
\APACrefYearMonthDay{2018}{}{}.
\newblock
{\BBOQ}\APACrefatitle {Phase curves of {WASP-33b} and {HD 149026b} and a new
  correlation between phase offset and irradiation temperature} {Phase curves
  of {WASP-33b} and {HD 149026b} and a new correlation between phase offset and
  irradiation temperature}.{\BBCQ}
\newblock
\APACjournalVolNumPages{The Astronomical Journal}{155}{}{83}.
\PrintBackRefs{\CurrentBib}

\bibitem [\protect \citeauthoryear {%
X.~Zhang%
}{%
X.~Zhang%
}{%
{\protect \APACyear {2020}}%
}]{%
Zhang:2020aa}
\APACinsertmetastar {%
Zhang:2020aa}%
\begin{APACrefauthors}%
Zhang, X.%
\end{APACrefauthors}%
\unskip\
\newblock
\APACrefYearMonthDay{2020}{}{}.
\newblock
{\BBOQ}\APACrefatitle {Atmospheric regimes and Trends on Exoplanets and Brown
  Dwarfs} {Atmospheric regimes and trends on exoplanets and brown
  dwarfs}.{\BBCQ}
\newblock
\APACjournalVolNumPages{Research in Astronomy and Astrophysics}{}{}{}.
\PrintBackRefs{\CurrentBib}

\bibitem [\protect \citeauthoryear {%
X.~Zhang%
\ \BBA {} Showman%
}{%
X.~Zhang%
\ \BBA {} Showman%
}{%
{\protect \APACyear {2017}}%
}]{%
Zhang:2016}
\APACinsertmetastar {%
Zhang:2016}%
\begin{APACrefauthors}%
Zhang, X.%
\BCBT {}\ \BBA {} Showman, A.%
\end{APACrefauthors}%
\unskip\
\newblock
\APACrefYearMonthDay{2017}{}{}.
\newblock
{\BBOQ}\APACrefatitle {Effects of bulk composition on the atmospheric dynamics
  of close-in exoplanets} {Effects of bulk composition on the atmospheric
  dynamics of close-in exoplanets}.{\BBCQ}
\newblock
\APACjournalVolNumPages{The Astrophysical Journal}{836}{}{73}.
\PrintBackRefs{\CurrentBib}

\bibitem [\protect \citeauthoryear {%
{Zhang}%
\ \BBA {} {Showman}%
}{%
{Zhang}%
\ \BBA {} {Showman}%
}{%
{\protect \APACyear {2018}}%
}]{%
Zhang18b}
\APACinsertmetastar {%
Zhang18b}%
\begin{APACrefauthors}%
{Zhang}, X.%
\BCBT {}\ \BBA {} {Showman}, A\BPBI P.%
\end{APACrefauthors}%
\unskip\
\newblock
\APACrefYearMonthDay{2018}{{\APACmonth{10}}}{}.
\newblock
{\BBOQ}\APACrefatitle {{Global-mean Vertical Tracer Mixing in Planetary
  Atmospheres. II. Tidally Locked Planets}} {{Global-mean Vertical Tracer
  Mixing in Planetary Atmospheres. II. Tidally Locked Planets}}.{\BBCQ}
\newblock
\APACjournalVolNumPages{\apj}{866}{1}{2}.
\newblock
\begin{APACrefDOI} \doi{10.3847/1538-4357/aada7c} \end{APACrefDOI}
\PrintBackRefs{\CurrentBib}

\end{thebibliography}
